\PassOptionsToPackage{table}{xcolor}

\documentclass[
11pt, 
english, 
onehalfspacing, 
headsepline, 
]{MastersDoctoralThesis} 

\usepackage{cite}
\usepackage[percent]{overpic}
\usepackage{float}
\usepackage{tcolorbox}
\usepackage{microtype}
\usepackage[breaklinks]{hyperref}
\usepackage{slashed}
\usepackage{wrapfig}
\usepackage{tabu}
\usepackage{diagbox}
\usepackage{mathrsfs,amsmath,amssymb,amsthm,amsfonts,tikz,graphicx,accents,hyperref, color}
\usepackage{dsfont,epiolmec, latexsym, stmaryrd, comment, cancel}
\usepackage{slashed,ccaption}
\usepackage{mathrsfs, calligra}
\usepackage{leftidx}
\usepackage{import}
\usepackage{multirow}
\usepackage{amsfonts}
\usepackage{pifont}
\usepackage{tabularx}
\usepackage[utf8]{inputenc}
\usetikzlibrary{intersections,calc}
\usepackage{tikz-3dplot}
\usepackage{ifthen}
\usetikzlibrary{arrows}
\usepackage{amsmath}
\usepackage{xcolor}
\usepackage{geometry}
\usepackage{setspace}
\usetikzlibrary{arrows}
\usepackage{array}
\usepackage[nottoc]{tocbibind}

\usepackage[autostyle=true]{csquotes}
\usepackage[utf8]{inputenc} 
\usepackage[T1]{fontenc} 

\usepackage{mathpazo} 

\newcommand{\di}{\text{d}}

\newcommand{\bmsf}{{$\mathfrak{bms}_4$}}
\newcommand{\hbmsf}{$\widehat{\mathfrak{bms}}_4$}
\newcommand{\wf}{${W(a,b;\bar{a},\bar{b})}$}
\newcommand{\wfh}{${\widehat{W}(a,b;\bar{a},\bar{b})}$}
\newcommand{\bms}{{$\mathfrak{bms}_3$}}
\newcommand{\w}{{$W(a,b)$}}
\newcommand{\wh}{{$\widehat{W(a,b)}$}}
\newcommand{\kac}{{$\mathfrak{KM}_{\mathfrak{u}(1)}$}}
\newcommand{\kach}{{$\widehat{\mathfrak{KM}_{\mathfrak{u}(1)}}$}}
\newcommand{\hbms}{$\widehat{\mathfrak{bms}}_3$}
\newcommand{\vir}{{$\mathfrak{vir}$}}
\newcommand{\wit}{{$\mathfrak{witt}$}}
\newcommand{\virtwo}{{$\mathfrak{vir}^2$}}
\newcommand{\Max}{$\mathfrak{Max}_{3}$}
\newcommand{\sltwo}{{\mathfrak{sl}(2,\mathbb{R})}}
\newcommand{\iso}{{\mathfrak{iso}(2,1)}}

\newcommand{\eps}{\varepsilon}

\newcommand{\be}{\begin{equation}}
\newcommand{\ee}{\end{equation}}

\newcommand\snote[1]{\textcolor{red}{\bf [Sh-J:\,#1]}}

\thesistitle{Deformation of Asymptotic Symmetry Algebras\\ \LARGE{and Their Physical Realizations} } 
\supervisor{Prof. M.M. \textsc{Sheikh-Jabbari}} 
\examiner{} 
\degree{Doctor of Philosophy} 
\author{Hamidreza \textsc{Safari}} 
\addresses{} 

\subject{Biological Sciences} 
\university{\href{http://ipm.ac.ir/}{Institute for Research in Fundamental Sciences }} 
\department{\href{http://physics.ipm.ac.ir/}{School of Physics}} 
\group{\href{http://physics.ipm.ac.ir/string.jsp}{High Energy Physics Group}} 
\faculty{\href{http://faculty.university.com}{Faculty Name}} 

\AtBeginDocument{
\hypersetup{pdftitle=\ttitle} 
\hypersetup{pdfauthor=\authorname} 
\hypersetup{pdfkeywords=\keywordnames} 
}

\begin{document}

\frontmatter 

\pagestyle{plain} 


\begin{titlepage}
\begin{center}

\vspace*{.001\textheight}
{\scshape\LARGE \univname\par}\vspace{0.5cm} 
	\begin{center}
		\includegraphics[width=0.17\textwidth]{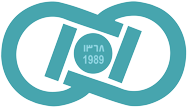}

			\end{center}

\textsc{\Large Doctoral Thesis}\\[0.5cm] 

\HRule \\[0.4cm] 
{\huge \ttitle\par}\vspace{0.4cm} 
\HRule \\[1.5cm] 
 
\begin{minipage}[t]{0.4\textwidth}
\begin{flushleft} \large
\emph{Author:}\\
\href{http://old.inspirehep.net/author/profile/H.R.Safari.1}{\authorname} 
\end{flushleft}
\end{minipage}
\begin{minipage}[t]{0.4\textwidth}
\begin{flushright} \large
\emph{Supervisor:} \\
\href{http://www.ipm.ac.ir/personalinfo.jsp?PeopleCode=IP0300097}{\supname} 
\end{flushright}
\end{minipage}\\[1.5cm]
 
\vfill

\large \textit{A thesis submitted in fulfillment of the requirements\\ for the degree of \degreename}\\[0.3cm] 
\textit{in the}\\[0.4cm]
\groupname\\\deptname\\[1.2cm] 
 
\vfill

{\large September 8, 2020}\\[2cm] 

\vfill
\end{center}
\end{titlepage}
\begin{figure}[hbt!]
    \centering
    \includegraphics[width=0.8\textwidth]{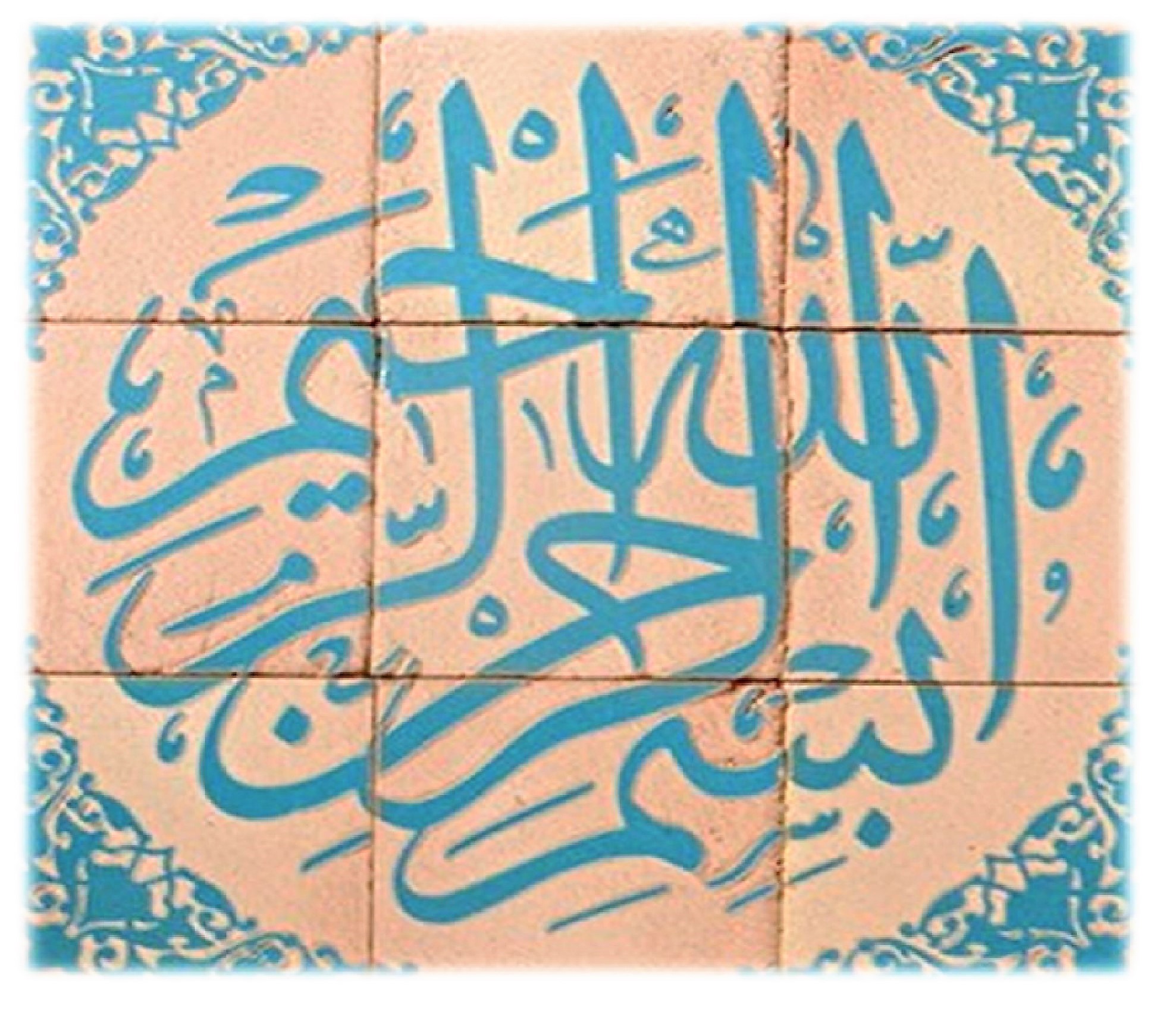}
\end{figure}

\newpage

\section*{PhD Thesis Referees Committee}

This thesis was defended on September 8nd, 2020 in front of the following jury at {\it Institute for Research in Fundamental Sciences (IPM)}:

\vspace{0.3cm}

\textbf{Dr. Hamidreza Afshar}, {\it Institute for Research in Fundamental Sciences (IPM), Tehran}

\vspace{0.3cm}

\textbf{Prof. Mohsen Alishahiha}, {\it Institute for Research in Fundamental Sciences (IPM), Tehran}
\vspace{0.3cm}

\textbf{Prof. Saeid Azam}, {\it University of Isfahan (Iran)}
\vspace{0.3cm}

\textbf{Prof. Glenn Barnich}, {\it Université Libre de Bruxelles (Belgium)}

\vspace{0.3cm}

\textbf{Prof. Stéphane Detournay}, {\it Université Libre de Bruxelles (Belgium)}
\vspace{0.3cm}

 \textbf{Prof. M.M. Sheikh-Jabbari} ({\it supervisor})

\vspace{0.3cm}

\textbf{Prof. Malihe Yousofzadeh}, {\it University of Isfahan (Iran)}


\begin{declaration}
\noindent I, \authorname, declare that this thesis titled, Deformation of Asymptotic Symmetry Algebras and Their Physical Realizations and the work presented in it are my own. I confirm that:

\begin{itemize} 
\item This work was done wholly or mainly while in candidature for a research degree at IPM.
\item Where any part of this thesis has previously been submitted for a degree or any other qualification at IPM or any other institution, this has been clearly stated.
\item Where I have consulted the published work of others, this is always clearly attributed.
\item Where I have quoted from the work of others, the source is always given. With the exception of such quotations, this thesis is entirely my own work.
\item I have acknowledged all main sources of help.
\item Where the thesis is based on work done by myself jointly with others, I have made clear exactly what was done by others and what I have contributed myself.
\end{itemize}
 
\noindent Signed: {\it Hamidreza Safari}\\
\rule[0.5em]{25em}{0.5pt} 
 
\noindent Date:  {\it September 8, 2020}\\
\rule[0.5em]{25em}{0.5pt} 
\end{declaration}

\newpage

\newpage

\begin{acknowledgements}
In this part I would like to acknowledge people and institutes who helped and supported me during my PhD period and emphasize that without them this work can not be completed. 

First of all, I would appreciate the referees who accepted being in my PhD committee: Prof. G. Barnich and Prof. S. Detournay from Physique Théorique et Mathématique, Université libre de Bruxelles and International Solvay Institutes,  Bruxelles, Prof. S. Azam and Prof. M. Yousofzadeh from Mathematics Department, University of Isfahan, Isfahan and also Prof. M. Alishahiha and Dr. H. Afshar from School of Physics, Institute for Research in Fundamental Sciences (IPM), Tehran. 

Then, I thank the partial support of Iranian NSF under grant No. 950124 and also ICTP network scheme NT-04. I should also thank F.R.S. FNRS which supported my scientific visit at ULB by fellowship "Bourse de séjour scientifique IN". 

 I would like to acknowledge my supervisor Prof. M. M. Sheikh-Jabbari. I am proud to have entered the world of hep-th with his fruitful courses. I thank him for his accurate guidance, compassionate advice, and exemplary patience which enabled me, as a newcomer to the hep-th field, to quickly learn the basics needed and enter the research course. During my research course, he accompanied me as an active colleague, a researcher with deep insight and a physicist with precise physical intuition
 until this moment. In addition, during my PhD period, he did not hesitate to help and support me in my scientific programs.
I am very grateful to him for all that he has taught me and for all his helps. 

Next I should thank people in my institute, IPM, who helped me during my PhD course training. At First, I should thank Prof. Yasaman Farzan for her very useful courses, active group meetings and helpful comments and hints for good presentation in particular in English. Then I am grateful to people who helped me specifically in early stages of my PhD. I thank so much R. Javadinejad, A. Rostami who explained difficult topics of PhD courses. I should thank Q. Exirifard, M. Vahidinia, H. Afshar and P. Bakhti who were TA of my PhD courses. I  acknowledge A. Mollabashi and  P. Mazloomi for useful discussions in training courses. I would like to specially thank M. Mohammadi Mozaffar who helped me during my PhD a lot with various discussions about the courses and also my research area. I am grateful to K. Hajian and J. Ebadi who helped me to train English language skills, introduced me Overleaf environment and  and for enjoyable discussion about different physical and non physical areas. 

Then I would like to appreciate people who were effective in my research works. First of all, I should thank Quantum Gravity group members in School of physics: H. Adami,  H. Afshar, S. Aghapour, A. Dehyadegari, E. Esmaili, K. Hajian, V. Hosseinzadeh, G. Jafari, M. J. Kazemi, M. Mohammadi Mozaffar, A. Mollabashi, M. Najafizadeh, S. Sadeghian, A. Seraj, H. Soltanpanahi and M. Vahidinia for fruitful discussions about different aspects of high energy issues and also my research topic during in our group meetings. We experienced beautiful moments with each other and I thank them very much for all the things I learned from them. 
I specifically acknowledge H. Adami for introduction of novel notions in asymptotic symmetry analysis and $3d$ massive gravity theories and also his helping to learn their computational details. I also thank A. Parsa from  School of Mathematics, IPM Isfahan branch who helped me to learn Mathematical concepts related to infinite dimensional Lie algebras. Thanks also to G. Jafari and H. Soltanpanahi for useful discussions and introduction of appropriate Mathematical packages for my problems.I am also grateful to my overseas collaborators and friends P. Concha and E. Rodriguez from Chile who assisted me to obtain more insight about Maxwell algebras and its physical importance. I had also some email exchanges with Alice Fialowski, Jose Figueroa-O’Farril and Claude Roger during my research. Thanks to all of them. Thanks also to B. Oblak for our friendly discussions during his visit at Tehran and through Skype. I appreciate also V. Hosseinzadeh, G. Jafari, H. Adami and A. Seraj for reading the thesis and their constructive comments on it.

During my PhD I had some scientific visits. First of all, I grateful very much Prof. M. Henneaux and Prof. S. Detournay who accepted to be my hosts and supervisors at ULB, Brussels, Belgium. During my visit at ULB, we had useful discussions and explored various problems specifically in $3d$ gravity. They also supported and helped me for other facilities. I should also thank H. Kajianis for her assistance before and during my visit. I am grateful to Prof. G. Barnich for our fruitful discussions about deformation and structure of BMS4 algebra. Thanks also to  S. Prohazka and A. Ranjbar who were my homemate in Brussels and for scientific discussions during and after my visit. I am also grateful to  L. Ciambelli, R. Ruzziconi, W. Merbis, N.Merino for useful discussions and specially thanks to A. Seraj for his kind helps during my visit. I acknowledge kind hospitality of Physique Théorique et Mathématique department of the Université Libre de Bruxelles (ULB). Next, I appreciate Prof. S. Azam at School of Mathematics, IPM Isfahan branch who accepted to be my host during my visit and for his friendly hospitality. He helped me to learn more about mathematical settings for deformation of Lie algebra and introduced his group members which led to some collaborations. I should also thank Prof. M. Yousofzadeh for her points and comments and useful discussion. I should also thank Prof. A. Nersessian for his kind hospitality in Yerevan Summer schools. 

There are some other people who assisted me not directly through scientific activities but their helps was very effective in my PhD. First of all, I am very grateful IPM's staffs N. Pileroudi, M. Babanzadeh, S. Jam, J. Aliabadi and H. Zarei for their activities which made IPM a pleasant research center. I also appreciate J. Yousefi, M. Khatiri,  E. Yousefi, M. Ahmadvand, M. Jamali and B. Eftekharnia for their guidance, advice and useful physical and non physical discussions specially during lunch times.

At last but not least, I appreciate my family for their kind unconditional supports during my PhD. In particular, I am grateful very much to my parents, my wife and my little daughter for their patience, encouragements and helps. 
\end{acknowledgements}

\newpage

{\LARGE Dedicated to}

\begin{center}
\Large{{\it my dear parents because of all the loving efforts they have made at different times in my life and they have kindly taught me how to live, }}
\end{center}

\begin{center}
\Large{{\it and to my kind wife for being with me throughout my education and for providing an environment full of health, security, peace and comfort for me,}}
\end{center}

\begin{center}
\Large{{\it and finally to my daughter, Maryam, who entered this world at the same time as I entered the new world of HEP-TH and accompanied me on a learning journey and taught me how to discover the mysteries of the new world.} }
\end{center}


\vspace*{0.2\textheight}
\chapter*{Preface}
\markboth{}{\small{\chaptername~\thechapter. Foreword}}

This thesis reviews and reports the results and thoughts of five years PhD period.

The original contributions of this thesis are based on the following 
publications:
\begin{itemize}
\item A.~Farahmand Parsa, H.~R.~Safari and M.~M.~Sheikh-Jabbari,
  ``On Rigidity of 3d Asymptotic Symmetry Algebras,''
  JHEP {\bf 2019}, 143 (2019)
  doi:10.1007/JHEP03(2019)143,
  \href{https://arxiv.org/abs/1809.08209}{\texttt{1809.08209}}.
\item H.~R.~Safari and M.~M.~Sheikh-Jabbari,
  ``BMS$_{4}$ algebra, its stability and deformations,''
  JHEP {\bf 2019}, 068 (2019)
  doi:10.1007/JHEP04(2019)068,
  \href{https://arxiv.org/abs/1902.03260}{\texttt{1902.03260}}.
\item P.~Concha and H.~R.~Safari,
  ``On Stabilization of Maxwell-BMS Algebra,''
  JHEP {\bf 2020}, 073 (2020)
  doi:10.1007/JHEP04(2020)073,
  \href{https://arxiv.org/abs/1909.12827}{\texttt{1909.12827}}.
  \item H.~Adami, P.~Concha, E.~Rodriguez and  H.~R.~Safari,
  ``Asymptotic Symmetries of Maxwell Chern-Simons Gravity with Torsion,'' Eur. Phys. J. C {\bf 2020}, C80 (2020) doi:10.1140/epjc/s10052-020-08537-z,
  \href{https://arxiv.org/abs/2005.07690}{\texttt{2005.07690}}.
  
\end{itemize}

\section*{How to read this thesis}

This thesis is organized in $7$ chapters and some appendices. In chapter $1$ we review some crucial concepts such as asymptotic symmetry algebra and deformation theory of Lie algebra and motivate  studying deformation of asymptotic symmetry algebra. Then we present the most important results obtained in this thesis. Chapter $2$ is devoted to reviewing the notion of asymptotic and surface charge algebra and then introduction of infinite dimensional Lie algebras which we will study their deformations. In chapter $3$ we have a detailed review on mathematical structure of contraction, deformation and stabilization of Lie algebra for finite and infinite dimensional cases. In particular, we review some important theorems in this context and mention which of them can not be applied for infinite dimensional algebras. Chapters $4,5$ and $6$ are tried to be written independently. So one can focus on each of them without losing any information. In chapter $4$ we consider deformations of $3d$ asymptotic/near horizon symmetry algebras \bms\ and Virasoro-Kac-Moody algebra, denoted as \kac, and their central extensions and comment on their physical interpretations. In chapter $5$ we compute deformations of \bmsf\ algebra as asymptotic symmetry algebra of $4d$ flat space time and its central extension. Then the physical realizations of deformation procedure and obtained algebras are discussed. In chapter $6$ we explore deformations of BMS-Maxwell algebra denoted by \Max\ which is an infinite dimensional enhancement of the $2+1$ Maxwell algebra. We then construct a Chern-Simons gravity theory containing torsion and obtain one of deformations of \Max\ algebra, $\mathfrak{bms}_{3}\oplus\mathfrak{wit}$ algebra, as its surface charge algebra. In last chapter, chapter $7$, we summarize our main results and discuss problems which can be considered as natural continuation of this thesis. In some Appendices we have gathered  some extra topics such as ``obtaining conserved surface charges for gauge theories'' and  some computational points which can be viewed as complementary of our analysis in the main text.


\begin{abstract}
\addchaptertocentry{\abstractname} 
This thesis is devoted to the study of the deformation and rigidity of infinite dimensional Lie algebras which are not subject to the Hochschild-Serre factorization theorem.
In particular, we consider \bms, Virasoro-Kac-Moody algebra  \kac\ and \bmsf\ and their central extensions which are respectively obtained as asymptotic and/or near horizon symmetry algebras for Einstein gravity on $3d$ flat, AdS$_{3}$ and $4d$ flat spacetimes. We also explore possible deformations of the Maxwell-BMS algebra \Max , which  is obtained as asymptotic symmetry algebra of the Chern-Simons gravity theory invariant under the $2+1$ dimensional Maxwell algebra. 

We find that these algebras are not rigid and can be deformed into new non isomorphic infinite dimensional (family of) algebras. We study these deformations by direct computations and also by cohomological analysis. We then classify all the algebras obtained through deformation of these algebras as well as all possible central extensions thereof. We propose/conjecture an extension of the Hochschild-Serre factorization theorem for infinite dimensional algebras as well as introducing a new notion of rigidity for family algebras obtained through deformation. 
We also explore physical realizations and significance of  the family of algebras we obtain through the deformation procedure.

\paragraph{Keywords:} Infinite dimensional Lie algebras, Asymptotic symmetry, Surface charge algebra, Deformation, Rigidity, BMS algebra, Maxwell-BMS algebra.
\end{abstract}


\tableofcontents 

\mainmatter 

\pagestyle{thesis} 

\chapter{Introduction} 

\label{ch1} 

\section*{ Symmetry and conservation laws}
       
      The notion of {\it symmetry} is the cornerstone of modern theoretical physics. In physics formulations symmetries are transformations on the space of degrees of freedom and/or parameters of the theory which keep the action or Hamiltonian invariant. While discrete symmetries play a crucial and important role in the structure of the standard model of particles, continuous symmetries appear ubiquitously in classical and quantum mechanics, quantum field theory and in general relativity. 
       
       The notion of {\it conservation laws} and constants of motion was well known among physicists, nonetheless they were given a renewed very insightful and fruitful meaning in the beginning of the 20th century by Emmy Noether. Noether's first theorem \cite{Noether:1918zz} states how to associate a {\it conserved charge} to each {\it continuous global symmetry}. A global symmetry is defined as a continuous symmetry associated with a transformation which has fixed given spacetime dependence. In this context, for instance, the conservation of energy and angular momentum of a system are respectively the result of invariance under time translation and rotation.
       
       Unlike global symmetries a {\it gauge symmetry} is induced by a transformation which is an arbitrary function of spacetime coordinates. Noether's second theorem for the gauge theories states that the gauge symmetries just lead to some identities known as Noether identities, and not a standard Noether charge. Therefore, gauge symmetries are often associated with the {\it redundancies} in the description of the theory; their role is to consistently remove the nonphysical (gauge) degrees of freedom.
       
       The above picture should, however, be revisited in the presence of spacetime boundaries. It may happen that we have gauge transformations which are {\it non-trivial} at the boundary in the sense that they change the field configurations at the boundary. One may then associate to these non-trivial gauge transformations charges which are given by some surface integrals. Then the ordinary ``bulk'' gauge transformations which have vanishing surface charges may be viewed as trivial gauge transformations, the redundancies of the system.

    \section*{  Lie group and Lie algebra in physics}
    
       {
Symmetries of a certain physical theory are associated to the mathematical notion of {\it groups}. Discrete symmetries have a discrete group structure while continuous symmetries deal with Lie groups and Lie algebras. Specifically, conserved charges of a theory play the role of generators of the Lie group associated to continuous symmetries and are endowed with the Lie algebra structure. For instance, the Hamiltonian and angular momentum of a system are generators of time translation and rotation which are elements of  Lie algebra of the spacetime isometries.   }

    {
   Depending on the details of the physical theories, the number of conserved charges may be finite or infinite. Therefore, we  may be dealing with finite or infinite dimensional Lie algebras. 
   The Poincar\'e group  consists of transformations which are generated by rotations, boosts, time and spatial translations which keep the metric of Minkowski spacetime intact. The Poincar\'e algebra in $d$ dimension has $\frac{d(d+1)}{2}$ generators which generate $\mathfrak{iso}(d-1,1)$ algebra \cite{weinberg1995quantum}. The Maxwell and Schr\"{o}dinger algebras are two other examples of finite dimensional Lie algebras which are obtained in the context of gravitational and statistical physics \cite{deAzcarraga:2010sw,Hoseinzadeh:2014bla,Alishahiha:2009nm,Henkel:1993sg}. }
    
    {
    There has been a growing interest in the study of infinite dimensional symmetries and algebras due to their applications in string theory, gravity theory, condensed matter physics, fluid mechanics and other areas of physics. Of particular interest are the symmetries of the {\it Virasoro} type. The Virasoro algebra is a prime example of an infinite dimensional Lie algebra with countable basis \cite{gel1968cohomologies, virasoro1970subsidiary}. Virasoro algebra is the {\it central extension} of the {\it Witt} algebra \cite{cartan1909groupes} which is the algebra of smooth vector fields on circle. It was first introduced in the context of  conformal field theory (CFT) \cite{francesco2012conformal} where it appears as the global symmetry of any physical system with conformal invariance in two dimensional space. There $2d$ CFTs appear in many different physical problems, like Ising model and percolation \cite{smirnov2010conformal,Cardy:2005kh} and also notably in string worldsheet theory \cite{polchinski1998string}. }
    
    {In the context of gauge theories such as Maxwell theory or gravity, as mentioned, one can associate a non zero {\it surface charge} to each non trivial gauge transformation and as gauge transformations are arbitrary smooth functions of on codimension one surfaces \cite{Grumiller:2020vvv}, the number of surface charges are countable infinite dimensional. So in this context we are tackling with infinite dimensional Lie algebras.}

    \section*{ Asymptotic symmetry group and algebra}
    
Infinite dimensional Lie algebras have appeared in the context of gauge theories (Maxwell or Yang-Mills) or in gravity theories, the diffeomorphisms. These, as discussed, are typically redundancies of description and are not ``physical''. However, in the presence of boundaries a measure zero subset of these lead to infinitely many physical surface charges, see e.g.  \cite{Brown:1986nw, Barnich:2001jy,  Strominger:2017zoo, Campoleoni:2017qot, Barnich:2017ubf, Compere:2007vx, Concha:2018zeb, Grumiller:2019fmp, Adami:2020amw}. These charges are given by integrals over spacelike, compact codimension two surfaces of a codimension two-form constructed from the action of the theory, see e.g. \cite{Seraj:2016cym, Compere:2018aar} for reviews. This codimension two surface is usually (but not necessarily) taken to be at the  asymptotic/near horizon region of spacetime and the charges are associated with gauge transformations which do not die off there. These surface charges are hence customarily called {\it asymptotic/near horizon symmetries} and the gauge transformations called (asymptotically) {\it large gauge transformations}.

The asymptotic symmetry algebra of nontrivial diffeomorphisms {has been of special interest}. The first example of such symmetries was constructed in early 1960's by Bondi-van der Burg-Metzner and Sachs \cite{Bondi:1962px, Sachs:1962zza, Sachs:1962wk} and usually goes under the name of the {\it BMS algebra}. The BMS algebra discussed in the early examples was associated with the symmetries of four dimensional ($4d$) asymptotic flat spacetime and is usually called {\it original} $\mathfrak{bms}_4$. Then Barnich and Troessaert showed that the Lorentz part of original \bmsf\ can be extended to the infinite dimensional ``{\it superrotation}'' algebra \cite{Barnich:2009se}. They also analyzed possible central extensions of \bmsf \cite{Barnich:2011ct}. There is a similar notion in three dimensional ($3d$) flat space \cite{Ashtekar:1996cd} which is denoted by \hbms. \hbms\ is central extension of \bms\ \cite{Barnich:2006av, Barnich:2011ct, Oblak:2016eij}. Another famous and classic example of asymptotic symmetries is the one discussed by Brown and Henneaux in 1986 \cite{Brown:1986nw}. The Brown-Henneaux analysis was carried out for asymptotically AdS$_3$ spacetime and revealed two copies of commuting Virasoro algebras (usually called by Left and Right sectors) at a given central charge; the Brown-Henneaux central charge is 3/2 of  AdS$_3$ radius measured in units of $3d$ Newton constant. It has been recently shown that \cite{Concha:2018zeb, Concha:2018jjj} the asymptotic symmetry algebras of Chern-Simons gravity theories invariant under $2+1$ Maxwell algebra and its deformation $\mathfrak{so}(2,2)\oplus\mathfrak{sl}(2,\mathbb{R})$ are {\it Maxwell-BMS algebra}, denoted by \Max\ , and three copies of Virasoro algebra, respectively. \Max \ is an infinite dimensional enhancement of $2+1$ Maxwell algebra.

{
The asymptotic symmetry analysis has been carried out for $3d$ and $4d$ asymptotically (locally) flat, de Sitter or AdS spacetimes with various boundary fall-off conditions recovering various classes of algebras, e.g. see \cite{Compere:2009zj, Compere:2014bia, Compere:2014cna, Troessaert:2013fma,  Afshar:2016wfy, Grumiller:2016pqb} for $3d$ and \cite{Ashtekar:1984zz, Henneaux:1985ey, henneaux1985asymptotically,  Ashtekar:1999jx} for $4d$. In specific $3d$ cases these
algebras typically contain a Virasoro algebra and other generators associated with primary fields in this Virasoro algebra, e.g. like a {\it Virasoro-Kac-Moody algebra}\footnote{{By Virasoro-Kac-Moody algebra, denoted as \kac, we mean the semi direct sum of a Virasoro and $\mathfrak{u}(1)$ current Kac-Moody algebra.}} \kac\ , \hbms\ or $2d$ Galilean-Conformal-Algebra (GCA). The most general in the family of AdS$_3$ algebras are $\sltwo\oplus \sltwo$ current algebras and the associated Virasoro algebras built through (twisted) {\it Sugawara construction} \cite{Grumiller:2016pqb}. One may then realize many subalgebras of this algebra as asymptotic symmetry algebras with appropriate {\it fall-off boundary conditions} \cite{Grumiller:2016pqb}. Similar analysis for asymptotically flat case has been carried out yielding $\mathfrak{iso}(2,1)$ current algebras  \cite{Grumiller:2017sjh}. There are also other similar algebras coming from surface charge analysis near the horizon of black holes (rather than the asymptotic charges) \cite{Afshar:2016wfy, Afshar:2016uax,Afshar:2017okz,Donnay:2015abr} and in the $3d$ higher spin theories \cite{Grumiller:2016kcp, Campoleoni:2017mbt}. This shows that the asymptotic symmetries and associated surface charge algebras crucially depend on the choice of fall-off and boundary conditions.}
    
    \section*{Contraction, deformation and rigidity of Lie algebras}

 While the {\it deformation} and {\it rigidity} of algebras in physics literature are not new \cite{levy1967deformation, levy1968first, Figueroa-OFarrill:1989wmj,mendes1994deformations,Chryssomalakos:2004gk} and have recently been studied in a series of papers {in} the context of ``kinematical algebras'' \cite{Figueroa-OFarrill:2017sfs, Figueroa-OFarrill:2017ycu, Figueroa-OFarrill:2017tcy, Andrzejewski:2018gmz, Figueroa-OFarrill:2018ygf, Figueroa-OFarrill:2018ilb}, physicists are more familiar with the reverse process; the {\it In\"on\"u-Wigner
(IW) contraction} \cite{Inonu:1953sp}. The IW contraction was primarily studied relating the Poincar\'e algebra to the Galilean symmetry algebra, through a non-relativistic limit; i.e.\ through sending speed of light $c$ to infinity and scaling the generators in an appropriate way \cite{levy1965nouvelle, mendes1994deformations}. A similar contraction is also known to relate the AdS or de Sitter symmetry algebras, respectively $\mathfrak{so}(d-2,2)$ or $\mathfrak{so}(d-1,1)$, to Poincar\'e algebra $\mathfrak{iso}(d-2,1)$ \cite{levy1967deformation}, see \cite{gilmore2012lie} for a comprehensive review. Geometrically, this latter may be understood as a particular large (A)dS$_{d-1}$ radius $\ell$ limit where it is expected to recover flat $d-1$ dimensional Minkowski spacetime. The IW contraction relates algebras with the same number of generators, but different structure constants and always results in a non-rigid (non-stable) algebra.

One may ask if the contraction process can be reversed in a systematic way. For example, one may deform the Galilean algebra order by order by $1/c$ corrections, or the Poincar\'e algebra by $1/\ell$ corrections and examine which algebra these order by order deformations will end up to. In this case, the expected result is of course to find the original algebra before the contraction. However, one may find other algebras too. Moreover, one may ask when does the process of deforming an algebra lead to a new algebra and when  it stops. There could be some algebras which do not admit deformations, i.e. deformations will map the algebra onto itself. In these cases the algebra is called rigid or stable. In other words, one would like to 
know, given an algebra whether the rigid algebra coming out of its deformations is unique; i.e. whether there are various algebras which upon IW contraction yield the same algebra. The theory of {\it cohomology} of algebras answers this question, unambiguously, for finite dimensional Lie algebras: The result of deforming/stabilizing an algebra is unique (up to signature of the metric of the algebra, and up to possible central extensions). As a rule of thumb, the (semi)simple algebra with the same number of generators is generically the answer. 
In the case of finite dimensional Lie algebras classification of deformations of a certain Lie algebra may be analyzed in the context of cohomology of algebras where the celebrated {\it Hochschild-Serre factorization theorem} sets the stage on how one can deform any given Lie algebra and which algebras are rigid. 

As mentioned, there are infinite dimensional algebras which are of interest from {physical viewpoint.} The most famous ones are Virasoro and (Kac-Moody)
current algebras, or algebras of diffeomorphisms on certain given manifolds. One may then pose the same question of the deformations here. To our knowledge the mathematical literature on deformations of these algebras, especially when they admit central extensions, is not as vast and established as the finite dimensional cases, see however, \cite{Fialowski:2001me, fialowski2012formal, gao2008derivations, gao2011low, Ecker:2017sen,Ecker:2018iqn}. Furthermore, it should be noticed that rigidity of the Witt and Virasoro algebras was first conjectured in \cite{fialowski1990deformations}, a sketch of the proof was provided in \cite{fialowski2003global} using cohomological point of view and finally a full proof of rigidity (stability)  was provided in \cite{fialowski2012formal, schlichenmaier2014elementary}. These proofs also point to how the usual algebraic cohomology arguments have shortcomings when applied to infinite dimensional algebras with central charges.

  \section*{Motivation to study deformation of asymptotic symmetry algebras}

In this thesis we  study infinite dimensional algebras appearing as asymptotic symmetry algebras and hopefully enrich the mathematical literature on this issue and provide physical implications and insights. The main motivations to consider deformation and stabilization of infinite dimensional algebras appearing in asymptotic/near horizon symmetry algebras are as follows:
 
(1) Physical theories and in particular quantum field theories are usually analyzed within perturbation theory: we deform the action and analyze the deformed/perturbed theory. These deformations may or may not respect symmetries of the original theory. A relevant question is then whether there is a similar notion as deformations at the level of (symmetry) algebras and if yes, whether there is any relation between the ``algebraic deformation theory'' and those of physical theories. As it is reviewed above there is a fairly well established notion of deformation of algebras in the mathematical settings. Moreover, in the context of quantum field theories we have the notion of {\it RG flows} and their {\it fixed points} in the space of deformation parameters. As a general question, one can ask whether there is a similar concept as fixed points in the space of algebraic deformations. As we discussed above the answer is affirmative and there is notion of rigid algebras which are stable against deformations.

(2) Asymptotic symmetry/near horizon algebras strongly depend on the choice of the boundary conditions. One may then ask how deformations in the algebra show up in the boundary conditions and  asymptotic/near horizon symmetry algebra. For instance, if one chooses boundary {conditions on near horizon} of a black hole instead of asymptotic infinity region, one obtains a completely different symmetry algebra as it has recently been shown in \cite{Grumiller:2019fmp} that the near horizon symmetry algebra  of $3d$ and $4d$ black holes may be obtained to be $W(0,b)$ or $W(a,a;a,a)$ algebras, see \cite{Parsa:2018kys, Safari:2019zmc} for the definition of the latter. Or, on the AdS$_3$ background there are many different choices of boundary conditions, see e.g. \cite{Grumiller:2016pqb} for discussions. Given that there are practically infinitely many possibilities for boundary conditions and infinitely many surfaces (like null infinity or horizon) to impose boundary conditions over, there seems to be  many different possibilities of such infinite dimensional algebras \cite{Adami:2020ugu}. Hence, a natural question is whether one can find/classify other such algebras? For instance it has been shown that the asymptotic symmetry algebra of AdS$_{4}$ is just $\mathfrak{so}(3,2)$. {One may ask whether there is a specific fall-off condition} which leads to obtain an infinite dimensional Lie algebra containing $\mathfrak{so}(3,2)$ as its global part? Our algebraic analysis has the advantage that is independent of such boundary condition possibilities and bypasses all those concerns. The result of our analysis is that there is no falloff conditions (for metric fluctuations) which may allow for a bigger symmetry algebra than $\mathfrak{so}(3,2)$ for AdS$_4$ asymptotic symmetry algebra.

(3) We note that the asymptotic symmetry algebras has been argued to be relevant for the formulation of holographic dual field theories, e.g. see \cite{Barnich:2010eb}. We will argue that the deformation of asymptotic symmetry algebras may be attributed to the holographic renormalization of the conformal weight (scaling dimension) of the operators. 

(4) Given the variety of infinite dimensional algebras and their ``geometric'' realization as asymptotic/near horizon symmetries, one may ask whether these algebras are rigid and if not, what is the result of the stabilization procedure. In fact, as discussed there are different physical realizations and interpretations for deformation procedure and obtained algebras. It may be interesting to explore the realization for deformation of infinite dimensional Lie algebras. To this end, we explore deformations of the infinite dimensional algebras obtained as asymptotic/near horizon symmetries of $3$ dimensional flat, AdS$_3$ and $4$ dimensional flat spacetimes and infinite enhancement of $2+1$ Maxwell algebra given by \Max\ as asymptotic symmetry of the Chern-Simons gravity theory invariant under the Maxwell algebra and their physical realizations. 

\section*{What is new in this thesis?}
The most important results obtained through this thesis are as follows

\subsection*{Deformation/stabilization of $3d$ asymptotic/near horizon symmetry algebras}

At first, we study deformation and stabilization of the \bms\ and \kac\ algebras and their central extensions. These are infinite dimensional Lie algebras  obtained as asymptotic/near horizon symmetry algebras in $3d$. We show that \bms\ in addition to the two \wit\ algebras, can be also deformed into {\it \w\ algebra family}. \w\  algebra family is obtained through deformation of commutators between superrotations and {\it supertranslations}, which is deviating from the Hochschild-Serre factorization theorem. The parameters $a$ and $b$ of \w\ algebra can be interpreted as change of periodicity and conformal weight of the primary fields (under the \wit\ algebra sector). In particular, $W(0,-1)$ and $W(0,0)$ in this family are respectively \bms\ and \kac\ algebras. We also examine deformations of \kac\ algebra and find that it can be deformed into \w\ algebra and a new algebra we dub as $\mathfrak{KM}(a,\nu)$ where $a$ and $\nu$ are deformation parameters. 

We also consider deformations of centrally extended \hbms\ and $\widehat{\mathfrak{KM}_{u(1)}}$. We show that the former can be deformed into either two \vir\ or \wh\ when $c_{JP}=0$ and the latter is rigid in the sense that it can not be deformed into a new algebra. 
We also study deformations of two \wit\ and its central extension two \vir\ algebra and show that they are rigid. We compute deformations of \w\ and its central extension \wh\ and show that for generic value of its parameters it  can only be deformed into another $W(\bar{a},\bar{b})$ algebra with shifted parameter. In this way, we introduce a new notion of rigidity for family algebras:

\begin{tcolorbox}[colback=red!3!white]
\emph{ The family algebra $W$ is rigid in the sense that for generic value of its parameters an element of family just can be deformed into another family algebra element with shifted parameters. }
\end{tcolorbox}

Based on the exhaustive analysis of various examples we propose a generalization of the Hochschild-Serre factorization theorem for infinite dimensional Lie algebras: 

\begin{tcolorbox}[colback=red!3!white]
\emph{For infinite dimensional algebras with countable basis the deformations may appear in ideal and non-ideal parts, however, the deformations are always by coefficients of terms in the ideal.}
\end{tcolorbox}

We provide several checks and  analysis  in support of this conjecture. Our results for deformation of $3d$ asymptotic symmetries algebra have been published in \cite{Parsa:2018kys}.

\subsection*{Deformation/stabilization of asymptotic symmetry algebra of $4d$ flat spacetime}

We also carry out the stability analysis of the \bmsf\ algebra which is asymptotic symmetry of $4d$ flat spacetime. We show that \bmsf\ can not be deformed into an infinite dimensional Lie algebra which includes $\mathfrak{so}(3,2)$ as the global part. This result 
rules out the possibility of finding an infinite dimensional asymptotic symmetry algebra of AdS$_{4}$ by choosing different boundary fall-off conditions. In addition, we explore all possible  deformations of \bmsf\ and find that it is not rigid and can be deformed into a new four parameters family algebra called {\it \wf\ algebra}. (Note that \bmsf\ is $W(-1/2,-1/2;-1/2,-1/2)$ algebra.) This example as another deviation of the Hochschild-Serre factorization theorem enforces our mentioned conjecture for extension of this theorem to infinite dimensional algebras. We also consider all possible deformations of centrally extended \hbmsf\ and show that it can be deformed into \wfh.  

We then explore deformations of \wf\ algebra and similar to \w\ algebra in $3d$ we find that for generic values of its parameters this family of algebras are stable, i.e. deformations typically move us within the family in the parameter space of the \wf\ algebra. The latter is another example of new notion of rigidity of family of algebras we introduce in the previous part. 
Then, we study some specific points in \wf\ parameters space and show that $W(0,0;0,0)$
and $W(0,-1;0,0)$ can be deformed into new algebras out of the \wf\ family. We next explore  possible central extensions of \wfh\ algebra and their deformations. Our results for deformation of \bmsf\ have been published in \cite{Safari:2019zmc}.

\subsection*{Deformation/stabilization infinite enhancement of $2+1$ Maxwell algebra and its physical realization}

As the last example we study possible deformation of Maxwell-BMS algebra, denoted by \Max , {which is an infinite enhancement} of the $2+1$ Maxwell algebra and can be obtained as asymptotic symmetry of a Chern-Simons gravity theory invariant under $2+1$ Maxwell algebra. We show that \Max\ algebra can be deformed in its ideal part to new algebras $\mathfrak{bms}_{3}\oplus\mathfrak{wit}$ and three copies of \wit. It has been shown that the latter can be obtained as asymptotic symmetry of a Chern-Simons gravity invariant under $\mathfrak{so}(2,2)\oplus\mathfrak{so}(2,1)$. We also find deformations in other commutators of \Max\ algebra which lead to  two new family algebras we coin them as $M(a,b;c,d)$ and $\bar{M}(\bar{\alpha},\bar{\beta};\bar{\nu})$ algebras. These two examples are new deviation from the Hochschild-Serre factorization theorem which also follow  our conjecture to extend this theorem. Then, we consider central extensions of the algebras obtained through deformation of \Max. Our results for deformation of \Max\ algebra have been published in \cite{Concha:2019eip}.

Finally, we present a Chern-Simons gravity theory with {\it torsion}, invariant under the algebra $\mathfrak{iso}(2,1)\oplus\mathfrak{so}(2,1)$ as a deformation of $2+1$ Maxwell algebra. By choosing appropriate fall-off boundary conditions we find gauge transformations which keep this fall-off condition intact. Then, we compute the surface charge algebra associated to these gauge transformations and find that the asymptotic symmetries algebra is $\mathfrak{bms}_{3}\oplus\mathfrak{wit}$ and the surface charge algebra is obtained as its central extension as $\widehat{\mathfrak{bms}_{3}}\oplus\mathfrak{vir}$ with three independent central charges. For more details about to construct a Chern-Simons theory based on $\mathfrak{iso}(2,1)\oplus\mathfrak{so}(2,1)$ and to find its surface charges algebra we refer the reader to our recent work \cite{Adami:2020xkm}.

\thispagestyle{empty}
\mbox{}



\chapter{Asymptotic symmetries, surface charges and infinite dimensional Lie algebras} \label{ch2}

In this chapter we  introduce various infinite dimensional Lie algebras obtained in the asymptotic symmetry analysis in gravitational physics. At first we  shortly review  notions of asymptotic symmetries and their associated charge algebra in gravity. We then study the structure of asymptotic symmetries algebra in $3d$ flat and AdS gravity and explore structure of \bmsf\ algebra which is asymptotic symmetry algebra of $4d$ flat spacetime and its central extensions. Finally we consider the $2+1$ Maxwell algebra and its infinite dimensional enhancement, \Max, and its central extensions. \Max\ algebra has appeared as asymptotic symmetry algebra of a Chern-Simons gravity theory invariant under $2+1$ Maxwell algebra.

 \section{Asymptotic symmetries and surface charges algebra}
Diffeomorphism invariance, invariance under general coordinate transformations, $x_{\mu}\rightarrow x_{\mu}+\xi_{\mu}$ where $\xi_{\mu}(x)$ are vector fields on spacetime manifold, in gravitational theories is a manifestation of the Equivalence Principle. Algebras of diffeomorphisms are infinite dimensional. This algebra is a (gauge) redundancy of the gravity theory. Nonetheless, certain measure zero subalgebras of diffeomorphisms, nontrivial diffeos, may appear as asymptotic symmetries of the theory. To these diffeos one can associate conserved charges given by surface integrals. The non-trivial diffeos certainly contain the isometry algebra of the manifold but may be a bigger set, respecting certain boundary conditions. In the context of asymptotic symmetries these are called asymptotic Killing vectors. While the algebra of diffeomorphisms, defined through Lie brackets, do not have central extensions, the algebra of corresponding charges in a given theory of gravity may admit central extensions.

Asymptotic symmetries has a longstanding history in gravitational physics and is recently extended to  other gauge theories such as Maxwell theory and $p$-form gauge theories \cite{Hosseinzadeh:2018dkh, Ortaggio:2014ipa, Compere:2007vx}. Historically, the first example of such an algebra was obtained by Bondi, van der Burg, Metzner and Sachs in 1962 \cite{Bondi:1962px,Sachs:1962wk,Sachs:1962zza} through a careful analysis of gravitational waves in the asymptotic region of $4d$ flat spacetime. The obtained algebra is called \bmsf\ which is a direct sum of $4d$ Lorentz algebra with infinite dimensional supertranslations as the ideal part. After that Brown and Henneaux considered asymptotic symmetries of AdS$_{3}$ with specific boundary fall-off conditions which lead to  two copies of Virasoro algebra as asymptotic symmetries algebra. Two Virasoro algebras can be also viewed as global symmetries of a $2d$ conformal field theory. So \cite{Brown:1986nw} can be viewed a precursor of the AdS/CFT. 
 
The more recent extended analysis carried out by Barnich and his collaborators led to obtain new asymptotic symmetry algebras for $3d$ and $4d$ flat spacetimes. One such analysis was the study of asymptotic symmetries of $3d$ flat spacetime, led to the \bms\ algebra. They also showed that \bms\ admits central extensions, the centrally extended version is denoted by \hbms. It has been shown that one can start with two copies of Virasoro algebra to reach \hbms\ algebra by using IW contraction. Barnich and Troessaert also showed that the old version of \bmsf\ in its Lorentz part can be extended to infinite dimensional superrotations. So the extended \bmsf\ is direct sum of infinite dimensional superrotations with infinite dimensional supertranslations as the ideal part. In this thesis, as it is common these days, we use \bmsf\ to refer to this extended version and use the term ``original \bmsf'' to denote the one obtained in \cite{Bondi:1962px,Sachs:1962wk,Sachs:1962zza}.

 
 \section*{Infinite dimensional algebras as asymptotic/near horizon algebra} 
 
 Now we review the structure of some infinite dimensional Lie algebras which are obtained as asymptotic/near horizon symmetry algebras and their central extensions. Although the algebras we  consider are obtained in diverse theories of gravity, their common feature is that all of them may be considered as extensions of Witt and Virasoro algebras. On the other hand, the first examples of infinite dimensional Lie algebra with extensive applications in physics are Witt and Virasoro algebras. We therefore first present a short review on the latter.  Then, we  review \bms\ and \bmsf\ algebras which are asymptotic symmetry algebras of $3d$ and $4d$ flat spacetimes, their commutation relations and  central extensions. Finally, we study the  \Max\ algebra and it central extension which is asymptotic symmetry algebra of a Chern-Simons gravity theory invariant under $2+1$ Maxwell algebra.

 \section{Witt algebra, its central extension and physical realization}
 In mathematical settings the Witt algebra, denoted by $\mathfrak{witt}$, is defined as the space of smooth vector fields on a circle and is the Lie algebra of group of diffeomorphisms on a circle $Diff(S^{1})$ \cite{Oblak:2016eij}. The {elements of the \wit\ algebra } as vector fields satisfy the Lie bracket as
 \begin{equation}
     [X,Y]=X(\varphi) Y^{\prime}(\varphi)- Y(\varphi) X^{\prime}(\varphi),
 \end{equation}
 where $X,Y\in \mathfrak{witt}$ and $\varphi$ is the coordinate on circle. One can then consider the Fourier expansion of the above and write the bracket in terms of Fourier expansion coefficients as 
 \begin{equation}
     [\mathcal{L}_{m},\mathcal{L}_{n}]=(m-n)\mathcal{L}_{m+n}.
 \end{equation}
  It is showed that the central extension of $\mathfrak{witt}$ algebra is Virasoro algebra denoted by \vir which is a one dimensional extension. 
  The central term can be viewed as a generator of algebra which commutes with all other generators. 
 The commutation relation of \vir\ algebra is as follows
 \begin{equation}
     [\mathcal{L}_{m},\mathcal{L}_{n}]=(m-n)\mathcal{L}_{m+n}+\frac{c}{12}(m^3-m)\delta_{m+n,0},
 \end{equation}
 where $c$ is an arbitrary constant, the central charge. In abstract mathematics, central extensions of $\mathfrak{g}$ are classified with the second real cohomology of the algebra  $\mathcal{H}^{2}(\mathfrak{g};\mathbb{R})$. The second real cohomology of \wit\ algebra is one dimensional.
 In the next chapter we return to the notion of cohomology with more details.
 
 One can show that the linear part of the central term can be absorbed by an appropriate redefinition of the generators so the non trivial central extension of \wit\ algebra may also be written as 
 \begin{equation}
     [\mathcal{L}_{m},\mathcal{L}_{n}]=(m-n)\mathcal{L}_{m+n}+\frac{c}{12}m^3\delta_{m+n,0}.
 \end{equation}
 Therefore, in our analysis we only include the $m^{3}$ term.
 It has been shown that the global symmetries of conformal field theory in $2d$ is two copies of the $\mathfrak{witt}$ algebra. 
 In fact the elements of the \wit\ algebra generate infinitesimal $2d$ conformal transformations. 

The central charge in \vir\ is a quantum feature of $2d$ CFT and appears in two-point function of energy-momentum tensor. 
 One can show that a necessary condition for the unitarity of the corresponding $2d$ CFT theory is the reality of the central charge.

 \section {Asymptotic symmetries of \textit{3d} spacetimes}\label{sec:2}
 
{In this section we review the structure of asymptotic symmetry algebras appearing in the context of $3d$ gravity.}
Depending on the asymptotic behavior of the metric and the chosen boundary fall-off conditions one can get different asymptotic symmetry algebras. The two set of ``standard'' fall-off behaviors are the Brown-Henneaux boundary conditions for AdS$_3$ which yields two copies of Virasoro \cite{Brown:1986nw} and the BMS fall-off behavior for $3d$ flat spacetime which leads to $\widehat{\mathfrak{bms}}_3$ \cite{Barnich:2006av,Barnich:2011ct, Ashtekar:1996cd}.  Here we briefly review these algebras. 


\subsection* { \texorpdfstring{AdS$_3$}{AdS} asymptotic symmetry algebra }\label{sec:2.1}
Centerless asymptotic symmetry algebra\footnote{{As discussed in Appendix \ref{appendix-A}, the asymptotic charge algebra is a central extension of asymptotic Killing vector field algebra. We refer to the latter as centerless asymptotic symmetry algebra.}} of $3$ dimensional AdS space with Brown-Henneaux boundary conditions is
\begin{equation} 
\begin{split}
 & i[\mathcal{L}_m,\mathcal{L}_n]=(m-n)\mathcal{L}_{m+n}, \\
  &i[\mathcal{L}_m,\Bar{\mathcal{L}}_n]=0, \\
 &i[\Bar{\mathcal{L}}_m,\Bar{\mathcal{L}}_n]=(m-n)\Bar{\mathcal{L}}_{m+n},
\end{split}\label{Centerless-AdS3}
\end{equation}
in which $m,\ n\ \in \mathbb{Z}$ and it is defined over the field of real numbers $\mathbb{R}$. It is obvious that the algebra \eqref{Centerless-AdS3} is direct sum of two different \wit\ algebras,
\begin{equation} 
\mathfrak{witt}^2= \mathfrak{witt}_L \oplus \mathfrak{witt}_R. 
\end{equation}

Centrally extended version of \eqref{Centerless-AdS3} is 
\begin{equation} 
\begin{split}
 & i[\mathcal{L}_m,\mathcal{L}_n]=(m-n)\mathcal{L}_{m+n} +\frac{c}{12} m^3\delta_{m+n,0}, \\
  &i[\mathcal{L}_m,\Bar{\mathcal{L}}_n]=0, \\
 &i[\Bar{\mathcal{L}}_m,\Bar{\mathcal{L}}_n]=(m-n)\Bar{\mathcal{L}}_{m+n}+\frac{\bar c}{12} m^3\delta_{m+n,0},
\end{split}\label{AdS3}
\end{equation}
which is the asymptotic symmetry algebra of asymptotically AdS$_3$ solutions to Einstein-$\Lambda$ theory with AdS radius $\ell$ and Newton constant $G_N$, $c=\bar c=\frac{3\ell}{2G_N}$ \cite{Brown:1986nw}. 
The central part, as mentioned, may also have a piece proportional to $m$ which can be absorbed in a proper redefinition of the generators.
The algebra \eqref{AdS3}  
will be denoted by \virtwo\ and is direct sum of two different Virasoro algebras,
\begin{equation} 
\mathfrak{vir}^2= \mathfrak{vir}_L \oplus \mathfrak{vir}_R.
\end{equation}
With our previous notation, Virasoro algebra is the centrally extended version of Witt algebra, i.e. \vir$=\widehat{\mathfrak{witt}}$. In other words, Virasoro algebra is Witt algebra which is extended by the Gel'fand-Fuks global 2-cocycle \cite{gel1969cohomologies}. The global part\footnote{{By the global part of an infinite dimensional Lie algebra $\mathfrak{g}$ we mean its maximal finite subalgebra. }} of \virtwo\ is either $\mathfrak{sl}(2,\mathbb{R})\oplus \mathfrak{sl}(2,\mathbb{R})\simeq \mathfrak{so}(2,2)$  which is the algebra of (global) isometries of AdS$_3$ \cite{Brown:1986nw}, or $\mathfrak{so}(3,1)\simeq \mathfrak{sl}(2;\mathbb{C})$ associated with global isometries of $3$ dimensional de Sitter space (without the factors of $i$ in the commutators, though) \cite{Compere:2014cna}. 

On the AdS$_3$ one can get other asymptotic symmetry algebras, e.g. if we relax the Brown-Henneaux boundary conditions to the ones introduced in \cite{Troessaert:2013fma} we get  two copies of centrally extended Virasoro-Kac-Moody algebras. One can also impose more restricted set of boundary conditions and get a single copy of Virasoro-Kac-Moody algebra \cite{Compere:2013bya}. The latter boundary conditions can be also used for the case of chiral gravity theory \cite{Li:2008dq}. In the most general case one can get two copies of $\mathfrak{sl}(2,\mathbb{R})$  Kac-Moody current algebras
\cite{Grumiller:2016pqb} which contains all the previous cases as subalgebras (the Virasoro algebras are related to this current algebra through the (twisted) Sugawara construction).

\subsection *{\textit{3d} flat space asymptotic symmetry algebra}\label{sec:2.2}

The centerless asymptotic symmetry  algebra of $3d$ flat spacetime is $\mathfrak{bms}_{3}$ \cite{Ashtekar:1996cd, Barnich:2006av}:
 \begin{equation} 
\begin{split}
 & i[\mathcal{J}_m,\mathcal{J}_n]=(m-n)\mathcal{J}_{m+n}, \\
 &i[\mathcal{J}_m,\mathcal{P}_n]=(m-n)\mathcal{P}_{m+n},\\
 &i[\mathcal{P}_m,\mathcal{P}_n]=0,
\end{split}\label{bms3}
\end{equation}
where $m,n\in \mathbb{Z}$ and it is defined over the field $\mathbb{R}$. {The $\mathfrak{bms}_{3}$ is an infinite dimensional algebra with countable basis} and has two sets of generators, ${\cal J}_n$'s and ${\cal P}_n$'s. The former generates the \wit\ subalgebra of \bms\ and is the algebra of smooth vector fields on a circle. ${\cal P}_n$'s  construct an adjoint representation of the \wit\ algebra and form the ideal part of \bms.  Eq.\eqref{bms3} makes it clear that $\mathfrak{bms}_{3}$ has a semi-direct sum structure:
\begin{equation} \label{bms3=witt+ideal}
\mathfrak{bms}_{3}= \mathfrak{witt}\inplus_{ad}\mathfrak{witt}_{ab},
\end{equation}
where the subscript $ab$ is to emphasize ${\cal P}_n$'s being abelian and $ad$ denotes the adjoint action.  
The global part of $\mathfrak{bms}_{3}$ is $3d$ Poincar\'{e} algebra, $\mathfrak{iso}(2,1)$, and is obtained by restricting to $m, n=\pm 1, 0$ in relation \eqref{bms3}. {In physics terminology the generators ${\cal J}_n$'s which are generalizations of the $3d$ Lorentz group are called superrotations and the generators ${\cal P}_n$'s which are generalizations of $3d$ momenta, supertranslations. }

The asymptotic symmetry analysis of $3d$ flat space actually leads to centrally extended version of the algebra, denoted by $\widehat{\mathfrak{bms}}_{3}$:
 \begin{equation}\label{BMS3-centrally-extended} 
\begin{split}
 & i[\mathcal{J}_m,\mathcal{J}_n]=(m-n)\mathcal{J}_{m+n}+\frac{c_{JJ}}{12}m^{3}\delta_{m+n,0}, \\
 &i[\mathcal{J}_m,\mathcal{P}_n]=(m-n)\mathcal{P}_{m+n}+\frac{c_{JP}}{12}m^{3}\delta_{m+n,0},\\
 &i[\mathcal{P}_m,\mathcal{P}_n]=0,
\end{split}
\end{equation}
in which $c_{JJ}$ and $c_{JP}$ are the central charges \cite{Barnich:2006av, Barnich:2014cwa}. 

For flat space there are of course ``more relaxed'' boundary conditions which yield algebras with more number of fields in the various representations of the Virasoro algebra, the most general being $\mathfrak{iso}(2,1)$  Kac-Moody current algebra 
\cite{Grumiller:2017sjh}. 

\subsection{Contraction of Virasoro to \texorpdfstring{\hbms}{HBMS3}}\label{sec:2.3-contraction}

The \hbms\ and \virtwo\ algebras are related by a IW contraction \cite{Barnich:2012rz}. To see this, let us introduce

\begin{equation}\label{Vir-BMS}
\mathcal{L}_n=\frac12(\mathcal{J}_n+\ell{\mathcal{P}_n}),\ \ \ {\bar{\mathcal{L}}}_{-n}=\frac12(\mathcal{J}_n-\ell{\mathcal{P}_n}),
\end{equation}
where $\ell$ is a parameter introduced to facilitate taking the contraction as an $\ell\to\infty$ limit, while keeping $\mathcal{J}_n, \mathcal{P}_n$ fixed. Plugging \eqref{Vir-BMS} into the Virasoro algebra we obtain 
\begin{align}
 & i[\mathcal{J}_{m},\mathcal{J}_{n}]=(m-n)\mathcal{J}_{m+n}+\frac{c+\bar c}{12}n^3\delta_{n+m,0}, \cr
 &i[\mathcal{J}_{m},\mathcal{P}_{n}]=(m-n)\mathcal{P}_{m+n}+\frac{c-\bar c}{12\ell}n^3\delta_{n+m,0},\\
 &i[\mathcal{P}_{m},\mathcal{P}_{n}]=\frac{1}{\ell^2}(m-n)\mathcal{J}_{m+n}+\frac{c+\bar c}{12\ell^2}n^3\delta_{n+m,0}.\nonumber
\end{align} 
If we keep $c_{JJ}=c+\bar c$ and $c_{JP}=(c-\bar c)/\ell$ finite while taking $\ell\to\infty$, the above reduces to the $\widehat{\mathfrak{bms}}_3$ \eqref{BMS3-centrally-extended}. In the above IW contraction $1/\ell$ may be viewed as the deformation parameter.

\subsection* {\textit{3d} asymptotic/near horizon symmetry algebras
}\label{sec:2.4}

On AdS$_3$ space we have the class of BTZ black holes \cite{Banados:1992wn} and/or their Virasoro descendants. It has been shown in \cite{Afshar:2016kjj,Afshar:2016wfy, Afshar:2017okz} that one can impose appropriate boundary (fall-off) conditions on the horizon of these black holes and find a ``near horizon'' (in contrast to asymptotic) symmetry algebra. This algebra has been shown to be two chiral copies of a $\mathfrak{u}(1)$ current algebra, i.e. a centrally extended Virasoro-Kac-Moody algebra, $\widehat{\mathfrak{KM}_{\mathfrak{u}(1)}}$, and is defined through 
\begin{align}\label{centrally-extended-u(1)-KM}
 & i[\mathcal{J}_{m},\mathcal{J}_{n}]=(m-n)\mathcal{J}_{m+n}+\frac{c_{JJ}}{12} n^3\delta_{n+m,0}, \cr
 &i[\mathcal{J}_{m},\mathcal{P}_{n}]=-n\mathcal{P}_{m+n}+\frac{c_{PJ}}{12}n^2\delta_{n+m,0},\\
 &i[\mathcal{P}_{m},\mathcal{P}_{n}]={\frac{c_{PP}}{12}} n\delta_{n+m,0}.\nonumber
\end{align} 
One should note that in the computations of the near horizon symmetry algebras the central charges $c_{JJ}, c_{JP}$ and $c_{PP}$ are not (parametrically) independent. 

On the other hand, it has been shown that with choosing a different fall-off condition on asymptotic region of AdS$_{3}$, one can obtain the same algebra which was obtained by Compère, Song and Strominger in \cite{Compere:2013bya}. In this case also there are just two independent central charges $c_{JJ}$ and $c_{PP}$. 
 
 \section{Introduction to \texorpdfstring{$\mathfrak{bms}_4$}{BMS4}  algebra}\label{sec:3.2}

The original \bmsf\ algebra introduced in \cite{Bondi:1962px, Sachs:1962zza, Sachs:1962wk} is a semi-direct sum of Lorentz algebra with Abelian ideal spanned by supertranslations 
\begin{equation*}
    (\mathfrak{bms}_{4})_{\text{old}}=\text{Lorentz}\,\inplus\,\text{Supertranslations},
\end{equation*}
and its global part is $4d$ Poincar\'e algebra. Barnich and Troessaert in \cite{Barnich:2009se, Barnich:2011ct} suggested the Lorentz part of the $\mathfrak{bms}_4$ might be replaced by a larger (infinite dimensional) algebra called superrotations. In this work, as it is common in the recent literature,  we use \bmsf\ to denote this extended version. \bmsf\ is hence semi-direct sum of superrotations and supertranslations,
\begin{equation*}
   \mathfrak{bms}_{4}= \text{Superrotations}\,\inplus\,\text{Supertranslations}.
\end{equation*}
Later, they also classified its central extensions $\widehat{\mathfrak{bms}_4}$ in \cite{Barnich:2011ct,Barnich:2017ubf}. Next, we review the structure of \bmsf\ algebra.

\subsection *{{4d} flat space asymptotic symmetry algebra}

The centerless asymptotic symmetry  algebra of 4d flat spacetime is $\mathfrak{bms}_{4}$:
 \begin{equation} 
\begin{split}
 & [\mathcal{L}_{m},\mathcal{L}_{n}]=(m-n)\mathcal{L}_{m+n}, \\
 & [\bar{\mathcal{L}}_{m},\bar{\mathcal{L}}_{n}]=(m-n)\bar{\mathcal{L}}_{m+n},\\
 &[{\mathcal{L}}_{m},\bar{\mathcal{L}}_{n}]=0,\\
 &[\mathcal{L}_{m},T_{p,q}]=(\frac{m+1}{2}-p)T_{p+m,q},\\
 &[\bar{\mathcal{L}}_{n},T_{p,q}]=(\frac{n+1}{2}-q)T_{p,q+n},\\
 &[T_{p,q},T_{r,s}]=0,
\end{split}\label{bms4}
\end{equation}
where $m,n,p,q,r,s\in \mathbb{Z}$ and it is defined over the field of real numbers $\mathbb{R}$. {The $\mathfrak{bms}_{4}$ is an infinite dimensional Lie algebra with countable basis} which is spanned by the generators $\mathcal{L}_m$, $\bar{\mathcal{L}}_{m}$ and $T_{p,q}$.  $\mathcal{L}_m$ and $\bar{\mathcal{L}}_{m}$ generate the direct sum of two \wit\  subalgebra of  $\mathfrak{bms}_{4}$ and are usually called ``superrotations''.  $T_{p,q}$, the ``supertranslations,'' form the ideal part of $\mathfrak{bms}_{4}$, $\mathfrak{T}_{ab}$. Eq.\eqref{bms4} makes it clear that $\mathfrak{bms}_{4}$ has a semi-direct sum structure
\begin{equation} \label{bms4=witt+ideal}
\mathfrak{bms}_{4}= \big(\mathfrak{witt}\oplus\mathfrak{witt}\big)\inplus \mathfrak{T}_{ab},
\end{equation}
where the subscript $ab$ is to emphasize the ideal part being abelian.  

The global part, i.e. the maximal finite subalgebra, of $\mathfrak{bms}_{4}$ is 4d Poincar\'{e} algebra $\mathfrak{iso}(3,1)$ and is generated by ${\cal L}_0, {\cal L}_{\pm1}$ and $\bar{\mathcal{L}}_0, \bar{\mathcal{L}}_{\pm1}$ (which form Lorentz algebra $\mathfrak{so}(3,1)$) and 
$T_{0,0}, T_{0,1}, T_{1,0}, T_{1,1}$ as the translations. In the next subsection we will make the connection to the more usual basis for Poincar\'e algebra  explicit. 

The above $\mathfrak{bms}_{4}$ admits central extensions in ${\cal L}_n$, $\bar{\cal L}_n$ sectors \cite{Barnich:2011ct,Barnich:2011mi}. The centrally extended algebra which will be denoted by $\widehat{\mathfrak{bms}}_{4}$ is
 \begin{equation} 
\begin{split}
 & [\mathcal{L}_{m},\mathcal{L}_{n}]=(m-n)\mathcal{L}_{m+n}+\frac{c_{\mathcal{L}}}{12}(m^{3})\delta_{m+n,0}, \\
 & [\bar{\mathcal{L}}_{m},\bar{\mathcal{L}}_{n}]=(m-n)\bar{\mathcal{L}}_{m+n}+\frac{c_{\bar{\mathcal{L}}}}{12}(m^{3})\delta_{m+n,0},\\
 &[{\mathcal{L}}_{m},\bar{\mathcal{L}}_{n}]=0,\\
 &[\mathcal{L}_{m},T_{p,q}]=(\frac{m+1}{2}-p)T_{p+m,q},\\
 &[\bar{\mathcal{L}}_{n},T_{p,q}]=(\frac{n+1}{2}-q)T_{p,q+n},\\
 &[T_{p,q},T_{r,s}]=0,
\end{split}\label{BMS4-centrally-extended}
\end{equation}
in which $c_{\mathcal{L}}$ and $c_{\bar{\mathcal{L}}}$ are called central charges. We note that these central charges do not arise in the computation of surface charges associated with the asymptotic symmetries of 4d flat space gravity \cite{Barnich:2011mi}. We also note that the global part of the algebra generated when the $m,n$ indices are restricted to  $0,\pm 1$ values, do not involve the central charges, as they can be absorbed in a shift of ${\cal L}_0, {\bar{\cal L}_0}$ respectively by $\frac{c_{\mathcal{L}}}{24}, \frac{c_{\bar{\mathcal{L}}}}{24}$.  Therefore, global part of $\widehat{\mathfrak{bms}}_{4}$ is also 4d Poincar\'{e}. We note that the second real cohomology of the $\mathfrak{bms}_{4}$ algebra ${\cal H}^2(\mathfrak{bms}_{4}; \mathbb{R})$, which  classifies (global) central extensions of the algebra, does not allow for any other central extension \cite{Barnich:2011ct}  than the $c_{\mathcal{L}}, c_{\bar{\mathcal{L}}}$.

\subsection{More on global part of the \texorpdfstring{$\mathfrak{bms}_4$}{BMS4} algebra}

As mentioned the $\mathfrak{bms}_4$ algebra should contain Poincar\'e algebra which is the isometry algebra of the flat space. Generators of the Poincar\'e algebra are usually written in the 4d tensorial basis, $J^{\mu \nu}$, the generators of Lorentz algebra, and $P^{\mu}$, the generators of translations, as it discussed in \cite{weinberg1995quantum} 
\begin{equation} \label{poincareJ-P}
\begin{split}
 & [J^{\mu \nu},J^{\rho \sigma}]=i(\eta^{\mu \rho} J^{\nu \sigma}+\eta^{\sigma \mu}J^{\rho \nu}-\eta^{\nu \rho}J^{\mu \sigma}-\eta^{\sigma \nu}J^{\rho \mu}), \\
  &[J^{\mu \nu},P^{\rho}]=i(\eta^{\rho  \mu}P^{\nu}-\eta^{\rho \nu}P^{\mu}), \\
 &[P^{\mu},P^{\nu}]=0,
\end{split}
\end{equation}
where $\mu, \nu=0,1,2,3$ and $\eta_{\mu\nu}=diag(-,+,+,+)$ is the Minkowski metric. To relate the above algebra to the global part of $\mathfrak{bms}_4$ one should decompose the Lorentz part into $\mathfrak{sl}(2)\oplus\mathfrak{sl}(2)$ basis:
\begin{equation*}
\begin{split}
    & \mathcal{L}_{\pm}\equiv iR^{1}\pm R^{2},\\
     &\bar{\mathcal{L}}_{\pm}\equiv iL^{1}\pm L^{2},\\
     & \mathcal{L}_{0} \equiv R^{3},\,\,\,\,\,\ \bar{\mathcal{L}}_{0} \equiv L^{3},
\end{split}
\end{equation*}
where 
\begin{equation}
     {L}^{i}\equiv \frac{1}{2}(\mathcal{J}^{i}+i\mathcal{K}^{i}),\quad
     {R}^{i}\equiv \frac{1}{2}(\mathcal{J}^{i}-i\mathcal{K}^{i}), \qquad i=1,2,3.
\end{equation}
Here ${\cal J}^i, {\cal K}^i$ are generators of rotation and boost: 
\begin{equation*}
    \mathcal{J}^{i}:=\frac{1}{2}\epsilon^{i}_{\ jk} J^{jk},\,\,\,\,\,\ \mathcal{K}^{i}:=J^{0i},
\end{equation*}
where $\epsilon^{i}_{\ jk}$ is an antisymmetric quantity with $\epsilon^{1}_{\ 23}=+1$. In other words, the generators of 4d Lorentz algebra $ J^{\mu\nu}$ can be decomposed as $({\bf 1},{\bf 3})\oplus ({\bf 3},{\bf 1})$ of $\mathfrak{sl}(2)\oplus\mathfrak{sl}(2)$ algebra. One may then readily show that
\begin{equation} 
\begin{split}
 & [\mathcal{L}_{m},\mathcal{L}_{n}]=(m-n)\mathcal{L}_{m+n}, \\
  &[\mathcal{L}_{m},\bar{\mathcal{L}}_{n}]=0, \\
 &[\bar{\mathcal{L}}_{m},\bar{\mathcal{L}}_{n}]=(m-n)\bar{\mathcal{L}}_{m+n},
\end{split}\label{PoincareL-L}
\end{equation}
where $m,n=\pm 1,0$. 

The translation generators, which are Lorentz four-vectors $P_\mu$ can be decomposed into $({\bf 2},{\bf 2})$ of $\mathfrak{sl}(2)\oplus\mathfrak{sl}(2)$ algebra, i.e. $P_\mu$ are linear combinations of $T_{m,n}, m,n=0,1$:
\begin{equation}\label{basisP-T}
\begin{split}
 & P^{0}\equiv H=(T_{1,0}-T_{0,1}), \\
 &P^{1}=(-i)(T_{1,1}+T_{0,0}),\\
 & P^{2}=(T_{1,1}-T_{0,0}),\\
 &P^{3}=(T_{1,0}+T_{0,1}).
\end{split}
\end{equation}


\subsection{\texorpdfstring{AdS$_4$}{4AdS} isometry, \texorpdfstring{$\mathfrak{so}(3,2)$}{AdS4} algebra as deformation of \texorpdfstring{$4d$}{4d} Poincar\'e algebra}

For our later use we also discuss AdS$_4$ isometry algebra generated by $J^{ab}, a,b=-1,0,1,2,3$
\begin{equation} 
\begin{split}
  [J^{a b},J^{c d}]=i(\eta^{a c} J^{b d}+\eta^{d a}J^{c b}-\eta^{b c}J^{a d}-\eta^{d b}J^{c a}), 
\end{split}\label{AdS-J}
\end{equation}
where $\eta^{ab}=(-1,-1,+1,+1,+1)$. $J^{ab}$'s which are in ${\bf 15}$ representation of $\mathfrak{so}(3,2)$ may be decomposed in terms of $\mathfrak{sl}(2)\oplus\mathfrak{sl}(2)$ as $({\bf 1},{\bf 3})\oplus ({\bf 3},{\bf 1})\oplus ({\bf 3},{\bf 3})$.
The first two are just ${\cal L}_m, \bar{\cal L}_m, m=0,\pm 1$, and the last one may be denoted by $T_{m,n}, m,n=0,\pm 1$ with the commutation relations:
 \begin{equation}\label{AdS4}
\begin{split}
 & [\mathcal{L}_{m},\mathcal{L}_{n}]=(m-n)\mathcal{L}_{m+n}, \\
 & [\bar{\mathcal{L}}_{m},\bar{\mathcal{L}}_{n}]=(m-n)\bar{{\mathcal L}}_{m+n},\\
 &[{\mathcal{L}}_{m},\bar{\mathcal{L}}_{n}]=0,\\
 &[\mathcal{L}_{m},T_{p,q}]=(\frac{m+1}{2}-p)T_{p+m,q},\\
 &[\bar{\mathcal{L}}_{n},T_{p,q}]=(\frac{n+1}{2}-q)T_{p,q+n},\\
 &[T_{m,n},T_{p,q}]=\frac{1}{\ell^2}\frac{1}{2}\bigg((q-n)\mathcal{L}_{m+p-1}+(p-m)\bar{\mathcal{L}}_{q+n-1}\bigg),
\end{split}
\end{equation}
where $m,n,p,q=0,\pm 1$ and $\frac{1}{\ell^2}$ is the AdS radius which may be viewed as a deformation parameter. 

It is known that $\mathfrak{iso}(3,1)$ is not a rigid (stable) algebra and may be deformed into $\mathfrak{so}(3,2)$ or $\mathfrak{so}(4,1)$, which are stable \cite{levy1967deformation}. In the $\mathfrak{sl}(2)\oplus\mathfrak{sl}(2)$ notation adopted above, one may readily use the Hochschild-Serre factorization theorem to argue that only the ideal part of $\mathfrak{iso}(3,1)$, the $[T,T]$ commutator, can be deformed such that only ${\cal L}_m, \bar{\mathcal{L}}_m$ appear in the right-hand-side of the commutator. This can be manifestly seen in the last equation in \eqref{AdS4}.\footnote{Conversely, one may view $\mathfrak{iso}(3,1)$ algebra as the In\"on\"u-Wigner contraction of the $\mathfrak{so}(3,2)$ algebra. Geometrically, this contraction corresponds to a large radius limit of an AdS$_4$ space yielding a 4d flat space.} As reviewed in the introduction, these theorems do not apply to the infinite dimensional algebras and one cannot extend the above result which is about the global part $\mathfrak{bms}_4$ to the whole algebra. We will show in the rest of this work that there is no infinite dimensional algebra in the family of $\mathfrak{bms}_4$ deformations which has $\mathfrak{so}(3,2)$ as its global part.

\section {Maxwell algebra and its infinite dimensional enhancement}\label{sec:4-2}

\subsection*{The Maxwell algebra}
The Maxwell algebra in $d$ dimension can be obtained as, first, an extension of the Poincar\'{e} algebra by an antisymmetric rank two tensor \cite{Bonanos:2008kr,Bonanos:2008ez},
as $\mathfrak{iso}(d-1,1)\inplus\mathfrak{M}_{ab}$, with
\begin{equation}
    [\mathcal{J}_{\mu\nu},\mathcal{M}_{\alpha\beta}]=-(\eta_{\alpha[\mu}\mathcal{M}_{\nu]\beta}-\eta_{\beta[\mu}\mathcal{M}_{\nu]\alpha}),
\end{equation}
where $\mathcal{J}_{\mu\nu}$ are generators of the Lorentz algebra $\mathfrak{so}(d-1,1)$ and $\mathcal{M}_{\alpha\beta}$ commute among themselves and also with generators of translations, $\mathcal{P}_{\mu}$'s, appearing in ideal part of Poincar\'{e} algebra $\mathfrak{iso}(d-1,1)$ {and bracket $[xy]$ denotes antisymmetrization. }

Furthermore, one can deform the commutator of translations so that it is no more zero but proportional to the new generators $\mathcal{M}$ to obtain Maxwell algebra as
\begin{equation}
\begin{split}
&[\mathcal{J}_{\mu\nu},\mathcal{J}_{\alpha\beta}]=-(\eta_{\alpha[\mu}\mathcal{J}_{\nu]\beta}-\eta_{\beta[\mu}\mathcal{J}_{\nu]\alpha}),\\
&[\mathcal{J}_{\mu\nu},\mathcal{P}_{\alpha}]=-(\eta_{\mu\alpha}\mathcal{P}_{\nu}-\eta_{\nu\alpha}\mathcal{P}_{\mu}),\\
   & [\mathcal{J}_{\mu\nu},\mathcal{M}_{\alpha\beta}]=-(\eta_{\alpha[\mu}\mathcal{M}_{\nu]\beta}-\eta_{\beta[\mu}\mathcal{M}_{\nu]\alpha}),\\
   &    [\mathcal{P}_{\mu},\mathcal{P}_{\nu}]=\varepsilon \mathcal{M}_{\mu\nu},
\end{split}
\end{equation}
where $\varepsilon$ may be viewed as the deformation parameter and can be absorbed by an appropriate redefinition of the generators. This algebra describes a relativistic particle which is coupled to a constant electromagnetic field \cite{schrader1972maxwell,beckers1983minimal} and has subsequently been studied in the gravity context in several papers \cite{Bonanos:2009wy,Izaurieta:2009hz,Lukierski:2010dy,Durka:2011nf,deAzcarraga:2012qj,Concha:2013uhq,deAzcarraga:2014jpa,Concha:2014tca,Concha:2014vka,Concha:2014zsa,Cebecioglu:2015jta,Concha:2016zdb,Caroca:2017izc,Ravera:2018vra,Aviles:2018jzw,Concha:2018jxx,Kibaroglu:2018mcn,Concha:2019icz,Barducci:2019fjc,Chernyavsky:2019hyp,Concha:2019lhn}. 

In three spacetime dimensions, the Poincar\'{e} algebra has six generators, three generators for rotation and boost and three generators for translation. In the $2+1$ Maxwell algebra, the Lorentz-covariant tensor adds three independent generators. Thus the Maxwell algebra in three spacetime dimensions has $9$ generators which can be written in an appropriate basis ($\mathfrak{sl}(2)$ basis) as
\begin{equation} 
\begin{split}\label{maxwell}
 & i[\mathcal{J}_m,\mathcal{J}_n]=(m-n)\mathcal{J}_{m+n}, \\
 &i[\mathcal{J}_m,\mathcal{P}_n]=(m-n)\mathcal{P}_{m+n},\\
 &i[\mathcal{J}_m,\mathcal{M}_n]=(m-n)\mathcal{M}_{m+n},\\
 &i[\mathcal{P}_m,\mathcal{P}_n]=(m-n)\mathcal{M}_{m+n},
\end{split}
\end{equation}
where $m,n=\pm1,0$. The $2+1$ Maxwell algebra can be enlarged to a new  infinite dimensional algebra with countable basis by extending the range of $m,n$,  $m,n\in \mathbb{Z}$ \cite{Caroca:2017onr}. In this work we shall denote the infinite dimensional version of the Maxwell algebra by $\mathfrak{Max}_{3}$. Interestingly, as was shown in \cite{Concha:2018zeb}, the latter can be obtained as the asymptotic symmetry of a $3d$ Chern-Simons gravity based on the Maxwell algebra.

\subsection* {Infinite dimensional \texorpdfstring{$3d$}{3d} Maxwell algebra through \texorpdfstring{$\mathfrak{bms}_{3}$}{bms3} algebra}

An infinite dimensional enhancement of $2+1$ Maxwell algebra $\mathfrak{Max}_{3}$ can be obtained as an extension and deformation of the \bms\ algebra. For review the properties of the \bms\ algebra, its commutation relations and central extension we refer the reader to section \ref{sec:2.2}.

\paragraph {$\mathfrak{Max}_{3}$ as an extension of \bms.}
We are interested in a particular extension of the \bms\ algebra, denoted by ${\widetilde{\mathfrak{bms}}}_{3}$, in which the additional generators have the same conformal weight as the \bms\ generators, $h=2$, where $h$ appears as $[\mathcal{J}_{m},\mathcal{P}_{n}]=((h-1)m-n)\mathcal{P}_{m+n}$. The non vanishing commutators of $\widetilde{\mathfrak{bms}}_{3}$ are given by
\begin{equation} 
\begin{split}
 & i[\mathcal{J}_m,\mathcal{J}_n]=(m-n)\mathcal{J}_{m+n}, \\
 &i[\mathcal{J}_m,\mathcal{P}_n]=(m-n)\mathcal{P}_{m+n},\\
 &i[\mathcal{J}_m,\mathcal{M}_n]=(m-n)\mathcal{M}_{m+n},
\end{split}\label{bms3-extended}
\end{equation}
in which $m,\ n\ \in \mathbb{Z}$, and is defined over the field $\mathbb{R}$. One can see that the algebra \eqref{bms3-extended} has a \wit\ subalgebra. In particular, the structure of $\widetilde{\mathfrak{bms}}_{3}$ is the semi direct sum of the $\mathfrak{bms}_{3}$ algebra with an abelian ideal part. Then, we have
\begin{equation} 
\widetilde{\mathfrak{bms}}_{3}= (\mathfrak{bms}_{3}) \inplus\mathfrak{M}_{ab},
\end{equation}
where the abelian ideal $\mathfrak{M}$ is spanned by $\mathcal{M}$ generators. 
One can show that $\widetilde{\mathfrak{bms}}_{3}$ admits only three independent central terms \cite{Caroca:2017onr} as 
\begin{equation} 
\begin{split}
 & i[\mathcal{J}_m,\mathcal{J}_n]=(m-n)\mathcal{J}_{m+n}+\frac{c_{JJ}}{12}m^{3}\delta_{m+n,0}, \\
 &i[\mathcal{J}_m,\mathcal{P}_n]=(m-n)\mathcal{P}_{m+n}+\frac{c_{JP}}{12}m^{3}\delta_{m+n,0},\\
 &i[\mathcal{J}_m,\mathcal{M}_n]=(m-n)\mathcal{M}_{m+n}+\frac{c_{JM}}{12}m^{3}\delta_{m+n,0},
\end{split}\label{bms3-central-extended}
\end{equation}

One can deform the  algebra in \eqref{bms3-extended} to obtain a new non isomorphic algebra with non vanishing commutators the same as \eqref{maxwell}.
Thus one can view the \Max\ algebra \eqref{maxwell} as an extension and deformation of the \bms\ algebra. 
 The \Max\ algebra as the centrally extended $\widetilde{\mathfrak{bms}}_{3}$ algebra \eqref{bms3-extended} admits only three independent central terms as
\begin{equation} 
\begin{split}\label{C-maxwell}
 & i[\mathcal{J}_m,\mathcal{J}_n]=(m-n)\mathcal{J}_{m+n}+\frac{c_{JJ}}{12}m^{3}\delta_{m+n,0}, \\
 &i[\mathcal{J}_m,\mathcal{P}_n]=(m-n)\mathcal{P}_{m+n}+\frac{c_{JP}}{12}m^{3}\delta_{m+n,0},\\
 &i[\mathcal{J}_m,\mathcal{M}_n]=(m-n)\mathcal{M}_{m+n}+\frac{c_{JM}}{12}m^{3}\delta_{m+n,0},\\
 &i[\mathcal{P}_m,\mathcal{P}_n]=(m-n)\mathcal{M}_{m+n}+\frac{c_{JM}}{12}m^{3}\delta_{m+n,0}.
\end{split}
\end{equation}
We denote the central extension of \Max\ by $\widehat{\mathfrak{Max}}_{3}$ with the commutators as \eqref{C-maxwell}.

Such an infinite-dimensional symmetry algebra in the presence of three central terms can also be obtained through the semi-group expansion method \cite{Caroca:2017onr}. This algebra describes the asymptotic symmetry of a three-dimensional Chern-Simons gravity theory invariant under the Maxwell algebra where the central charges $c_{JJ}$, $c_{JP}$ and $c_{JM}$ are proportional to the couplings of three terms of the Chern-Simons Maxwell gravity action \cite{Concha:2018zeb}.

\section{Summary of the chapter}
In this chapter we reviewed the structures of some infinite dimensional Lie algebras appearing as asymptotic and or near horizon symmetry algebras. These algebras contain a \wit\ or \vir\ subalgebra, which  we reviewed first. Next, we explored the asymptotic symmetry algebras in $3d$ flat and AdS spacetimes. In particular, we explored the structure of \bms, \kac\ and two copies of \wit\ algebra and their central extensions. Then we considered the commutation relations of asymptotic symmetry algebra of $4d$ flat spacetime denoted \bmsf\ and its central extensions. Finally, we introduced $2+1$ Maxwell algebra and its infinite dimensional enhancement called \Max\ which is obtained as asymptotic symmetry algebra of a Chern-Simons gravity theory based on $2+1$ Maxwell algebra. After that we studied commutation relations of \Max\ algebra and its central extension.  

 


\chapter{Contraction and deformation theory of Lie algebras} \label{ch3}

\section{Contraction of Lie algebras}\label{3.1}
The notion of contraction group and algebra in physics has been introduced from 1950s when Erdal İnönü and Eugene Wigner discussed possibility of obtaining a new non isomorphic Lie group (algebra) by a continuous limiting of a given Lie group (algebra) \cite{Inonu:1953sp}. 
At the level of Lie algebra, the IW contraction procedure leads to  a new non isomorphic algebra with different structure constants. To see this consider the rotation algebra $\mathfrak{so}(3)$ with commutation relations as
\begin{equation}
    [L,M]=N,\qquad[M,N]=L, \qquad[N,L]=M,
\end{equation}
where $M,N$ and $L$ are generators of $\mathfrak{so}(3)$. Suppose one redefines the generators as $L=\ell \tilde{L}$, $N=\ell\tilde{N}$ and $M=\tilde{M}$ so the commutation relations are changed as 
\begin{equation}
    [\tilde{L},\tilde{M}]=\tilde{N},\qquad[\tilde{M},\tilde{N}]=\tilde{L}, \qquad[\tilde{N},\tilde{L}]=\frac{1}{\ell^2}\tilde{M},
\end{equation}
by sending $\ell$ to infinity one obtains 
\begin{equation}
    [\tilde{L},\tilde{M}]=\tilde{N},\qquad[\tilde{M},\tilde{N}]=\tilde{L}, \qquad[\tilde{N},\tilde{L}]=0,
\end{equation}
which is $\mathfrak{iso}(2)$, isometries of Euclidean $2d$ plane. The new algebra $\mathfrak{iso}(2)$ is not isomorphic with $\mathfrak{so}(3)$ in the sense that the latter can not be obtained by a redefinition, change of the basis, of generators of the former. Geometrically, if $\mathfrak{so}(3)$ and $\mathfrak{iso}(2)$ are viewed as isometries of a $2$ sphere and plane, $\ell^2$ which is the contraction parameter may be viewed as radius of the $2$ sphere. 
This result can be generalized: We know that $SO(n+1)$ is the isometry group of $n$ dimensional sphere $S^{n}$ of an arbitrary radius $\ell$. This sphere may be embedded into the $n+1$ dimensional flat space $\mathbb{R}^{n+1}$. The commutation relations of its algebra are
\begin{equation}
    [s_{ij},s_{kl}]=s_{il}\delta_{jk}+s_{jk}\delta_{il}-s_{ik}\delta_{jl}-s_{jl}\delta_{ik},\label{so(n)}
\end{equation}
where $0\leq i,j,k,l\leq n+1$. In the limit of $\ell\to\infty$, as (IW) contraction, the above algebra reduces to the algebra of Euclidean group $E(n)$ or $ISO(n)$, with the associated algebra:
\begin{equation}
\begin{split}
   & [r_{ij},r_{kl}]=r_{il}\delta_{jk}+r_{jk}\delta_{il}-r_{ik}\delta_{jl}-r_{jl}\delta_{ik},\\
  & [r_{ij},p_{k}]=p_{i}\delta_{jk}-p_{j}\delta_{ik},\\
  & [p_{i},p_{j}]=0,\label{iso(n)}
\end{split}
\end{equation}
where one has replaced $s_{ij}->r_{ij}$ and $s_{i,n+1}->(\frac{1}{\ell}) p_{i}$ for $0\leq i,j,k,l \leq n$. 
The next example is the isometry group of  $n$ dimensional hyperboloid $H^{n}$ radius $\ell$, $SO(n,1)$. The commutation relations of $\mathfrak{so}(n,1)$ is the same as \eqref{so(n)}. A similar limit again brings us to Euclidean group $E(n)$ or $ISO(n)$ with the same commutation relations as $\mathfrak{iso}(n)$. By the Weyl unitarity trick two algebras $\mathfrak{so}(n+1)$ and $\mathfrak{so}(n,1)$ are mapped to each other. Therefore, both of $\mathfrak{so}(n+1)$ and $\mathfrak{so}(n,1)$ algebras in the  $\ell\to \infty$ limit can be contracted to $\mathfrak{iso}(n)$. Conversely, this means that by deformation procedure one can reach  $\mathfrak{so}(n+1)$ or $\mathfrak{so}(n,1)$, starting from $\mathfrak{iso}(n)$. We return to the concept of deformation in the next section. This is very similar to the problem  considered by Levy-Nahas in \cite{levy1967deformation} where it is shown that the only algebras which can be contracted to the Poincar\'{e} algebra $\mathfrak{iso}(3,1)$, are $\mathfrak{so}(4,1)$ and $\mathfrak{so}(3,2)$ which are, respectively the algebras corresponding to dS$_4$ and AdS$_4$ isometries.  

As mentioned, the notion of contraction was first used to obtain Galilean algebra from Poincar\'e algebra and Poincar\'e algebra from isometries algebra of AdS/dS spacetimes by sending the speed of light and the AdS/dS radius to infinity respectively  \cite{weinberg1995quantum}. 

Although the IW contraction is well known among physicists, it should be noticed that the concept of contraction was first introduced by Segal in \cite{segal1951class}. In 1960 Saletan introduced another contraction approach \cite{saletan1961contraction}. To review different methods of contraction and comparison between them we refer the reader to  \cite{gilmore2012lie}.

\section{Deformation as inverse procedure of contraction}\label{3.2}
{\it{Deformation}} of a certain Lie algebra $\mathfrak{g}$ is a modification of its structure constants. Some of such deformations could just be a change of basis which are called trivial deformations. {Nontrivial deformations modify/deform a Lie algebra $\mathfrak{g}$ to another Lie algebra with the same vector space structure. In the case of finite {dimensional} Lie algebras the latter implies that deformation does not change the dimension of the algebra.} 
The concept of deformation of rings and algebras was first introduced in a series of papers by Gerstenhaber \cite{gerstenhaber1964deformation, gerstenhaber1966deformation, gerstenhaber1968deformation, gerstenhaber1974deformation} and by Nijenhuis and Richardson for Lie algebras in \cite{nijenhuis1967deformations}. 
Deformations introduced by Nijenhuis and Richardson are known as 'formal' deformations where a Lie algebra is deformed by a formal power series of some variables (deformation parameters). {If {one considers the deformation only up to the linear term in the power series it is called `infinitesimal' deformation}}. 
While other kinds of generalized deformations of Lie algebras, like 'versal', 'global' and 'analytic' deformations {have} also been introduced and studied \cite{fialovski1986deformations, fialowski1988example, fialowski2003global, fialowski2005global, guerrini1998formal}, in this work we only focus on infinitesimal and formal deformations.

A Lie algebra $\mathfrak{g}$ is called rigid or stable if it does not admit any deformation or equivalently if its deformed algebra $\mathfrak{g}_{\varepsilon}$, $\varepsilon$ being the deformation parameter, is isomorphic to the initial algebra $\mathfrak{g}$. As deformations, there are some different notions of rigidity. Here, we will focus on the `formal' and `infinitesimal' rigidity. One may consider `analytic,' `global or `versal' rigidity of Lie algebras associated with similar deformations, as e.g. done in \cite{guerrini1999formal, fialowski2005global}. 

{Our main focus in this section will be on infinite dimensional Lie algebras, nonetheless for setting the stage and building an intuitive picture we review the more studied and established case of finite dimensional Lie algebras. In {the} following sections (up to section \ref{sec:3.4}) we assume that the Lie algebra is finite dimensional and in section \ref{sec:3.4} discuss which notions are not extendable to infinite dimensional cases.}

\paragraph{Lie algebra structure and $\mathcal{S}_{n}$ space.}
 We call $(\mathfrak{g},[,])$ a Lie algebra in which $\mathfrak{g}$ is a vector space over a field $\mathbb{F}$ with characteristic zero (for example $\mathbb{R}$ or $\mathbb{C}$) equipped with a Lie bracket, a bilinear and antisymmetric product function $[,]$,
\begin{equation}
    [,]: \mathfrak{g}\times \mathfrak{g}\longrightarrow \mathfrak{g}.
\end{equation}
The Lie bracket $[,]$ must also satisfy the Jacobi identity,
\begin{equation}
    [g_{i},[g_{j},g_{k}]]+[g_{j},[g_{k},g_{i}]]+[g_{k},[g_{i},g_{j}]]=0,\qquad \forall g_{i} \in \mathfrak{g}. \label{GJ}
\end{equation}
We denote the Lie algebra $(\mathfrak{g},[,])$ by $\mathfrak{g}$ and $\{g_{i}\}$ are the basis elements of $n$ dimensional Lie algebra $\mathfrak{g}$ where $i=1,...,n=dim \mathfrak{g}$ which satisfy the Lie bracket $[,]$ as
 \begin{equation}
     [g_{i},g_{j}]=f^{k}_{i,j}g_{k},
 \end{equation}
where $f^{k}_{i,j}=-f^{k}_{j,i}$'s are the $\mathbb{F}$-valued structure constants. The Jacobi identity in terms of structure constants is 
 \begin{equation}
   f^{l}_{i,j}f^{m}_{k,l}+f^{l}_{j,k}f^{m}_{i,l}+f^{l}_{k,i}f^{m}_{j,l}=0,\label{stj} 
 \end{equation}
where summation is over the repeated indices. 
For an $n$ dimensional Lie algebra $\mathfrak{g}$ the maximum number of independent structure constants are $\frac{1}{2}n^{2}(n-1)$. The space of all independent structure constants can be considered as a $\frac{1}{2}n^{2}(n-1)$ dimensional space, $\mathbb{F}^{\frac{1}{2}n^{2}(n-1)}$, each axis is labeled by one independent structure constant $f^{k}_{i,j}$ with values in field $\mathbb{F}$. The space of all $n$ dimensional Lie algebras over the same field, which is specified with the constrains \eqref{stj} and denoted by $\mathcal{S}_{n}$, is an algebraic subset  
of this space \cite{onishchik41lie}. Intuitively each point on the algebraic subset
$\mathcal{S}_{n}$ corresponds to a certain class of Lie algebras isomorphic to $\mathfrak{g}$ with specific structure constants.    

\paragraph{Formal deformation of Lie algebras.}
{A formal one parameter deformation of a Lie algebra ($\mathfrak{g},{[,]}_{0}$), abbreviated as $\mathfrak{g}$, is defined as a skew symmetric bilinear map $\mathfrak{g}\times\mathfrak{g}\rightarrow\mathfrak{g}[[\varepsilon]]$ which satisfies Jacobi identity in all orders of $\varepsilon$. Here $\mathfrak{g}[[\varepsilon]]$ is the space of formal power series in $\varepsilon$ with coefficients in $\mathfrak{g}$ \cite{Figueroa-OFarrill:1989wmj,fox1993introduction}.} This means that the commutation relations of $\mathfrak{g}$ are modified as :
 \begin{equation}\label{eq:defdefor}
     {[g_{i},g_{j}]}_\varepsilon:=\Psi(g_{i},g_{j};\varepsilon)=\Psi(g_{i},g_{j};\varepsilon=0)+{{\psi }_1(g_{i},g_{j})}\varepsilon^{1}+{{\psi }_2(g_{i},g_{j})}\varepsilon^{2}+...,
 \end{equation}
where $ \Psi(g_{i},g_{j};\varepsilon=0)={[g_{i},g_{j}]}_0$, $g_{i},g_{j}$ are basis elements of $\mathfrak{g}$, $\varepsilon \in \mathbb{F}$ (for example $\mathbb{R}$ or $\mathbb{C}$) is the {\it{deformation parameter}} and functions ${\psi }_i:\ \mathfrak{g}\times \mathfrak{g}\longrightarrow \mathfrak{g}$ are bilinear antisymmetric functions, the 2-\textit{cochains}. 
Intuitively, a formal one parameter deformation can be seen as a continuous curve on the algebraic subset $\mathcal{S}_{n}$ parametrized by the deformation parameter $\varepsilon$. A smooth and analytic curve corresponds to a smooth and analytic deformation \cite{onishchik41lie}.  For every $\varepsilon$, the new Lie algebra ($\mathfrak{g},{[,]}_{\varepsilon}$) should satisfy the Jacobi identity,
\be
[g_{i},[g_{j},g_{k}]_{\varepsilon}]_{\varepsilon}+\text{cyclic permutations of}\ (g_{i},g_{j},g_{k}) =0,\label{Jepsilon}
\ee
 order by order in $\varepsilon$, which leads to infinite sequence of equations among $\psi_{i}$.

 For small $\varepsilon$ the leading deformation is given by ${\psi }_{1}(g_{i},g_{j})$-term and the associated Jacobi identity leads to
\be\label{coe}
  {[g_{i},{\psi }_1(g_{j},g_{k})]}_0+{\psi }_1(g_{i},{[g_{j},g_{k}]}_0)+\text{cyclic permutations of}\ (g_{i},g_{j},g_{k}) =0.
\ee
This relation is known as the $2-$cocycle condition. Its solution, the  $2-$cocycle ${\psi }_1$,  specifies an {\it infinitesimal} deformation of Lie algebra $\mathfrak{g}$. The Jacobi identity for higher orders of $\varepsilon$ should also be checked as integrability conditions of ${\psi }_1$ and may lead to obstructions, which will be discussed later in this section. From now on we denote the deformed algebra ($\mathfrak{g},{[,]}_{\varepsilon}$) just by $\mathfrak{g}_{\varepsilon}$. Conversely, given an algebra $\mathfrak{g}$ one can take $\varepsilon \to 0$ limit and obtain $\mathfrak{g}_{0}$.
This procedure is known as {\it{contraction}} of Lie algebras. The contraction and deformation are hence inverse of each other.

Deformation by a $2-$cocycle is trivial if the deformed algebra is isomorphic to the initial algebra, i.e. the deformed and original algebras are related by a redefinition of generators or equivalently a change of basis. If two algebras $\mathfrak{g}$ and $\mathfrak{g}_{\varepsilon}$ are isomorphic to each other, they are related by \begin{equation}
    {[g_{i},g_{j}]}_{\varepsilon}=M_{\varepsilon}{[M^{-1}_\varepsilon (g_{i}),M^{-1}_\varepsilon(g_{j})]}_0,\label{change-basis}
\end{equation}
in which invertible operator $M$ is a linear transformation on the vector space $\mathfrak{g}$. 
The operator $M_\varepsilon$ can be expanded as\footnote{{For finite dimensional Lie algebras, operator $M$ can be considered as a square matrix which can be expanded in terms of unit matrix $I$ and invertible square matrix ${\varphi }_1$.}}
\begin{equation*}
     M_\varepsilon=I+\varepsilon {\varphi }_1.
\end{equation*}
By inserting this relation into \eqref{change-basis} one gets
\begin{equation*}
    {[g_{i},g_{j}]}_{\varepsilon}={(I+\varepsilon {\varphi }_1)[(I-\varepsilon{\varphi }_1)g_{i}, (I-\varepsilon {\varphi }_1)g_{j}]}_0,
\end{equation*}
yielding
\begin{equation*}
    {[g_{i},g_{j}]}_\varepsilon={[g_{i},g_{j}]}_0+\bigg( {{\varphi }_1([g_{i},g_{j}])}_0-{[{\varphi }_1(g_{i}),g_{j}]}_0-{[g_{i},{\varphi }_1(g_{j})]}_0\bigg)\varepsilon+O(\varepsilon^2).
\end{equation*}
Next consider
\begin{equation}
    {\psi }_1(g_{i},g_{j})={{\varphi }_1([g_{i},g_{j}])}_0-{[{\varphi }_1(g_{i}),g_{j}]}_0-{[g_{i},{\varphi }_1(g_{j})]}_0.\label{cob}
\end{equation}
One can readily check that this satisfies the 2-cocycle condition \eqref{coe}. In fact \eqref{cob} shows ${\psi}_1$ is a 2-coboundary if ${\varphi }_1$ is a 1-cochain. When ${\psi }_1$ is a 2-coboundary the deformation \eqref{eq:defdefor} is called trivial, meaning that the deformation is just a redefinition of basis elements.
\subsection{Relation of deformation theory and cohomology of Lie algebra}\label{sec:3.3}

We start with the definition of the Chevalley-Eilenberg complex and differential.
A vector space $\mathbb{V}$ is called a $\mathfrak{g}$-module if {there exists}
a bilinear map $\omega:\ \mathfrak{g}\times \mathbb{V}\longrightarrow \mathbb{V}$  
for all $x\in \mathbb{V}$ and $g_{1},g_{2} \in \mathfrak{g}$ with the property $\omega([g_{1},g_{2}],x)=\omega(g_{1},\omega(g_{2},x))-\omega(g_{2},\omega(g_{1},x))$,  \cite{fuks2012cohomology}. In this setting, the Jacobi identity of the Lie bracket implies that a Lie algebra $\mathfrak{g}$ with the adjoint action is a $\mathfrak{g}$-module. 
{A $p$-cochain $\psi$ is a $\mathbb{V}$-valued (as $\mathfrak{g}$-module),  
 bilinear and completely antisymmetric function which is defined as:
\begin{align*}
    \psi: \underbrace{\mathfrak{g}\times\cdots\times\mathfrak{g}}_{p \, \ times}&\longrightarrow \mathbb{V}\\
    (g_1,\cdots ,g_p)&\longmapsto \psi(g_1,\cdots ,g_p);\,\,\,\,\,\ 0 \leq p \leq dim(\mathfrak{g}).
\end{align*}
Suppose ${\mathcal{C}}^p(\mathfrak{g};\mathbb{V})$ is the space of $\mathbb{V}$-valued $p$-cochains on $\mathfrak{g}$. One can then define the cochain complex ${\mathcal{C}}^*(\mathfrak{g};\mathbb{V})=\oplus^{dim(\mathfrak{g})}_{p=o}{\mathcal{C}}^p(\mathfrak{g};\mathbb{V})$ which is known as the \textbf{\textit{Chevalley-Eilenberg complex}}.

The \textbf{\textit{Chevalley-Eilenberg differential}} or equivalently \textbf{\textit{coboundary operator}} ``$d$'' is a linear map defined as \cite{ChevalleyEilenberg, MR0054581}
\begin{align*}
d_{p}:\ {\mathcal{C}}^{p} (\mathfrak{g} ;\mathfrak{g})&\longrightarrow{\mathcal{C}}^{p+1}\left(\mathfrak{g};\mathfrak{g}\right),\\
     \psi &\longmapsto d_{\mathrm{p}}\psi,
\end{align*}
and $p+1$-cochain $d_{{p}}\psi$ is given by:
   \begin{align}\label{p-cochain}
     (d_{p}\psi )\left(g_0,\dots ,g_{p}\right)\equiv&\sum_{0\leq {i}{<}{j}\leq {p}}{({{1}})^{{i}{+}{j}{-}{1}}} {\psi }\left(\left[{{g}}_{{i}},{{g}}_{{j}}\right],{{g}}_{{0}},{\dots },\widehat{{{g}}_{{i}}},{\dots },\widehat{{{g}}_{{j}}},{\dots },{{g}}_{{p}{+1}}\right)\cr {+}&
    \sum_{{1}\leq {i}\leq {p}{+1}}{({{-}{1}})^{{i}}}\left[{{g}}_{{i}},{\psi }\left({{g}}_{0},{\dots },\widehat{{{g}}_{{i}}}, {\dots },{{g}}_{p}\right)\right],
\end{align}
the hat denotes omission. One can check that $d_{p}\circ d_{{p}{-}{1}}{=0.\ }$
A $p$-cochain ${\psi}$  is called \textbf{\textit{$p$-cocycle}} if  ${d}_p{\psi}=0$, and a \textbf{\textit{$p$-coboundary}} if $\ {\psi}={d }_{p-1}{\varphi}$.
By the property $d_{{p}}\circ d_{{p}{-}{1}}{=0\ }$ one concludes that every $p$-coboundary is also a $p$-cocycle. With this definition one can check that 2-cocycle condition \eqref{coe} is just $d_{2} \psi_{1}=0$ where $\psi$ is a $\mathfrak{g}$-valued $2-$cochain and $d_{2}$  given in \eqref{p-cochain}, and the relation \eqref{cob} is $2-$coboundary condition $\psi_{1}=d_{1}\varphi_{1}$ where $\varphi_{1}$ is a $\mathfrak{g}$-valued $1-$cochain.

{One defines $Z^{p}(\mathfrak{g};\mathbb{V})$ as space of $p$-cocycles which is kernel of differential $d$ as
\begin{equation}
    Z^{p}(\mathfrak{g};\mathbb{V})=\{\psi\in\mathcal{C}^p(\mathfrak{g};\mathbb{V})|d_{p}\psi=0 \}.
\end{equation}
$Z^{2}(\mathfrak{g};\mathfrak{g})$ is hence the space of all $\mathfrak{g}-$valued $2-$cocycles which satisfy the relation \eqref{coe}. 
One also defines $B^{p}(\mathfrak{g};\mathbb{V})$ as space of $p$-coboundaries as
\begin{equation}
    B^{p}(\mathfrak{g};\mathbb{V})=\{\psi\in\mathcal{C}^p(\mathfrak{g};\mathbb{V})|\psi=d_{p-1}\varphi \,\,\, \text{for some}\ \ \varphi\ \text{in}\ \  \mathcal{C}^{p-1}(\mathfrak{g};\mathbb{V}) \}.
\end{equation}
$B^{2}(\mathfrak{g};\mathfrak{g})$ is therefore the space of all $\mathfrak{g}-$valued $2-$cocycles which are also $2-$coboundary which means its elements satisfy both relations \eqref{coe} and \eqref{cob}.  $p^{th}\ cohomology$ space of $\mathfrak{g}$ with coefficients in $\mathbb{V}$ is then defined as quotient of the space of  $p$-cocycles $ Z^{p}(\mathfrak{g};\mathbb{V})$ to the space of  $p$-coboundaries $ B^{p}(\mathfrak{g};\mathbb{V})$ as:
 \begin{equation}
     {\mathcal{H}}^p(\mathfrak{g};\mathbb{V}):=Z^{p}(\mathfrak{g};\mathbb{V})/B^{p}(\mathfrak{g};\mathbb{V})=\text{Ker}({\ d }_p)/\text{Im}({\ d }_{p-1}).
 \end{equation}
 
 It should be noticed that isomorphic Lie algebras has the same cohomology spaces. 
 
 \paragraph{Interpretations of cohomology spaces}
 \begin{itemize}
     \item  ${\mathcal{H}}^{0}(\mathfrak{g};\mathbb{V})$ is defined as: 
  \begin{equation}
     \mathcal{H}^{0}(\mathfrak{g};\mathbb{V})=\text{Inv}_{\mathfrak{g}}(\mathbb{V})=\{x\in \mathbb{V}|g.x=0,\,\, \forall g\in \mathfrak{g}\},\label{H01}
 \end{equation}
and is the space of invariants \cite{Roger:2006rz}.
\item ${\mathcal{H}}^{1}(\mathfrak{g};\mathbb{V})$ can be interpreted as exterior derivations of Lie algebra $\mathfrak{g}$ with values in $\mathbb{V}$. In the case ${\mathcal{H}}^{1}(\mathfrak{g};\mathfrak{g})$ the map $\varphi: \mathfrak{g} \longrightarrow \mathfrak{g}$ is called a derivation if it satisfies the Leibniz rule $\varphi([g_{1},g_{2}])=[\varphi(g_{1}),g_{2}]+[g_{1},\varphi(g_{2})]$.
\item ${\mathcal{H}}^{2}(\mathfrak{g};\mathbb{V})$ classifies deformations or isomorphic classes of extensions of the Lie algebra $\mathfrak{g}$ with respect to $\mathbb{V}$. Two particular cases of interest are 
 \begin{itemize}
     \item[i.]  $\mathbb{V}=\mathbb{R}$ (or $\mathbb{C}$) where ${\mathcal{H}}^{2}(\mathfrak{g};\mathbb{R})$ classifies (global) central extensions of algebra $\mathfrak{g}$. 
      \item[ii.] $\mathbb{V}=\mathfrak{g}$ where ${\mathcal{H}}^{2}(\mathfrak{g};\mathfrak{g})$, the second adjoint cohomology, classifies all infinitesimal deformations of algebra $\mathfrak{g}$.  \end{itemize}
 \end{itemize}
 
Therefore,  to classify infinitesimal deformations of a given Lie algebra $\mathfrak{g}$ one has to compute the second adjoint cohomology $\mathcal{H}^2(\mathfrak{g};\mathfrak{g})$. Not all infinitesimal deformations integrate to a formal (finite) deformation; there could be obstructions.  We will return to the concept of integrability and obstructions later in this section.

\begin{tcolorbox}
\paragraph{Definition \cite{fialowski2012formal}.} \emph{A Lie algebra ($\mathfrak{g},{[,]}_{0}$) is formally rigid, if and only if its every formal deformation is a trivial deformation.} 
\end{tcolorbox}

Intuitively rigidity of Lie algebra $\mathfrak{g}$ means each Lie algebra ${\mathfrak{g}}_{\varepsilon}$ which is close to $\mathfrak{g}$, is isomorphic to it. In the physics literature rigid algebras are also called stable algebras.

\begin{tcolorbox}
 \paragraph{Theorem 3.1.}\label{2.1} \emph{ If $\mathcal{H}^2(\mathfrak{g};\mathfrak{g})=0$ then Lie algebra ($\mathfrak{g},{[,]}_{0}$) is infinitesimally and formally rigid .}
 \end{tcolorbox}

 For more details we refer to \cite{nijenhuis1966cohomology, goze2006lie, richardson1967rigidity}.
In fact, for finite dimensional Lie algebras the condition $\mathcal{H}^2(\mathfrak{g};\mathfrak{g})=0$ means that these Lie algebras are rigid in any sense {e.g. infinitesimally, formally, analytically, \dots } \cite{gerstenhaber1964deformation, gerstenhaber1966deformation, gerstenhaber1968deformation,gerstenhaber1974deformation, nijenhuis1967deformations, fialowski-Mc:2005}. For infinite dimensional Lie algebras, however,  $\mathcal{H}^2(\mathfrak{g};\mathfrak{g})=0$ means that these are just infinitesimally and formally rigid \cite{fialowski1988example}.
That is, the second adjoint cohomology $\mathcal{H}^{2} (\mathfrak{g};\mathfrak{g})$ classifies infinitesimal deformations of Lie algebra $\mathfrak{g}$; if it is zero from the last theorem one concludes that $\mathfrak{g}$ is rigid and does not admit a nontrivial deformation. Otherwise, one has found nontrivial infinitesimal deformations and then verify possible obstructions to make sure that these deformations are also formal deformations.

As an intuitive picture, recall the  $\mathcal{S}_{n}$, the set of all $n$ dimensional Lie algebras which is an algebraic subset in the space of structure constants; each point of this set denotes (an isomorphism class of) a certain Lie algebra. Consider the point $\mathfrak{g}$ with the coordinate $(f^{1}_{1,2},f^{2}_{1,2},...,f^{n}_{n-1,n})$. So the space of all $\mathfrak{g}$-valued $2-$cocycles, $\mathcal{Z}^2(\mathfrak{g};\mathfrak{g})$, is the tangent space to $\mathcal{L}_{n}$ in point $\mathfrak{g}$. The space of all $\mathfrak{g}$-valued $2-$coboundaries, $\mathcal{Z}^2(\mathfrak{g};\mathfrak{g})$, is a subspace of the tangent space on point $\mathfrak{g}$. If $\mathcal{H}^2(\mathfrak{g};\mathfrak{g})=0$, the elements of $\mathcal{Z}^2(\mathfrak{g};\mathfrak{g})$ and $\mathcal{B}^2(\mathfrak{g};\mathfrak{g})$ are the same, so the point or equivalently $\mathfrak{g}$ is rigid. However, if $\mathcal{H}^2(\mathfrak{g};\mathfrak{g})\neq 0$, there is at least one direction in the tangent space which can take the algebra $\mathfrak{g}$ to another Lie algebra $\mathfrak{g}_{\varepsilon}$ which is not isomorphic to $\mathfrak{g}$. In fact, nontrivial infinitesimal deformations which are elements of $\mathcal{H}^2(\mathfrak{g};\mathfrak{g})$ are directions where the algebra $\mathfrak{g}$ along them can be deformed to another algebra $\mathfrak{g}_\eps$. 

\paragraph{Integrability conditions and obstructions.} \label{Integrability}
As pointed out earlier, to have a  formal deformation \eqref{eq:defdefor}, we need the  corresponding nontrivial infinitesimal deformation to be integrable, to be valid to all orders in the deformation parameter. To first few orders in $\varepsilon$, \eqref{Jepsilon}  leads to 
\begin{subequations}\label{infs}
\begin{align}   &[g_{i},[g_{j},g_{k}]_{0}]_{0}+\text{cyclic permutation of}\ (g_{i},g_{j},g_{k})=0,\label{infs-a}\\
    &d_{2}\psi_{1}=0,\label{infs-b}\\
    &d_{2}\psi_{2}=-\frac{1}{2}[\![\psi_{1},\psi_{1}]\!],\label{infs-c}\\
   & d_2\psi_3=-[\![\psi_1,\psi_2]\!],\label{infs-d}
\end{align}
\end{subequations}
where we used the definition of the {Chevalley-Eilenberg differential} $d_{2}$ in \eqref{p-cochain} and the double-bracket is the Nijenhuis and Richardson bracket \cite{nijenhuis1967deformations} defined as 
$$\frac{1}{2}[\![\psi_{r},\psi_{s}]\!](g_{i},g_{j},g_{k}):=\psi_{r}(g_{i},\psi_{s}(g_{j},g_{k}))+\text{cyclic permutation of}\ (g_{i},g_{j},g_{k}).
$$
The zeroth order in $\varepsilon$, \eqref{infs-a}, is nothing but the Jacobi relation for the undeformed algebra and is hence satisfied by definition. The second equation \eqref{infs-b} is the $2-$cocycle condition \eqref{coe} for $\psi_{1}$ and its solutions provides nontrivial \emph{infinitesimal} deformations. Eq.\eqref{infs-c} would then guarantee that there are no obstructions in viewing $\psi_1(g_i,g_j)$ as the first order term of a formal deformation $\Psi(g_i,g_j;\eps)$ which admits a power series expansion in $\eps$. Of course one should continue the reasoning in higher orders of $\eps$. It is readily seen that for $\eps^3$ level one should satisfy \eqref{infs-d} and so on. The sequence of relations will be stopped if there is an obstruction. 

From cohomological point of view, one can verify that all obstructions are in the space ${\mathcal{H}}^{3}(\mathfrak{g};\mathfrak{g})$.  If ${\mathcal{H}}^{3}(\mathfrak{g};\mathfrak{g})=0$ then there are no obstructions \cite{nijenhuis1967deformations}. The latter is a sufficient condition for integrability of ${\mathcal{H}}^{2}(\mathfrak{g};\mathfrak{g})$'s elements. However, the converse is not true and absence of obstructions does not mean ${\mathcal{H}}^{3}(\mathfrak{g};\mathfrak{g})$ is vanishing.  As a result if ${\mathcal{H}}^{2}(\mathfrak{g};\mathfrak{g})\neq 0$ while ${\mathcal{H}}^{3}(\mathfrak{g};\mathfrak{g})=0$ we have a formal  (formal power series) deformation and the deformation parameter $\varepsilon$ need not be taken to be small. 

Given a nontrivial 2-cocycle $\psi_{1}$ or equivalently given ${\mathcal{H}}^{2}(\mathfrak{g};\mathfrak{g})\neq 0$, some different situations may happen:
\begin{itemize}
\item [I.] While there is no obstruction, all the other functions $\psi_{r}$ ($r\geq 2$) have trivial solution $0$. This means that there is a formal deformation which has only linear term in power series. {As we will see except some specific cases, which are examples of case V. below, the deformations we will find are examples of this case}.
\item [II.] There are no obstructions but other functions $\psi_{r}$ ($r\geq 2$) also have nontrivial solutions. This means that there is a formal power series in which $\psi_{r}$'s can be seen as Taylor coefficients of a function $\Psi(g_{i},g_{j};\varepsilon)$.
\item [III.] There is no obstruction but for the set of functions $\psi_{r}$ ($r\geq 2$), \eqref{infs} lead to different solution sets. That is, there are at least two different curves in the $\mathcal{L}_{n}$ space which have the same  first derivative with respect to $\varepsilon$ in the initial point. 
\item [IV.] There is no obstruction up to a specific order $s$. It means that there are only solutions for functions $\psi_{r}$ where $r=2,...,s$ and for $r>s$ equations do not have any solution. We then call $\psi_{1}$ is integrable up to order $s$ \cite{goze2006lie}.
\item [V.] Obstructions start from second order and there is no solution for $\psi_{2}$ and other $\psi_{r}$ where $r\geq2$. As we saw in theorem \ref{2.1} the condition ${\mathcal{H}}^{2}(\mathfrak{g};\mathfrak{g})=0$ is a sufficient condition for the algebra $\mathfrak{g}$ to be rigid. We can also see the meaning of the latter from the relations \eqref{infs}. The condition ${\mathcal{H}}^{2}(\mathfrak{g};\mathfrak{g})=0$ means \eqref{infs-a} does not have any solution. Therefore, other equations which use $\psi_1$ as an input do not have any solution either. However, there are cases  where $\psi_{1}$ exists but not other $\psi_{r}$ where $r\geq 2$. These deformations are not integrable and do not admit a power series as \eqref{eq:defdefor}. An example of such rigid algebra with nontrivial linear (infinitesimal) deformation can be found in \cite{richardson1967rigidity}. {In our analysis of \bms\ or \hbms\ algebra deformations we will find some other examples of this case.}
\end{itemize}

\paragraph{Relation between deformation and contraction.}
As mentioned  the contraction procedure is inverse of deformation. In fact by taking the $\varepsilon\to 0$ limit one can return  to the  original algebra $\mathfrak{g}$ from the  deformed algebra. Physically, deformation can be interpreted as reaching to a ``corrected'' theory from a nonexact one while contraction is getting a certain limit of this corrected theory in which deformation parameter appears as some physical parameter or in some specific examples as  a fundamental constant of nature, e.g. see \cite{mendes1994deformations}. All finite dimensional semisimple Lie algebras are rigid in the sense that they do not admit any deformation so they can be viewed as symmetry algebras of a more fundamental or an undeformed physical theory. On the other hand a non-semisimple Lie algebra (like Galilean algebra) as the symmetry algebra of a physical theory admits deformations (to Poincar\'e algebra, where speed of light $c$ is the deformation parameter and a constant of nature). Conversely, a relativistic theory can be reduced to a non-relativistic theory through a contraction (obtained through sending $c$ to infinity). There are  different approaches to contract a certain Lie algebra. The concept of contraction as limiting process on a Lie algebra was first introduced by Segal in \cite{segal1951class} and then by In\"on\"u and Wigner in \cite{Inonu:1953sp}, which is more known among physicists. Further discussions may be found in  \cite{patera1992graded, weimar1995contractions, gilmore2012lie}.

\subsection{Whitehead's lemma and Hochschild-Serre theorem for finite Lie algebras}

Here we apply theorem \ref{2.1} to finite dimensional Lie algebras. First, we consider simple and semi-simple Lie algebras. In this case a theorem which has been proven by Whitehead has a very important role

\begin{tcolorbox}
  \paragraph{Lemma 3.1.}\label{2.2} ({\it Whitehead's lemma}) {\it Let $M$ be a (finite-dimensional) simple module over a finite dimensional semisimple
Lie algebra $\mathfrak{g}$ with nontrivial $\mathfrak{g}$-action. Then $\mathcal{H}^p(\mathfrak{g};M)=0$ for all $p \geq 0$.} 
\end{tcolorbox}

 As $\mathfrak{g}$ is also a $\mathfrak{g}$-module, the above theorem states that $\mathcal{H}^p(\mathfrak{g};\mathfrak{g})=0$ for all $p \geq 0$. This lemma and theorem \ref{2.1} lead to,
 
 \begin{tcolorbox}
    \paragraph{Corollary 3.1.}\label{2.3}  {\it All  semisimple finite dimensional Lie algebras are rigid.} 
 \end{tcolorbox}

 As examples all $\mathfrak{so}(n)\,\,n\geq 3$, $\mathfrak{su}(n)$ and $\mathfrak{sl}(n,\mathbb{R}\,\,\text{or}\,\,\mathbb{C})$ are rigid in the sense that they cannot be deformed to another non-isomorphic algebra. 
 
 \paragraph{Semi-direct sum structure.} There are also some powerful mathematical methods for specific finite dimensional  Lie algebra structures. Many of finite dimensional Lie algebras which appear in physics have semi-direct sum structure in the sense that $\mathfrak{g}$ is a semi-direct sum of two (or possibly more) parts as $\mathfrak{g}=\mathfrak{g}_{1}\inplus\mathfrak{g}_{2}$ in which $\mathfrak{g}_{2}$ is ideal part of $\mathfrak{g}$, e.g. Poincar\'{e} algebra in general $d$ dimensions has this form (actually any non-solvable finite dimensional Lie algebra over a field of characteristic zero can be written as a semi direct sum of a semisimple Lie algebra and its radical ideal. This is known as the Levi decomposition). Let us look at the commutation relations of Poincar\'{e} Lie algebra in $3d$,
  \begin{equation} 
\begin{split}
 & i[\mathcal{J}_m,\mathcal{J}_n]=(m-n)\mathcal{J}_{m+n}, \\
 &i[\mathcal{J}_m,\mathcal{P}_n]=(m-n)\mathcal{P}_{m+n},\\
 &i[\mathcal{P}_m,\mathcal{P}_n]=0,
\end{split}\label{poincare-3D}
\end{equation}
where $m,n=0,\pm1$. This algebra is a semi-direct sum of two parts as: Poincar\'{e}$_{3}=\iso= \sltwo\inplus_{ad}\mathfrak{h}$ in which $\sltwo$ is spanned by $\mathcal{J}$'s and $\mathfrak{h}$ is ideal part which is spanned by $\mathcal{P}$'s.
For algebras with semi-direct sum structure, there is the classic {\it Hochschild-Serre factorization theorem}  proven in \cite{MR0054581} which makes the calculations more convenient. 

\begin{tcolorbox}
  \paragraph{Theorem 3.2.}({\it Hochschild-Serre factorization theorem})\label{2.4} {\it Let $\mathfrak{g}$ be a finite dimensional Lie algebra over the field $\mathbb{F}$ of characteristic 0, and let $M$ be a finite dimensional $\mathfrak{g}$-module. Suppose that $\mathfrak{h}$ is ideal
 of $\mathfrak{g}$ such that $\mathfrak{g}/\mathfrak{h}$ is semi-simple. Then
 $\mathcal{H}^p(\mathfrak{g};M)\cong \sum_{r+s=p}\mathcal{H}^r(\mathfrak{g}/\mathfrak{h};\mathbb{F})\otimes \mathcal{H}^s(\mathfrak{h};M)$.}
 \end{tcolorbox}

As we mentioned $\mathfrak{g}$ is a $\mathfrak{g}$-module and the characteristic of real numbers $\mathbb{R}$ (and also complex number $\mathbb{C}$) is zero. Then, Lemma \ref{2.2} and Theorem \ref{2.4} imply that
 \begin{equation}
     \mathcal{H}^2(\mathfrak{g};\mathfrak{g})\cong  \mathcal{H}^2(\mathfrak{h};\mathfrak{g}).
 \end{equation}
This means that all nontrivial infinitesimal deformations of $\mathfrak{g}$ are located in the ideal part of $\mathfrak{g}$. In other words, it is not necessary to analyze the deformation of the full algebra, the deformation of the ideal part already yields the most general deformations of $\mathfrak{g}$. In the case of  the Poincar\'{e} Lie algebra in $3d$ the ideal part is the third line in \eqref{poincare-3D}. According to the {\it Hochschild-Serre factorization theorem} this algebra is deformed to  
\begin{equation} 
\begin{split}
 & i[\mathcal{J}_m,\mathcal{J}_n]=(m-n)\mathcal{J}_{m+n}, \\
 &i[\mathcal{J}_m,\mathcal{P}_n]=(m-n)\mathcal{P}_{m+n},\\
 &i[\mathcal{P}_m,\mathcal{P}_n]=\sigma^2 (m-n)\mathcal{J}_{m+n},\qquad m,n=0,\pm 1,\ \sigma^2=1,
\end{split}\label{satbilized-3d-Poincare}
\end{equation}
which is the $\mathfrak{so}(2,2)$ algebra, isometry of AdS$_3$ for $\sigma=+1$ or the $\mathfrak{so}(3,1)$ algebra, isometry group of dS$_3$, for $\sigma=-1$. See \cite{Figueroa-OFarrill:2017sfs, Figueroa-OFarrill:2017ycu, Figueroa-OFarrill:2017tcy} for more discussions and examples.

\subsection{Hochschild-Serre spectral sequence}\label{sec:HS-seq}

Although in the case of infinite dimensional Lie algebras the Hochschild-Serre factorization theorem does not work, one can still use Hochschild-Serre spectral sequence method {which works for both finite and infinite cases} and extract information about deformations from that. We focus on certain split abelian extensions of Lie algebras. For a Lie algebra $(\mathfrak{g},[,])$ {with a semi-direct sum structure as $\mathfrak{g}=\mathfrak{g}_{0}\inplus \mathfrak{h}$ where $\mathfrak{h}$ is an abelian ideal and $\mathfrak{g}_{0}\cong \mathfrak{g}/\mathfrak{h}$ is its quotient Lie algebra}, we have the following short exact sequence
\begin{equation}
    0\longrightarrow \mathfrak{h}\longrightarrow \mathfrak{g} \longrightarrow \mathfrak{g}/\mathfrak{h}\cong \mathfrak{g}_{0}\longrightarrow 0,\label{short-exact}
\end{equation}
where arrows show {Lie algebra morphisms. Short exactness means the image of each morphism is equal to the kernel of the next  . For this sequence one obtains the Hochschild-Serre spectral sequence of cochain complexes whose  first terms are
    \begin{equation*} 
\begin{split}\label{chain-complex}
 E_{0}^{p,q}=\mathcal{C}^{q}(\mathfrak{g}_{0},\mathcal{C}^{p}(\mathfrak{h},M)),\ E_{1}^{p,q}=\mathcal{H}^{q}(\mathfrak{g}_{0};\mathcal{C}^{p}(\mathfrak{h},M)),\ 
   E_{2}^{p,q}=\mathcal{H}^{p}(\mathfrak{g}_{0};\mathcal{H}^{q}(\mathfrak{h};M)),\ ...,\ E_{n}^{p,q},... 
\end{split}
\end{equation*}
in which $M$ is a $\mathfrak{g}$-module, $\mathcal{C}^{p}$ as we introduced earlier is the space of $p$-cochains and $E$'s are related to each other by the differential operator $d_{n}^{p,q}:E_{n}^{p,q}\longrightarrow E_{n}^{p+n,q-n+1}$ \cite{MR0054581,fuks2012cohomology}. In some specific cases the differential function becomes trivial for $n\geq n_0$ (for certain $n_0$) and $E_{n}^{p,q}, \forall n\geq n_0$ are isomorphic to each other and therefore, $E_{n}^{p,q}\cong E_{\infty}^{p,q}$. So for the latter we have \footnote{Note that,  in general, this equality is true modulo extensions but all the terms in our cases are vector spaces and hence those extensions do not appear.}
\begin{equation}
    \mathcal{H}^{n}(\mathfrak{g};M)=\oplus_{p+q=n}E_{\infty}^{p,q}.\label{dec}
\end{equation}
{In this setting by the Hochschild-Serre spectral theorem \cite{MR0054581} we have} 
\begin{equation}
E_{2}^{p,q}=
    \mathcal{H}^{p}(\mathfrak{g}_{0}; \mathcal{H}^{q}(\mathfrak{h},M)).\label{E2-dec}
\end{equation}
 This theorem works for both finite and  infinite  dimensional Lie algebras.
{For those split abelian extensions with the property that the ideal action on $M$ is trivial, Theorem 1.2 in \cite{degrijse2009cohomology} states that 
we always have} $n_{0}=2$ and {therefore} $E_{2}^{p,q}\cong E_{\infty}^{p,q}$. So, combining \eqref{dec} and \eqref{E2-dec} one finds 
\begin{equation}
    \mathcal{H}^{2}(\mathfrak{g};M)=\oplus_{p+q=2}E_{2}^{p,q}.\label{dec2}
\end{equation}

{Note that $\mathfrak{h}$ is an ideal of $\mathfrak{g}$ and hence a $\mathfrak{g}$-module and because it is abelian, as a $\mathfrak{g}$-module its action on itself is trivial. Using the short exact sequence \eqref{short-exact} we consider 
 $\mathfrak{g}_{0}$ as a $\mathfrak{g}$-module as well. In this way the action of $\mathfrak{h}$ on $\mathfrak{g}_{0}$ is trivial. We conclude that via the above arguments, $\mathfrak{g}_{0}$ and $\mathfrak{h}$ are both $\mathfrak{g}$-modules satisfying conditions of theorem 1.2 in \cite{degrijse2009cohomology}, and one can compute the spaces $\mathcal{H}^{2}(\mathfrak{g};\mathfrak{g}_{0})$ and $\mathcal{H}^{2}(\mathfrak{g};\mathfrak{h})$.} 
 
{The short exact sequence \eqref{short-exact} induces the long exact sequence at the level of cohomologies 
\begin{equation}
    \begin{split}
    & \cdots \longrightarrow \mathcal{H}^{1}(\mathfrak{g};\mathfrak{g}_{0})\longrightarrow \mathcal{H}^{2}(\mathfrak{g};\mathfrak{h}) \longrightarrow \mathcal{H}^{2}(\mathfrak{g};\mathfrak{g})\longrightarrow \mathcal{H}^{2}(\mathfrak{g};\mathfrak{g}_{0})
     \longrightarrow \mathcal{H}^{3}(\mathfrak{g};\mathfrak{g}_{0}) \longrightarrow \cdots \label{long-exact}
    \end{split}
\end{equation}
One may use the above sequence to get information about  $\mathcal{H}^{2}(\mathfrak{g};\mathfrak{g})$ or even compute it. The long exact sequence \eqref{long-exact} is true for both finite and infinite dimensional Lie algebras with the semi-direct sum structure. In finite dimensional cases as a consequence of Hochschild-Serre factorization theorem we have $ \mathcal{H}^{2}(\mathfrak{g};\mathfrak{g})\cong \mathcal{H}^{2}(\mathfrak{h};\mathfrak{g})$. 

 In the case of infinite dimensional Lie algebras we can still use \eqref{long-exact}. While the sequence has in general infinite terms, in some specific cases one finds that some of the terms in \eqref{long-exact} are equal to zero, leading to another  short exact sequence. In such situations we can infer some information about lower cohomologies. As we will discuss in the next sections, from \eqref{dec2} and \eqref{long-exact} we can learn about cohomological structure of asymptotic symmetry algebras introduced in chapter 2.

\section{Deformations and rigidity of infinite dimensional Lie algebras}\label{sec:3.4}

{So far we considered deformation theory of Lie algebras mainly for finite dimensional Lie algebras. One can readily extend the main results and notions discussed earlier, except for the the Hochschild-Serre factorization theorem, to infinite dimensional Lie algebras. Here we will be interested in a specific class of infinite dimensional algebras with countable bases. Such algebras, like \wit\ which is the Lie algebra of smooth vector fields on circle $S^{1}$ and its universal central extension, Virasoro algebra, have very crucial role in {the} context of $2d$ quantum field theory and string theory.}

Rigidity and deformation analysis of infinite dimensional Lie algebras are more complicated than finite dimensional Lie algebras. For instance {in spite of existence of semi-direct} sum structures for certain infinite dimensional algebras, like $\mathfrak{bms}_3$, we cannot use the Hochschild-Serre factorization theorem.  Rigidity of Witt and Virasoro algebras has been considered in \cite{fialowski1990deformations, guerrini1999formal, fialowski2003global,  fialowski2012formal, schlichenmaier2014elementary} where it is shown that Witt algebra and its universal central extension are formally rigid. Later Fialowski and Schlichenmaier showed that the Witt algebra is not globally rigid and global deformation of Witt and Virasoro algebra lead to a specific family of algebras  known as Krichever-Novikov type \cite{fialowski2003global}. Fialowski  {has} also worked on formal deformations of some other infinite dimensional Lie algebras such as vector fields on the line \cite{fialowski1988example}.   
 
\emph{Affine Kac-Moody} Lie algebras are another class of infinite dimensional Lie algebras of interest in theoretical physics. Roger in \cite{hazewinkel2012deformation} has given a theorem stating that affine Kac-Moody Lie algebras are (formally) rigid in the sense that their second adjoint cohomology is zero. 
 
\emph{Schr\"{o}dinger-Virasoro} algebras are also infinite dimensional algebras which appear in nonequilibrium statistical physics \cite{Christe:1993ij, Henkel:2012zz}. Deformations of these families have been studied by Unterberger and Roger. They have found three nontrivial formal deformations for a specific kind of Schr\"{o}dinger-Virasoro algebra known as twisted Schr\"{o}dinger-Virasoro algebra. We refer the reader to \cite{unterberger2011schrodinger} for further analysis of Schr\"{o}dinger-Virasoro algebras and their deformations.    
 
 The concept of contraction has also been considered for some infinite dimensional algebras. For instance, contractions of Kac-Moody and affine Kac-Moody Lie algebras {have} been studied in \cite{majumdar1993inonu, daboul2008gradings}. The second example which was discussed earlier in section \ref{sec:2} is contraction of asymptotic symmetry algebras of (A)dS$_3$ spacetime in $3d$ to asymptotic symmetry algebra of flat spacetime in the same dimension. 
 
 \section{Summary of the chapter}
 In this chapter we reviewed concepts of contraction, deformation and stabilization of Lie algebras. 
 We  introduced deformation procedure by using the notion of cohomology of Lie algebras. We mentioned Whitehead's lemma and Hochschild-Serre factorization theorem which are crucial to compute deformations of finite dimensional Lie algebras. We exemplified deformations and contractions of some important physical finite algebras like Poincar\'{e}. 
We then showed that deformation of infinite dimensional Lie algebras are not subject to Hochschild-Serre factorization theorem so we should compute it explicitly. Finally, we introduced some Mathematical tools such as Hochschild-Serre long exact sequence which may help us to find deformation of infinite dimensional Lie algebras.



\chapter{Deformation of \texorpdfstring{$3d$}{3d} asymptotic symmetry algebras} \label{ch4}
Here we study infinitesimal and formal deformations of \bms\ and \kac\ algebras and classify them as shown in fig. \ref{Fig-abnu}. We then present our results in the cohomology language. Next we analyze deformations of central extensions of these algebras \hbms\ and \kach. Finally, we explore possible deformations of algebras we obtain as a result of these deformations and introduce the notion of rigidity of family of algebras. The computation details may be found in appendix \ref{appendix-B} and the summary of our results in this chapter are presented in Tables \hyperlink{table1}{4.1} and \hyperlink{table1}{4.2}.

\section{Deformation of \texorpdfstring{$\mathfrak{bms}_{3}$}{bms3} algebra}\label{sec:4.1}
In this section we consider deformations of  $\mathfrak{bms}_{3}$ defined in \eqref{bms3}. As discussed the Hochschild-Serre factorization theorem is not applicable for infinite dimensional Lie algebras and working with them is more complicated than finite dimensional cases. Here, we first analyze possible deformations of $\mathfrak{bms}_{3}$ algebra by deforming each commutation relation of the $\mathfrak{bms}_{3}$ algebra separately. One should then make sure that in this way we do not miss any possible deformation which may involve two or more sets of commutators. This will be discussed in section \ref{sec:4.5}. Finally, we study obstructions, which infinitesimal deformations yield formal deformations and which rigid algebras are obtained from deformations of $\mathfrak{bms}_{3}$. In subsection \ref{sec:cohomology-bms} we establish and reinforce our results of the previous subsections through the algebraic cohomology analysis. We have summarized all possible formal deformations of the \bms\ and $\mathfrak{KM}_{\mathfrak{u}(1)}$ algebras in fig. \ref{Fig-abnu}.

\subsection*{Deformation of commutators of two \texorpdfstring{$\mathcal{P}$'s}{PP} }

For the first step we construct  all deformations of the ideal part of $\mathfrak{bms}_{3}$. As we can see from \eqref{bms3} the ideal part of $\mathfrak{bms}_{3}$ algebra is spanned by supertranslation generators $\mathcal{P}$'s. We deform infinitesimally this ideal part as
\begin{equation} 
i[\mathcal{P}_{m},\mathcal{P}_{n}]=\varepsilon\psi_{1}^{PP}(\mathcal{P}_{m},\mathcal{P}_{n}),\label{deform of ideal}
\end{equation}
where $\varepsilon$ is deformation parameter and  $\psi_{1}^{PP}(\mathcal{P}_{m},\mathcal{P}_{n})$ is a $\mathfrak{bms}_{3}$-valued 2-cocycle and hence admits the expansion as
\begin{equation}\label{psi1-PP}
\psi_{1}^{PP}(\mathcal{P}_{m},\mathcal{P}_{n})=F(m,n)\mathcal{J}_{u(m,n)}+G(m,n)\mathcal{P}_{v(m,n)},
\end{equation}
in which coefficients $F(m,n)$ and $G(m,n)$ are antisymmetric functions while indices $u(m,n)$ and $v(m,n)$ are symmetric functions. For convenience, we extract the antisymmetric part of $F(m,n)$ (or  $G(m,n)$) as
\begin{equation} 
F(m,n)=(m-n)f(m,n),\qquad f(m,n)=f(n,m).
\end{equation}

The main goal is to find the explicit form of $f, g, u$ and $v$. One may show that the general form of functions $u(m,n)$ and $v(m,n)$ can be chosen as $u(m,n)=v(m,n)=m+n$.
To this end, we recall \eqref{bms3=witt+ideal} and that $\mathfrak{witt}$ is related to the adjoint action of vectors on two-tensors on an S$^1$.  Consider tensor densities on circle  $\mathcal{M}(\phi)(d\phi)^{\lambda}$ in which $\lambda\in \mathbb{Z}$ is called the degree of tensor density and $\mathcal{M}(\phi)$ is a periodic function on circle (e.g. see \cite{Oblak:2016eij} and references therein). In this way $\mathcal{J}(\phi)(d\phi)^{-1}$ is the vector field on circle. So one finds the Fourier expansion of $\mathcal{M}(\phi)(d\phi)^{\lambda}$ as
\begin{equation}
    \mathcal{M}(\phi)(d\phi)^{\lambda}=\sum M_{n}e^{in\phi}(d\phi)^{\lambda},\label{Fourier}
\end{equation}
where $e^{in\phi}(d\phi)^{\lambda}$ are the basis. In the terminology of 2d CFT's, if $\mathcal{J}$ is associated with conformal transformations, then $\lambda$ is the conformal weight of operator ${\cal{M}}$. As an example the basis of Witt algebra are $e^{in\phi}\partial_{\phi}=\mathcal{J}_{n}$. The adjoint action of Witt generators on tensor density $\mathcal{M}(\phi)$ is obtained as
$adj_{\mathcal{J}(\phi_1)}\mathcal{M}(\phi_2)=[\mathcal{J}(\phi_1),\mathcal{M}(\phi_2)]=\big(\mathcal{M}^{'}(\phi_2)\mathcal{J}(\phi_1)+\lambda \mathcal{M}(\phi_2)\mathcal{J}^{'}(\phi_1)\big)\delta(\phi_1-\phi_2)$ \cite{Oblak:2016eij}. One can then write the above in terms of its Fourier basis and see that the final result has the form 
$i[\mathcal{J}_{m},\mathcal{M}_{n}]=(\lambda m-n)\mathcal{M}_{m+n}$. That is, dealing with periodic tensors on a circle $u(m,n), v(m,n)$ appearing in \eqref{psi1-PP} should be $m+n$.

\paragraph{Remark.} By the above explanation, from now on we just consider deformations which are induced by generators with the fixed index as $m+n$. 

To determine functions $f, g$ one has to consider the Jacobi identities (the 2-cocycle conditions \eqref{coe} or equivalently \eqref{infs-b}). The only Jacobi identities which can lead to some relations for $f, g$, are $[\mathcal{P}_{m},[\mathcal{P}_{n},\mathcal{P}_{l}]]+\text{cyclic permutations}=0$ and  $[\mathcal{P}_{m},[\mathcal{P}_{n},\mathcal{J}_{l}]]+\text{cyclic permutations}=0$. The former just contains a relation with linear terms in $\varepsilon$ for $f(m,n)$, yielding
\begin{equation}
(n-l)(m-n-l)f(n,l)+ (l-m)(n-l-m)f(l,m)+ (m-n)(l-m-n)f(m,n)=0, \label{f-first}
\end{equation}
while from the second, one gets
\begin{equation}\label{fg-first}
\begin{split}
&(n-l)(m-n-l)f(m,l+n)+ (l-m)(n-l-m)f(n,l+m)+\\
&(m-n)(l-m-n)f(m,n)=0,  \\ 
&(n-l)(m-n-l)g(m,l+n)+ (l-m)(n-l-m)g(n,l+m)+\\
&(m-n)(l-m-n)g(m,n) =0.
\end{split}
\end{equation}
As we will show below the equations for $f$ in \eqref{fg-first} and \eqref{f-first} {are the same} and $f,g$ should satisfy the same equation. So, we only focus on $f$ and on \eqref{fg-first}.

We now tackle \eqref{fg-first}. Our goal is to find the most general form of $f(m,n), g(m,n)$. 
As the first step we will prove a very important proposition.
\paragraph{Proposition.}
The most general solution of \eqref{fg-first} is
  \begin{equation}
f(m,n)=\text{constant}.
 \end{equation}
We prove this proposition by proving some lemmas.
  \paragraph{Lemma 1.}
 For $m\neq n$
 \begin{equation}
 f(m,n)=f(m,n+m).
 \end{equation}
  \begin{proof}
 One puts $l=m$ into the relation \eqref{fg-first} and finds:
 \begin{equation}
 (n-m)f(m,n+m)+(m-n)f(m,n)=0,
 \end{equation}
 then choosing $m\neq n$ one finds :
 \begin{equation}
   f(m,n)=f(m,n+m).  
 \end{equation}
 \end{proof}
 
 \paragraph{Lemma 2.}
 For $n\neq 0$
 \begin{equation}
 f(0,n)=f(0,1).
 \end{equation}
  \begin{proof}
 By insertion of $l=1, m=0$ into the relation \eqref{fg-first} one finds:
  \begin{equation}
 (-n-1)f(0,n+1)+f(n,1)+(n)f(0,n)=0.
   \end{equation}
 From lemma 1 one can show that $f(1,n)=f(1,0)$ (by $n$ times using lemma 1). So we have:
  \begin{equation}
 (-n-1)f(0,n+1)+f(0,1)+(n)f(0,n)=0.
 \end{equation}\label{f-induction}
We  prove the lemma by induction. For $n=1$ the statement holds $f(0,1)=f(0,1)$. We suppose for $n=k$ the statement holds which means $f(0,1)=f(0,k)$. We use the latter assumption and \eqref{f-induction} to reach
 \begin{equation}
(k+1)f(0,k+1)=(k+1)f(0,1)\rightarrow f(0,k+1)=f(0,1).
\end{equation}
 \end{proof}
  \paragraph{Lemma 3.} For $m\neq n$ we have
  \begin{equation}
 f(m,n)=f(0,n).
 \end{equation}
 \begin{proof}
By inserting $l=-m$ in \eqref{fg-first} we get:
\begin{equation}
(n+m)(2m-n)f(m,n-m)+(-2m)(n)f(n,0)+(m-n)(-2m-n)f(m,n)=0.\label{lemma 3}
\end{equation}
From lemma 1 we infer that $f(m,n)=f(m,n-m)$. By putting the latter into \ref{lemma 3} we get
\begin{equation}
 f(m,n)=f(n,0).
\end{equation}
\end{proof}
Combining results of lemmas 2 and 3 we conclude $f(m,n)=f(0,1)=\text{constant}=f$ and proof of the proposition is completed. It is immediate to see that this solution also solves \eqref{f-first}. So, by solving the $2-$cocycle condition  we have found that $\psi_{1}^{PP}(\mathcal{P}_{m},\mathcal{P}_{n})=f(m-n)\mathcal{J}_{m}+g(m-n)\mathcal{P}_{m}$ in which $f,g$ are some arbitrary (complex or real) constants, i.e.
\begin{equation}\label{mostgeneral-ideal-bms3}
 i[\mathcal{P}_{m},\mathcal{P}_{n}]=\varepsilon_{1}(m-n)\mathcal{J}_{m+n}+\varepsilon_{2}(m-n) \mathcal{P}_{m+n},
\end{equation}
where $\varepsilon_{1}=\varepsilon f$ ($\varepsilon_{2}=\varepsilon g$). 
We should note that special $\varepsilon_1=0$ or $\varepsilon_2=0$ cases in \eqref{mostgeneral-ideal-bms3}, corresponding to
$i[\mathcal{P}_{m},\mathcal{P}_{n}]=\varepsilon_2(m-n)\mathcal{P}_{m+n}$ or $i[\mathcal{P}_{m},\mathcal{P}_{n}]=\varepsilon_1(m-n)\mathcal{J}_{m+n}$ are both isomorphic to the general case of \eqref{mostgeneral-ideal-bms3}. This means that  by a proper redefinition of the generators they can be mapped onto each other and the algebra becomes the direct sum of two Witt algebras as $\mathfrak{witt}_{L}\oplus \mathfrak{witt}_{R}$.
So we have the theorem
\begin{tcolorbox}[colback=red!3!white]
\paragraph{Theorem 4.1}{\it The most general infinitesimal deformation of $\mathfrak{bms}_{3}$ ideal part is 
 \begin{equation} 
i[\mathcal{P}_{m},\mathcal{P}_{n}]=\varepsilon_{2}(m-n) \mathcal{P}_{m+n}.\label{theorem 4.1}
\end{equation}}
\end{tcolorbox}

 Since the cases $\varepsilon_{1}=0$,  $\varepsilon_{2}=0$ and \eqref{mostgeneral-ideal-bms3} are isomorphic, without any loss of generality only consider
 $i[\mathcal{P}_{m},\mathcal{P}_{n}]=\varepsilon_1(m-n)\mathcal{J}_{m+n}$. 
$|\varepsilon_1|$ may be absorbed into the normalization of ${\cal P}_n$ but its sign will remain: $i[\mathcal{P}_{m},\mathcal{P}_{n}]=\sigma (m-n)\mathcal{J}_{m+n}, \sigma^2=1$. Therefore, there are two choices for this deformation. These two choices parallels the two choices for the stabilization of $\mathfrak{iso}(2,1)$ into $\mathfrak{so}(2,2)$ and $\mathfrak{so}(3,1)$ discussed in \eqref{satbilized-3d-Poincare}.

So far we have found a nontrivial infinitesimal deformation of $\mathfrak{bms}_{3}$ ideal part, showing that the $\mathfrak{bms}_{3}$ algebra is not, at least, ``infinitesimally rigid'' and $\mathcal{H}^2(\mathfrak{bms}_{3};\mathfrak{bms}_{3})\neq 0$. One can show that in an appropriate basis the deformed algebra is just the direct sum of two Witt algebras $\mathfrak{witt}_{L}\oplus \mathfrak{witt}_{R}$ and so not isomorphic to $\mathfrak{bms}_{3}$. This is of course expected  from contraction procedure which was introduced in section \ref{sec:2.3-contraction}. Here, we have shown there is a unique deformation in $\mathfrak{bms}_{3}$ ideal part corresponding to that contraction. 
To establish that $\mathfrak{bms}_{3}$ is not also formally rigid one should check the integrability conditions or probably obstructions for this infinitesimal deformation, which will be discussed in section \ref{sec:integrability-bms}.

\subsection*{ Deformation of commutators of \texorpdfstring{$[\mathcal{J},\mathcal{P}]$}{JP} }

Now, we consider  deformations of commutator of superrotations and supertranslations which is the second line in \eqref{bms3} without changing other commutators. To this end and as in the previous subsection, we add a $2-$cocycle function:
\begin{equation} 
 i[\mathcal{J}_{m},\mathcal{P}_{n}]=(m-n)\mathcal{P}_{m+n}+\zeta \psi_{1}^{JP}(\mathcal{J}_{m},\mathcal{P}_{n}).\label{JP-deformation}
\end{equation}
The 2-cocycle $\psi_{1}^{JP}(\mathcal{J}_{m},\mathcal{P}_{n})$ is a linear combination of generators as
\begin{equation}\label{eqdefIK}
\psi_{1}^{JP}(\mathcal{J}_{m},\mathcal{P}_{n})=I(m,n)\mathcal{J}_{m+n}+K(m,n)\mathcal{P}_{m+n},
\end{equation}
where we have fixed the indices of $\mathcal{J}$ and $\mathcal{P}$ to be $m+n$ (\emph{cf.}  discussion around \eqref{Fourier}) and
the coefficients $I(m,n)$ and $K(m,n)$ are arbitrary functions.

To find the explicit form of functions $I(m,n)$ and $K(m,n)$ we check Jacobi identities. Two different Jacobi identities put constraints on $I(m,n)$ and $K(m,n)$. The first Jacobi identity is  $[\mathcal{P}_{m},[\mathcal{P}_{n},\mathcal{J}_{l}]]+\text{cyclic permutations}=0$. Keeping up to first order in $\zeta$ and using the fact that $[\mathcal{P}_{m},\mathcal{P}_{n}]=0$, we get
\begin{equation} 
 -(m-n-l)I(l,n)+(n-l-m)I(l,m)=0.\label{firsteqI}
\end{equation}
 This relation is exactly the same as $2-$cocycle condition \eqref{coe}. It is obvious that this specific Jacobi does not put any constraint on $K(m,n)$. By some easy steps we show $I(m,n)=0$. Let us put $m=l+n$ into \eqref{firsteqI} to reach
\begin{equation} 
 (2l)I(l,l+n)=0,
\end{equation}
and for $l\neq 0$, one gets
\begin{equation} 
 I(l,q)=0,
\end{equation}\label{result1}
 where $q=l+n$. By insertion of $l=0$ into \eqref{firsteqI} we get
 \begin{equation} 
 I(0,n)=-I(0,m),\label{result2}
\end{equation}
for $m\neq n$. Suppose $I(0,n)=F(n)$ so from \eqref{result2} we have $F(n)=-F(m)$ which means that $I(0,n)=0 ,\,\,\,\text{for}\,\,\,\, n\neq 0$. Finally one concludes that
\begin{equation}
    I(m,n)=0 ,\,\,\,\forall\,\, m,n\in \mathbb{Z}.
\end{equation}

We next examine the Jacobi identity $[\mathcal{J}_{m},[\mathcal{J}_{n},\mathcal{P}_{l}]]+\text{cyclic permutations}=0$ and up to first order in $\zeta$ we obtain,
\begin{multline}
(n - l) K(m, l + n) + ( m-n-l) K(n, l) + (l-m) K(
   n, l+m) +\\  +
   (l+m-n) K(m,l)+(n-m) K(m+n,l)=0.\label{eq-K3}
\end{multline}
One puts $l=m=0$ into \eqref{eq-K3} to get
\begin{equation} 
 n (K(0, n) - K(0,0))=0.
\end{equation}
The above then yields 
\begin{equation} 
  K(0, n)=\text{constant}.\label{first contraint on K-3}
\end{equation}
To solve \eqref{eq-K3} we note that it is linear in $K$ and hence linear combination of any two solutions is also a solution. Since the coefficients of the $K$'s is first order in $l,m$ or $n$, the solutions should be homogeneous functions of a given degree $N$, i.e. $K(m,n)=\sum_{r=1}^N A_r m^r n^{N-r}$ where we already used \eqref{first contraint on K-3}. For $N=0,1$ one may readily check that $K(m,n)=\alpha+\beta m $ in which $\alpha,\beta$ are arbitrary real constants. So, let us focus on $N\geq 2$ case. For $N=2$ one may show that there is a solution of the form
\begin{equation} 
 K(m,n)=\gamma m(m-n),\label{Ksolution}
\end{equation}
where $\gamma$ is an arbitrary constant. This term, however, is not representing a nontrivial deformation and can be absorbed into normalization of ${\cal P}$. To see this consider redefining ${\cal P}$ as 
\begin{equation}\label{rescale-P}
   \mathcal{P}_{n}:= N(n)\tilde{\mathcal{P}}_{n},
\end{equation}
for a function $N$ we are free to choose. Replacing this into \eqref{bms3} one gets
\begin{equation} 
\begin{split}
 & i[\mathcal{J}_{m},\mathcal{J}_{n}]=(m-n)\mathcal{J}_{m+n}, \\
 &i[\mathcal{J}_{m},\tilde{\mathcal{P}}_{n}]=(m-n)\frac{N(m+n)}{N(n)}\tilde{\mathcal{P}}_{m+n},\\
 &i[\tilde{\mathcal{P}}_{m},\tilde{\mathcal{P}}_{n}]=0.
\end{split}\label{Q-normalization}
\end{equation}
If we choose $N$ as
\begin{equation*}
   N(n)=1+\zeta\gamma n+\mathcal{O}(\zeta^2),
\end{equation*}
then the $\gamma$ term can be absorbed into redefinition of generators. 
\paragraph{Higher $N$ and uniqueness of solutions of \eqref{eq-K3}.} An explicit verification with straightforward algebra for generic $N$ shows that there is no other solution.

The most general deformation of \bms\ through deformation of $[\mathcal{J},\mathcal{P}]$ commutator is hence
\begin{equation} 
\begin{split}
 & i[\mathcal{J}_{m},\mathcal{J}_{n}]=(m-n)\mathcal{J}_{m+n}, \\
 &i[\mathcal{J}_{m},\mathcal{P}_{n}]=-(n+bm+a)\mathcal{P}_{m+n},\\
 &i[\mathcal{P}_{m},\mathcal{P}_{n}]=0,
\end{split}\label{W(a,b)}
\end{equation}
where $a,b \in \mathbb{R}$ are two independent deformation parameters. The Lie algebra \eqref{W(a,b)} is known as $W(a,b)$ algebra and some of its properties such as its central extensions has been analyzed in  
\cite{Roger:2006rz}.\footnote{Since  the Witt algebra is rigid any extension of the Witt algebra by its abelian representations is a semi-direct sum $\mathfrak{witt}\inplus \mathfrak{witt}_{ab}$ which is the case for $\mathfrak{bms}_{3}$. This also indicates that we cannot expect exotic extensions of the Witt algebra. Furthermore, in deformations of $[\mathcal{J},\mathcal{P}]$ commutator only elements from the abelain part can appear.
Despite the similarity between the \bms\ case and the kinematical algebras, e.g. see \cite{Figueroa-OFarrill:2017sfs, Figueroa-OFarrill:2017ycu, Figueroa-OFarrill:2017tcy}, this result already shows the difference between the infinite and finite dimensional algebras and that the former does not obey Hochschild-Serre factorization theorem.}  

\paragraph{More on $W(a,b)$ algebras.}
As we mentioned before, ${W}(a,b)$ algebra which is an extension of the Witt algebra, has been studied in different papers \cite{gao2011low,ovsienko1996extensions}. It has a semi-direct sum structure as $W(a,b)\cong \mathfrak{witt}\inplus \mathcal{P}(a,b)$ in which $\mathcal{P}(a,b)$, with $a, b$ being arbitrary real constants, is called tensor density module. Here we briefly discuss two interesting questions about these algebras: 1. What is the physical interpretations of parameter $a,b$? 2.  Are there  special points in $(a,b)$ parameter space? 

To answer the first question we rewrite the algebra as 
\begin{equation} 
\begin{split}
 & i[\mathcal{J}_{m},\mathcal{J}_{n}]=(m-n)\mathcal{J}_{m+n}, \\
 &i[\mathcal{J}_{m},\mathcal{P}_{n}]=-\left(n+(1-h)m+a\right)\mathcal{P}_{m+n},\\
 &i[\mathcal{P}_{m},\mathcal{P}_{n}]=0,\label{cwh}
\end{split}
\end{equation}
where $h=1-b$. These commutation relations are very familiar in the context of $2d$ conformal field theories. Consider a primary field ${\cal P}(\phi)$ of conformal weight $h$, with Fourier modes ${\cal P}_n$ and assume that $P(\phi)$ has the quasi-periodicity property
\be\label{Fourier-expand-a}
{\cal P}(\phi+2\pi)=e^{2\pi i a} {\cal P}(\phi),\qquad {\cal P}(\phi)=\sum_n {\cal P}_n e^{i(n+a)\phi}.
\ee
The usual conformal transformation for this conformal primary field can then be recast as $[{\cal J},{\cal P}]$ commutator in \eqref{cwh}. Here we are assuming that ${\cal P}_n$'s satisfy a commuting (abelian) algebra . The above argument makes it clear that the range of $a$ parameter which yields independent algebras is $a\in[-1/2,1/2]$. However, one may show that the negative and positive values of $a$ parameter are related by a $\mathbb{Z}_2$ parity transformation which takes $\phi\to 2\pi-\phi$, i.e. a $\mathbb{Z}_2$ inner automorphism of the $W(a,b)$ algebra: ${\cal J}_n\to -{\cal J}_{-n},\ {\cal P}_n\to -{\cal P}_{-n}$. Therefore, the independent range for $a$ parameter is $[0,1/2]$, as depicted in fig. \ref{Fig-abnu}.   

The deformation which moves us along the $b$ direction in the parameter space, given the physical interpretation above, can be understood as an RG flow in the presumed $2d$ CFT dual to the $3d$ flat space (which realizes \bms \ as its symmetry algebra). A flow from $b=-1$ $(h=2)$ to $b=0$ $(h=1)$ is what is expected from comparing the asymptotic symmetry and the ``near horizon'' symmetry analysis. We shall comment on this latter further in the discussion section.


As for the second question, the definition of $W(a,b)$ algebra implies that \bms$=W(0,-1)$ and that the Virasoro-Kac-Moody algebra is $W(0,0)$. These two points are special in the sense that one can deform the algebra in two different ways while in a generic point of $a,b$ space allowed deformations move us only within the $W(a,b)$ family.\footnote{By $W(a,b)$ we  mean a specific algebra for a given $a,b$ parameters and by the ``$W(a,b)$ family,''  all algebras for generic but different values of $a,b$.\label{footnote3}} As discussed, one can deform the ideal part of \bms\ to obtain $\mathfrak{witt}\oplus \mathfrak{witt}$, while one can also keep the ideal part intact and change $(a,b)$ from $(0,-1)$ value. As for the Virasoro-Kac-Moody case, one can check that one cannot deform the ideal part. However, besides deforming the $[{\cal J},{\cal P}]$ part and moving in $(a,b)$ plane, one is also allowed to deform the algebra in the $[{\cal J},{\cal J}]$ part (by adding a $(m-n) {\cal P}_{m+n}$ 2-cocyle). We have depicted the parameter space of $W(a,b)$ algebras in fig. \ref{Fig-abnu}, which shows as far as the infinitesimal deformations are concerned, there are no other special points.\footnote{We note that when the central extensions of $W(a,b)$ are concerned and as discussed in \cite{gao2011low}, for generic $(a,b)$ there is always one central extension in the Witt part of the algebra. There are however, other special points in $a,b$ space with more possibilities for central extensions. Points $(0,-1)$, $(0,0)$, $(0,1)$ and $(1/2,0)$ are special in the sense that they admit other central extensions \cite{gao2011low}.}

\subsection* {Deformation of commutators of two \texorpdfstring{$\mathcal{J}$'s}{JJ} }\label{sec:4.3}

We finally consider deformations of the $[\mathcal{J},\mathcal{J}]$ part of $\mathfrak{bms}_{3}$. As mentioned the Witt algebra is rigid in the sense that it cannot be deformed by a $\mathcal{J}$-valued 2-cocycle. We can however add a $\mathcal{P}$-valued 2-cocycle as
\begin{equation}
i[\mathcal{J}_{m},\mathcal{J}_{n}]=(m-n)\mathcal{J}_{m+n}+\eta\psi_{1}^{JJ}(J_{m},J_{n})=
(m-n)\mathcal{J}_{m+n}+\eta(m-n)h(m,n)\mathcal{P}_{m+n}\label{psi-JJ}
\end{equation} 
in which $\psi_{JJ}(J_{m},J_{n})$ is a 2-cocycle and $h(m,n)$ is a symmetric function. We then insert \eqref{psi-JJ} into the Jacobi identity $[\mathcal{J}_{m},[\mathcal{J}_{n},\mathcal{J}_{l}]]+\text{cyclic permutations}=0$ which in first order in $\varepsilon$ yields
\begin{multline} 
(n-l)(m-n-l)[h(m,l+n)+h(n,l)]+ (l-m)(n-l-m)[h(n,l+m)+h(l,m)]+\\
 (m-n)(l-m-n)[h(l,m+n)+h(m,n)]=0.\label{h-eq}
\end{multline}
One can then readily check that any $h$ of the form
\be\label{JJ-h-Z}
h(m,n)=Z(m)+Z(n)-Z(m+n),
\ee
for any arbitrary $\mathbb{R}$-valued function $Z$, provides a solution to \eqref{h-eq}. One may argue that this is in fact the most general solution. To this end, we note that \eqref{h-eq} is linear in $h$ and has quadratic coefficients in $l,m$ or $n$. So, a generic solution for $h(m,n)$ is expected to be a polynomial of homogeneous degree $N$:
\begin{equation*}
    h(m,n)=\sum_{r=0}^N A_{r} m^r n^{N-r},\qquad A_r=A_{N-r}.\label{power series}
\end{equation*}
Subtracting the general solution \eqref{JJ-h-Z}, without loss of generality, one can use the ansatz $h(m,n)=mn \sum_{r=1}^{N-1} A_r m^r n^{N-r}$. Plugging this into \eqref{h-eq} one can see that the equation is only satisfied for $A_r=0$.  

One can, however, show deformations of the form  \eqref{JJ-h-Z} are trivial deformations, as they can be reabsorbed into the redefinition of generators:
\begin{equation} 
\begin{split}
 &  \mathcal{J}_{m}:=\tilde{\mathcal{J}}_{m}+Z(m)\tilde{\mathcal{P}}_{m}, \\
 &\mathcal{P}_{m}:=\tilde{\mathcal{P}}_{m},
\end{split}\label{Z-redefinition}
\end{equation}
where $\tilde{\mathcal{J}}_{m}$ and $\tilde{\mathcal{P}}_{m}$ satisfy  $\mathfrak{bms}_{3}$ commutation relations \eqref{bms3}.

\paragraph{Virasoro-Kac-Moody algebra, ${\mathfrak{KM}}_{\mathfrak{u}(1)}$; an example of deformation of $[\mathcal{J},\mathcal{J}]$ part.}
The $\mathfrak{u}(1)$ Kac-Moody algebra, ${\mathfrak{KM}}_{\mathfrak{u}(1)}$, is defined through 
\begin{equation} 
\begin{split}
 & i[\mathcal{J}_{m},\mathcal{J}_{n}]=(m-n)\mathcal{J}_{m+n}, \\
 &i[\mathcal{J}_{m},\mathcal{P}_{n}]=-n\mathcal{P}_{m+n},\\
 &i[\mathcal{P}_{m},\mathcal{P}_{n}]=0.
\end{split}\label{Kac-Moody}
\end{equation}
One may verify that the $[\mathcal{J},\mathcal{J}]$ part can be deformed by the term $\nu (m-n)\mathcal{P}_{m+n}$. One can then check that the latter is infinitesimal and also formal deformation. This deformation has been discussed in \cite{Roger:2006rz}. In the next subsections, we discuss the most general deformations of Virasoro-Kac-Moody algebra.


\subsection{ Integrability conditions and obstructions}\label{sec:integrability-bms}

So far we found all infinitesimal deformations of $\mathfrak{bms}_{3}$ which consist of three independent classes: two of them deform the $[\mathcal{J},\mathcal{P}]$ part of the algebra while keeping the ideal part $[\mathcal{P},\mathcal{P}]=0$. This leads to $W(a,b)$ algebra. The last case deforms the ideal part and leads to $\mathfrak{witt}\oplus \mathfrak{witt}$. Here we study integrability conditions of these infinitesimal deformations.    

There are three different approaches to check integrability conditions: 
\begin{itemize}
\item[1.] Direct method: One can consider the entire infinite sequence of relations \eqref{infs} and directly verify their solutions or probable obstructions. 
\item[2.] ${\cal H}^3$ method: We mentioned that all obstructions are located in $\mathcal{H}^{3}(\mathfrak{bms}_{3};\mathfrak{bms}_{3})$. It can be computed e.g. using the Hochschild-Serre spectral sequence, \emph{cf}. subsection \ref{sec:HS-seq}. If $\mathcal{H}^{3}(\mathfrak{bms}_{3};\mathfrak{bms}_{3})$ vanishes there is no obstruction. For further discussions we refer the reader to subsection \ref{Integrability}.
\item[3.] ``A quick test'': One may examine if an infinitesimal deformation is a formal one by promoting the linear (infinitesimal) deformation $\psi_{1}(g_{i},g_{j})$ to $\Psi(g_{i},g_{j};\varepsilon)$ and check whether it satisfies the Jacobi identity or not. If one finds that the linear term in the Taylor expansion of $\Psi(g_{i},g_{j};\varepsilon)$ satisfies the Jacobi \eqref{GJ} one concludes it is also a formal deformation of algebra. This is basically checking whether we are dealing with case \texttt{I.} discussed in subsection \ref{Integrability}.
\end{itemize}
Since in most of the known cases we are dealing with case \texttt{I} of subsection \ref{Integrability} and since it is very convenient and handy to check this, we will focus on ``quick test'' and we will comment on the cohomology considerations  in section \ref{sec:cohomology-bms}.

Let us start with the deformation of ideal part of $\mathfrak{bms}_{3}$ which is deformed by $2-$cocycle
$\psi^{PP}_{1}=(m-n)\mathcal{P}_{m+n}$. As discussed in the end of first part of section \ref{sec:4.1}, one may ignore the $(m-n)\mathcal{J}_{m+n}$ deformation as it does not lead to a nonisomorphic infinitesimal deformation. Two Jacobi identities put constraints on the form of the deformation function $\Psi(g_{i},g_{j};\varepsilon)$. The first Jacobi is  
\begin{equation*}
    [\mathcal{P}_{m},[\mathcal{P}_{n},\mathcal{P}_{l}]]+\text{cyclic permutations}=0,
\end{equation*}
which in terms of $\Psi(\mathcal{P}_{m},\mathcal{P}_{n};\varepsilon)=(m-n)\tilde{g}(m,n;\varepsilon)\mathcal{P}_{m+n}$ leads to
\begin{equation}
  \begin{split}
       &(n-l)(m-n-l)\tilde{g}(m,n+l;\varepsilon)\tilde{g}(n,l;\varepsilon)+(l-m)(n-l-m)\tilde{g}(n,l+m;\varepsilon)\tilde{g}(l,m;\varepsilon)\\
      & +(l-n-m)(m-n)\tilde{g}(m,n;\varepsilon)\tilde{g}(l,m+n;\varepsilon)=0. 
  \end{split}
\end{equation}
One can check that the solution of linear term $\psi^{PP}_{1}(\mathcal{P}_{m},\mathcal{P}_{n})=g(m-n)\mathcal{P}_{m+n}$ ($g$ is an arbitrary constant) satisfies the above equation. The second Jacobi $[\mathcal{P}_{m},[\mathcal{P}_{n},\mathcal{J}_{l}]]+\text{cyclic permutations}$\\$=0$ leads to the constraint 
\begin{multline}
       (n-l)(m-n-l)\tilde{g}(m,l+n;\varepsilon)+(l-m)(n-l-m)\tilde{g}(n,m+l;\varepsilon)\\
       +(l-n-m)(m-n)\tilde{g}(m,n;\varepsilon)=0, 
\end{multline}   
which is exactly the same as \eqref{fg-first}, so it has the same solution $\tilde{g}(m,n;\varepsilon)=\text{constant}=g$. Therefore,  $\psi^{PP}_{1}(\mathcal{P}_{m},\mathcal{P}_{n})=g(m-n)\mathcal{P}_{m+n}$ is also a formal deformation and moreover, as argued in previous subsection, this deformation is unique. 

The same procedure may be repeated for linear (infinitesimal) deformation $\psi^{JP}_{1}(\mathcal{J}_{m},\mathcal{P}_{n})=(\alpha + \beta m){\cal P}_{m+n}$. The only Jacobi to consider is  $[\mathcal{J}_{m},[\mathcal{J}_{n},\mathcal{P}_{l}]]+\text{cyclic permutations}=0$ which for the formal deformation function $\Psi(\mathcal{J}_{m},\mathcal{P}_{n};\varepsilon)=\tilde{K}(m,n;\varepsilon)\mathcal{P}_{m+n}$ leads to,
\begin{equation}\label{eq-Ktild}
(m-l) X(l+m,n;\varepsilon) - X(l,n;\varepsilon) X(m, n+l;\varepsilon)+ X(l,m+n;\varepsilon) X(m,n;\varepsilon)=0,
\end{equation}
where $X(m,n;\varepsilon)=(m-n)+\tilde{K}(m, n;\varepsilon)$. One can check that 
\begin{equation}
\tilde{K}(m,n;\varepsilon)=\alpha +\beta m,\label{K-tilde-generic}
\end{equation}
provides a solution to \eqref{eq-Ktild}, and that, if $X(m,n)$ is a solution to \eqref{eq-Ktild}, so as 
$$
Y(m,n)=X(m,n)\frac{N(m+n)}{N(n)},
$$
for an arbitrary function $N(n)$. We had argued in the previous subsection that the unique solution to the linearized equation (up to the rescaling by $N(n)$) is of the form \eqref{K-tilde-generic}. Therefore, \eqref{K-tilde-generic} is also a unique all orders solution to  \eqref{eq-Ktild}, if we require solutions to be smoothly connected to the linear solution. 
In summary,  the infinitesimal deformation $\psi^{JP}_{1}(\mathcal{J}_{m},\mathcal{P}_{n})=(\alpha + \beta m){\cal P}_{m+n}$ is also a formal deformation. In this way we have found that all three infinitesimal deformations are integrable and there is no obstruction. 

\subsection{Classification of \texorpdfstring{$\mathfrak{bms}_3$}{BMS3} and 
\texorpdfstring{$\mathfrak{KM}_{\mathfrak{u}(1)}$}{KMu(1)} deformations}\label{sec:4.5}

In section \ref{sec:4.1},  we classified all ``infinitesimal'' deformations of the $\mathfrak{bms}_3$ algebra, by deforming each of $[{\cal P}, {\cal P}]$, $[{\cal P}, {\cal J}]$ and $[{\cal J}, {\cal J}]$ separately. This led to only two nontrivial deformations. Then in section \ref{sec:integrability-bms} we showed that
both of these two cases are integrable and one can obtain two classes of ``formal'' deformations, the $W(a,b)$ algebra and the $\mathfrak{witt}\oplus \mathfrak{witt}$ algebra.

One may wonder if in this process we could have missed cases which involve simultaneous deformations  of the three classes of the commutators of the algebra. With the experience gained in the previous analysis here we tackle this question. The most general possible deformation of the $\mathfrak{bms}_3$ algebra takes the form:
\begin{equation} 
\begin{split}
 & i[\mathcal{J}_{m},\mathcal{J}_{n}]=(m-n)\mathcal{J}_{m+n}+ (m-n)h(m,n)\mathcal{P}_{m+n}, \\
 &i[\mathcal{J}_{m},\mathcal{P}_{n}]=(m-n)\mathcal{P}_{m+n}+ K(m,n)\mathcal{P}_{m+n}+ I(m,n) {\cal J}_{m+n},\\
 &i[\mathcal{P}_{m},\mathcal{P}_{n}]=(m-n)f(m,n)\mathcal{J}_{m+n}+(m-n)g(m,n)\mathcal{P}_{m+n},\label{general-deformation}
\end{split}
\end{equation}
in which we absorbed the expansion deformation parameters into the functions. 
The Jacobi identities in the first order in functions for $f(m,n)$, $K(m,n)$ and $h(m,n)$ yield exactly equations as we found in the previous subsections. For $I(m,n)$ one finds a relation similar to \eqref{eq-K3} and for $g(m,n)$ we find
\begin{multline}\label{g-I-mixed}
(n-l)(m-n-l)g(m,l+n)+ (l-m)(n-l-m)g(n,l+m)+(m-n)(l-m-n)g(m,n) \\ +\left[(n+l-m)I(l,n)- (m+l-n) I(l,m)\right]=0.
\end{multline}
The separate equation we have for $I(m,n), g(m,n)$, which yields $I(m,n)=\tilde{\alpha}+\tilde{\beta}m$ for some constants $\tilde{\alpha},\tilde{\beta}$. We note the above equation has the following structure: it is linear in $I, g$ and the $I$ terms come with coefficients first order in $m,n,l$ while the $g$ terms come with second order coefficients. Since $g, I$ are expected to be polynomials of positive powers in their arguments, if $I$ is a monomial of degree $p$, $g$ should be a monomial of degree $p+1$. On the other hand we know $p$ is either 0 or 1 and we may check these two cases directly in  \eqref{g-I-mixed} and verify that it does not have any solution expect for $I(m,n)=0$. Therefore, $g(m,n)$, also, satisfies the same equation as \eqref{fg-first} and hence our method of deforming each commutator separately, to linear order, captures all possible infinitesimal deformations.

One may discuss integrability of the most general deformation \eqref{general-deformation}. Since $I(m,n)=0$ at linear order, it will remain zero to all orders. If we for instance turn on $g(m,n)$ and $K(m,n)$ simultaneously, we observe that although there can be an infinitesimal deformation, higher order Jacobi equations for this deformation are not integrable. The same feature happens when we consider the deformation induced by $f(m,n)$ and $K(m,n)$; the Jacobi allows one to make simultaneous  infinitesimal deformations in  the $[\mathcal{J},\mathcal{P}]$  part by $K(m,n)$ and ideal part by $f(m,n)$ and $g(m,n)$. One should hence analyze integrability of these infinitesimal deformations. The answer, as mentioned above, is that the only integrable deformations  are constructed from the  infinitesimal deformations  induced by $K(m,n)$ or $g(m,n)$ and $f(m,n)$. One then realizes that although the simultaneous deformation induced by  $K(m,n)$ and $g(m,n)$ is an infinitesimal deformation, it does not integrate to a formal deformation. There is a similar structure when we have simultaneous infinitesimal deformation induced by  $K(m,n)$ and $f(m,n)$, which is isomorphic to the infinitesimal deformation induced by $K(m,n)$ and $g(m,n)$, and is not integrable.To summarize, all infinitesimal deformations of $\mathfrak{bms}_{3}$ are those induced by $K(m,n)$ and $g(m,n)$.

In fact the latter can be considered as an example of the case V in subsection \ref{Integrability} where we have an infinitesimal deformation which is not integrable in higher order analysis ($r\geq 0$).}
We can summarize the above analysis in the following  theorem: 
\begin{tcolorbox}[colback=red!3!white]
\paragraph{Theorem 4.2} {\it The most general formal deformations of \bms\ are either $\mathfrak{witt} \oplus \mathfrak{witt}$ or $W(a,b)$ algebras}. 
\end{tcolorbox}
We note that, as discussed earlier, the $\mathfrak{witt} \oplus \mathfrak{witt}$ has two options: the subalgebra generated by ${\cal P}_{0,\pm1},\ {\cal J}_{0,\pm 1}$ can be either $\mathfrak{so}(2,2)$ or $\mathfrak{so}(3,1)$.

\paragraph{Formal deformations of Virasoro-Kac-Moody algebra ($\mathfrak{KM}_{\mathfrak{u}(1)}$).}
In this part we would like to discuss the most general infinitesimal and formal deformations of Virasoro-Kac-Moody algebra $\mathfrak{KM}_{\mathfrak{u}(1)}$ \eqref{Kac-Moody}. The most general infinitesimal deformation of $\mathfrak{KM}_{\mathfrak{u}(1)}$ is of the same form as \eqref{general-deformation} (but with $m-n$ in the first term on the right-hand-side of the second line  is replaced by $-n$). The Jacobi identities then yield $f(m,n)=0$, $g(m,n)=constant=g$, $I(m,n)=gm$, $K(m,n)=\alpha+\beta m$ and $h(m,n)=constant=h$. One can then show that the infinitesimal deformation induced by $g(m,n)$ and $I(m,n)$ can be absorbed by a proper redefinition of generators and hence 
these are just three different nontrivial infinitesimal deformations of $\mathfrak{KM}_{\mathfrak{u}(1)}$. (None of the other combinations of the above leads to a nontrivial deformation.) One can then also show that the three parameter family of infinitesimal deformations have no obstructions and can be integrated into a formal deformation. These results can be summarized in the following theorem:
\begin{tcolorbox}[colback=red!3!white]
\paragraph{Theorem 4.3 } {\it The most general formal deformations of $\mathfrak{KM}_{\mathfrak{u}(1)}$ algebra are either $W(a,b)$ algebras or the $\mathfrak{KM}(a,\nu)$ algebra defined as}
\begin{equation} 
\begin{split}
 & i[\mathcal{J}_{m},\mathcal{J}_{n}]=(m-n)(\mathcal{J}_{m+n}+\nu \mathcal{P}_{m+n}), \\
 &i[\mathcal{J}_{m},\mathcal{P}_{n}]=-(n+a)\mathcal{P}_{m+n},\\
 &i[\mathcal{P}_{m},\mathcal{P}_{n}]=0.
\end{split}\label{Kac-Moody-nu}
\end{equation}
\end{tcolorbox}
For more details we refer to \cite{Roger:2006rz}.

\begin{figure}[hbt!]
\centering
		\includegraphics[width=0.9\textwidth]{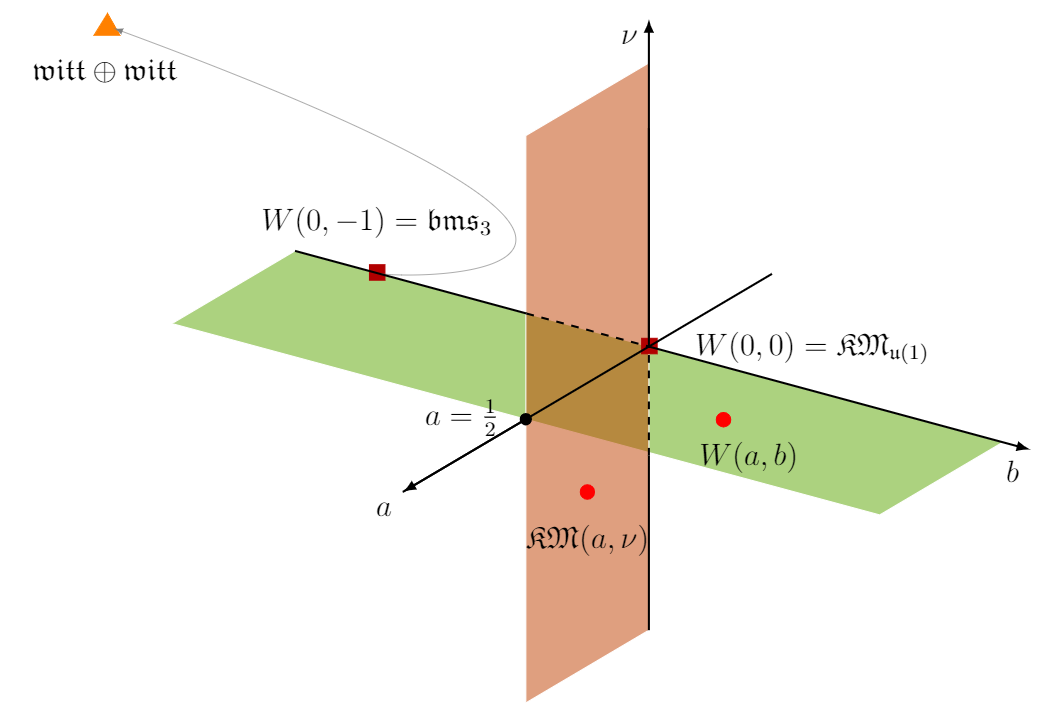}
		\caption{\small Parametric space of formal deformations of $\mathfrak{bms}_{3}$ and $\mathfrak{KM}_{\mathfrak{u}(1)}$. $W(a,b)$ algebra appears as the formal deformation of both $\mathfrak{bms}_{3}$ and $\mathfrak{KM}_{\mathfrak{u}(1)}$. As discussed below \eqref{Fourier-expand-a} parameter $a$ ranges in $[0,1/2]$ while $b,\nu $ can be any real number. Besides the $W(a,b)$,  $\mathfrak{KM}_{\mathfrak{u}(1)}$
can be deformed to  $\mathfrak{KM}(a,\nu)$ which is spanned by $b=0$ plane. The
\bms\ algebra  can be  deformed  to $\mathfrak{witt}\oplus \mathfrak{witt}$ which again is out of the $(a,b,\nu)$ space displayed above. The original version of this figure is presented in \cite{Parsa:2018kys}.}
\label{Fig-abnu}
			\end{figure}
\subsection{Algebraic cohomology considerations}\label{sec:cohomology-bms}
In the previous subsections we classified all nontrivial infinitesimal and formal deformations of the \bms\ algebra by an explicit checking of Jacobi identity and then verifying nontriviality of the deformation by checking if it can be removed by a change of basis. As discussed in section \ref{sec:3.3}, a similar question can be addressed and analyzed by algebraic cohomology considerations through computing $\mathcal{H}^{2}(\mathfrak{bms}_{3};\mathfrak{bms}_{3})$. The main tools to study the cohomology are the Hochschild-Serre spectral sequence \eqref{E2-dec} and the long and short exact sequences \eqref{long-exact} discussed in subsection \ref{sec:HS-seq}. By the former we obtain  information  independently about $\mathcal{H}^{2}(\mathfrak{bms}_{3};\mathfrak{witt})$ and $\mathcal{H}^{2}(\mathfrak{bms}_{3};\mathcal{P})$ where $\mathcal{P}$ and $\mathfrak{witt}$ respectively denote the ideal part and the Witt subalgebra of  $\mathfrak{bms}_{3}$. Note that since $\mathfrak{witt}$ is not a \bms\ module by the adjoint action, $\mathcal{H}^{2}(\mathfrak{bms}_{3};\mathfrak{witt})$ is defined by the action used in the short exact sequence \eqref{short-exact-bms3} below, as discussed in subsection \ref{sec:HS-seq}. Note also that given the semi-direct sum structure of the \bms\ algebra from \eqref{bms3=witt+ideal} we should not expect 
$\mathcal{H}^{2}(\mathfrak{bms}_{3};\mathfrak{bms}_{3})=\mathcal{H}^{2}(\mathfrak{bms}_{3};\mathfrak{witt}) \oplus \mathcal{H}^{2}(\mathfrak{bms}_{3};\mathcal{P})$. 

One can however get information about $\mathcal{H}^{2}(\mathfrak{bms}_{3};\mathfrak{bms}_{3})$ from the long exact sequence \eqref{long-exact} as one can show that some terms therein  are zero and hence it reduces to a short exact sequence or an isomorphism. In fact as it has been shown in \cite{gao2011low} in the case of $W(a,b)$ algebras, especially for $\mathfrak{bms}_{3}$ which is $W(0,-1)$, $\mathcal{H}^{1}(W(a,b);\mathfrak{witt})$ is equal to zero. One can, however, show that $\mathcal{H}^{2}(\mathfrak{bms}_{3};\mathfrak{witt})$ is not zero and therefore, the long exact sequence does not give useful information about $\mathcal{H}^{2}(\mathfrak{bms}_{3};\mathfrak{bms}_{3})$. Nonetheless, as we will see, direct analysis of $\mathcal{H}^{2}(\mathfrak{bms}_{3};\mathfrak{witt})$ and $ \mathcal{H}^{2}(\mathfrak{bms}_{3};\mathcal{P})$ reveals the structure of $\mathcal{H}^{2}(\mathfrak{bms}_{3};\mathfrak{bms}_{3})$.    
To this end, we follow the Hochschild-Serre spectral sequence method (\emph{cf}. subsection \ref{sec:HS-seq}) and consider the following short exact sequence of $\mathfrak{bms}_{3}$  
\begin{equation}
    0\longrightarrow \mathcal{P}\longrightarrow \mathfrak{bms}_{3} \longrightarrow \mathfrak{bms}_{3}/\mathcal{P}\cong \mathfrak{witt}\longrightarrow 0.\label{short-exact-bms3}
\end{equation}

\paragraph{Computation of  $\mathcal{H}^{2}(\mathfrak{bms}_{3};\mathcal{P})$.}
From \eqref{dec2} and \eqref{E2-dec} we have 
\begin{equation}
\begin{split}
    \mathcal{H}^{2}(\mathfrak{bms}_{3};\mathcal{P})&=\oplus_{p+q=2}E_{2;\mathcal{P}}^{p,q}=E_{2;\mathcal{P}}^{2,0}\oplus E_{2;\mathcal{P}}^{1,1}\oplus E_{2;\mathcal{P}}^{0,2}\\
    &=\mathcal{H}^{2}(\mathfrak{witt};\mathcal{H}^{0}(\mathcal{P};\mathcal{P}))\oplus \mathcal{H}^{1}(\mathfrak{witt};\mathcal{H}^{1}(\mathcal{P};\mathcal{P}))\oplus \mathcal{H}^{0}(\mathfrak{witt};\mathcal{H}^{2}(\mathcal{P};\mathcal{P})),\label{bmsp}
\end{split}
\end{equation}
where the subscript $\mathcal{P}$ in $E_{2;\mathcal{P}}^{p,q}$ denotes we are computing $\mathcal{H}^{2}(\mathfrak{bms}_{3};\mathcal{P})$. We compute the three terms above separately. 
$\mathcal{H}^{2}(\mathfrak{witt};\mathcal{H}^{0}(\mathcal{P};\mathcal{P}))$ contains $\mathcal{H}^{0}(\mathcal{P};\mathcal{P})$ which by the definition \eqref{H01} and the fact that the action of $\mathcal{P}$ on $\mathcal{P}$ is trivial, one concludes that $\mathcal{H}^{0}(\mathcal{P};\mathcal{P})=\mathcal{P}$ then $\mathcal{H}^{2}(\mathfrak{witt};\mathcal{H}^{0}(\mathcal{P};\mathcal{P}))=\mathcal{H}^{2}(\mathfrak{witt};\mathcal{P})$.
 On the other hand, as it is obvious from commutators of $\mathfrak{bms}_{3}$, the adjoint action of its $\mathfrak{witt}$ subalgebra on itself is exactly the same as adjoint action of its $\mathfrak{witt}$ subalgebra on the ideal part $\mathcal{P}$. This can be seen in the first and second lines in \eqref{bms3}. For this reason one concludes that $\mathfrak{witt}\cong \mathcal{P}$ as {$\mathfrak{witt}$-modules}. So one gets $\mathcal{H}^{2}(\mathfrak{witt};\mathcal{P})\cong \mathcal{H}^{2}(\mathfrak{witt};\mathfrak{witt})$, and since the Witt algebra is rigid, i.e. $\mathcal{H}^{2}(\mathfrak{witt};\mathfrak{witt})=0$, we conclude that $\mathcal{H}^{2}(\mathfrak{witt};\mathcal{P})=0$.
 Therefore the first commutator in \eqref{bms3} remains intact by the deformation procedure.

Next we  analyze $\mathcal{H}^{1}(\mathfrak{witt};\mathcal{H}^{1}(\mathcal{P};\mathcal{P}))$. It  is constructed by $1-$cocycle $\varphi_{1}$ which is defined as a function $ \varphi_{1}: \mathfrak{witt}\longrightarrow \mathcal{H}^{1}(\mathcal{P};\mathcal{P})$. So
the expression of $\varphi_{1}(\mathcal{J}_{m})(\mathcal{P}_{n})$ can be expanded in terms of $\mathcal{P}$'s as $\varphi_{1}(\mathcal{J}_{m})(\mathcal{P}_{n})=\tilde{K}(m,n)\mathcal{P}_{m+n},\label{Cjpp}$
where $\tilde{K}(m,n)$ is an arbitrary function. The deformation of $[\mathcal{J},\mathcal{P}]$ part corresponding to $\varphi_{1}$ is $i[\mathcal{J}_m,\mathcal{P}_n]=(m-n)\mathcal{P}_{m+n}+\tilde{K}(m,n){\mathcal{P}}_{m+n}$.
The Jacobi identity for the above bracket  restricts  $\tilde{K}(m,n)$ exactly as $K(m,n)$ in \eqref{eqdefIK}, so one finds the same result as $\tilde{K}(m,n)=\alpha +\beta m+ \gamma m(m-n)$. As we mentioned earlier the $\gamma$ term can be absorbed in choosing a proper normalization factor.

The last term we need to consider is  $\mathcal{H}^{0}(\mathfrak{witt};\mathcal{H}^{2}(\mathcal{P};\mathcal{P}))$. We use the definition of $\mathcal{H}^{0}$ 
\begin{equation}
  \mathcal{H}^{0}(\mathfrak{witt};\mathcal{H}^{2}(\mathcal{P};\mathcal{P}))=\{ \psi\in \mathcal{H}^{2}(\mathcal{P};\mathcal{P})| \mathcal{J}\circ \psi=0,\,\,\, \forall \mathcal{J}\in \mathfrak{witt} \}.\label{H0P}
\end{equation}
 where $\psi$ is a $\mathcal{P}$-valued  $2-$cocycle. 
 The action ``$\circ$'' of $\mathcal{J}$ on a 2-cocycle $\psi$ is defined as \cite{MR0054581}
\begin{equation}
  (\mathcal{J}_{l}\circ \psi)(\mathcal{P}_{m},\mathcal{P}_{n})=[\mathcal{J}_{l},\psi(\mathcal{P}_{m},\mathcal{P}_{n})]-\psi([\mathcal{J}_{l},\mathcal{P}_{m}],\mathcal{P}_{n})-\psi(\mathcal{P}_{m},[\mathcal{J}_{l},\mathcal{P}_{n}]),\label{hjacobi}
\end{equation}
Expanding $\psi$ in terms of $\mathcal{P}$s as $\psi(\mathcal{P}_{m},\mathcal{P}_{n})=(m-n)g(m,n)\mathcal{P}_{m+n}$, we get the same relation as \eqref{fg-first} which has the solution $g(m,n)=constant$. 
The above discussion leads to 
\begin{equation}
\begin{split}
    \mathcal{H}^{2}(\mathfrak{bms}_{3};\mathcal{P})&=\mathcal{H}^{1}(\mathfrak{witt};\mathcal{H}^{1}(\mathcal{P};\mathcal{P}))\oplus \mathcal{H}^{0}(\mathfrak{witt};\mathcal{H}^{2}(\mathcal{P};\mathcal{P})),\label{bmsp2}
\end{split}
\end{equation}
which means that turning on deformations with coefficients in $\mathcal{P}$, we can only  deform the $[\mathcal{J},\mathcal{P}]$ part by $K(m,n)$ and  the ideal part $[\mathcal{P},\mathcal{P}]$  by $g(m,n)$ or the combination of these two, when we turn on both of them simultaneously. This is exactly in agreement of our results of direct and explicit calculations in the previous subsections.

\paragraph{Computation of ${\mathcal{H}}^{2}(\mathfrak{bms}_{3};\mathfrak{witt})$.} Again considering the short exact spectral sequence \eqref{short-exact-bms3} we have 
\begin{equation}
\begin{split}
     \mathcal{H}^{2}(\mathfrak{bms}_{3};\mathfrak{witt})&=\oplus_{p+q=2}E_{2;\mathfrak{w}}^{p,q}=E_{2;\mathfrak{w}}^{2,0}\oplus E_{2;\mathfrak{w}}^{1,1}\oplus E_{2;\mathfrak{w}}^{0,2}\\
    &=\mathcal{H}^{2}(\mathfrak{witt};\mathcal{H}^{0}(\mathcal{P};\mathfrak{witt}))\oplus \mathcal{H}^{1}(\mathfrak{witt};\mathcal{H}^{1}(\mathcal{P};\mathfrak{witt}))\oplus \mathcal{H}^{0}(\mathfrak{witt};\mathcal{H}^{2}(\mathcal{P};\mathfrak{witt})),\label{bmsj}
\end{split}
\end{equation}
where the subscript $\mathfrak{w}$ denotes we are computing $  \mathcal{H}^{2}(\mathfrak{bms}_{3};\mathfrak{witt})$.

To compute  $\mathcal{H}^{2}(\mathfrak{witt};\mathcal{H}^{0}(\mathcal{P};\mathfrak{witt}))$, we recall the action  of $\mathcal{P}$ on $\mathfrak{witt}$, which is induced via the short exact sequence  \eqref{short-exact-bms3}, is trivial and  hence $\mathcal{H}^{0}(\mathcal{P};\mathfrak{witt})\cong \mathfrak{witt}$. We then conclude $\mathcal{H}^{2}(\mathfrak{witt};\mathcal{H}^{0}(\mathcal{P};\mathfrak{witt}))\cong  \mathcal{H}^{2}(\mathfrak{witt};\mathfrak{witt})\cong 0,$ where in the last step we used the fact that Witt algebra is rigid \cite{fialowski2012formal, schlichenmaier2014elementary}.

Next, we consider the second term in \eqref{bmsj}, $\mathcal{H}^{1}(\mathcal{J};\mathcal{H}^{1}(\mathcal{P};\mathcal{J}))$ which is constructed by 1-cocycle $\varphi_{2}$ as $\varphi_{2}: \mathfrak{witt}\longrightarrow \mathcal{H}^{1}(\mathcal{P};\mathfrak{witt})$, where 
$\varphi_{2}(\mathcal{J}_{m})(\mathcal{P}_{n})=\tilde{I}(m,l)\mathcal{J}_{m+l}\label{Cjp}$, in which $\tilde{I}(m,l)$ is an arbitrary function. $\varphi_2$ deforms the commutator $[\mathcal{J},\mathcal{P}]$ part as $i[\mathcal{J}_m,\mathcal{P}_n]=(m-n)\mathcal{P}_{m+n}+\tilde{I}(m,n){\mathcal{J}}_{m+n}$. Jacobi identity for the above bracket implies that $\tilde{I}(m,n)$ should solve \eqref{eqdefIK}, so one finds the same result as $\tilde{I}(m,n)=0$. This means that the $[\mathcal{J},\mathcal{P}]$ commutator cannot be deformed by the terms with coefficients {in} $\mathcal{J}$.

We finally compute the last term in \eqref{bmsj} which is $\mathcal{H}^{0}(\mathfrak{witt};\mathcal{H}^{2}(\mathcal{P};\mathfrak{witt}))$.  One can repeat the procedure exactly the same as the previous case to get 
\begin{equation}
  (\mathcal{J}_{l}\circ \psi)(\mathcal{P}_{m},\mathcal{P}_{n})=[\mathcal{J}_{l},\psi(\mathcal{P}_{m},\mathcal{P}_{n})]-\psi([\mathcal{J}_{l},\mathcal{P}_{m}],\mathcal{P}_{n})-\psi(\mathcal{P}_{m},[\mathcal{J}_{l},\mathcal{P}_{n}]),\label{hjacobi-1}
\end{equation}
so we can expand $\psi$ in terms of $\mathcal{J}$s as $\psi(\mathcal{P}_{m},\mathcal{P}_{n})=(m-n)f(m,n)\mathcal{J}_{m+n}$ where $f(m,n)$ is an arbitrary symmetric function. By inserting the latter into \eqref{hjacobi-1} we get the same relation as 
 \eqref{fg-first} which has the solution $f(m,n)=constant$. 

With the above discussion we conclude that 
\begin{equation}
    \mathcal{H}^{2}(\mathfrak{bms}_{3};\mathfrak{witt})\cong \mathcal{H}^{0}(\mathfrak{witt};\mathcal{H}^{2}(\mathcal{P};\mathfrak{witt})),\label{bms-witt}
\end{equation}
i.e. deformations of $\mathfrak{bms}_{3}$ with coefficients in $\mathcal{J}$ are just in ideal part of the algebra and this is in agreement with our results in {the} previous subsection.  
{So far we have  separately computed  $\mathcal{H}^{2}(\mathfrak{bms}_{3};\mathcal{P})$ and  $\mathcal{H}^{2}(\mathfrak{bms}_{3};\mathfrak{witt})$. Combination of the first term in RHS of \eqref{bmsp2} and \eqref{bms-witt} are just isomorphic with the first and second terms in \eqref{bmsp2} and the second term in RHS of \eqref{bmsp2} and \eqref{bms-witt} are  isomorphic. One can then conclude that \eqref{bmsp2} contains the whole  content of \eqref{bms-witt}.}

As summary of the above discussions one may conclude that 
\begin{equation}
    \mathcal{H}^{2}(\mathfrak{bms}_{3};\mathfrak{bms}_{3})\cong \mathcal{H}^{2}(\mathfrak{bms}_{3};\mathcal{P})\cong \mathcal{H}^{1}(\mathfrak{witt};\mathcal{H}^{1}(\mathcal{P};\mathcal{P}))\oplus \mathcal{H}^{0}(\mathfrak{witt};\mathcal{H}^{2}(\mathcal{P};\mathcal{P})).
\end{equation}

One may also ask about $\mathcal{H}^{3}(\mathfrak{bms}_{3};\mathfrak{bms}_{3})$.  One can show that all obstructions are located in the space $\mathcal{H}^{3}$ \cite{nijenhuis1967deformations}. As we discussed, nontrivial infinitesimal deformations lead to formal deformations if there is no obstruction. However, as discussed above when we turn on deformations $K(m,n)$ and $g(m,n)$ together, there are obstructions to integrability of infinitesimal deformation.  So since we have found a specific example of nontrivial infinitesimal deformation in the case of $\mathfrak{bms}_{3}$ which is not integrable, we conclude that $\mathcal{H}^{3}(\mathfrak{bms}_{3};\mathfrak{bms}_{3})\neq 0$.

\paragraph{Virasoro-Kac-Moody algebra, $\mathfrak{KM}_{\mathfrak{u}(1)}$.} We can repeat the above cohomological analysis for $\mathfrak{KM}_{\mathfrak{u}(1)}$ algebra. The results are discussed in \cite{Roger:2006rz}.  
One can consider the short exact sequence for this algebra as 
 \begin{equation}
    0\longrightarrow \mathcal{P}_{\mathfrak{u}(1)} \longrightarrow \mathfrak{KM}_{\mathfrak{u}(1)} \longrightarrow \mathfrak{KM}_{\mathfrak{u}(1)}/\mathcal{P}_{\mathfrak{u}(1)} \cong \mathfrak{witt}\longrightarrow 0,\label{short-exact-Kac}
\end{equation}
where $\mathcal{P}_{\mathfrak{u}(1)}$ denotes the ideal part and $\mathfrak{witt}$ the Witt subalgebra of $\mathfrak{KM}_{\mathfrak{u}(1)}$. Our goal here is to compute $\mathcal{H}^{2}(\mathfrak{KM}_{\mathfrak{u}(1)};\mathfrak{KM}_{\mathfrak{u}(1)})$.
Based on the above sequence 
and the structure of the algebra, as in the \bms\ case, we should not expect $\mathcal{H}^{2}(\mathfrak{KM}_{\mathfrak{u}(1)};\mathfrak{KM}_{\mathfrak{u}(1)})$ to be equal to  $\mathcal{H}^{2}(\mathfrak{KM}_{\mathfrak{u}(1)};\mathfrak{witt})\oplus \mathcal{H}^{2}(\mathfrak{KM}_{\mathfrak{u}(1)};\mathcal{P}_{\mathfrak{u}(1)}).$ Nonetheless, our arguments of the \bms\ case readily extend to the Virasoro-Kac-Moody case yielding
\begin{equation}
    \mathcal{H}^{2}(\mathfrak{KM}_{\mathfrak{u}(1)};\mathfrak{witt})=0.\label{KM-witt}
\end{equation}
We recall that $\mathcal{H}^{1}(W(a,b);\mathfrak{witt})=0$ \cite{gao2011low} which is also true for $\mathfrak{KM}_{\mathfrak{u}(1)}$ as $W(0,0)$.  From the long exact sequence \eqref{long-exact}, and that $\mathcal{H}^{n}(\mathfrak{KM}_{\mathfrak{u}(1)};\mathfrak{witt})=0$ for $n=1,2$, one concludes that
\begin{equation}
\mathcal{H}^{2}(\mathfrak{KM}_{\mathfrak{u}(1)};\mathfrak{KM}_{\mathfrak{u}(1)})= \mathcal{H}^{2}(\mathfrak{KM}_{\mathfrak{u}(1)};\mathcal{P}_{\mathfrak{u}(1)}).    
\end{equation}

Next, we note that $\mathcal{H}^{2}(\mathfrak{KM}_{\mathfrak{u}(1)};\mathcal{P}_{\mathfrak{u}(1)})$  can be decomposed as 
\begin{equation}
\begin{split}
    \hspace*{-5mm}\mathcal{H}^{2}(\mathfrak{KM}_{\mathfrak{u}(1)};\mathcal{P}_{\mathfrak{u}(1)})
     \cong
    & \mathcal{H}^{2}(\mathfrak{witt};\mathcal{H}^{0}(\mathfrak{u}(1);\mathcal{P}_{\mathfrak{u}(1)}))\oplus \mathcal{H}^{1}(\mathfrak{witt};\mathcal{H}^{1}(\mathcal{P}_{\mathfrak{u}(1)};\mathcal{P}_{\mathfrak{u}(1)}))\\\oplus &\mathcal{H}^{0}(\mathfrak{witt};\mathcal{H}^{2}(\mathfrak{u}(1);\mathcal{P}_{\mathfrak{u}(1)})).\nonumber
\end{split}
\end{equation}
As in the \bms\ case, the last term can be argued to be zero and hence
\begin{equation}
\begin{split}
     \mathcal{H}^{2}(\mathfrak{KM}_{\mathfrak{u}(1)};\mathfrak{KM}_{\mathfrak{u}(1)})\cong 
     \mathcal{H}^{2}(\mathfrak{witt};\mathcal{H}^{0}(\mathcal{P}_{\mathfrak{u}(1)};\mathcal{P}_{\mathfrak{u}(1)}))\oplus \mathcal{H}^{1}(\mathfrak{witt};\mathcal{H}^{1}(\mathcal{P}_{\mathfrak{u}(1)};\mathcal{P}_{\mathfrak{u}(1)})),
\end{split}
\end{equation}
where the first and second term in the right hand side are respectively associated to deformations of $[\mathcal{J},\mathcal{P}]$  and $[\mathcal{P},\mathcal{P}]$ parts of $\mathfrak{KM}_{\mathfrak{u}(1)}$ algebra. The former leads to $W(a,b)$ and the latter to \eqref{Kac-Moody-nu}.

One can then show that all nontrivial deformations of $\mathfrak{KM}_{\mathfrak{u}(1)}$ algebra discussed above have no obstructions and are integrable. However, as mentioned in subsection \ref{Integrability}, absence of obstructions does not mean $\mathcal{H}^{3}(\mathfrak{KM}_{\mathfrak{u}(1)};\mathfrak{KM}_{\mathfrak{u}(1)})=0$.


\section{Deformations of \texorpdfstring{$\widehat{\mathfrak{bms}_3}$}{BMS3} and  \texorpdfstring{$\widehat{\mathfrak{KM}_{\mathfrak{u}(1)}}$}{KM-u1} centrally extended algebras}\label{sec:4.2}

 The second real cohomology 
 of $\mathfrak{bms}_{3}$ algebra, ${\mathcal{H}}^2(\mathfrak{g};\mathbb{R})$, is two dimensional \cite{gao2008derivations} in the sense that it admits two independent nontrivial Gelfand-Fucks real 2-cocycles which are \cite{Barnich:2011ct}:
 \begin{equation}
     \psi_{1}({\mathcal{J}_{m},\mathcal{J}_{n}})=\frac{c_{JJ}}{12}m^{3}\delta_{m+n,0},\,\,\,\  \psi_{2}({\mathcal{J}_{m},\mathcal{P}_{n}})=\frac{c_{JP}}{12}m^{3}\delta_{m+n,0}.\,\,\,\
 \end{equation}
Having two independent Gelfand-Fucks 2-cocycles means that we are extending the algebra by two independent central identity elements \cite{Oblak:2016eij}. We denote central extension of $\mathfrak{bms}_{3}$ algebra with $\widehat{\mathfrak{bms}}_{3}$.
The commutation relations of $\widehat{\mathfrak{bms}}_{3}$ are (see \cite{Oblak:2016eij} and references therein):
\begin{equation} 
\begin{split}
 & i[\mathcal{J}_{m},\mathcal{J}_{n}]=(m-n)\mathcal{J}_{m+n}+\frac{c_{JJ}}{12}m^{3}\delta_{m+n,0}, \\
 &i[\mathcal{J}_{m},\mathcal{P}_{n}]=(m-n)\mathcal{P}_{m+n}+\frac{c_{JP}}{12}m^{3}\delta_{m+n,0},\\
 &i[\mathcal{P}_{m},\mathcal{P}_{n}]=0.\label{eq:5.2}
\end{split}
\end{equation}
The central charges $c_{JJ}$ and $c_{JP}$ are  arbitrary real numbers. Algebras with different nonzero values of the central charges $c_{JJ}$ and $c_{JP}$, are cohomologous, i.e they are isomorphic to each other. Here we take the viewpoint that the global central extensions are deformations of the algebra by addition of the unit elements to the algebra, one unit element for each 2-cocycle. The latter means that if deformation procedure leads to $\widehat{\mathfrak{bms}}_{3}$ algebra with different central charges than we started with, we do not view it as a nontrivial deformation of the initial algebra.

Deformations of $\widehat{\mathfrak{bms}}_{3}$ algebra
can be analyzed as we did in the previous section, by  deforming  each commutator of $\widehat{\mathfrak{bms}}_{3}$ algebra separately. As we will see, however, presence of central extensions makes some of the deformations which were trivial become nontrivial, and conversely some of the nontrivial deformations may become trivial.

\subsection{Classification of 2-cocycles of \texorpdfstring{$\widehat{\mathfrak{bms}}_3$}{BMS3} algebra}\label{sec:5.1}

\paragraph{Deformation of commutators of two $\mathcal{P}$’s.}  Since we added new generators as central terms to $\mathfrak{bms}_{3}$ algebra, the most general deformations of the algebra also involves the central term. So the commutation relations of deformed $\widehat{\mathfrak{bms}}_{3}$ algebra are:
\begin{equation} 
\begin{split}
 & i[\mathcal{J}_{m},\mathcal{J}_{n}]=(m-n)\mathcal{J}_{m+n}+\frac{c_{JJ}}{12}m^{3}\delta_{m+n,0}, \\
 &i[\mathcal{J}_{m},\mathcal{P}_{n}]=(m-n)\mathcal{P}_{m+n}+\frac{c_{JP}}{12}m^{3}\delta_{m+n,0},\\
 &i[\mathcal{P}_{m},\mathcal{P}_{n}]=\varepsilon\tilde{\psi}_{1}^{PP}(\mathcal{P}_{m},\mathcal{P}_{n}),\label{deform of ideal-ce2}
\end{split}
\end{equation}
where $\varepsilon$ is the deformation parameter and $\tilde{\psi}_{1}^{PP}(\mathcal{P}_{m},\mathcal{P}_{n})$ is a $2-$cocycle which may be expanded as a linear combination of generators,
\begin{equation}
   \tilde{\psi}_{1}^{PP}(\mathcal{P}_{m},\mathcal{P}_{n})= (m-n)g(m,n)\mathcal{P}_{m+n}+(m-n)f(m,n)\mathcal{J}_{m+n}+\frac{{c}_{PP}}{12} X(m)\delta_{m+n,0},\label{hat-psiPP}
\end{equation}
in which $X(m)$ is an arbitrary function and ${c}_{PP}$ is a new central charge. As in the case of centerless $\mathfrak{bms}_{3}$, we should consider two Jacobi identities $[\mathcal{P}_{m},[\mathcal{P}_{n},\mathcal{P}_{l}]]+\text{cyclic permutations}=0$ and $[\mathcal{P}_{m},[\mathcal{P}_{n},\mathcal{J}_{l}]]+\text{cyclic permutations}=0$. The former leads to two independent relations one with coefficients in $\mathcal{J}$, which is the same as \eqref{fg-first}, and another for the central part. 
The second Jacobi identity leads to two independent equations for $f(m,n)$ and $g(m,n)$ which are the same as \eqref{fg-first}, and a new equation for central part which is
\begin{equation}\label{C_JJ+C_JP+C_PP}
\big[{c}_{PP}\big((l-m)X(n)+(n-l)X(m)\big)+l^3(m-n)\big(c_{JJ}f(m,n)+c_{JP}g(m,n)\big)\big]\delta_{m+n+l,0}=0.
    \end{equation} 

As discussed in the first part of section \ref{sec:4.1}, solutions to \eqref{fg-first} are $f(m,n),\,\, g(m,n)=\text{constant}$, and \eqref{C_JJ+C_JP+C_PP} yields 
\begin{equation}
{c}_{PP}\big((2n+m)X(m)-(2m+n)X(n)\big)=
 (f {c}_{JJ} + g{c}_{JP})(m-n)(m+n)^{3}.\label{tildec(JJ)}
\end{equation}
The only nontrivial solution of \eqref{tildec(JJ)} is
\begin{equation}
c_{PP}=f {c}_{JJ}+g {c}_{JP},\qquad X(m)=m^{3}. \label{PP-solutions}
\end{equation}
So the most general deformations of $\widehat{\mathfrak{bms}}_{3}$ ideal part are 
\begin{equation} 
\begin{split}
 & i[\mathcal{J}_{m},\mathcal{J}_{n}]=(m-n)\mathcal{J}_{m+n}+\frac{c_{JJ}}{12}m^{3}\delta_{m+n,0}, \\
 &i[\mathcal{J}_{m},\mathcal{P}_{n}]=(m-n)\mathcal{P}_{m+n}+\frac{c_{JP}}{12}m^{3}\delta_{m+n,0},\\
 &i[\mathcal{P}_{m},\mathcal{P}_{n}]=(m-n) \varepsilon(f\mathcal{P}_{m+n}+g\mathcal{J}_{m+n})+\varepsilon \frac{(f {c}_{JJ} +g {c}_{JP})}{12}m^{3}\delta_{m+n,0},\label{deform-ideal-central}
\end{split}
\end{equation}
where $f,g$ are two arbitrary (deformation) constants. {One can then show that by a proper redefinition of the generators the relation \eqref{deform-ideal-central} is exactly the same as \eqref{AdS3} which has only two independent central charges.}

As a special case, we may study deformations when we do not allow for the central term $c_{PP}$. In this case, we can have $f(m,n)=f,\ g(m,n)=g$ constants, but they are not arbitrary any more, as they are related by $f c_{JJ}+ g c_{JP}=0$. In the special case when $c_{JP}$ or $c_{JJ}$ are zero, this respectively yields $f=0$ or $g=0$. 

\paragraph{Deformation of commutators of $[\mathcal{J},\mathcal{P}]$.}

We then consider deformation of second commutation relation of $\widehat{\mathfrak{bms}}_{3}$ algebra \eqref{eq:5.2} as:
\begin{equation} 
\begin{split}
 & i[\mathcal{J}_{m},\mathcal{J}_{n}]=(m-n)\mathcal{J}_{m+n}+\frac{c_{JJ}}{12}m^{3}\delta_{m+n,0}, \\
 &i[\mathcal{J}_{m},\mathcal{P}_{n}]=(m-n)\mathcal{P}_{m+n}+\frac{c_{JP}}{12}m^{3}\delta_{m+n,0}+\zeta\tilde{\psi}^{JP}_{1}(\mathcal{J}_{m},\mathcal{P}_{n}),\\
 &i[\mathcal{P}_{m},\mathcal{P}_{n}]=0,\label{c_(JP)commutation}
\end{split}
\end{equation}
where $\tilde{\psi}^{JP}_{1}(\mathcal{J}_{m},\mathcal{P}_{n})$ is a $2-$cocycle and $\zeta$ is the deformation parameter. One can write $\tilde{\psi}^{JP}_{1}(\mathcal{J}_{m},\mathcal{P}_{n})$ in terms of linear combination of generators as 
\begin{equation}
    \tilde{\psi}^{JP}_{1}(\mathcal{J}_{m},\mathcal{P}_{n})=K(m,n)\mathcal{P}_{m+n}+I(m,n)\mathcal{J}_{m+n}+\frac{\hat{c}_{JP}}{12}Y(m).\label{hat-psiJP}
\end{equation}
We have to check two Jacobi identities
$[\mathcal{P}_{m},[\mathcal{P}_{n},\mathcal{J}_{l}]]+\text{cyclic permutations}=0$ and \\$ [\mathcal{J}_{m},[\mathcal{J}_{n},\mathcal{P}_{l}]]+\text{cyclic permutations} =0$. From the first Jacobi one gets a relation for $I(m,n)$ which is exactly the same as \eqref{firsteqI}, when we just keep linear term in $\zeta$, with the only solution $I(m,n)=0$. The second Jacobi leads to two relations for $K(m,n)$ up to first order in $\zeta$. The first one is exactly the same as \eqref{eq-K3} with the only nontrivial solutions $K(m,n)=\alpha+\beta m$. The second equation obtained from the above Jacobi relates $K(m,n)$ to $Y(m)$ as
 \begin{equation}
 \begin{split}
\big[{c_{JP}K(n,-m-n)m^{3}-\hat{c}_{JP}}\big((2m+n)Y(n)+m Y(m+n)\big)\big]-(m\leftrightarrow n)=0.\label{y and k}
 \end{split}
\end{equation}
The above should be solved for $Y$ with $K(m,n)=\alpha+\beta m$. One can immediately see that for $\alpha,\beta, c_{JP}\neq 0$ the above has no solutions. If $c_{JP}=0$, then \eqref{y and k} admits a solution of the form $Y(n)=a(n^3-n)\delta_{m+n,0}$ where $a$ is a real number. This solution, however, as we will see is not integrable. If $\alpha,\beta=0$, i.e. when we turn off the $K$ deformation, then again we can have $Y(n)=a(n^3=n)\delta_{m+n,0}$ as a solution. This solution is not a new deformation, it is just the same as the central extension already turned on. To summarize, for $c_{PJ}\neq 0$ case we do not have any nontrivial deformation and for $c_{JP}=0$ as have $K(m)=\alpha+\beta m$ deformations, as discussed in the second part of section \ref{sec:4.1}. That is, the $W(a,b)$ algebra does not admit a $c_{JP}$ central extension, in accord with \cite{gao2008derivations,gao2011low}.

 \paragraph{Deformation of commutators of two $\mathcal{J}$’s.}\label{JJhat}
 
We finally consider deformations of third commutation relations of $\widehat{\mathfrak{bms}}_{3}$ algebra as:
\begin{equation} 
\begin{split}
 & i[\mathcal{J}_{m},\mathcal{J}_{n}]=(m-n)\mathcal{J}_{m+n}+\frac{c_{JJ}}{12}m^{3}\delta_{m+n,0}+\eta \tilde{\psi}^{JJ}_{1}(\mathcal{J}_{m},\mathcal{J}_{n}) \\
 &i[\mathcal{J}_{m},\mathcal{P}_{n}]=(m-n)\mathcal{P}_{m+n}+\frac{c_{JP}}{12}m^{3}\delta_{m+n,0},\\
 &i[\mathcal{P}_{m},\mathcal{P}_{n}]=0,\label{JJ-h-center}
\end{split}
\end{equation}
in which $\eta$ is a deformation parameter and $\tilde{\psi}^{JJ}_{1}(\mathcal{J}_{m},\mathcal{J}_{n})$ is a $2-$cocycle which can be written in terms of linear combination of generators as
\begin{equation}
    \tilde{\psi}^{JJ}_{1}(\mathcal{J}_{m},\mathcal{J}_{n})=(m-n)h(m,n)\mathcal{P}_{m+n}+\frac{1}{12} U(m)\delta_{m+n,0},\label{hat-psiJJ}
\end{equation}
where $U(m)$ is an arbitrary function. We did not include coefficients in $\mathcal{J}$ since the Virasoro algebra is proved to be a rigid algebra \cite{fialowski2012formal}. 

To find the constraints on $h(m,n)$ and $U(m)$ we  consider the Jacobi identity $ [\mathcal{J}_{m},[\mathcal{J}_{n},\mathcal{J}_{l}]]$\\$+\text{cyclic permutations}=0$. This Jacobi for \eqref{JJ-h-center} leads to two different equations. One of them is exactly the same as \eqref{h-eq} and its only solution is \eqref{JJ-h-Z}. The second one is related to the central part as
 \begin{equation}
 \begin{split}
      \big[(n-l)\big(c_{JP}h(n,l)m^{3}+U(m)\big)+\text{cyclic permutations}
      \big]\delta_{m+n+l,0}=0,
      \label{h-center}
 \end{split}
\end{equation}
or equivalently
\begin{equation}
\begin{split}
   & \big[c_{JP}\big((2n+m)m^3 h(n,-m-n)+n(m+n)^3 h(m,n)\big)+(2n+m)U(m)-nU(-m-n)\big]\\
   & - (m\leftrightarrow n)=0.  
    \end{split}
\end{equation}
One can readily verify that $h(m,n)=h=\text{constant}$, $U(m)=a+bm+cm^3$ for arbitrary constants $a,b,c$, provides a general solution. One may also show that these are the most general solutions for $c_{JP}\neq 0$. The $a,b$ terms in $U(m)$ may be reabsorbed in a shift of ${\cal J}_m$ and the $c$ term is nothing but a shift of central charge $c_{JJ}$. So, let us focus on the $h$-deformation:
\begin{equation} 
\begin{split}
 & i[\mathcal{J}_{m},\mathcal{J}_{n}]=(m-n)(\mathcal{J}_{m+n}+\nu P_{m+n})+ \frac{c_{JJ}}{12}m^{3}\delta_{m+n,0}, \\
 &i[\mathcal{J}_{m},\mathcal{P}_{n}]=(m-n)\mathcal{P}_{m+n}+\frac{c_{JP}}{12}m^{3}\delta_{m+n,0},\\
 &i[\mathcal{P}_{m},\mathcal{P}_{n}]=0,\label{JJ-JP-nu}
\end{split}
\end{equation}
where $\nu=\eta h$. This algebra has three parameters, $c_{JJ},\ c_{JP}$ and $\nu$. However, if $c_{JP}\neq 0$ one can remove $c_{JJ}$ or $\nu$ by a simple redefinition of generators. To this end consider
\begin{equation} \label{J-shift-P}
\begin{split}
 &  \mathcal{J}_{m}:=\tilde{\mathcal{J}}_{m}+Z\tilde{\mathcal{P}}_{m}, \\
&\mathcal{P}_{m}:=\tilde{\mathcal{P}}_{m}.
\end{split}
\end{equation}
By choosing $Z=\nu$ we can remove $\nu$ term. This does not change the $c_{JP}$ while shifts $c_{JJ}$ to $\tilde c_{JJ}=c_{JJ}-2\nu c_{JP}$.
Alternatively, one may choose $Z=\frac{c_{JJ}}{2c_{JP}}$ and remove $c_{JJ}$ term in favor of the $\nu$ term, by shifting $\nu$ to $\tilde{\nu}=\nu-\frac{c_{JJ}}{2c_{JP}}$. It is not possible to remove both central charges in the $\nu$ term.  

If $c_{JP}=0$, \eqref{h-center} does not put any new constraint on $h(m,n)$ and we get $U(m)=\hat{c}_{JJ}m^{3}$. In this case by redefinition \eqref{J-shift-P} with $\nu=Z$, $\eta h=\nu$ deformation can be absorbed and hence there is no nontrivial deformation of $\widehat{\mathfrak{bms}}_{3}$ in the $[{\cal J},{\cal J}]$ sector.

Although \bms\  admits two independent central terms, only $c_{JP}$ is obtained through gravitational computations \cite{Barnich:2006av}. On the other hand, we showed that turning on the $\nu$ deformation is equivalent to turning on $c_{JJ}$. It is worth noting that in a recent work by W. Merbis and M. Riegler \cite{Merbis:2019wgk} it has been shown that the second central charges $c_{JJ}$ of \bms\ can be obtained as a quantum correction. This guides us toward considering $\nu$ deformation as quantum correction of $3d$ flat Einstein gravity.


\subsection{Integrability conditions and obstructions}

After classifying nontrivial infinitesimal deformations we now discuss integrability of these infinitesimal deformations and their possible obstructions. As pointed out there are different approaches to the latter. Here we follow the same approach as in section \ref{sec:4.5} and check if the $2-$cocycle $\psi_{1}$ provides a formal deformation. 

We first consider integrability conditions of deformation of the ideal part of $\widehat{\mathfrak{bms}}_{3}$, $\tilde{\psi}_{1}^{PP}(\mathcal{P}_{m},\mathcal{P}_{n})$. The Jacobi identity $ [\mathcal{P}_{m},[\mathcal{P}_{n},\mathcal{J}_{l}]]+\text{cyclic permutations} =0$ leads to a linear relation for each function $f(m,n)$, $g(m,n)$ and $X(m)$ separately with the same solutions we mentioned before and hence the linear order solution is also a solution to higher order equations. The Jacobi $ [\mathcal{P}_{m},[\mathcal{P}_{n},\mathcal{P}_{l}]]+\text{cyclic permutations} =0$  leads to some other relations. The first one is  the same as \eqref{C_JJ+C_JP+C_PP} which is again linear in terms of deformation parameter. Two other relations are 
\begin{align*}
(n-l)(m-n-l)f(n,l)f(m,n+l)+\text{cyclic permutations}&=0, \\ (n-l)(m-n-l)f(n,l)g(m,n+l)+\text{cyclic permutations}&=0,
\end{align*}
which are quadratic in  the deformation parameter. One can check that both of these relations are satisfied with the solutions $f(m,n), g(m,n)=\text{constant}$ and hence the 
nontrivial deformation of ideal part is integrable.

The next infinitesimal deformation is obtained through deforming $[\mathcal{J},\mathcal{P}]$ by $\tilde{\psi}_{1}^{JP}(\mathcal{J}_{m},\mathcal{P}_{n})$. To analyze integrability we  study the Jacobi identity $ [\mathcal{J}_{m},[\mathcal{J}_{n},\mathcal{P}_{l}]]+\text{cyclic permutations} =0$. The latter leads to two different relations.  The first relation is 
\eqref{eq-Ktild} with solution $K(m,n)=\alpha+\beta m$, therefore there is no obstruction on integrability of $K(m,n)$. The second relation is related to the central part, 
\begin{equation}
 \begin{split}
       &\big[\frac{c_{JP}}{12}\big({K}(n,l;\zeta)m^{3}-{K}(m,l;\zeta)n^{3}\big) +{K}(n,l;\zeta){Y}(m;\zeta)-K(m,l;\zeta){Y}(n;\zeta)-\\
       &\big((l-n){Y}(m;\zeta)+(m-l){Y}(n;\zeta)+(m-n){Y}(m+n;\zeta)\big)\big]\delta_{m+n+l,0}=0.\label{yandk-tilde}
 \end{split}
\end{equation}
As discussed before, when  $c_{JP}\neq 0$ the above has no solutions. For $c_{JP}=0$ it can be seen from \eqref{yandk-tilde} that, although in the infinitesimal level $K(m,n)$ and $Y(m)$ can admit nonzero values, in the higher order (e.g. in second order in $\zeta$), there are obstructions to integrability. Therefore, to have a nontrivial formal deformation one can only turn on $K(m,n)$ or $Y(m)$, and not both. To summarize, all  deformations which we have found for $\widehat{\mathfrak{bms}}_{3}$ are integrable. We note that although we deformed the commutators of $\widehat{\mathfrak{bms}}_{3}$ separately, one can show that simultaneous deformations do not lead to new relations. So, there are only three integrable nontrivial infinitesimal deformations. As the $\mathfrak{bms}_{3}$ case, infinitesimal deformations induced by $f(m,n)$ or by simultaneous deformations  $f(m,n)$ and $K(m,n)$ are equivalent to deformations induced by $g(m,n)$ and $K(m,n)$. One can also show that, similarly to the $\mathfrak{bms}_{3}$ case, simultaneous $g(m,n)$ and $K(m,n)$ infinitesimal deformation is not integrable; this latter is another example of the case v. discussed in subsection \ref{Integrability}.

\begin{tcolorbox}[colback=red!3!white]
\paragraph{Theorem 5.1} {\it The most general formal deformations of \hbms\ are either $\mathfrak{vir} \oplus \mathfrak{vir}$ or $\widehat{ W(a,b)} $ algebras,  (the latter is a formal deformation when $c_{JP}=0$).}
\end{tcolorbox}

We note that the $\mathfrak{vir} \oplus \mathfrak{vir}$ has two options: the subalgebra generated by ${\cal P}_{0,\pm1},\ {\cal J}_{0,\pm 1}$ can be either $\mathfrak{so}(2,2)$ or $\mathfrak{so}(3,1)$.

\subsection*{Deformations of \texorpdfstring{$\widehat{\mathfrak{KM}_{\mathfrak{u}(1)}}$}{KMu1}}
The commutation relations of  $\widehat{\mathfrak{KM}_{\mathfrak{u}(1)}}$ are 
\begin{equation} 
\begin{split}
 & i[\mathcal{J}_{m},\mathcal{J}_{n}]=(m-n)\mathcal{J}_{m+n}+\frac{c_{JJ}}{12}m^{3}\delta_{m+n,0}, \\
 &i[\mathcal{J}_{m},\mathcal{P}_{n}]=(-n)\mathcal{P}_{m+n}+\frac{c_{JP}}{12}m^{2}\delta_{m+n,0},\\
 &i[\mathcal{P}_{m},\mathcal{P}_{n}]=\frac{c_{PP}}{12}m\delta_{m+n,0}.\label{KM-hat}
\end{split}
\end{equation}
As it is obvious from the above the second real cohomology of {$\mathfrak{KM}_{\mathfrak{u}(1)}$} algebra, {${\mathcal{H}}^2(\mathfrak{KM}_{\mathfrak{u}(1)};\mathbb{R})$}, is three dimensional \cite{gao2008derivations} in the sense that it admits three independent nontrivial 2-cocycles which are:
 \begin{equation}
     \psi_{1}({\mathcal{J}_{m},\mathcal{J}_{n}})=\frac{c_{JJ}}{12}m^{3}\delta_{m+n,0},\,\,\,\  \psi_{2}({\mathcal{J}_{m},\mathcal{P}_{n}})=\frac{c_{JP}}{12}m^{2}\delta_{m+n,0},\,\,\,\ \psi_{3}({\mathcal{P}_{m},\mathcal{J}_{n}})=\frac{c_{PP}}{12}m\delta_{m+n,0}.
 \end{equation}
The most general deformations of the $\widehat{\mathfrak{KM}}_{\mathfrak{u}(1)}$ may be parametrized as
\begin{equation} 
\begin{split}
 & i[\mathcal{J}_{m},\mathcal{J}_{n}]=(m-n)\mathcal{J}_{m+n}+\frac{c_{JJ}}{12}m^{3}\delta_{m+n,0}+\eta (m-n)h(m,n)\mathcal{P}_{m+n}+\frac{\hat{c}_{JJ}}{12}U(m)\delta_{m+n,0}, \\
 &i[\mathcal{J}_{m},\mathcal{P}_{n}]=-n\mathcal{P}_{m+n}+\frac{c_{JP}}{12}m^{2}\delta_{m+n,0}+\zeta K(m,n)\mathcal{P}_{m+n}+\kappa I(m,n) {\mathcal J}_{m+n}+\frac{\hat{c}_{JP}}{12}Y(m)\delta_{m+n,0},\\
 &i[\mathcal{P}_{m},\mathcal{P}_{n}]=\frac{c_{PP}}{12}m\delta_{m+n,0}+(m-n)(\varepsilon_{1}f(m,n)\mathcal{J}_{m+n}+\varepsilon_{2}g(m,n)\mathcal{P}_{m+n})+\frac{\hat{c}_{PP}}{12}X(m)\delta_{m+n,0}.\label{general-deformation-KM-hat}
\end{split}
\end{equation}
Jacobi identities to the first order in deformation parameters lead to two different family of relations for the deformation functions. The first set  are exactly the same relations of the $\mathfrak{KM}_{\mathfrak{u}(1)}$ case discussed in section \ref{sec:4.5}. The second set come from central terms. These new relations put constraints on central functions and also new constraints on the first family solutions. Here we do not present the details and  just summarize the results:

The Jacobi identity $[\mathcal{P}_{m},[\mathcal{P}_{n},\mathcal{P}_{l}]]+\text{cyclic permutations} =0$ does not put any new constraints on functions at the infinitesimal deformation level. The central part of the Jacobi $[\mathcal{P}_{m},[\mathcal{P}_{n},\mathcal{J}_{l}]]+\text{cyclic permutations} =0$ gives two relations from power analysis point of view as
\begin{equation}
\begin{split}
    \{ &\hat{c}_{PP}(nX(m)- mX(n))+c_{PP}(nK(l,m)-m K(l,n))\\
     & +c_{JP}((n+l)^{2}I(l,n)-(m+l)^{2}I(l,m)+(m-n)(l)^{2}g(m,n))\}\delta_{m+n+l,0}=0.
\end{split}
\end{equation}
One can then show that, taking the solutions $I(m,n)=gm$ and $g(m,n)=g$, the term with the coefficient $c_{JP}$ is equal to zero  independently. The first term, by a power series analysis, leads to $X(m)=m$ and $K(m,n)=0$. 
So one then concludes that there is no simultaneous deformation by $I(m,n)$, $g(m,n)$ and $K(m,n)$. Also the above result leads to the conclusion that there is no deformation by $K(m,n)$ independently and by $K(m,n)$ and $h(m,n)$ simultaneously. 

The Jacobi $[\mathcal{J}_{m},[\mathcal{J}_{n},\mathcal{P}_{l}]]+\text{cyclic permutations} =0$ leads to another relation as
\begin{equation}
\begin{split}
     \big[&\hat{c}_{JP}(-lY(m)+lY(n)-(m-n)Y(m+n)) + c_{JJ}(m^{3}I(n,l)-n^{3}I(m,l))\\
     &+c_{PP}l(m-n)h(m,n)\big]\delta_{m+n+l,0}=0,
\end{split}
\end{equation}
in which we used the previous result $K(m,n)=0$. By a power series analysis, one can show that the above is satisfied when $h(m,n)=0$ and $Y(m)=m^{3}$ when $\hat{c}_{JP}=c_{JJ}$. As a result one concludes that the central part does not allow the deformations by $h(m,n)$ individuallyly and by $h(m,n)$, $I(m,n)$ and $g(m,n)$ simultaneously. 
 The final Jacobi $[\mathcal{J}_{m},[\mathcal{J}_{n},\mathcal{J}_{l}]]+\text{cyclic permutations} =0$ just gives the solution $U(m)=m^{3}$ as we expected. 
 
To summarize, we have found that the only nontrivial infinitesimal deformations of $\widehat{\mathfrak{KM}_{\mathfrak{u}(1)}}$ are those induced by $I(m,n)$, $g(m,n)$ and central functions $X(m)=m$, $Y(m)=m^{3}$ and $U(m)=m^{3}$. The central functions $X(m)=m$ and $U(m)=m^{3}$ do not lead to  a nontrivial deformation.  The Jacobi $[\mathcal{P}_{m},[\mathcal{P}_{n},\mathcal{J}_{l}]]+\text{cyclic permutations} =0$ for higher order deformations yields $I$, $g$ and $Y$ deformations are not integrable. {So, we conclude that
\begin{tcolorbox}[colback=red!3!white]
\paragraph{Theorem 5.2} {\it Centrally extended Virasoro-Kac-Moody algebra $\widehat{\mathfrak{KM}_{\mathfrak{u}(1)}}$ is formally rigid.}
\end{tcolorbox}

 {One may verify that this is another example of the case V in subsection \ref{Integrability}.}\footnote{While we have shown  $\widehat{\mathfrak{KM}_{\mathfrak{u}(1)}}$ with generic central charges does not admit any formal deformation, for special case of $c_{PP}=c_{JP}=0$,  $\widehat{\mathfrak{KM}_{\mathfrak{u}(1)}}$ can be formally deformed to $\widehat{W(a,b)}$. } 
 
 \subsubsection*{Proposal}
 We have shown that deformation of \bms\ and \kac\ algebras and their central extensions, obviously, deviate from the Hochschild-Serre factorization theorem since they can be deformed on their non ideal parts. However, we can propose an extension of it which is applicable for infinite dimensional Lie algebras. This new version is\\ 
 \emph{For infinite dimensional algebras with countable basis the deformations may appear in ideal and non-ideal parts, however, the deformations are always by coefficients of terms in the ideal part.}\\
 As we will see, our results in the next chapters reinforce this proposal. 
 

\subsection{Algebraic cohomology considerations}\label{sec:cohomology-bms-hat}

We  can extend our discussion in the subsection \ref{sec:cohomology-bms} to  $\widehat{\mathfrak{bms}}_{3}$. {As in the $\mathfrak{bms}_{3}$ case, $\mathfrak{vir}$ is not a $\widehat{\mathfrak{bms}}_{3}$ module by the adjoint action and $\mathcal{H}^{2}(\widehat{\mathfrak{bms}}_{3};\mathfrak{vir})$ is defined by the action induced from the short exact sequence \eqref{short-exact-vir}, as it was discussed in \ref{sec:HS-seq}. Moreover, as in the case without central extension, we should not expect
$\mathcal{H}^{2}(\widehat{\mathfrak{bms}}_{3};\widehat{\mathfrak{bms}}_{3})$ to be equal to $\mathcal{H}^{2}(\widehat{\mathfrak{bms}}_{3};\mathfrak{vir})\oplus \mathcal{H}^{2}(\widehat{\mathfrak{bms}}_{3};\mathfrak{vir}_{ab})$.} As in the 
\bms\ case, computation of each of these two factors would be helpful in computing the former.
To this end we employ the Hochschild-Serre spectral sequence procedure. 
The $\widehat{\mathfrak{bms}}_{3}$ has semi direct 
sum structure as $\widehat{\mathfrak{bms}}_{3}\cong\mathfrak{vir}\inplus_{ad}\mathfrak{vir}_{ab}$ where $\mathfrak{vir}$ is spanned by  $\mathcal{J}$ generators plus a unit central generator and $\mathfrak{vir}_{ab}$ is the ideal part which is spanned by $\mathcal{P}$ plus another unit central generator.
The short exact sequence for the above is  
\begin{equation}
    0\longrightarrow \mathfrak{vir}_{ab}\longrightarrow \widehat{\mathfrak{bms}}_{3} \longrightarrow \widehat{\mathfrak{bms}}_{3}/\mathfrak{vir}_{ab}\cong \mathfrak{vir}\longrightarrow 0.\label{short-exact-vir}
\end{equation}

\paragraph{Computation of $\mathcal{H}^{2}(\widehat{\mathfrak{bms}}_{3};\mathfrak{vir}_{ab})$.}
Using theorem 1.2 in \cite{degrijse2009cohomology} and Hochschild-Serre spectral sequence method we get
\begin{multline}
         \mathcal{H}^{2}(\widehat{\mathfrak{bms}}_{3};\mathfrak{vir}_{ab})=\oplus_{p+q=2}E_{2;\mathfrak{vir}_{ab}}^{p,q}=E_{2;\mathfrak{vir}_{ab}}^{2,0}\oplus E_{2;\mathfrak{vir}_{ab}}^{1,1}\oplus E_{2;\mathfrak{vir}_{ab}}^{0,2}
     \\= \mathcal{H}^{2}(\mathfrak{vir};\mathcal{H}^{0}(\mathfrak{vir}_{ab};\mathfrak{vir}_{ab}))\oplus \mathcal{H}^{1}(\mathfrak{vir};\mathcal{H}^{1}(\mathfrak{vir}_{ab};\mathfrak{vir}_{ab}))\oplus\mathcal{H}^{0}(\mathfrak{vir};\mathcal{H}^{2}(\mathfrak{vir}_{ab};\mathfrak{vir}_{ab})),\nonumber
\end{multline}
where we defined $E_{2;\mathfrak{vir}_{ab}}^{p,q}=\mathcal{H}^{p}(\mathfrak{vir};\mathcal{H}^{q}(\mathfrak{vir}_{ab};\mathfrak{vir}_{ab}))$.
 
The first term we should consider is $E_{2;\mathfrak{vir}_{ab}}^{2,0}=\mathcal{H}^{2}(\mathfrak{vir};\mathcal{H}^{0}(\mathfrak{vir}_{ab};\mathfrak{vir}_{ab}))$. From the definition of $\mathcal{H}^{0}$ (subsection \ref{sec:3.3}) one gets $\mathcal{H}^{0}(\mathfrak{vir}_{ab};\mathfrak{vir}_{ab})=\mathfrak{vir}_{ab}$ because the action of $\mathfrak{vir}_{ab}$ on $\mathfrak{vir}_{ab}$ is trivial. Then one has to compute $E_{2}^{2,0}=\mathcal{H}^{2}(\mathfrak{vir};\mathfrak{vir}_{ab})$. The latter is just the same deformation as \eqref{JJ-h-center}. As we mentioned earlier the latter just leads to a shift in $c_{JJ}$ so it does not lead to a nontrivial infinitesimal deformation and therefore $E_{2}^{2,0}=\mathcal{H}^{2}(\mathfrak{vir};\mathfrak{vir}_{ab})=0$. 
 
 $E_{2;\mathfrak{vir}_{ab}}^{1,1}=\mathcal{H}^{1}(\mathfrak{vir};\mathcal{H}^{1}(\mathfrak{vir}_{ab};\mathfrak{vir}_{ab}))$ is constructed by the map    $\hat{\varphi_{1}}: \mathfrak{vir}\longrightarrow \mathcal{H}^{1}(\mathfrak{vir}_{ab},\mathfrak{vir}_{ab})$.
The expression of $\hat{\varphi_{1}}(\hat{\mathcal{J}}_{m})(\hat{\mathcal{P}}_{n})$ can be written in terms of $\hat{\mathcal{P}}$ as $ \hat{\varphi_{1}}(\hat{\mathcal{J}}_{m})(\hat{\mathcal{P}}_{n})=\tilde{K}(m,l)\mathcal{P}_{m+l}+\tilde{Y}(m),\label{phi-jp}$ in which $\tilde{K}(m,l)$ and $\tilde{Y}(m)$ are arbitrary functions; here we have used the hatted generators to emphasize that they also include the central element. The deformation corresponding to $\hat{\varphi_{1}}$ is similar to the relation \eqref{c_(JP)commutation} when $I(m,n)=0$. The Jacobi identity constraints on $\hat{K}(m,n)$  are \eqref{eq-K3} and  \eqref{y and k}, whose solutions has been discussed below \eqref{y and k}.  Moreover,  we 
cannot deform the $\mathcal{J}\mathcal{P}$ part for $c_{JP}\neq 0$ and hence  $E_{2;\mathfrak{vir}_{ab}}^{1,1}=\mathcal{H}^{1}(\mathfrak{vir};\mathcal{H}^{1}(\mathfrak{vir}_{ab};\mathfrak{vir}_{ab}))=0$.
 
The last term we must compute is $E_{2}^{0,2}=\mathcal{H}^{0}(\mathfrak{vir};\mathcal{H}^{2}(\mathfrak{vir}_{ab},\mathfrak{vir}_{ab}))$. One can extend the definition of $\mathcal{H}^{0}$ in \eqref{H0P} for the above and verify that its elements are solutions of the equation
\begin{equation}
  (\hat{\mathcal{J}}_{l}\circ \hat{\psi}_{1})(\hat{\mathcal{P}}_{m},\hat{\mathcal{P}}_{n})=[\hat{\mathcal{J}}_{l},\hat{\psi}_{1}(\hat{\mathcal{P}}_{m},\hat{\mathcal{P}}_{n})]-\hat{\psi}_{1}([\hat{\mathcal{J}}_{l},\hat{\mathcal{P}}_{m}],\hat{\mathcal{P}}_{n})-\hat{\psi}_{1}(\hat{\mathcal{P}}_{m},[\hat{\mathcal{J}}_{l},\hat{\mathcal{P}}_{n}]),\label{psi-hat-1}
\end{equation}
where $\hat{\psi}_{1}(\hat{\mathcal{P}}_{m},\hat{\mathcal{P}}_{n})$ is a $2-$cocycle.  
The linear expansion of $\hat{\psi}_{1}$ in terms of generators is $\hat{\psi}_{1}(\hat{\mathcal{P}}_{m},\hat{\mathcal{P}}_{n})=(m-n)g(m,n)\mathcal{P}_{m+n}+X(m)\delta_{m+n,0}$. So by inserting the expansion of $\hat{\psi}_{1}$ into \eqref{psi-hat-1} we reach to two relations. The first one only involves $g(m,n)$ and is exactly the same as \eqref{fg-first} with the solution $g(m,n)=constant$. The second equation is exactly \eqref{C_JJ+C_JP+C_PP} when $f(m,n)=0$ and its solution is \eqref{PP-solutions}. By a proper redefinition of generators, one can show that the deformed algebra  is just the direct sum of two Virasoro algebras $\mathfrak{vir}\oplus \mathfrak{vir}$. This means that one can deform the ideal part of the $\widehat{\mathfrak{bms}}_{3}$ with coefficients in $\mathcal{P}$ into direct sum of two Virasoro algebras:
\begin{equation}
    \mathcal{H}^{2}(\widehat{\mathfrak{bms}}_{3};\mathfrak{vir}_{ab})\cong \mathcal{H}^{0}(\mathfrak{vir};\mathcal{H}^{2}(\mathfrak{vir}_{ab},\mathfrak{vir}_{ab})). \label{H2-vir(ab)}
\end{equation}

\paragraph{Computation of $\mathcal{H}^{2}(\widehat{\mathfrak{bms}}_{3};\mathfrak{vir})$.}

As the second case we consider $\mathcal{H}^{2}(\widehat{\mathfrak{bms}}_{3};\mathfrak{vir})$ which can be decomposed as 
\begin{equation}
\begin{split}
     \hspace*{-5mm}\mathcal{H}^{2}(\widehat{\mathfrak{bms}}_{3};\mathfrak{vir})&=\oplus_{p+q=2}E_{2;\mathfrak{vir}}^{p,q}=E_{2;\mathfrak{vir}}^{2,0}\oplus E_{2;\mathfrak{vir}}^{1,1}\oplus E_{2;\mathfrak{vir}}^{0,2}\\
     &= \mathcal{H}^{2}(\mathfrak{vir};\mathcal{H}^{0}(\mathfrak{vir}_{ab},\mathfrak{vir}))\oplus \mathcal{H}^{1}(\mathfrak{vir};\mathcal{H}^{1}(\mathfrak{vir}_{ab},\mathfrak{vir}))\oplus\mathcal{H}^{0}(\mathfrak{vir};\mathcal{H}^{2}(\mathfrak{vir}_{ab},\mathfrak{vir})),
\end{split}
\end{equation}
where $E_{2;\mathfrak{vir}}^{p,q}\equiv\mathcal{H}^{p}(\mathfrak{vir};\mathcal{H}^{q}(\mathfrak{vir}_{ab},\mathfrak{vir}))$.
 
The first term we have to consider is $E_{2;\mathfrak{vir}}^{2,0}=\mathcal{H}^{2}(\mathfrak{vir};\mathcal{H}^{0}(\mathfrak{vir}_{ab},\mathfrak{vir}))$. From the definition of $\mathcal{H}^{0}$ (subsection \ref{sec:cohomology-bms}) one gets $\mathcal{H}^{0}(\mathfrak{vir}_{ab},\mathfrak{vir})=\mathfrak{vir}$ because the action of $\mathfrak{vir}_{ab}$ as an ideal part of the algebra, on $\mathfrak{vir}$, {induced by the short exact sequence \eqref{short-exact-vir},} is trivial. Then, recalling the fact that $\mathfrak{vir}$ algebra is rigid \cite{fialowski2012formal, schlichenmaier2014elementary} one concludes that $E_{2;\mathfrak{vir}}^{2,0}=\mathcal{H}^{2}(\mathfrak{vir};\mathfrak{vir})=0$.

$E_{2;\mathfrak{vir}}^{1,1}=\mathcal{H}^{1}(\mathfrak{vir};\mathcal{H}^{1}(\mathfrak{vir}_{ab},\mathfrak{vir}))$ is constructed by $\hat{\varphi_{2}}$ as 
   $\hat{\varphi_{2}}: \mathfrak{vir}\longrightarrow \mathcal{H}^{1}(\mathfrak{vir}_{ab},\mathfrak{vir})$.
The expression of $\hat{\varphi_{2}}(\hat{\mathcal{J}}_{m})(\hat{\mathcal{P}}_{n})$ can be written in terms of $\hat{\mathcal{J}}$ as $ \hat{\varphi_{2}}(\hat{\mathcal{J}}_{m})(\hat{\mathcal{P}}_{n})=\tilde{I}(m,l)\mathcal{J}_{m+l}+\tilde{Y}(m),$ in which $\hat{I}(m,l)$ and $\hat{Y}(m)$ are arbitrary functions. 
 This deformation  is similar to the relation \eqref{c_(JP)commutation} when $K(m,n)=0$. 
 The Jacobi identity yields to \eqref{eqdefIK} for $\hat{I}(m,n)$, so one finds  $\hat{I}(m,n)=0$ and $\hat{Y}(m)=m^{3}$. The latter does not lead to a nontrivial deformation. This means that the $[\mathcal{J},\mathcal{P}]$ commutator cannot be deformed by the terms with coefficient of $\hat{\mathcal{J}}$, hence $E_{2}^{1,1}=0$. 

The last term we compute is $E_{2;\mathfrak{vir}}^{0,2}=\mathcal{H}^{0}(\mathfrak{vir};\mathcal{H}^{2}(\mathfrak{vir}_{ab},\mathfrak{vir}))$. One can extend the definition of $\mathcal{H}^{0}$ in \eqref{H0P} for the above and observe that its elements are solutions to
\begin{equation}
  (\hat{\mathcal{J}}_{l}\circ \hat{\psi}_{2})(\hat{\mathcal{P}}_{m},\hat{\mathcal{P}}_{n})=[\hat{\mathcal{J}}_{l},\hat{\psi}_{2}(\hat{\mathcal{P}}_{m},\hat{\mathcal{P}}_{n})]-\hat{\psi}_{2}([\hat{\mathcal{J}}_{l},\hat{\mathcal{P}}_{m}],\hat{\mathcal{P}}_{n})-\hat{\psi}_{2}(\hat{\mathcal{P}}_{m},[\hat{\mathcal{J}}_{l},\hat{\mathcal{P}}_{n}]),\label{psi-hat}
\end{equation}                         
where $\hat{\psi}_{2}(\hat{\mathcal{P}}_{m},\hat{\mathcal{P}}_{n})$ is a $2-$cocycle and the hatted objects  denote generators of the Viraroso algebra, i.e. Witt algebra plus central element. The linear expansion of $\hat{\psi}_{2}$ in terms of generators is $\hat{\psi}_{2}(\hat{\mathcal{P}}_{m},\hat{\mathcal{P}}_{n})=(m-n)f(m,n)\mathcal{J}_{m+n}+X(m)\delta_{m+n,0}$. Inserting the expansion of $\hat{\psi}$ into \eqref{psi-hat} we reach to two relations, the first one just contains $f(m,n)$ exactly the same as \eqref{fg-first} with the solution $f(m,n)=constant$, and the second equation is \eqref{C_JJ+C_JP+C_PP} when $g(m,n)=g=0$ which has solutions as \eqref{PP-solutions}. So we conclude that $\mathcal{H}^{2}(\widehat{\mathfrak{bms}}_{3};\mathfrak{vir})\cong \mathcal{H}^{0}(\mathfrak{vir};\mathcal{H}^{2}(\mathfrak{vir}_{ab},\mathfrak{vir}))$. The deformed algebra is, however, nothing but the direct sum of two Virasoro algebras $\mathfrak{vir} \oplus \mathfrak{vir}$. Note also that the deformation induced by $g(m,n)$ and $X(m)$, $f(m,n)$ and $X(m)$ or $g(m,n)$, $f(m,n)$ and $X(m)$, are not independent deformations and can be mapped to each other by a proper redefinition of generators. The content of $\mathcal{H}^{2}(\widehat{\mathfrak{bms}}_{3};\mathfrak{vir})$ is hence the same as $\mathcal{H}^{2}(\widehat{\mathfrak{bms}}_{3};\mathfrak{vir}_{ab})$ or their combination. In summary, we have shown that 
\begin{equation}
    \mathcal{H}^{2}(\widehat{\mathfrak{bms}}_{3};\widehat{\mathfrak{bms}}_{3})
    \cong \mathcal{H}^{2}(\widehat{\mathfrak{bms}}_{3};\mathfrak{vir}_{ab}),\label{cent-bms-H2}
\end{equation}
where in the last equality we used the fact that $\mathcal{H}^{2}(\widehat{\mathfrak{bms}}_{3};\mathfrak{vir})$, $\mathcal{H}^{2}(\widehat{\mathfrak{bms}}_{3};\mathfrak{vir}_{ab})$ are isomorphic.\footnote{In the above analysis we have assumed generic case where none of central charges $c_{JP}$ or $c_{JJ}$ are zero. As we observed from direct calculations in the previous section, for $c_{JP}=0$ case we can also deform $\widehat{\mathfrak{bms}}_{3}$ to $\widehat{W(a,b)}$. The cohomological analysis also confirms direct calculations of  section \ref{sec:5.1}. {We should also mention that cohomological considerations for $\widehat{\mathfrak{KM}_{\mathfrak{u}(1)}}$ case needs more analysis than what we did for other cases since the conditions of theorem 1.2 in \cite{degrijse2009cohomology} do not hold.}}





\section{Physical realization of obtained algebras through deformation}
Similar to the finite dimensional Lie algebra, deformations of a certain infinite dimensional Lie algebra may have various physical interpretations which highly depends on the physical context these algebras have appeared. Here we discuss  possible physical realizations of the deformation procedure and also infinite dimensional algebras obtained through it. 

The first case which we would like to consider is deformations of \bms\ algebra and its centrally extended \hbms. We showed that \bms\ algebra can be deformed either into two copies of \wit\ algebra which is obtained as asymptotic symmetries of AdS$_{3}$ or $W(a,b)$. Geometrically, deformation of $d$ dimensional Poincare algebra to $\mathfrak{so}(d-1,2)$ ($\mathfrak{so}(d,1)$) which is isometries algebra of AdS$_{d}$ (dS$_{d}$) is interpreted as going from a flat spacetime with zero curvature to another spacetime with a certain curvature. In fact, as mentioned, the deformation parameter in this picture is interpreted as AdS (dS) radius. The same interpretation is considered for deformation from \bms\ to two copies of \wit. 

\bms\ algebra can also be deformed into $W(a,b)$ family algebra. Here there are different arguments
\begin{itemize}
    \item As mentioned, \kac algebra which is exactly the same as $W(0,0)$, is obtained as asymptotic symmetry algebra of AdS$_{3}$ by choosing a specific boundary fall-off conditions \cite{Compere:2013bya}. This guides us again to interpret the deformation of \bms\ to \kac\ as procedure which takes us from asymptotic region of $3d$ flat spacetime to asymptotic region of AdS$_{3}$. But unlike the two copies of \wit, here the deformation parameter can not be considered as AdS radius. On the other hand, although \hbms\ can be deformed into two copies of \vir\ with two independent central term, it can be deformed into \wh\ when $c_{JP}=0$. The latter shows that during deformation of \hbms\ to \wh\ one of central terms will survive while in \cite{Compere:2013bya} it has been shown that two central terms are obtained. 
    
    \item $W(0,0)$ is also obtained \cite{Afshar:2016kjj,Afshar:2016wfy, Afshar:2017okz} as the near horizon symmetry of BTZ black holes in AdS$_{3}$. In fact, it has been shown that two copies of $\widehat{W(0,0)}$ with appropriate fall-off condition with two independent central terms (two central terms for each copy of $W(0,0)$) is obtained as the near horizon symmetry algebras. In this picture, deformation of \bms\ to $W(0,0)$ can be interpreted as procedure which takes us from the near horizon to asymptotic region. On the other hand, deformation of the symmetry algebra may then be associated with symmetries of the perturbed field theory deformed by a relevant deformation. Recalling analysis of \cite{Bagchi:2012xr} one may then associate the former to an irrelevant deformation taking us to a UV fixed point and the latter as an IR deformation connecting us to the near horizon. To be more specific, consider a $3d$ flat space cosmology background. On the asymptotic null infinity we find \hbms\ ($\widehat{W(0,-1)}$) algebra, one can then model moving in toward the horizon inside the spacetime, where we find $\widehat{W(0,0)}$ algebra. As discussed in second part of section \ref{sec:4.1} the $b$ parameter of the $W(a,b)$ family can be associated with the scaling dimension of the operator ${\cal P}$ with which we have deformed the dual theory. Then moving in the $b$ direction corresponds to the RG flow for this operator. In short, the idea is to make a direct connection between algebra and (holographic) dual (conformal) field theory deformations. This viewpoint may prove fruitful in analyzing field theory duals of $3d$ flat spaces. 
    
    \item In a recent work \cite{Grumiller:2019fmp} the family algebra $W(0,b)$ is obtained as the near horizon symmetry algebra of $3d$ black holes. This again reinforces our previous interpretation. On the other hand, as $W(0,-1)$ is \bms\ one may interprets \bms\ as the near horizon symmetry algebra of $3d$ black holes. Deformation of \bms , as the near horizon symmetry algebra, into \w family algebra can be interpreted as procedure which circulates us among $b$ parameter space. In the mentioned work, the parameter $b$ is obtained as spin of the fields living  near the horizon. 
    
    \item In the context of GCA, the $W(0,b)$ algebra can be appeared as infinite dimensional version of the symmetries known as Galilei $\ell$-conformal symmetry. The $\ell$-conformal Galilei symmetry at first was discussed in \cite{negro1997nonrelativistic} and its realizations are studied in \cite{Gomis:2011dw, Galajinsky:2012rp, Chernyavsky:2016mnw, Andrzejewski:2013ioa, Chernyavsky:2015vav}. One can also show that this finite dimensional algebra can be enhanced to infinite dimensional as it is discussed in \cite{Galajinsky:2011iz,Martelli:2009uc}. One can also check that $\ell=1$ leads to GCA algebra (or equivalently the asymptotic symmetry of $3d$ flat spacetimes \cite{Bagchi:2010eg}) discussed in 
\cite{Bagchi:2009my} (the only difference is in the latter case generators of spatial rotations $SO(d)$ commute with infinite dimensional generators of translation. See the relation 3.5 in \cite{Bagchi:2009my}). The specific case of infinite dimensional version of the $\ell$-conformal Galilei  algebra is obtained when we restrict ourselves to $d=1$ \cite{aizawa2016aspects} which is exactly the same as $W(0,b)$ and can be interpreted as two-dimensional Galilean field theories \cite{Chen:2019hbj}. This is the next physical interpretation of deformations of \bms. In fact by deformation procedure \bms, as $W(0,-1)$ or $\ell=1$-conformal Galilei algebra is deformed into $\ell$-family algebra with arbitrary value for parameter $\ell$.   

    \item We explored that \hbms\ in its $\mathcal{J}\mathcal{J}$ part can be deformed by the $\nu$ term \eqref{JJ-JP-nu}. We showed that this deformation can be absorbed, leading to shift in $c_{JJ}$ central charge; or conversely, $c_{JJ}$ can be dropped by a shift in $\nu$. This shows that the deformation procedure and the central extension in some specific cases may play the same role. On the other hand, it has been shown that the $c_{JJ}$ can not be obtained through gravity computations. In the recent work \cite{Merbis:2019wgk} it has been shown that the central charge $c_{JJ}$ can be recovered as a quantum correction to $3d$ gravity. Thus,  one may interpret the deformation of $\mathcal{J}\mathcal{J}$ commutators of \hbms\ as a quantum correction. 
\end{itemize}

\section{Summary of the chapter}
We have summarized our discussion and results in this chapter in the Tables \hyperlink{table1}{4.1} and \hyperlink{table1}{4.2}. In \hyperlink{table1}{4.1} we have listed infinitesimal and formal deformations of \bms\ and \kac\ algebras and their central extensions \hbms\ and \kach. We showed that \bms\ can be deformed into either two copies of \wit\ algebra, or \w\ family algebra. We also showed that \kac\ algebra can be deformed onto the \w\ family algebra and a new algebra we called $\mathfrak{KM}(a,\nu)$. We explored deformations of centrally extended \hbms\ and \kach. We concluded \hbms\ can be deformed into two copies of \vir\ algebra or centrally extended \wh\ algebra when $c_{JP}=0$.

\begin{figure} [hbt!]
    	\includegraphics[width=1\textwidth]{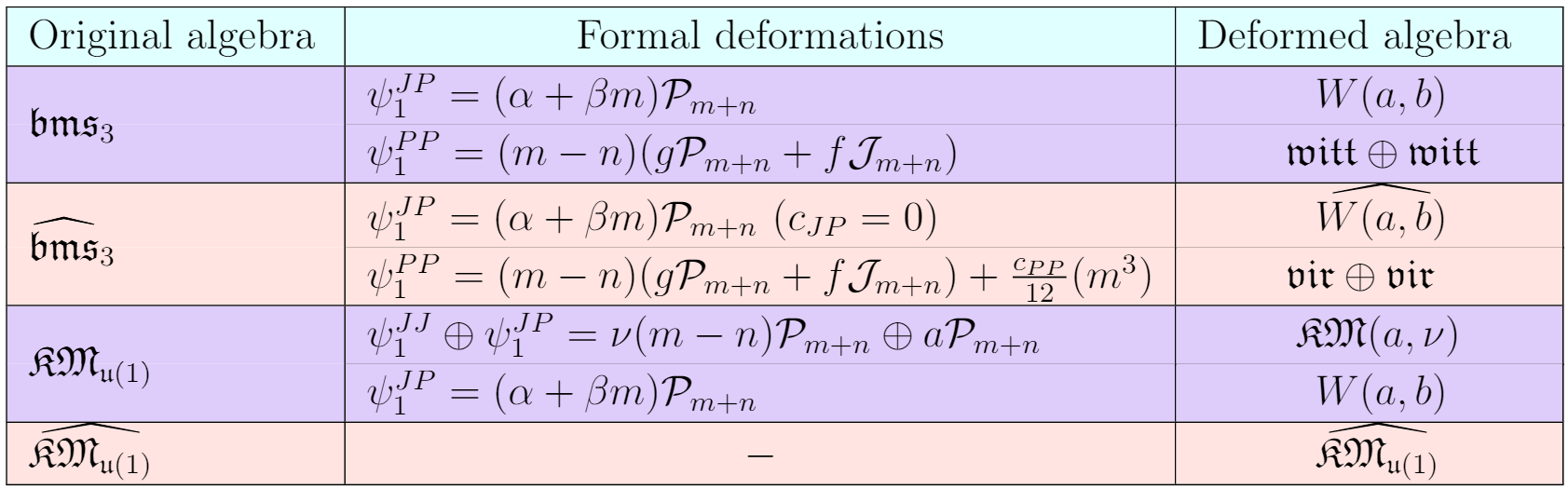}
    {\small \textbf{Table 4.1}: Infinitesimal and formal deformations and the 2-cocycles of \bms, \hbms, $\mathfrak{KM}_{\mathfrak{u}(1)}$ and $\widehat{\mathfrak{KM}_{\mathfrak{u}(1)}}$ algebras. This table is taken from \cite{Parsa:2018kys}.}
    \hypertarget{table1}{}
\end{figure}

We also examined deformations of \w\ and two \wit\ algebras and their central extensions \wh\ and two \vir. We showed that for generic values in parameters space of \w algebra, it just can be deformed into a new \w\ algebra with shifted parameters. In this sense we introduced a new notion of rigidity: The \w\ family algebra is rigid for generic value of its parameters in the sense that it only can be deformed into another $W(\tilde{a},\tilde{b})$ algebra with shifted parameters.

We also proved that two copies of \wit\ algebra and two copies of \vir\ algebra as its central extension are rigid in the sense that it can not be deformed to any new non isomorphic algebra. For details of computation we refer the reader to appendix \ref{appendix-B}.

\begin{figure}[hbt!]
		\includegraphics[width=1\textwidth]{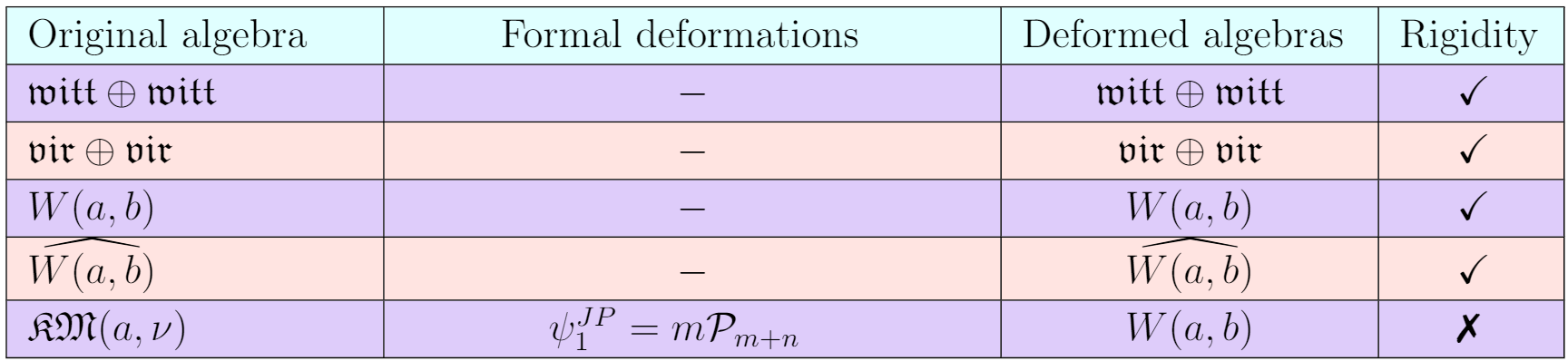}
		
	{\small \textbf{Table 4.2}: Rigidity of the algebras appearing as formal deformations of algebras in table \hyperlink{table1}{4.1}. We note that here by $W(a,b)$ or $\widehat{W(a,b)}$ algebra we mean family of $W$-algebras with generic values of the $a,b$ parameters. Deformations can move us within this family on the $a,b$ plane. {For specific points $(a=0,b=0)$ or $(a=0,b=-1)$ which are respectively corresponding to $\mathfrak{KM}_{u(1)}$ and \bms, the algebra admits non-trivial deformations. This table is taken from \cite{Parsa:2018kys}.}}
	\hypertarget{table2}{}
			\end{figure}

\chapter{Deformation of \texorpdfstring{$\mathfrak{bms}_{4}$}{bms4} algebra} \label{ch5}

There are many physical or mathematical motivations to carry out the stability analysis of the \bmsf. To state our main motivation, let us review some facts:
\begin{itemize}
    \item[(1)] As mentioned, the asymptotic symmetries of AdS$_3$ space is two Virasoro algebra at Brown-Henneaux central charge \cite{Brown:1986nw}. The seminal Brown-Henneaux analysis was a precursor of the celebrated AdS/CFT. 
    \item[(2)] This algebra upon the In\"on\"u contraction goes over to the \hbms\ \cite{Barnich:2012rz}, which is asymptotic symmetry group of 3d flat space. This contraction is geometrically the large AdS radius $\ell$ limit under which the AdS$_3$ goes over to 3d flat space.
    \item[(3)] Under a similar large radius limit, geometrically,   AdS$_d$ space yields a $d$ dimensional flat Minkowski space for any $d$. 
    \item[(4)] It has been  argued that the asymptotic symmetry group of AdS$_d,\ d>3$ is nothing more than isometries of the spacetime $\mathfrak{so}(d-1,2)$ \cite{Ashtekar:1984zz, Henneaux:1985ey, henneaux1985asymptotically,  Ashtekar:1999jx}.
    \item[(5)] Asymptotic symmetry algebra analysis depends very much on the choice of boundary falloff behavior on metric fluctuations and there could always be a question whether the results mentioned in item (4) above could some how be evaded by a more relaxed boundary condition.
    \item[(6)] The asymptotic symmetry group of 4d flat space is known to be the infinite dimensional \bmsf\ algebra.\footnote{
Note that the notion of BMS algebra, which includes superrotations plus supertranslations, does not seemingly exist in dimensions higher than 4  \cite{Kapec:2015vwa,Hollands:2016oma}.}
    \end{itemize}
    
    Therefore, it is natural to wonder if the \bmsf\ may come from contraction of an infinite dimensional ``asymptotic symmetry algebra of AdS$_4$''. In this work we  confirm these earlier results, in the sense that we show \bmsf\ algebra cannot be deformed into an algebra which has $\mathfrak{so}(3,2)$ as its subalgebra.
Our algebraic analysis and results has the advantage that it rules out possibility of existence of falloff conditions (for metric fluctuations) which may allow for a bigger symmetry algebra than $\mathfrak{so}(3,2)$ for AdS$_4$ asymptotic symmetry algebra.

\section{Deformation of \texorpdfstring{$\mathfrak{bms}_{4}$}{bms4} algebra}\label{sec:1.5}
In this section we consider deformations of  $\mathfrak{bms}_{4}$ defined in \eqref{bms4}.
As discussed the Hochschild-Serre factorization theorem is not applicable for infinite dimensional Lie algebras and working with them is more complicated than finite dimensional cases. Here, we first analyze possible deformations of $\mathfrak{bms}_{4}$ algebra by deforming  each commutation relation of $\mathfrak{bms}_{4}$ algebra separately. Of course one should check that in this way we do not miss any possible deformation which may involve more than one set of commutators. Finally, we study obstructions, which infinitesimal deformations yield formal deformations and  what are the rigid algebras obtained from deformations of $\mathfrak{bms}_{4}$. 

\paragraph{Remark} From now on, for simplicity, we will drop deformation parameters and will not write the deformation terms in terms of 2-cocycles. We only write the generators and their function coefficients which the form of the functions will be determined through Jacobi considerations.

\subsection*{ Deformation in the two Witt subalgebras}\label{Witt-deformations-sec}

The Witt algebra is known to be rigid and hence there is no 2-cocycle which deforms $[\mathcal{L}_{m},\mathcal{L}_{n}]$ by coefficients of ${\cal L}_p$ \cite{fialowski2012formal, schlichenmaier2014elementary}. Similarly, we cannot deform $\mathfrak{witt}\oplus\mathfrak{witt}$ algebra as we showed that in appendix  \ref{witt+witt-rigid}. Therefore, in this sector the only option is deforming $\mathfrak{witt}\oplus\mathfrak{witt}$ sector by coefficients of $T_{m,n}$ generators:\footnote{Here we are allowing for $d,\bar{d}$ to take arbitrary  values. However, as the discussions in the appendix \ref{appendix-C} indicates one could have fixed them by the requirement that the generators are functions on the $S^2$. This, however, does not affect our analysis and results in this subsection as all these deformations happen to be trivial and may be absorbed into redefinition of generators.} 
\begin{equation} \label{h-barh-H-deformation}
\begin{split}
    &[\mathcal{L}_{m},\mathcal{L}_{n}]=(m-n)\mathcal{L}_{m+n}+(m-n)\sum_{d,\bar{d}} h^{d,\bar{d}}(m,n)T_{m+n+d,\bar{d}},\\
    &[\bar{\mathcal{L}}_{m},\bar{\mathcal{L}}_{n}]=(m-n)\bar{\mathcal{L}}_{m+n}+(m-n)\sum_{d,\bar{d}} \bar{h}^{d,\bar{d}}(m,n)T_{d,m+n+\bar{d}},\\
    &[\mathcal{L}_{m},\bar{\mathcal{L}}_{n}]=\sum_{d,\bar{d}} H^{d,\bar{d}}(m,n)T_{m+d,n+\bar{d}},
\end{split}
\end{equation}
where $h,\bar{h}$ are symmetric and $H$ is arbitrary functions and $d$  and $\bar{d}$ are arbitrary numbers but we should note that the indices $d, \bar d$  should be equal with each other in all three relations.

The Jacobi $[\mathcal{L}_{m},[\mathcal{L}_{n},\mathcal{L}_{l}]]+cyclic\,\,\,permutation=0$ leads to 
\begin{equation} 
\begin{split}
    &\sum_{d,\bar{d}}\big((n-l)(m-n-l)h^{d,\bar{d}}(m,n+l)+(n-l)(\frac{m+1}{2}-n-l-d)h^{d,\bar{d}}(n,l)+\\
    &(l-m)(n-m-l)h^{d,\bar{d}}(n,m+l)+(l-m)(\frac{n+1}{2}-m-l-d)h^{d,\bar{d}}(l,m)+\\
    &(m-n)(l-m-n)h^{d,\bar{d}}(l,m+n)+(m-n)(\frac{l+1}{2}-m-n-d)h^{d,\bar{d}}(m,n)\big)T_{m+n+l+d,\bar{d}}=0,
\end{split}
\end{equation}
which its solution is $h^{d,\bar{d}}(m,n)=constant=h^{d,\bar{d}}$. The same relation can be obtained from the Jacobi $[\bar{\mathcal{L}}_{m},[\bar{\mathcal{L}}_{n},\bar{\mathcal{L}}_{l}]]+cyclic\,\,\,permutation=0$ for $\bar{h}^{d,\bar{d}}(m,n)$ which its solution is $\bar{h}(m,n)=constant=\bar{h}^{d,\bar{d}}$. 

The next two Jacobi identities to analyze are  $[\mathcal{L}_{m},[\mathcal{L}_{n},\bar{\mathcal{L}}_{l}]]+cyclic\,permutation=0$ and $[\bar{\mathcal{L}}_{m},[\bar{\mathcal{L}}_{n},\mathcal{L}_{l}]]+cyclic\,permutation=0$ which yield
\begin{align}
\label{H-h}   (\frac{m+1}{2}-n-d)H^{d,\bar{d}}(n,l)&-(\frac{n+1}{2}-m-{d})H^{d,\bar{d}}(m,l)-\cr
&-(m-n)\left(H^{d,\bar{d}}(m+n,l)-
(\frac{l+1}{2}-\bar{d})h^{d,\bar{d}}\right) =0,\\
\label{H-barh}   (\frac{n+1}{2}-m-\bar{d})H^{d,\bar{d}}(l,m)&-(\frac{m+1}{2}-n-\bar{d})H^{d,\bar{d}}(l,n)+\cr &+(m-n)\left(H^{d,\bar{d}}(l,m+n)+
    (\frac{l+1}{2}-{d})\bar{h}^{d,\bar{d}}\right)=0.
\end{align} 
Let us first consider the special case of $h^{d,\bar{d}}=\bar {h}^{d,\bar{d}}=0$. In this case one can easily see that 
\be\label{H0dd}
H_0^{d,\bar{d}}(m,n)=H^{d,\bar{d}}_0 (m+1-2d)(n+1-2\bar{d}),
\ee
where $H^{d,\bar{d}}_0$ is an arbitrary coefficient. 
Next, let us consider the generic case where $h^{d,\bar{d}},\bar {h}^{d,\bar{d}}\neq 0$. In this case the solution is of the form 
$$
H^{d,\bar{d}}(m,n)=H_0^{d,\bar{d}}(m,n)+\tilde{H}^{d,\bar{d}}(m,n),
$$
where $\tilde{H}^{d,\bar{d}}(m,n)$ is a solution to \eqref{H-h} and \eqref{H-barh} which vanishes as $h, \bar{h}=0$. The form of equations \eqref{H-h} and \eqref{H-barh} suggests that the most general solution is of the form $H(m,n)=a mn+ bm+c n+d$. The $mn$ term, however, can be absorbed in the ``homogeneous solution'' part $H_0^{d,\bar{d}}(m,n)$. 
Therefore, we consider the solution ansatz 
$\tilde{H}^{d,\bar{d}}(m,n)=Am+Bn+C$. Plugging this into \eqref{H-h} and \eqref{H-barh} yields
$A=\bar{h},\,\,\,B=-h,\,\,\,C=h(2\bar{d}-1)+\bar{h}(1-2d)$. To summarize, 
\begin{equation}
    H^{d,\bar{d}}(m,n)=H^{d,\bar{d}}_0 (m+1-2d)(n+1-2\bar{d})+\bar{h}(m+1-2d)-h(n+1-2\bar{d}).
\end{equation}


\paragraph{On triviality of these deformations.} One may examine whether the  $h,\bar h$ and $H(m,n)$ deformations are nontrivial or may be absorbed in the redefinition of generators. To this end let us consider redefined generators as
\begin{equation}\label{redefinitionXY}
\begin{split}
    &\Tilde{\mathcal{L}}_{m}\equiv \mathcal{L}_{m}+\sum_{d,\bar{d}}\ X^{d,\bar{d}}(m)T_{m+d,\bar{d}} ,\\
    &\Tilde{\bar{\mathcal{L}}}_{m}\equiv \bar{\mathcal{L}}_{m}+\sum_{d,\bar{d}}\ Y^{d,\bar{d}}(m)T_{d,m+\bar{d}},\\
    &\Tilde{T}_{m,n}\equiv T_{m,n},
\end{split}
\end{equation}
 where $X^{d,\bar{d}}(m)$ and $Y^{d,\bar{d}}(m)$ are  functions to be determined upon requirement of removing $h^{d,\bar{d}},{\bar h}^{d,\bar{d}}$ and $H^{d,\bar{d}}$ terms in \eqref{h-barh-H-deformation}. Removal of $h,\bar{h}$, i.e.  
requiring $[\Tilde{\mathcal{L}}_{m},\Tilde{\mathcal{L}}_{n}]=(m-n)\Tilde{\mathcal{L}}_{m+n}$ and performing the same analysis for $\Tilde{\bar{\mathcal{L}}}_{m}$ yields $X^{d,\bar{d}}(m)=A(m+1-2d)-2h^{d,\bar{d}},\ Y^{d,\bar{d}}(m)=\bar{A}(m+1-2\bar{d})- 2\bar{h}^{d,\bar{d}}$. 
Requiring $[\Tilde{\mathcal{L}}_{m},\bar{\tilde{\mathcal{L}}}_{n}]=0$, yields $A-\bar{A}=2H_0^{d,\bar{d}}$. One may take $A=-\bar{A}=H_0^{d,\bar{d}}$ and hence
\be\label{Redef-XY}
X^{d,\bar{d}}(m)=H_0^{d,\bar{d}}(m+1-2d)-2h^{d,\bar{d}},\qquad Y^{d,\bar{d}}(m)=-H_0^{d,\bar{d}}(m+1-2\bar{d})- 2\bar{h}^{d,\bar{d}}
\ee
would remove the deformations. Therefore, the deformations in \eqref{h-barh-H-deformation} are all trivial.


\subsection*{ Deformation of  \texorpdfstring{$[\mathcal{L},T]$}{LT} commutators}

The deformations in this sector could be with coefficients of $T_{m,n}$ or ${\cal L}_m$. We consider these two cases separately. 

\paragraph{With coefficients in \texorpdfstring{$T$}{T}.}
Consider the deformations of commutator of superrotations and supertranslations which is the fourth line in \eqref{bms4} without changing other commutators. To this end as in the previous subsection we add a 2-cocycle function:
\begin{equation} 
 [\mathcal{L}_{m},T_{p,q}]=(\frac{m+1}{2}-p)T_{p+m,q}+ K(m,p)T_{p+m,q}.\label{LT-Tdeformation}
\end{equation}
We have fixed the first index of $T$ on the right-hand-side to be $m+p$, see appendix \ref{appendix-C} for further discussions. Here, $K(m,n)$ is an unknown function to be determined through closure of algebra requirements.

To find the explicit form of function $K(m,n)$, there are two Jacobi identities to check. The first one is  $[\mathcal{L}_{m},[\mathcal{L}_{n},T_{p,q}]]+{cyclic\ permutations}=0$, which to first order in the deformation parameter yields
\begin{multline}
(\frac{n+1}{2} - p) K(m, p + n) + ( \frac{m+1}{2}-p-n) K(n, p) + (p-\frac{m+1}{2}) K(
   n, p+m) +\\  +
  (p+m - \frac{n+1}{2}) K(m,p)+ (n-m) K(m+n,p)=0.\label{eq-K}
\end{multline}
For $p,m=0$ we get
\begin{equation} 
 (\frac{n+1}{2}) (K(0, n) - K(0, 0))=0,\nonumber
\end{equation}
and hence
\begin{equation} 
  K(0, n)=\text{constant}.\label{first contraint on K}
\end{equation}
To solve \eqref{eq-K} we note that it is linear in $K$ and hence linear combination of any two solutions is also a solution. One may then check that 
\be\label{K-linear}
K(m,n)= \alpha+\beta m,
\ee
is a solution for any $\alpha,\beta$. This equation has solutions which involve higher powers of $m,n$. One may then examine a  degree $N$, i.e. $K(m,n)=\sum_{r=1}^N A_{rs} m^r n^{s}$ ansatz. At $N=2$ we obtain
a solution of the form
\begin{equation} 
 K(m,n)=\gamma m(\frac{m+1}{2}-n),\label{Ksolution1}
\end{equation}
where $\gamma$ is an arbitrary constant and we have added the 1/2 factor for later convenience. This solution, however, is a trivial deformation as it can be absorbed in rescaling of $T_{m,n}$ generators:
\be\label{T-redef}
T_{m,n}\to \Tilde{T}_{m,n}=M(m) T_{m,n}
\ee
with $M(m)=1+\gamma m$. In general, one can show that the most general solution to \eqref{eq-K} is 
\be\label{K-redef}
K(m,n)=(\frac{m+1}{2}-n)\left(\frac{M(m+n)}{M(n)}-1\right),
\ee
which again can be absorbed in a redefinition of $T$ of the form \eqref{T-redef}. (Note that here $K(m,n)$ is to be viewed as an infinitesimal function. In subsection \ref{Integrability-Wab} we discuss finite deformations.\footnote{A similar pattern was also found in the 3d case, the second part of subsection \ref{sec:4.1}.})

The other Jacobi to be checked is $[\bar{{\cal L}}_{m},[{\cal L}_{n},T_{p,q}]]+{cyclic\ permutations}=0$, which  does not yield a new constraint on $K(m,p)$. So the most general solutions of  \eqref{eq-K} are those we have derived. Deformations in  $[\bar{{\cal L}}_{n},T_{p,q}]$ can be analyzed in a similar manner, yielding similar results.

To summarize, the only non-trivial deformations are those generated by \eqref{K-linear} which yields $W(a,b;\bar{a},\bar{b})$ algebra defined through  commutation relations,
 \begin{equation} 
\begin{split}
 & [\mathcal{L}_{m},\mathcal{L}_{n}]=(m-n)\mathcal{L}_{m+n}, \\
 & [\bar{\mathcal{L}}_{m},\bar{\mathcal{L}}_{n}]=(m-n)\bar{\mathcal{L}}_{m+n},\\
 &[{\mathcal{L}}_{m},\bar{\mathcal{L}}_{n}]=0,\\
 &[\mathcal{L}_{m},T_{p,q}]=-(p+bm+a)T_{p+m,q},\\
 &[\bar{\mathcal{L}}_{n},T_{p,q}]=-(q+\bar{b}n+\bar{a})T_{p,q+n},\\
 &[T_{p,q},T_{r,s}]=0.
\end{split}\label{W4-algebra}
\end{equation}
The above is a 4d extension of the $3$ dimensional $W(a,b)$ family algebra, obtained in chapter $3$, which is a deformation of \bms\ \eqref{W(a,b)}.

One may wonder if the index of $T$ generator appearing in the RHS of $[{\cal L}_n,T_{p,q}]$ is restricted to be $T_{n+p,q}$. As discussed in the previous part around \eqref{h-barh-H-deformation}, the answer is no, at least as far as the Jacobi identity and the closure of the algebra is concerned. Explicitly, let us consider the following deformation,
\begin{equation} 
 [\mathcal{L}_{m},T_{p,q}]=(\frac{m+1}{2}-p)T_{p+m,q}+K(m,p)T_{p+m+d_0,q+\bar{d}_0},
\end{equation}
where $d_0,\bar{d}_0$ are two arbitrary constants.  The Jacobi identity then leads to
\begin{multline}\label{K-d}
(\frac{n+1}{2} - p) K(m, p + n) + ( \frac{m+1}{2}-p-n-d_0) K(n, p) + (p-\frac{m+1}{2}) K(
   n, p+m) +\\  +
  (p+m+d_0- \frac{n+1}{2}) K(m,p)+ (n-m) K(m+n,p)=0.
\end{multline}
It can be readily seen that for $d\neq 0$ the only solution to \eqref{K-d} is $K(m,n)=K=constant$. Nonetheless, this is trivial deformation, as it can be absorbed in the redefinition of $T_{m,n}$ as follows:
\[
\tilde{T}_{m,n}=\sum_{d} C_d T_{m+d,n}
\]
where $C_d$ are coefficients to be fixed upon request that $[{\cal L}_n,\tilde{T}_{p,q}]=(\frac{n+1}{2}-p)\tilde{T}_{n+p,q}$. This requirement yields $K C_{d-d_0}=d C_d$. 

The deformations discussed above and also those of \eqref{h-barh-H-deformation} involve an index structure which has a shift (by $d, \bar{d}$). In all of these cases, as we explicitly showed, such deformations are trivial ones and can be absorbed in the redefinition of generators. One may show that all such shifts in the indices are trivial deformations. This may be understood  geometrically recalling that the $\mathfrak{bms}_4$ algebra is associated with asymptotic symmetry algebra of 4d flat space and the generators are functions on the 2d celestial sphere \cite{Barnich:2011mi, Barnich:2011ct}. The deformations with the shifted indices are then an inner automorphism of the asymptotic symmetry generating diffeomorphisms; see appendix \ref{appendix-C} for more discussions on this point. Therefore, from now on we only consider deformations with appropriately fixed indices; we do not consider the extra shifts.

\paragraph{With coefficients in \texorpdfstring{$\mathcal{L}$}{L} and \texorpdfstring{$\bar{\mathcal{L}}$}{bar-L}.}

As the \bms\ case, we can consider deformations of the $[\mathcal{L},T]$ (or $[\mathcal{L},T]$) by $\mathcal{L}$ and $\bar{\mathcal{L}}$  terms:
\begin{equation} 
\begin{split}\label{LT-Ldeform}
 [\mathcal{L}_{m},T_{p,q}] &=(\frac{m+1}{2}-p)T_{p+m,q}+ f(m,p)\mathcal{L}_{p+m-1}\delta_{q,0}+ g(m)\bar{\mathcal{L}}_{q-1}\delta_{m+p,0}, \\
 [\bar{\mathcal{L}}_{n},T_{p,q}] &=(\frac{n+1}{2}-q)T_{p,n+q}+ \bar{f}(n)\mathcal{L}_{p-1}\delta_{n+q,0}+ \bar{g}(n,q)\bar{\mathcal{L}}_{n+q-1}\delta_{p,0},
\end{split}\end{equation}
where functions $f,g,\bar{f}$ and $\bar{g}$ are functions to be fixed upon the requirement of closure of the algebra. The index structure of the deformations has been fixed recalling the discussions in last part of the previous subsubsection. 

To find the explicit form of the functions we should consider three different Jacobi identities. The Jacobi $[\mathcal{L}_{m},[\bar{\mathcal{L}}_{n},T_{p,q}]]+{cyclic\ permutation}=0$ leads to one relation for each of $\mathcal{L}$ and $\bar{\mathcal{L}}$ coefficients as
\begin{equation} 
 \delta_{n+q,0}\big( (m-p+1)\bar{f}(n)+(p-\frac{m+1}{2})\bar{f}(n)+(\frac{n+1}{2}-q)f(m,p) \big)=0,\label{barf-f}
\end{equation}
and 
\begin{equation} 
 \delta_{m+p,0}\big( -(n-q+1)g(m)+(\frac{n+1}{2}-q)g(m)+(p-\frac{m+1}{2})\bar{g}(n,q)\big)=0.\label{barg-g}
\end{equation}
From the first relation we have 
\begin{equation} 
(\frac{m+1}{2})\bar{f}(n)=-(\frac{3n+1}{2})f(m,p), \label{barf-f1}
\end{equation}
which suggests that $f(m,n)=a(1+m)$ and $\bar{f}(n)=-a(3n+1)$ and similarly for $\bar{g}(m,n)$ and $g(n)$. 

The next Jacobi we should consider is $[\mathcal{L}_{m},[\mathcal{L}_{n},T_{p,q}]]+{cyclic\ permutation}=0$ which leads to, 
\begin{equation} 
    \delta_{m+p+n,0}\big((\frac{n+1}{2}-p)g(m)+(p-\frac{m+1}{2})g(n)+(n-m)g(m+n)\big) \bar{\mathcal{L}}_{q-1}=0,\label{LT-g}
\end{equation}
and
\begin{equation} 
\begin{split}
   &\delta_{q,0}\big((m-n-p+1)f(n,p)+(\frac{n+1}{2}-p)f(m,p+n)-(n-m-p+1)f(m, p)\\
   &+(p-\frac{m+1}{2})f(n,p+m)+(n-m)f(m+n,p)\big) \mathcal{L}_{m+p-1}=0.\label{LT-f}  
\end{split}
\end{equation}
One may readily verify that $g(n)=-a(1+3n)$ and $f(m,n)=a(1+m)$ respectively solve \eqref{LT-g} and \eqref{LT-f}, as also implied from our previous analysis. 

The last Jacobi we should consider is 
$[T_{p,q},[T_{r,s},\mathcal{L}_{m}]]+{cyclic\ permutation}=0$ which leads to
\begin{equation} 
\begin{split}
   &\delta_{s,0}(\frac{m+r}{2}-p)f(m,r)T_{m+r+p-1,q}+\delta_{r+m,0}(\frac{s}{2}-q)g(m)T_{p,q+s-1}+\\
   &\delta_{q,0}(r-\frac{m+p}{2})f(m,p)T_{m+r+p-1,s}+\delta_{p+m,0}(s-\frac{q}{2})g(m)T_{r,q+s-1}=0.\label{LTT-fg}
\end{split}
\end{equation}
There is a similar equation for $\bar{f}$ and $\bar{g}$ from $[T_{p,q},[T_{r,s},\bar{\mathcal{L}}_{m}]]+{cyclic\ permutation}=0$. The terms with coefficients $g$ should be equal to zero as they are coefficients of different $T_{m,n}$'s. In a similar way the $f(m,n)$ terms should be zero.
So, $[{\cal L}, T]$ cannot be deformed with coefficients in ${\cal L}, \bar{{\cal L}}$.

To summarize this subsection, \bmsf\ algebra can be deformed to a four parameter family of $W(a,b;\bar{a},\bar{b})$ algebras; where $\mathfrak{bms}_4=W(-\frac12,-\frac12;-\frac12,-\frac12)$. $W(a,b;\bar{a},\bar{b})$ for any value of parameters $a,b;\bar{a},\bar{b}$ 
share a $\mathfrak{witt}\oplus\mathfrak{witt}$ subalgebra spanned by 
${\cal L}_n$ and ${\bar{\cal L}}_n$. In section \ref{sec:5.3} we will study this family of algebras, its stability and  deformations in more detail.


\subsection*{ Deformations of commutator of  \texorpdfstring{$[T,T]$}{TT} }\label{ideal-bms}

$[T,T]$ commutator may be deformed in terms involving $T$ or  ${\cal L}$ and  ${\bar{\cal L}}$. In what follows we consider these cases separately. 

\paragraph{With coefficients in \texorpdfstring{$T$}{T}.}
As general case, we can consider the deformation of $[T,T]$ as
\begin{equation} 
 [T_{m,n},T_{p,q}]=G(m,n;p,q)T_{m+p,n+q},\label{TT-T}
\end{equation}
in which $G$ is an  antisymmetric function under the replacements $m\leftrightarrow p$ and $n\leftrightarrow q$. One must check the Jacobi identity $[\mathcal{L}_{r},[T_{m,n},T_{p,q}]]+cyclic\,\,permutations=0$, which yields
\begin{equation}\label{TTT}
    (p-\frac{r+1}{2})G(m,n;p+r,q)+(\frac{r+1}{2}-m)G(p,q;m+r,n)+
    (\frac{r+1}{2}-m-p)G(m,n;p,q)=0.
\end{equation}
For $r=0$, and recalling  $G(m,n;p,q)=-G(p,q;m,n)$, we get
\begin{equation} 
    \big((p-\frac{1}{2})-(\frac{1}{2}-m)+(\frac{1}{2}-m-p)\big)G(m,n;p,q)=0,
\end{equation}
which means that $G(m,n;p,q)=0$. In this way, we have shown that the ideal part of \bmsf\ cannot be deformed by terms with coefficients in $T$, when other commutators are untouched.

\paragraph{With coefficients in \texorpdfstring{$\mathcal{L}$}{L} and \texorpdfstring{$\bar{\mathcal{L}}$}{bar-L}.}\label{L-barL-ideal}
We next consider  deformation of the $[T,T]$ by terms with coefficients in $\mathcal{L}$ and $\bar{\mathcal{L}}$ as 
\begin{equation}\label{A-B--TTL}
 [T_{m,n},T_{p,q}]=A(m,n;p,q)\mathcal{L}_{m+p-1}+B(m,n;p,q)\bar{\mathcal{L}}_{n+q-1}
 \end{equation}
in which the coefficients $A(m,n;p,q)$ and $B(m,n;p,q)$ are antisymmetric under the replacement $m,n\leftrightarrow p,q$.\footnote{Note that the global part of this deformed algebra is always a deformation of 4d Poincar\'e which is  not (necessarily) AdS$_{4}$ algebra $\mathfrak{so}(3,2)$.} 
The index structure of ${\cal L}$ and ${\bar{\mathcal{L}}}$ in \eqref{A-B--TTL} is chosen recalling discussions of appendix \ref{appendix-C} that the generators may be viewed as fields (operators) on an $S^2$ in Poincar\'e coordinates. Moreover, the Jacobi $[\mathcal{L},[T,T]]+cyclic\,\,permutations=0$ restricts the index of $\mathcal{L}$ to be a linear function of the first indices of $T$. The same argument is obtained for $\bar{\mathcal{L}}$.

One should examine  the Jacobi identities $[\mathcal{L}_{r},[T_{m,n},T_{p,q}]]+cyclic\,\,permutations=0$ and $[T_{r,s},[T_{m,n},T_{p,q}]]+cyclic\,\,permutations=0$. From the first identity for the terms with coefficients $\bar{\mathcal{L}}$  one gets
\begin{equation} 
 \big((p-\frac{r+1}{2})B(m,n;p+r,q)+(\frac{r+1}{2}-m)B(p,q;m+r,n)\big)\bar{\mathcal{L}}_{n+q-1}=0.\label{B-eq0}
 \end{equation}
For $r=0$ and recalling the antisymmetry of $B$ function, we find $(m+p-1) B(m,n;p,q)=0$ and therefore, $B(m,n;p,q)=B_0(n,q) \delta_{m+p,1},$ where $B_0(n,q)=-B_0(q,n)$. Next we plug this form of $B$ back into \eqref{B-eq0} to obtain
$-2r B_0(n,q)\delta_{m+p+r,1}=0$ which implies $B_0(n,q)=0$ and hence $B$ should vanish.
A similar argument works for $A(m,n;p,q)$ when we consider the Jacobi $[\bar{\mathcal{L}}_{r},[T_{m,n},T_{p,q}]]+cyclic\,\,permutations=0$ and hence $A=B=0$.

To summarize this section, we have shown that $\mathfrak{bms}_4$ algebra admits non-trivial infinitesimal deformation only in $[{\cal L}, T]$ and $[\bar{\cal L}, T]$ parts of the algebra by coefficients in $T$. Therefore, the only allowed infinitesimal deformations of the $\mathfrak{bms}_4$ algebra is $W(a,b;\bar{a},\bar{b})$.

\section{Most general formal deformations of \texorpdfstring{$\mathfrak{bms}_4$}{BMS4}  algebra}\label{sec:4}

Here we complete the analysis of previous section by showing that (1) the infinitesimal deformations of \bmsf\ into $W(a,b;\bar{a},\bar{b})$ are also formal deformations and (2) there are no other deformations possible when we consider simultaneous deformations of two or more commutators. To this end, let us consider the schematic form of the most general deformations of  $\mathfrak{bms}_4$ in which all deformations are turned on simultaneously 
\begin{equation} 
\begin{split}
 & [\mathcal{L},\mathcal{L}]=\mathcal{L}+hT, \\
 & [\bar{\mathcal{L}},\bar{\mathcal{L}}]=\bar{\mathcal{L}}+\bar{h}T,\\
 &[{\mathcal{L}},\bar{\mathcal{L}}]=HT,\\
 &[\bar{\mathcal{L}},T]=T+\bar{K}T+\bar{f}\mathcal{L}+\bar{g}\bar{\mathcal{L}},\\
 &[\mathcal{L},T]=T+KT+f\mathcal{L}+g\bar{\mathcal{L}},\\
 &[T,T]=GT+A\bar{\mathcal{L}}+B\mathcal{L},
\end{split}\label{most-deform}
\end{equation}

The Jacobi $[\mathcal{L},[\mathcal{L},\mathcal{L}]]+cyclic\,\,permutations=0$ (and $[\bar{\mathcal{L}},[\bar{\mathcal{L}},\bar{\mathcal{L}}]]+cyclic\,\,permutations=0$) leads to some relations just for $h$ (and $\bar{h}$), in accord with the analysis of section \ref{Witt-deformations-sec},  have solution $h,\bar{h}=constant$ up to first order in functions (infinitesimal deformation). 
The Jacobi $[\mathcal{L},[\mathcal{L},\bar{\mathcal{L}}]]+cyclic\,\,permutations=0$ (and $[\bar{\mathcal{L}},[\bar{\mathcal{L}},\mathcal{L}]]+cyclic\,\,permutations=0$) up to first order just leads to \eqref{H-h} and \eqref{H-barh}; deformations in the $[{\cal L}, T], [\bar{\cal L}, T], [T, T]$ parts do not alter the equations on $h,\bar{h}$ and $H$. Therefore, there are no non-trivial deformations coming from this sector.

The Jacobi $[\mathcal{L},[\bar{\mathcal{L}},T]]+cyclic\,\,permutations=0$ up to first order just leads to the constraints \eqref{barf-f} and \eqref{barg-g}. 
  
The Jacobi $[\mathcal{L},[\mathcal{L},T]]+cyclic\,\,permutations=0$ (and $[\bar{\mathcal{L}},[\bar{\mathcal{L}},T]]+cyclic\,\,permutations=0$) up to first order just leads to the constraints \eqref{eq-K}, \eqref{LT-g} and \eqref{LT-f}.

The Jacobi $[T,[T,T]]+cyclic\,\,permutations=0$ does not lead to any constraints up to first order in the deformation parameter.

Finally, the Jacobi $[T,[T,\mathcal{L}]]+cyclic\,\,permutations=0$ (and $[T,[T,\bar{\mathcal{L}}]]+cyclic\,\,permutations=0$) up to first order leads to the following  three independent relations
\begin{equation} 
\begin{split}
  &\delta_{s,0}(\frac{m+r}{2}-p)f(m,r)T_{m+r+p-1,q}+\delta_{r+m,0}(\frac{s}{2}-q)g(m)T_{p,q+s-1}+\\
   &\delta_{q,0}(r-\frac{m+p}{2})f(m,p)T_{m+r+p-1,s}+\delta_{p+m,0}(s-\frac{q}{2})g(m)T_{r,q+s-1}+ \\
  &\big((r-\frac{m+1}{2})G(p,q;r+m,s)+(\frac{m+1}{2}-p)G(r,s;p+m,q)+\\
    &(\frac{m+1}{2}-(r+p+d))G(p,q;r,s)\big)T_{m+p+r,s+q}=0,
\end{split}\label{most-deform-LTT1}
\end{equation}
\begin{equation} 
\begin{split}
 & \big[(r-\frac{m+1}{2})A(p,q;m+r,s)+(\frac{m+1}{2}-p)A(r,s;p+m,q)+\\
  +& (m-p-r+1)A(p,q;r,s)\big]\mathcal{L}_{m+p+r-1}=0, 
\end{split}\label{most-deform-LTT2}
\end{equation}
and 
 \begin{equation} \label{most-deform-LTT3}
  \big((r-\frac{m+1}{2})B(p,q;m+r,s)+(\frac{m+1}{2}-p)B(r,s;p+m,q)\big)\bar{\mathcal{L}}_{s+q-1}=0. 
\end{equation}
As discussed, \eqref{most-deform-LTT3} leads to $B(p,q;r,s)=0$. A similar argument (analyzing \eqref{most-deform-LTT2} for $m=0$) yields $A(p,q;r,s)=0$. So we should just analyze \eqref{most-deform-LTT1}. Since $T_{m,n}$ for different $m,n$ are linearly independent, a careful analysis of the indices of $T$ generators in \eqref{most-deform-LTT1} reveals that  $f(m,n),g(m)$ and $G(m,n;p,q)$ should all vanish. To summarize, turning on deformations simultaneously, up to the first order, does not yield any new deformation other than $W(a,b;\bar{a},\bar{b})$ algebra. 

\subsection{Integrability, obstructions and formal deformation.}\label{Integrability-Wab}

We have shown in the previous section that the most general infinitesimal nontrivial deformation of \bmsf\ is $W(a,b;\bar{a},\bar{b})$. Now, we would like to explore integrability of these deformations and check if they are formal deformations. As in the case of \bms\ discussed in section \ref{sec:integrability-bms} one only needs to consider the relation
  \begin{multline}
(\frac{n+1}{2} - p) K(m, p + n) + ( \frac{m+1}{2}-p-n) K(n, p) + K(n,p)K(m,n+p)+ \\
+(p-\frac{m+1}{2}) K( n, p+m)   + (p+m - \frac{n+1}{2}) K(m,p)-K(m,p)K(n,p+m)+\\
+(n-m) K(m+n,p)=0.\label{formal-W}
\end{multline} 
which is satisfied by $K(m,n)=\alpha+\beta m$.\footnote{We note that the most general solution of \eqref{formal-W} is $\alpha+\beta m$ plus the solution given in \eqref{K-redef}. However, the latter is trivial deformation and may be absorbed in redefinition of $T$ as in \eqref{T-redef}. A similar pattern was also found in the 3d case, \emph{cf.} in second part of section \ref{sec:4.1}.} (Considering the Jacobi $[\bar{{\cal L}}_{m},[{\cal L}_{n},T_{p,q}]]+\text{cyclic permutations}\, \\ \text{permutations}=0$ does not change this result.) This means that the obtained infinitesimal deformation is integrable and can be extended to formal deformation.

Analysis of previous and this section may be summarized in the  the following  theorem: 
\begin{tcolorbox}[colback=red!3!white]
\paragraph{Theorem 5.1} {\it The most general formal deformation of \bmsf\ is $W(a,b;\bar{a},\bar{b})$ algebra}. 
\end{tcolorbox}


\section{On \texorpdfstring{$W(a,b;\bar{a},\bar{b})$}{W4} algebra, its subalgebras and  deformations}\label{sec:5.3}

We have introduced $W(a,b;\bar{a},\bar{b})$ which appears as formal deformation of \bmsf\ and here we would like to study this algebra a bit more. We first analyze its global subalgebras and then consider possible deformations of $W(a,b;\bar{a},\bar{b})$, particularly for special values of $a,b,\bar{a},\bar{b}$ parameters. Before starting we note that, as in the \bms\ and $W(a,b)$ algebra cases, \emph{cf.} section \ref{sec:4.1}, 
\begin{itemize}
\item $W(a,b;\bar{a}, \bar{b})$ and  $W(\bar{a}, \bar{b}; a,b)$ algebras are isomorphic. 
\item the range of $a, \bar{a}$ parameters may be limited to $[-1/2,1/2)$, as   $a=k+r, k\in\mathbb{Z}$  and $a=r$ cases can be related  by just a shift in the index of the associated $T_{m,n}$ generator, $T_{m,n}\to T_{m-k,n}$, and simiarly for the $\bar{a}$. 
\item $W(a,b;\bar{a},\bar{b})$ and $W(-a,b;\bar{a},\bar{b})$ algebras are isomorphic, as renaming ${\cal L}_m\to -{\cal L}_{-m}$ and $T_{p,q}\to T_{-p,q}$ relates these two algebras. Therefore, one may restrict the range of $a$ and $\bar{a}$ parameters to $[0,1/2]$. 
\end{itemize}

\subsection*{Subalgebras of \texorpdfstring{$W(a,b;\bar{a},\bar{b})$}{W4}}
 Irrespective of the values of $a,b;\bar{a},\bar{b}$ parameters, all $W(a,b;\bar{a},\bar{b})$ algerbas share a $\mathfrak{witt}\oplus\mathfrak{witt}$ subalgebra spanned by 
${\cal L}_n$ and $\bar{\mathcal{L}}_n$. This subalgebra in turn has a Lorentz subalgebra $\mathfrak{so}(3,1)=\mathfrak{sl}(2\mathbb{R})_L\oplus \mathfrak{sl}(2\mathbb{R})_R$ spanned by $\mathcal{L}_{0},\mathcal{L}_{\pm1},\ \bar{\mathcal{L}}_{0}, \bar{\mathcal{L}}_{\pm 1}$. Depending on the values of the four parameters, some $T_{m,n}$ generators may also be a part of this global subalgebra, e.g. as discussed in the previous section for $\mathfrak{bms}_4=W(-\frac12,-\frac12;-\frac12,-\frac12)$,  $T_{0,0}, T_{0,1}, T_{1,0}, T_{1,1}$ are the other four generators which turn the global subalgebra to 4d Poincar\'e $\mathfrak{iso}(3,1)$. To verify which of $T_{m,n}$ appear in the global subalgebra, we consider the relevant commutators:
\begin{equation} 
\begin{split}
 &[\mathcal{L}_{+1},T_{p,q}]=-(p+b+a)T_{p+1,q}, \\
 & [\mathcal{L}_{0},T_{p,q}]=-(p+a)T_{p,q},\\
 &[\mathcal{L}_{-1},T_{p,q}]=-(p-b+a)T_{p-1,q},
\end{split}\label{L-Tglobal}
\end{equation}
and 
\begin{equation} 
\begin{split}
 & [\bar{\mathcal{L}}_{+1},T_{p,q}]=-(q+\bar{b}+\bar{a})T_{p,q+1}, \\
 & [\bar{\mathcal{L}}_{0},T_{p,q}]=-(q+\bar{a})T_{p,q},\\
 &[\bar{\mathcal{L}}_{-1},T_{p,q}]=-(q-\bar{b}+\bar{a})T_{p,q-1}.
\end{split}\label{barL-Tglobal}
\end{equation}
The above commutations close off for finite number of $T_{m,n}$ generators only for three cases, $a=b=0$, $a=0, b=-1$ and $a=b=-1/2$, and similarly for $\bar{a},\bar{b}$. For generic values of $a,b$ other than these two cases the global part (finite subalgebra) of $W(a,b;\bar{a},\bar{b})$ is just $\mathfrak{so}(3,1)$. We have therefore, nine independent special cases for which the global part of the algebra is bigger than the Lorentz algebra $\mathfrak{so}(3,1)$:
\begin{enumerate}
\item $W(-1/2,-1/2;-1/2,-1/2)$ which is nothing but the \bmsf\ and its global part is 4d Poincar\'e algebra $\mathfrak{iso}(3,1)$.
\item $W(0,0;0,0)$, where $T_{0,0}$ falls into the global part and hence we have $\mathfrak{so}(3,1)\oplus \mathfrak{u}(1)$ global algebra.
\item $W(0,-1;0,0)$ (or $W(0,0;0,-1)$), where $T_{r,0},\ r=0,\pm 1$ (or $T_{0,r}$) fall into the global part and the global subalgebra is $\mathfrak{iso}(2,1)\oplus \mathfrak{sl}(2,\mathbb{R})_R$ (or $\mathfrak{iso}(2,1)\oplus \mathfrak{sl}(2,\mathbb{R})_L$). Generators of this algebra may be represented in usual Minkowski tensors: ${J}_{\mu\nu}, F^+_{\mu\nu}$, where $F^+_{\mu\nu}$ is a self-dual (anti-self-dual) anti-symmetric object, representing the $T_{r,0}$:
\be\begin{split}
[J_{\mu\nu}, J_{\alpha\beta}] &= i(\eta_{\mu \alpha} J_{\nu \beta}+\eta_{\beta \mu}J_{\alpha \nu}-\eta_{\nu \alpha}J_{\mu \beta}-\eta_{\beta \nu}J_{\alpha \mu}),\\
[{J}_{\mu\nu}, F^+_{\alpha\beta}] &={i(\eta_{\mu \alpha} F^+_{\nu \beta}+\eta_{\beta \mu}F^+_{\alpha\nu}-\eta_{\nu \alpha}F^+_{\mu \beta}-\eta_{\beta \nu}F^+_{\alpha\mu}),} \\
[F^+_{\mu\nu},F^+_{\alpha\beta}] &= 0. 
\end{split}
\ee
\item $W(0,-1;0,-1)$ where the global subalgebra involves nine generators $T_{r\bar{s}}, r,\bar{s}=0,\pm 1$. This 15 dimensional algebra which generated by ${\cal L}_r, {\bar{\mathcal{L}}}_{\bar s}, T_{r\bar{s}}$, $r,\bar{s}=0,\pm 1$. These generators may be gathered in a traceless Lorentz two-tensor $K_{\mu\nu}$, where its antisymmetric part is Lorentz generators $J_{\mu\nu}$ and its symmetric part captures $T_{r\bar{s}}$ and satisfy the algebra
\begin{equation}
   i [K_{\mu\nu}, K_{\alpha\beta}]=\eta_{\mu\beta}K_{\alpha\nu}-\eta_{\alpha\nu}K_{\mu\beta}.
\end{equation}
The above algebra is $\mathfrak{so}(3,1)\inplus_{ad} \mathfrak{T}$ subalgebra where $\mathfrak{T}$ denotes the ideal part which is spanned by $T_{r\bar{s}}$; the $T_{r\bar{s}}$ are in the bi-adjoint of the $\mathfrak{sl}(2,\mathbb{R})_L\oplus \mathfrak{sl}(2,\mathbb{R})_R$.

\item $W(0,0;-1/2,-1/2)$ (or $W(-1/2,-1/2;0,0)$), where $T_{0,\alpha}, \alpha=0,1$ (or $T_{\alpha,0}$) are also in the global algebra which is eight dimensional. This global algebra is  $\mathfrak{sl}(2,\mathbb{R})\oplus \mathfrak{sch}_2$ algebra where $\mathfrak{sch}_2$ denotes the $2d$ Schr\"{o}dinger algebra without central element.

\item $W(0,-1;-1/2,-1/2)$ (or $W(-1/2,-1/2;0,-1)$) where its global subalgebra is $\mathfrak{so}(3,1) $\\$\inplus_{ad} \mathfrak{T}_{\alpha r}$, where $\mathfrak{T}_{\alpha r}$ is  spanned by $T_{\alpha r}, \alpha=0,1, \ r=0,\pm 1$
which are in the vector and spinor representation of $\mathfrak{sl}(2,\mathbb{R})_L\oplus \mathfrak{sl}(2,\mathbb{R})_R$ subalgebra.

\end{enumerate}

\paragraph{Some other infinite dimensional subalgebras of $W(a,b;\bar{a},\bar{b})$.} Besides the above examples, one may consider other infinite dimensional subalgebras of $W(a,b;\bar{a},\bar{b})$. The simplest of these subalgebras are $\mathfrak{witt}\oplus\mathfrak{witt}$ or $W(a,b)_p$ algebras generated by ${\cal L}_n, T_{m, p}$ where $p$ is a fixed (but arbitrary) number.  As a subalgebra of $W(a,b; 0, -1)$ one may consider the one generated by ${\cal L}_n, T_{m, r},\bar{\mathcal{L}}_r\  (r=0,\pm 1)$. In special case $a=b=0$ one may view this as a $\mathfrak{u}(1)^3$ Kac-Moody algebra\footnote{{By $\mathfrak{u}(1)^3$ Kac-Moody algebra we mean semi direct sum of Virasoro algebra with $\mathfrak{u}(1)^3$  Kac-Moody current algebra.}} where the three currents fall into triplet representation  of the $\mathfrak{sl}(2,\mathbb{R})_R$ generated by  ${\bar{\mathcal{ L}}}_r$. This latter may be viewed as an ``internal symmetry'' of the Kac-Moody part. 

\subsection*{Deformations of generic \texorpdfstring{$W(a,b;\bar{a},\bar{b})$}{W4}  algebra}

As discussed formal deformations of \bmsf\ algebra yields the four parameter family $W(a,b;\bar{a},\bar{b})$ algebra. As such, one expects this algebra to be rigid. However, a more careful look into the analysis of previous section also reveals that deformations may move us in the $a,b;\bar{a},\bar{b}$ plane. This is very similar to the 3d example of $W(a,b)$ discussed in chapter \ref{ch4}, and is what we will explore more explicitly in this section. Here we use $W(a,b;\bar{a},\bar{b})$ in two different meanings, which will hopefully be clear from the context: (1) $W(a,b;\bar{a},\bar{b})$ for a given (but generic) value of the four parameter. This latter denotes a specific algebra; (2) $W(a,b;\bar{a},\bar{b})$ as ``family'' of algebras for different values of the parameters. 

While the $W(a,b;\bar{a},\bar{b})$ family is expected it to be rigid, as in the 3d example of $W(a,b)$ discussed in chapter \ref{ch4}, there could be special values of parameters where one can deform the algebra to other families of algebras. These special points, as we will see, correspond to cases with larger global part discussed in the previous subsection.
To make a formal analysis of $W(a,b;\bar{a},\bar{b})$ as in the previous section we consider all possible deformations of its commutators and check the closure conditions.

\paragraph{Deformations of $\mathfrak{witt}\oplus \mathfrak{witt}$ part.}\label{witt-witt-T} One can deform this sector of  $W(a,b;\bar{a},\bar{b})$ algebra as
\begin{equation} 
\begin{split}
     &[\mathcal{L}_{m},\mathcal{L}_{n}]=(m-n)\mathcal{L}_{m+n}+(m-n)\sum_{d,\bar{d}} h^{d,\bar{d}}(m,n)T_{m+n+d,\bar{d}},\\
    &[\bar{\mathcal{L}}_{m},\bar{\mathcal{L}}_{n}]=(m-n)\bar{\mathcal{L}}_{m+n}+(m-n)\sum_{d,\bar{d}} \bar{h}^{d,\bar{d}}(m,n)T_{d,m+n+\bar{d}},\\
    &[\mathcal{L}_{m},\bar{\mathcal{L}}_{n}]=\sum_{d,\bar{d}} H^{d,\bar{d}}(m,n)T_{m+d,n+\bar{d}},\label{w-w-W4-deform}
\end{split}
\end{equation}
in which $h^{d,\bar{d}}(m,n),\bar{h}^{d,\bar{d}}(m,n)$ are symmetric and $H^{d,\bar{d}}(m,n)$ an arbitrary functions. 
As the first step, one considers the Jacobi $[\mathcal{L}_{m},[\mathcal{L}_{n},\mathcal{L}_{l}]]+cyclic\,\,permutations=0$ which leads to 
\begin{multline}
        \sum_{d,\bar{d}}\{\big((n-l)(m-n-l)h^{d,\bar{d}}(m,n+l)+(n-l)(-bm-a-n-l-d)h^{d,\bar{d}}(n,l)+\\
    +(l-m)(n-m-l)h^{d,\bar{d}}(n,m+l)+(l-m)(-bn-a-m-l-d)h^{d,\bar{d}}(l,m)+\\
    +(m-n)(-bl-a-m-n-d)h^{d,\bar{d}}(m,n)+\\+(m-n)(l-m-n)h^{d,\bar{d}}(l,m+n)\big)T_{m+n+l+d,\bar{d}}\}=0,\label{h-a,b}
\end{multline} 

and the same relation for $\bar{h}^{d,\bar{d}}(m,n)$. The only solution to the above equations for generic  $a,b$ is $h^{d,\bar{d}}(m,n)=constant$ (and $\bar{h}^{d,\bar{d}}(m,n)=constant$).

The next two Jacobi identities to analyze are $[\mathcal{L}_{m},[\mathcal{L}_{n},\bar{\mathcal{L}}_{l}]]+cyclic\,\,permutations=0$ and 
$[\bar{\mathcal{L}}_{m},[\bar{\mathcal{L}}_{n},\mathcal{L}_{l}]]+cyclic\,\,permutations=0$, which yield
\begin{align}
     &\bigg(-(a+bm+n+d)H^{d,\bar{d}}(n,l)+(a+bn+m+d)H^{d,\bar{d}}(m,l)+(n-m)H^{d,\bar{d}}(m+n,l)+\cr
     &(n-m)(\bar{a}+\bar{b}l+\bar{d})h^{d,\bar{d}}(m,n)\bigg)T_{m+n+d,l+\bar{d}}=0,\label{H-h-ab}\\ \;\;\;\cr
   &\bigg((\bar{a}+\bar{b}m+n+\bar{d})H^{d,\bar{d}}(l,n)-(\bar{a}+\bar{b}n+m+\bar{d})H^{d,\bar{d}}(l,m)+(m-n)H^{d,\bar{d}}(l,m+n)+\cr
   &(n-m)(a+bl+d)\bar{h}^{d,\bar{d}}(m,n)\bigg)T_{l+d,m+n+\bar{d}}=0.\label{H-barh-ab}  
\end{align}
As in the \bmsf\  case one can obtain the most general form of the $H^{d,\bar{d}}(m,n)$ as
\begin{equation}
    H^{d,\bar{d}}(m,n)=H^{d,\bar{d}}_0 (bm+a+d)(\bar{b}n+\bar{a}+\bar{d})+\frac{\bar{h}}{\bar{b}}(bm+a+d)-\frac{h}{b}(\bar{b}n+\bar{a}+\bar{d}).\label{general-H-w4}
\end{equation}

One can then show that for generic $a,b,\bar{a},\bar{b}$ through the redefinitions \eqref{Redef-XY} in which $X^{d,\bar{d}}(m)$, $Y^{d,\bar{d}}(m)$ are changed to
\begin{equation}
    X^{d,\bar{d}}(m)=H_0^{d,\bar{d}}(bm+a+d)+\frac{h^{d,\bar{d}}}{b},\,\,\, Y^{d,\bar{d}}(m)=-H_0^{d,\bar{d}}(\bar{b}m+\bar{a}+\bar{d})+\frac{\bar{h}^{d,\bar{d}}}{\bar{b}},\label{Redef-XY-w4}
\end{equation}
these deformations can be reabsorbed and hence they are all trivial. However, as we see for the special case of $b=0$ or $\bar{b}=0$ these redefinitions are not well-defined and hence there remains a non-trivial deformation for  $W(a,b;0,0)$, $W(0,0;\bar{a},\bar{b})$ and  $W(0,0;0,0)$ cases which we discuss below separately.

 \paragraph{Deformations of \texorpdfstring{$[\mathcal{L},T]$}{LT-W} and \texorpdfstring{$[\bar{\mathcal{L}},T]$}{barLT-W} commutators.}\label{K-f-g-w4}
The most general deformations in this sector of  $W(a,b;\bar{a},\bar{b})$ algebra is 
\begin{equation} 
\begin{split}
    &[\mathcal{L}_{m},T_{p,q}]=-(a+bm+p)T_{p+m,q}+K(m,p)T_{p+m,q}++\eta f(m,p)\mathcal{L}_{p+m-1}\delta_{q,0}+\sigma g(m)\bar{\mathcal{L}}_{q-1}\delta_{m+p,0},\\
 &[\bar{\mathcal{L}}_{n},T_{p,q}]=-(\bar{a}+\bar{b}n+q)T_{p,q+n}+\bar{K}(n,q)T_{p,n+q}+\bar{\eta} \bar{f}(n)\mathcal{L}_{p-1}\delta_{n+q,0}+\bar{\sigma} \bar{g}(n,q)\bar{\mathcal{L}}_{n+q-1}\delta_{p,0},\label{L-barL-T-W4-deform}
\end{split}
\end{equation}
in which $K,\bar{K},f,g,\bar{f}$ and $\bar{g}$ are arbitrary functions. 
As the first step, we considers the Jacobi $[\mathcal{L}_{m},[\mathcal{L}_{n},T_{p,q}]]+cyclic\,\,permutations=0$, leading to 
\begin{multline}\label{K-w4}
(-a-bn - p) K(m, p + n) + ( -a-bm-p-n) K(n, p) + (p+a+bm) K(
   n, p+m) +\\  +
  (p+m+a+bn) K(m,p)+ (n-m) K(m+n,p)=0.
\end{multline}
and the same relation for $\bar{K}(m,n)$. One can solve the latter to get $K(m,n)=\alpha+\beta m$ and similar result for $\bar{K}(m,n)$. From previous Jacobi one also obtains two other relations for $f(m,n)$ and $g(m,n)$ as 
\begin{equation} 
    \delta_{m+p+n,0}\big((-a-bn-p)g(m)+(p+a+bm)g(n)+(n-m)g(m+n)\big) \bar{\mathcal{L}}_{q-1}=0,\label{LT-g-w4}
\end{equation}
and
\begin{equation} 
\begin{split}
   &\delta_{q,0}\big((m-n-p+1)f(n,p)+(-a-bn-p)f(m,p+n)-(n-m-p+1)f(m, p)\\
   &+(p+a+bm)f(n,p+m)+(n-m)f(m+n,p)\big) \mathcal{L}_{m+p-1}=0,\label{LT-f-w4}  
\end{split}
\end{equation}
and similar relations for $\bar{f}(m,n)$ and $\bar{g}(m,n)$ can be obtained from the Jacobi $[\bar{\mathcal{L}}_{m},[\bar{\mathcal{L}}_{n},T_{p,q}]]+cyclic\,\,permutations=0$.

On the other hand, the Jacobi $[\mathcal{L}_{m},[\bar{\mathcal{L}}_{n},T_{p,q}]]+cyclic\,\,permutations=0$ leads to one relation for each of $\mathcal{L}$ and $\bar{\mathcal{L}}$ coefficients as
\begin{equation} 
 \delta_{n+q,0}\big( (m-p+1)\bar{f}(n)+(p+a+bm)\bar{f}(n)+(-a-bn-q)f(m,p) \big)=0,\label{barf-f-w4}
\end{equation}
and 
\begin{equation} 
 \delta_{m+p,0}\big( -(n-q+1)g(m)+(-a-bn-q)g(m)+(p+a+bm)\bar{g}(n,q)\big)=0.\label{barg-g-w4}
\end{equation}
From the first relation we have 
\begin{equation} 
(m(b+1)+a+1)\bar{f}(n)=(a+n(b-1))f(m,p), \label{barf-f1-w4}
\end{equation}
which suggests that $f(m,n)=c(m(b+1)+a+1)$ and $\bar{f}(n)=-c(a+n(b-1))$ where $c=constant$ and similarly for $\bar{g}(m,n)$ and $g(n)$ which is in agreement with \eqref{barf-f1-w4} obtained for $a=b=\frac{-1}{2}$ case.

The last Jacobi we should consider is 
$[T_{p,q},[T_{r,s},\mathcal{L}_{m}]]+\text{cyclic permutation}=0$ which leads to
\begin{equation} 
\begin{split}
   &\delta_{s,0}(a+b(m+r-1)+p)f(m,r)T_{m+r+p-1,q}+\delta_{r+m,0}(\bar{a}+\bar{b}(s-1)+q)g(m)T_{p,q+s-1}-\\
   &\delta_{q,0}(a+b(m+p-1)+r)f(m,p)T_{m+r+p-1,s}-\delta_{p+m,0}(\bar{a}+\bar{b}(q-1)+s)g(m)T_{r,q+s-1}=0.\label{LTT-fg-w4}
\end{split}
\end{equation}
There is a similar equation for $\bar{f}$ and $\bar{g}$ from $[T_{p,q},[T_{r,s},\bar{\mathcal{L}}_{m}]]+\text{cyclic permutations}=0$. The terms with coefficients $g$ should be equal to zero as they are coefficients of different $T_{m,n}$'s. In a similar way $f(m,n)$ terms should be zero.

\paragraph*{Deformations of  \texorpdfstring{$W(a,b;\bar{a},\bar{b})$}{W4}'s ideal part: }
\begin{equation} 
 i[T_{m,n},T_{p,q}]=G(m,n;p,q)T_{m+p,n+q}+A(m,n;p,q)\mathcal{L}_{m+p-1}+B(m,n;p,q)\bar{\mathcal{L}}_{n+q-1},\label{TT-L}
\end{equation}
in which $G,A$ and $B$ are  antisymmetric functions under  $m\leftrightarrow p$ and $n\leftrightarrow q$. The Jacobi identities $[\mathcal{L}_{r},[T_{m,n},T_{p,q}]]+cyclic\,\,permutations=0$ and  $[\bar{\mathcal{L}}_{r},[T_{m,n},T_{p,q}]]+cyclic\,\,permutations=0$ yield two relations for $G$,
\begin{equation}\label{TTT1-w4}
    (p+a+br)G(m,n;p+r,q)-(a+br+m)G(p,q;m+r,n)-
    (a+br+m+p)G(m,n;p,q)=0,
\end{equation}
and
 \begin{equation} 
   (\bar{a}+\bar{b}r+q)G(m,n;p,q+r)-
    (\bar{a}+\bar{b}r+n)G(p,q;m,n+r)-(\bar{a}+\bar{b}r+n+q)G(m,n;p,q)=0,\label{TTT2-w4}
 \end{equation}
and two relations for $A$,
\begin{multline}\label{A-eq0-w4}
     \big((r-m-p+1)A(m,n;p,q)+(p+a+br)A(m,n;p+r,q)+\\
     (m+a+br)A(m+r,n;p,q)\big)\mathcal{L}_{m+p+r-1}=0, 
\end{multline} 
and 
\begin{equation} 
 \big((q+\bar{a}+\bar{b}r)A(m,n;p,q+r)+(-\bar{a}-\bar{b}r-n)A(p,q;m,n+r)\big)\mathcal{L}_{m+p-1}=0.\label{A-eq1-w4}
 \end{equation}
The same relation can be obtained from these two Jacobi for $B(m,n;p,q)$.

One can check that our argument in section \ref{ideal-bms} also works for deformations of $W(a,b;\bar{a},\bar{b})$'s in this sector, in the sense that for generic $a,b,\bar{a}$ and $\bar{b}$ it does not admit any non-trivial deformation in its ideal part. 

\paragraph{The most general deformations of \texorpdfstring{$W(a,b;\bar{a},\bar{b})$}{W4} algebra.}
When we consider the most general deformations of $W(a,b;\bar{a},\bar{b})$ algebra simultaneously as in the \bmsf\ case of \eqref{most-deform}, one can verify that except one case, similar to \bmsf\ case, all Jacobi identities lead to the relations obtained in the above. Similarly the \bmsf\ case, the only case we must study is the Jacobi $[T,[T,\mathcal{L}]]+cyclic\,\,permutations=0$ and $[T,[T,\bar{\mathcal{L}}]]+cyclic\,\,permutations=0$ which leads to the relations \eqref{A-eq0-w4}, \eqref{A-eq1-w4} and their analogue for $B$ and sum of two relations \eqref{TTT1-w4} and \eqref{LTT-fg-w4}. One can then show that these two latter are independently equal to zero. In this way the most general deformations of $W(a,b;\bar{a},\bar{b})$ are restricted to that induces by $K(m,n)$ and takes the $W(a,b;\bar{a},\bar{b})$ to $W({a'},{b'};{\bar{a}'},{\bar{b}'})$ with shifted parameters.  

The following theorem summarizes our above discussion as: 
\begin{tcolorbox}[colback=red!3!white]
\paragraph{Theorem 5.2} {\it The family of $W(a,b;\bar{a},\bar{b})$ algebra for generic values of the four parameters is stable (rigid) algebra}.
\end{tcolorbox}


\subsection*{Deformations of special \texorpdfstring{$W$}{W} algebras}

As discussed while in generic points of the parameter space of $W$ algebra they are rigid, there are special points in the parameter space where the algebra is not rigid and may admit other deformations. In this subsection we discuss such special cases.

\paragraph{Deformations of \texorpdfstring{$W(a,b;0,0)$}{u(1)Kac-Moody}.}
As discussed for $a=b=0$ case the $h$-deformations (\emph{cf}. \eqref{w-w-W4-deform}) become non-trivial and cannot be absorbed into a redefinition of generators. One can go through the Jacobi's of the previous subsection and verify allowed deformations. Here we do not repeat the analysis and just present the final result. The most general \emph{new} non-trivial deformation of  $W(a,b;0,0)$ is:
\begin{equation} 
\begin{split}
    &[\mathcal{L}_{m},\mathcal{L}_{n}]=(m-n)\mathcal{L}_{m+n},\\
    &[\bar{\mathcal{L}}_{m},\bar{\mathcal{L}}_{n}]=(m-n)\bar{\mathcal{L}}_{m+n}+\bar{\nu} \ (m-n) T_{0,m+n},\\
    &[\mathcal{L}_{m},\bar{\mathcal{L}}_{n}]=0,\\
    &[\mathcal{L}_{m},T_{p,q}]=-(p+bm+a)T_{p+m,q},\\
 &[\bar{\mathcal{L}}_{n},T_{p,q}]=-(q)T_{p,q+n},\\
 &[T_{m,n},T_{p,q}]=0.\label{u(1)-u(1)}
\end{split}
\end{equation}
Some comments are in order:
\begin{itemize}
    \item Among the ${\bar h}^{d,\bar{d}}$ deformations only $\bar{d}=0$ terms remain. The others are still trivial or do not satisfy Jacobi identity.
    \item The possibility of moving in the $(a,b, \bar{a},\bar{b})$ parameter space via the deformation still exists and the above is the new non-trivial deformation which did not exist for $\bar{a},\bar{b}\neq 0$ point. That is, the $\bar{\nu}$ deformation and moving in $\bar{a},\bar{b}$ plane are not mutually inclusive.
\item   $[T_{m,n},T_{p,q}]$ cannot be deformed. Once again, $[\mathcal{L}_{m},T_{p,q}],\ [\bar{\mathcal{L}}_{n},T_{p,q}]$ can only be deformed into 
those of $W(a,b;\bar{a},\bar{b})$ algebra, but in that case we need to set $\nu=0$.

\item Here we are considering $a,b\neq 0$ case. The special case of $W(0,0;0,0)$ will be discussed next.

\end{itemize}

\paragraph{The special case of $W(0,0;0,0)$.} In this case one can deform the algebra into a generic point in the  $(a,b, \bar{a},\bar{b})$ parameter space or alternatively deform the $[\mathcal{L}_{m},\mathcal{L}_{n}]$ (or $[\bar{\mathcal{L}}_{m},\bar{\mathcal{L}}_{n}]$) with the coefficients in $T_{m+n,0}$ (or in $T_{0,m+n})$ while moving in $\bar{a},\bar{b}$ (or $a,b$) plane; or turn on the two $[\mathcal{L}_{m},\mathcal{L}_{n}]$ and $[\bar{\mathcal{L}}_{m},\bar{\mathcal{L}}_{n}]$ deformations without moving in the parameter space. These possibilities are mutually exclusive and the last deformation has the explicit form:
\begin{equation} 
\begin{split}
    &[\mathcal{L}_{m},\mathcal{L}_{n}]=(m-n)\mathcal{L}_{m+n}+\nu\ (m-n)T_{m+n,0},\\
    &[\bar{\mathcal{L}}_{m},\bar{\mathcal{L}}_{n}]=(m-n)\bar{\mathcal{L}}_{m+n}+\bar{\nu}\ (m-n)T_{0,m+n},\\
    &[\mathcal{L}_{m},\bar{\mathcal{L}}_{n}]=H_{0}(\alpha+\beta m)(\bar{\alpha}+\bar{\beta}n)T_{m,n},\\
    &[\mathcal{L}_{m},T_{p,q}]=(-p)T_{p+m,q},\\
 &[\bar{\mathcal{L}}_{n},T_{p,q}]=(-q)T_{p,q+n},\\
 &[T_{m,n},T_{p,q}]=0.\label{W4(00)-deformed}
\end{split}
\end{equation}
 
To summarize, $W(0,0;0,0)$ can be deformed  to the four parameter family $W(a,b;\bar{a},\bar{b})$ or (exclusively) by three independent formal deformations parametrized by $\nu, \bar\nu, H_0$. Also, if one chooses to move in $(\bar{a},\bar{b})$ or $(a,b)$ planes, we are left with $\bar{\nu}$ or $\nu$ deformations, respectively, \emph{cf.} the  $W(a,b;0,0)$ case discussed above.

\paragraph{Deformations of \texorpdfstring{$W(0,-1;0,0)$}{u(1)Kac-Moody-bms3}. }
The next special case we study is $W(0,-1;0,0)$:
\begin{equation} 
\begin{split}
    &[\mathcal{L}_{m},\mathcal{L}_{n}]=(m-n)\mathcal{L}_{m+n},\\
    &[\bar{\mathcal{L}}_{m},\bar{\mathcal{L}}_{n}]=(m-n)\bar{\mathcal{L}}_{m+n},\\
    &[\mathcal{L}_{m},\bar{\mathcal{L}}_{n}]=0,\\
    &[\mathcal{L}_{m},T_{p,q}]=(m-p)T_{p+m,q},\\
 &[\bar{\mathcal{L}}_{n},T_{p,q}]=(-q)T_{p,q+n},\\
 &[T_{m,n},T_{p,q}]=0,\label{u(1)-bms3}
\end{split}
\end{equation}
which is obtained from \eqref{W4-algebra} when we put $a=\bar{a}=\bar{b}=0$ and $b=-1$. As discussed $W(0,-1;0,0)$ can be considered as combination of a Virasoro-Kac-Moody algebra (on the right sector) and a \bms\ (on the left sector). The global part of $W(0,-1;0,0)$ is $\mathfrak{iso}(2,1)\oplus \mathfrak{sl}(2,\mathbb{R})$ spanned respectively by $\mathcal{L}_{r}, T_{r,0}$ and $\bar{\mathcal{L}}_{\bar{r}}$, $ r,{\bar{r}}=\pm1,0$.

Inspired by the discussions of previous subsection and recalling the results of chapter \ref{ch4} for deformations of \bms, we expect to be able to turn on a $T_{0,m+n}$ deformation in $[\bar{\mathcal{L}}_{m},\bar{\mathcal{L}}_{n}]$ and also be able to deform the ideal part; this is of course besides deforming by moving in the $(a,b;\bar{a},\bar{b})$ parameter space. The two allowed deformations are hence
\begin{equation} 
\begin{split}
 & [\mathcal{L}_{m},\mathcal{L}_{n}]=(m-n)\mathcal{L}_{m+n}, \\
 & [\bar{\mathcal{L}}_{m},\bar{\mathcal{L}}_{n}]=(m-n)\bar{\mathcal{L}}_{m+n}+\bar{\nu}\ (m-n)T_{0,m+n},\\
 &[{\mathcal{L}}_{m},\bar{\mathcal{L}}_{n}]=0,\\
 &[\mathcal{L}_{m},T_{p,q}]=(m-p)T_{p+m,q},\\
 &[\bar{\mathcal{L}}_{n},T_{p,q}]=(-q)T_{p,q+n},\\
 &[T_{m,n},T_{p,q}]=\varepsilon(m-p)T_{m+p,n+q}.
\end{split}\label{W4(0-100)-deformed}
\end{equation}
One can readily verify that the above deformations are formal. 

For the $\bar{\nu}=0$ the global part of the above algebra is $\mathfrak{sl}(2,\mathbb{R})\oplus \mathfrak{sl}(2,\mathbb{R}) \oplus \mathfrak{sl}(2,\mathbb{R})$, which is generated by ${\cal L}_r-\frac{1}{\varepsilon}T_{r,0}, \frac{1}{\varepsilon}T_{r,0}, \bar{\mathcal{L}}_{\bar r}$, $(r,\bar{r}=0,\pm 1)$. The first two $\mathfrak{sl}(2,\mathbb{R})$'s may be viewed as $\mathfrak{so}(2,2)$, the isometry of  AdS$_{3}$ space and the last $\mathfrak{sl}(2,\mathbb{R})$ factor as an ``internal symmetry'' for the AdS$_3$ space. One can then observe that the above algebra for $\bar{\nu}=0$ has a $\mathfrak{witt}\oplus \mathfrak{witt}\oplus \mathfrak{witt}$ subalgebra generated by ${\cal L}_n-\frac{1}{\varepsilon}T_{n,0}, \frac{1}{\varepsilon}T_{n,0}, \bar{\mathcal{L}}_{n}$. This latter has been studied as deformation of Maxwell algebra \cite{salgado2014so,gomis2009deformations, Concha:2018zeb,Caroca:2017onr}. Moreover, this algebra has a $\mathfrak{witt}\oplus$ Virasoro-Kac-Moody subalgebra generated by  ${\cal L}_n,  \bar{\mathcal{L}}_{n}, T_{0,n}$.

To summarize, one can deform $W(0,-1;0,0)$ to a generic $W(a,b;\bar{a},\bar{b})$ by moving in $(a,b;\bar{a},\bar{b})$ plane, or by turning on  $\nu$ or, exclusively, $\varepsilon(m-p)$ deformations.\footnote{Note that while $\nu$ and $\varepsilon$ deformations can be turned on simultaneously at infinitesimal level, they cannot both be elevated to a formal deformation at the same time.} If we move in $(\bar{a},\bar{b})$ plane we cannot turn on $\nu$ or $\varepsilon$ deformations and if we move in $(a,b)$ plane we cannot turn on $\varepsilon$ while $\nu$ deformation is possible. 

\section{Deformation of \texorpdfstring{$\widehat{W}(a,b;\bar{a},\bar{b})$}{hat-w4} centrally extended algebra }\label{sec:6}

Global central extensions (which in short are usually called central extensions) of an algebra $\mathfrak{g}$ are classified by its second real cohomology ${\cal H}^2(\mathfrak{g};\mathbb{R})$. Central extensions may hence be viewed as a special class of deformations  given by Gel'fand-Fucks 2-cocycles \cite{gel1969cohomologies}. One may show, following analysis of \cite{Barnich:2011ct}, that $W(a,b;\bar{a},\bar{b})$ for generic values of the parameters admits two independent central extensions which may be associated with deforming the algebra by two independent unit elements added to the algebra. The centrally extended $W$-algebra which will be  denoted by $\widehat{W}(a,b;\bar{a},\bar{b})$ is given by:
\begin{equation} 
\begin{split}
    &[\mathcal{L}_{m},\mathcal{L}_{n}]=(m-n)\mathcal{L}_{m+n}+\frac{C_{\mathcal{L}}}{12}m^{3}\delta_{m+n,0},\\
    &[\bar{\mathcal{L}}_{m},\bar{\mathcal{L}}_{n}]=(m-n)\bar{\mathcal{L}}_{m+n}+\frac{C_{\bar{\mathcal{L}}}}{12}m^{3}\delta_{m+n,0},\\
    &[\mathcal{L}_{m},\bar{\mathcal{L}}_{n}]=0,\\
    &[\mathcal{L}_{m},T_{p,q}]=-(p+bm+a)\ T_{p+m,q},\\
 &[\bar{\mathcal{L}}_{n},T_{p,q}]=-(q+\bar{b}n+\bar{a})\ T_{p,q+n},\\
 &[T_{m,n},T_{p,q}]=0,\label{central-w4}
\end{split}
\end{equation}
where $C_{\mathcal{L}}$ and $C_{\bar{\mathcal{L}}}$ are central charges. Algebras with different nonzero values of central charges $C_{\mathcal{L}}$ and $C_{\bar{\mathcal{L}}}$, are cohomologous, i.e they are isomorphic to each other. As we can see for the generic $a,b,\bar{a}$ and $\bar{b}$, in the centrally extended $W(a,b;\bar{a},\bar{b})$ algebra the $\mathfrak{witt}$ subalgebras are turned to two Virasoro algebras and other commutators are untouched. 

As in the case of $W(a,b)$ algebras discussed in chapter \ref{ch4} and \cite{gao2011low}, there may be special points in the $(a,b,\bar{a},\bar{b})$ parameter space  which admit other central terms.
 As the first case let us consider \bmsf$=W(-\frac{1}{2},-\frac{1}{2},-\frac{1}{2},-\frac{1}{2})$.  This algebra  admits only two independent central terms in its two $\mathfrak{witt}$ algebras \cite{Barnich:2011ct}.

Recalling that in the $W(a,b)$ case there is a possibility of a central extension in ${\cal L}, T$ sector $a=0, b=1$ case \cite{gao2011low}, we examine if there is  a possibility of central extension in $[\mathcal{L}_{m},T_{p,q}]$ or $[\bar{\mathcal{L}}_{m},T_{p,q}]$ for specific values of $a,b,\bar{a}, \bar{b}$ parameters. Explicitly, consider

{
\begin{equation}
    [\mathcal{L}_{m},T_{p,q}]=-(a+bm+p)T_{m+p,q}+f(m,p)\delta_{q,0},
\end{equation}
and 
\begin{equation}
[\bar{\mathcal{L}}_{n},T_{p,q}]=-(\bar{a}+\bar{b}n+p)T_{p,n+q}+\bar{f}(n,q)\delta_{p,0},
\end{equation}
where $f(m,n)$ and $\bar{f}(m,n)$ are arbitrary functions. $[\mathcal{L}_{m},[\mathcal{L}_{n},T_{p,q}]]+cyclic\,\,permutations=0$ and $[\bar{\mathcal{L}}_{m},[\bar{\mathcal{L}}_{n},T_{p,q}]]+cyclic\,\,permutations=0$ Jacobi relations lead to 
\begin{equation}
   -(a+bn+p)f(m,p+n)+(a+bm+p)f(n,p+m)+(n-m)f(m+n,p)=0,\label{central-ab-LT}
\end{equation}
and 
\begin{equation}
   -(\bar{a}+\bar{b}n+p)\bar{f}(m,p+n)+(\bar{a}+\bar{b}m+p)\bar{f}(n,p+m)+(n-m)\bar{f}(m+n,p)=0.\label{central-bar-ab-LT}
\end{equation}
Let us now examine the equations for $m$ (or $n$)  or $p$ equal to zero. For $p=0$ \eqref{central-ab-LT}  yields
\begin{equation}
   -(a+bn)f(m,n)+(a+bm)f(n,m)+(n-m)f(m+n,0)=0.
\end{equation}
We can consider two cases, either $f$ is symmetric $f(m,n)=f(n,m)$, or it is antisymmetric $f(m,n)=-f(n,m)$. For the symmetric case we get $ bf(m,n)=f(m+n,0)$. So, either $b=0$ which leads to $f(m+n,0)=0$, or $b\neq 0$ for which $f(m,n)=\frac{1}{b}f(m+n,0)=F(m+n)$. For the former one learns that the only solution is $f(m,n)=m^{2}\delta_{m+n,0}$ as we expected for the Virasoro-Kac-Moody algebra, \emph{cf.} section \ref{sec:4.1}. Plugging the solution $f(m,n)=F(m+n)$ into \eqref{central-ab-LT} restricts us to $b=1$ and arbitrary $a$.

For the antisymmetric $f(m,n)$, putting $p=0$ in \eqref{central-ab-LT} one finds 
\begin{equation}
   (2a+b(m+n))f(n,m)=(m-n)f(m+n,0),
\end{equation}
which for $m=0$ yields, either $b=-1, a=0$ or $f(n,0)=0$. One may then examine these two possibilities in \eqref{central-ab-LT} to find that $b=-1, a=0$ and that $f(m,n)=m^3\delta_{m+n,0}$ is the only non-trivial solution. This latter is of course expected recalling the \bms\ analysis of chapter \ref{ch4}. Similar analysis goes through for \eqref{central-bar-ab-LT}.} To summarize so far, the $f,\bar{f}$ type central terms are allowed only for 
    $a=b=0, f(m,n)=m^2\delta_{m+n,0}$; 
$b=1, a=arbitrary, f(m,n)=F(m+n)$; 
$b=-1, a=0, f(m,n)=m^3\delta_{m+n,0}$.

We should now verify if the central terms in special points  obtained above satisfy the Jacobi 
$[\mathcal{L}_{r},[\bar{\mathcal{L}}_{s},T_{m,n}]]+cyclic\,\,permutations=0$.
For generic $a,b,\bar{a}$ and $\bar{b}$ one obtains
\begin{equation}
    (a+br+p)\bar{f}(s,q)-(\bar{a}+\bar{b}s+q)f(r,p)=0. \label{C-f-barf-LLT}
\end{equation}
For the special point $a=b=\bar{a}=\bar{b}=0$ corresponding to  $W(0,0;0,0)$ algebra,
we obtained $\bar{f}(m,n)=f(m,n)=m^{2}\delta_{m+n,0}$  which does not fulfill \eqref{C-f-barf-LLT}. The next point is $W(0,-1;0,0)$, which can be viewed as combination of \bms\ and Virasoro-Kac-Moody, we obtained $f(m,n)=m^{3}\delta_{m+n,0}$ and $\bar{f}(m,n)=m^{2}\delta_{m+n,0}$. This too, does not satisfy \eqref{C-f-barf-LLT}. One therefore concludes that $W(0,0;0,0)$ and $W(0,-1;0,0)$ do not admit a central term in its $[\mathcal{L}_{m},T_{p,q}]$ or $[\bar{\mathcal{L}}_{m},T_{p,q}]$ commutators. But one can consider $W(0,1;0,0)$ and $W(0,1;0,-1)$ which admit central terms as $\bar{f}(m,n)=m^{2}\delta_{m+n,0}$ and $\bar{f}(m,n)=m^{3}\delta_{m+n,0}$ respectively. For the special case $W(0,1;0,1)$ one learns that it admits two independent central terms as $f(m,n)=(\alpha +\beta m)\delta_{m+n,0}$ and $\bar{f}(m,n)=(\bar{\alpha}+\bar{\beta }m)\delta_{m+n,0}$.

The specific case $W(0,0;a,b)$ algebra which can be seen as combination of  $W(a,b)$ and Virasoro-Kac-Moody algebras may also admit central terms in its ideal part which can be parametrized as
\begin{equation}
    [T_{m,n},T_{p,q}]= c_1m\delta_{m+p,0}\delta_{n,q}+c_2n\delta_{m,p}\delta_{n+q,0}.
\end{equation}
This structure guarantees antisymmetry w.r.t. $m\leftrightarrow p$ and $n\leftrightarrow q$. 
The Jacobi $[\mathcal{L}_{r},[T_{m,n},T_{p,q}]]+cyclic\,\,permuutation=0$ leads to
\begin{equation}
   -p\big((\bar{a}+\bar{b}r+q)\delta_{n,q+r}+(\bar{a}+\bar{b}r+n)\delta_{q,n+r}\big)=0,
\end{equation}
which cannot be satisfied for any values of $\bar{a}$ and $\bar{b}$. Therefore, one finds that $W(0,0;\bar{a},\bar{b})$ admits only central terms in its two Witt algebras (unless the case $\bar{a}=0$ and $\bar{b}=1$), just as \bmsf\ and $W(a,b;\bar{a},\bar{b})$ for generic $a,b,\bar{a}$ and $\bar{b}$. 

\subsection*{Most general deformations of centrally extended \texorpdfstring{$W(a,b;\bar{a},\bar{b})$}{w4} algebra}

Let us now consider the most general deformations of the $
\widehat{W}(a,b;\bar{a},\bar{b})$ algebra. As we checked in the previous part $
\widehat{W}(a,b;\bar{a},\bar{b})$ admits two nontrivial central charges in its two Witt algebras. We start with this algebra and schematically consider its most general deformations as
\begin{equation} 
\begin{split}
 & [\mathcal{L},\mathcal{L}]=\mathcal{L}+C_{\mathcal{L}}+hT+X, \\
 & [\bar{\mathcal{L}},\bar{\mathcal{L}}]=\bar{\mathcal{L}}+\bar{C}_{\bar{\mathcal{L}}}+\bar{h}T+\bar{X},\\
 &[{\mathcal{L}},\bar{\mathcal{L}}]=HT+U,\\
 &[\mathcal{L},T]=T+KT+fL+g\bar{L}+Y,\\
 &[\bar{\mathcal{L}},T]=T+\bar{K}T+\bar{f}L+\bar{g}\bar{L}+\bar{Y},\\
 &[T,T]=GT+AL+B\bar{L}+Z.
\end{split}\label{most general-hatW4-algebra}
\end{equation}
in which we dropped the indices of generators and arguments of functions to simplify the notation. The functions $X$, $U$, $Y$ and $Z$ and their analogues in barred sector, are deformations by terms with coefficients in unit generators (central terms). As we have seen in the case of $W(a,b)$ algebra which has been studied in previous chapter \ref{ch4}, the Jacobi analysis leads to two different family of relations. The first family is exactly the same as relations analyzed in the previous section for the functions related to non-central parts while the second set of relations include linear combinations of central and non-central functions. 
In this way, for the generic values of $a,b,\bar{a}$ and $\bar{b}$  we obtained that the only nontrivial solutions are $K(m,n)=\alpha+\beta m$ and $\bar{K}(m,n)=\bar{\alpha}+\bar{\beta}m$ which lead to a new $\widehat{W}(a,b;\bar{a},\bar{b})$ with shifts in the four parameters and none of the other functions, central or non-central, cannot be turned on. So, one concludes that the family of $\widehat{W}(a,b;\bar{a},\bar{b})$ algebras are rigid (or stable), in the sense that it can just be deformed to another $\widehat{W}(a,b;\bar{a},\bar{b})$ in the same family. As we discussed, however, there are  special points in the space $(a,b,\bar{a},\bar{b})$ which can admit some other deformations. Now, we are going to review the results of the most general 
deformations of $\widehat{W}(a,b;\bar{a},\bar{b})$ in special points.  
 
 \subsection{The most general deformations of \texorpdfstring{$\widehat{\mathfrak{bms}}_4$}{hbmsf} algebra}
As the first case we consider $\widehat{W}(-\frac{1}{2},-\frac{1}{2};-\frac{1}{2},-\frac{1}{2})$ which is the central extension of \bmsf\ denoted by \hbmsf . As mentioned, from the first family of relations, the only nontrivial functions are $K(m,n)=\alpha+\beta m$ and $\bar{K}(m,n)=\bar{\alpha}+\bar{\beta}m$ and other non-central functions are zero, as we have shown in section \ref{sec:4}. The Jacobi $[\mathcal{L},[\bar{\mathcal{L}},T]]+cyclic\,\,permutation=0$ leads to $Y=\bar{Y}=0$. The Jacobi $[\bar{\mathcal{L}},[\mathcal{L},\mathcal{L}]]+cyclic\,\,permutation=0$ leads to $U=0$. The Jacobi $[\mathcal{L},[T,T]]+cyclic\,\,permutation=0$ leads to $Z=0$. The Jacobi $[\mathcal{L},[\mathcal{L},\mathcal{L}]]+cyclic\,\,permutation=0$ leads to $X(m)=m^{3}\delta_{m+n,0}$ and a similar result for $\bar{X}(m)$ which just lead to a shift of $C_{\mathcal{L}}$ and $C_{\bar{\mathcal{L}}}$. However, as mentioned, the algebras with different values of $C_{\mathcal{L}}$ and $C_{\bar{\mathcal{L}}}$ are cohomologous and isomorphic to each other. In this way, we have found that the most general deformations of \hbmsf$=\widehat{W}(-\frac{1}{2},-\frac{1}{2};-\frac{1}{2},-\frac{1}{2})$  is $\widehat{W}(a,b;\bar{a},\bar{b})$ with the shifted parameters. 

\subsection*{Most general deformations of specific points in \texorpdfstring{$(a,b;\bar{a},\bar{b})$}{abab} space}
The next special point is $\widehat{W}(0,0;0,0)$. We showed in the  previous section that non-central functions with nontrivial solutions are  $h(m,n)=constant$, $K(m,n)=\alpha+\beta m$ and the same result for $\bar{h}(m,n)$ and $\bar{K}(m,n)$ and $H(m,n)=H_{0}(\alpha+\beta m)(\bar{\alpha}+\bar{\beta }n)$. 
The Jacobi $[\mathcal{L},[\bar{\mathcal{L}},T]]+cyclic\,\,permutation=0$ leads to $Y=\bar{Y}=0$. The Jacobi $[\bar{\mathcal{L}},[\mathcal{L},\mathcal{L}]]+cyclic\,\,permutation=0$ leads also to $U=0$. The Jacobi $[\mathcal{L},[T,T]]+cyclic\,\,permutation=0$ leads to $Z=0$. The Jacobi $[\mathcal{L},[\mathcal{L},\mathcal{L}]]+cyclic\,\,permutation=0$ leads to $X(m)=m^{3}\delta_{m+n,0}$ and a similar result for $\bar{X}(m)$ which just shifts the value of $C_{\mathcal{L}}$ and $C_{\bar{\mathcal{L}}}$.
We hence recover exactly the same results as the infinitesimal single deformation case. 

The next special point is $\widehat{W}(0,-1;0,0)$. Non-central functions with nontrivial solutions are  $\bar{h}(m,n)=constant$, $K(m,n)=\alpha+\beta m$, $\bar{K}(m,n)=\bar{\alpha}+\bar{\beta} n$ and $G(m,n;p,q)=(m-p)$ and other functions are zero. The Jacobi $[\mathcal{L},[\bar{\mathcal{L}},T]]+cyclic\,\,permutation=0$ leads to $Y=\bar{Y}=0$. The Jacobi $[\bar{\mathcal{L}},[\mathcal{L},\mathcal{L}]]+cyclic\,\,permutation=0$ leads to $U=0$. The Jacobi $[\mathcal{L},[\mathcal{L},\mathcal{L}]]+cyclic\,\,permutation=0$ leads to $X(m)=m^{3}\delta_{m+n,0}$ and a similar result for $\bar{X}(m)$. Finally, when we deform the ideal part as $[T_{m,n},T_{p,q}]=(m-p)T_{m+p,n+q}+Z(m,n;p,q)$, the Jacobi $[\mathcal{L}_{m},[T_{p,q},T_{r,s}]]+cyclic\,\,permutation=0$ leads to $Z(m,n;p,q)=0$. 

One may consider the specific subalgebra of $\widehat{W}(0,-1;0,0)$ generated by $\mathcal{L}^1_n\equiv T_{n,0}, \mathcal{L}^2_n=\mathcal{L}_n- T_{n,0}, \mathcal{L}^3_n=\bar{\mathcal{L}}_n$ for which, $$[\mathcal{L}^a_n,\mathcal{L}^b_n]=\delta^{ab}\left( (n-m) \mathcal{L}^a_{n+m}+\frac{1}{12} C^a n^3\delta_{m+n,0}\right),\quad a,b=1,2,3.$$ For this subalgebra,  $Z(m,n;p,q)=m^{3}\delta_{m+p,0}\delta_{n,0}\delta_{q,0}$ is an allowed central extension (as well as a formal deformation), denoted by $C^a$ in the above algebra. This subalgebra hence admits three central charges and is therefore, direct sum of three Virasoro algebras.

The next specific point is $\widehat{W}(0,1;0,0)$.  Non-central functions with nontrivial solutions are  $\bar{h}(m,n)=constant$, $K(m,n)=\alpha+\beta m$ and $\bar{K}(m,n)=\bar{\alpha}+\bar{\beta}n$. The Jacobi $[\bar{\mathcal{L}},[\mathcal{L},\mathcal{L}]]+cyclic\,\,permutation=0$ leads to $U=0$. The Jacobi $[\mathcal{L},[\mathcal{L},\mathcal{L}]]+cyclic\,\,permutation=0$ leads to $X(m)=m^{3}\delta_{m+n,0}$ and a similar result  for $\bar{X}(m)$. The Jacobi $[\mathcal{L},[T,T]]+cyclic\,\,permutation=0$ leads to $Z=0$. The Jacobi $[\mathcal{L},[\bar{\mathcal{L}},T]]+cyclic\,\,permutation=0$ yields $Y=0$. Unlike the previous cases, however, we obtained $\bar{Y}(m)=m^{2}\delta_{m+n,0}$. Although the latter is a formal deformation, when we turn on $\bar{K}$, $\bar{h}$ and $\bar{Y}$ simultaneously they cannot satisfy the Jacobi in higher order in deformation parameter, so they should be considered as independent formal deformations.

The next specific case is $\widehat{W}(0,1;0,-1)$.  Non-central functions with nontrivial solutions are  $K(m,n)=\alpha+\beta m$ and $\bar{K}(m,n)=\bar{\alpha}+\bar{\beta}n$. The Jacobi $[\bar{\mathcal{L}},[\mathcal{L},\mathcal{L}]]+cyclic\,\,permutation=0$ leads to $U=0$,  $[\mathcal{L},[\mathcal{L},\mathcal{L}]]+cyclic\,\,permutation=0$  to $X(m)=m^{3}\delta_{m+n,0}$, and a similar result for $\bar{X}(m)$. The Jacobi $[\mathcal{L},[\bar{\mathcal{L}},T]]+cyclic\,\,permutation=0$ leads to $Y=0$ and $\bar{Y}(m)=m^{3}\delta_{m+n,0}$. As in the previous cases one can show that the functions $\bar{K}$ and $\bar{Y}$ cannot be turned on simultaneously, implying that we do not have formal deformations induced with both $\bar{K}$ and $\bar{Y}$. 

The next specific point is $\widehat{W}(0,1;0,1)$. Non-central functions with nontrivial solutions are  $K(m,n)=\alpha+\beta m$ and $\bar{K}(m,n)=\bar{\alpha}+\bar{\beta}n$.The Jacobi $[\bar{\mathcal{L}},[\mathcal{L},\mathcal{L}]]+cyclic\,\,permutation=0$ leads to $U=0$, and $[\mathcal{L},[\mathcal{L},\mathcal{L}]]+cyclic\,\,permutation=0$ to $X(m)=m^{3}\delta_{m+n,0}$, and similarly for $\bar{X}(m)$. The Jacobi $[\mathcal{L},[\bar{\mathcal{L}},T]]+cyclic\,\,permutation=0$ leads to $Y(m)=(\tilde{\alpha} +\tilde{\beta} m)\delta_{m+n,0}$ and  $\bar{Y}(m)=(\tilde{\bar{\alpha}} +\tilde{\bar{\beta}} m)\delta_{m+n,0}$. One can show that the functions $K, Y$ and $\bar{K}, \bar{Y}$ cannot be turned on simultaneously.

  
 \section{ Cohomological consideration of  \texorpdfstring{$\mathfrak{bms}_{4}$}{BMS4} and \texorpdfstring{$W(a,b;\bar{a},\bar{b})$}{w4} algebras and their central extensions }\label{sec:5.5}
 
The direct and explicit verification of Jacobi identities for deformations may be presented in the language of algebraic cohomology. Following our discussions for  \bms\ in chapter \ref{ch4}, here we  study second adjoint cohomology of \bmsf\ and its central extension \hbmsf\ as well as the $W(a,b;\bar{a},\bar{b})$ and its central extension $\widehat{W}(a,b;\bar{a},\bar{b})$ for generic $a,b,\bar{a}$ and $\bar{b}$. 
The main tools to this cohomological analysis is the Hochschild-Serre spectral sequence which has been reviewed in subsection \ref{sec:HS-seq}. {Similar analysis for infinite dimensional Schr\"{o}dinger-Virasoro type algebras may be found in \cite{Unterberger:2011yya}.}

\subsection*{Cohomological consideration of \texorpdfstring{$\mathfrak{bms}_{4}$}{BMS4}  algebra}

As we know $\mathcal{H}^{2}(\mathfrak{bms}_{4};\mathfrak{bms}_{4})$ classifies all infinitesimal deformations of $\mathfrak{bms}_{4}$ algebra which may be computed using 
the spectral sequence \eqref{E2-dec} and the long exact sequence \eqref{long-exact}. By the former we obtain  information about $\mathcal{H}^{2}(\mathfrak{bms}_{4};(\mathfrak{witt}\oplus\mathfrak{witt}))$ and $\mathcal{H}^{2}(\mathfrak{bms}_{4};\mathfrak{T})$ independently where $\mathfrak{T}$ and $(\mathfrak{witt}\oplus\mathfrak{witt})$ respectively denote the ideal part and the Witt subalgebra of the $\mathfrak{bms}_{4}$. Note that since $(\mathfrak{witt}\oplus\mathfrak{witt})$ algebra is not a \bmsf\ module {by the adjoint action,} $\mathcal{H}^{2}(\mathfrak{bms}_{4};(\mathfrak{witt}\oplus\mathfrak{witt}))$ is defined by the action used in the short exact sequence \eqref{short-exact-bms} below, as discussed in section \ref{sec:HS-seq}. Note also that given the semi-direct sum structure of the \bmsf\ algebra \eqref{bms4=witt+ideal}, one should not expect 
$\mathcal{H}^{2}(\mathfrak{bms}_{4};\mathfrak{bms}_{4})$ to be equal to $\mathcal{H}^{2}(\mathfrak{bms}_{4};(\mathfrak{witt}\oplus\mathfrak{witt})) \oplus \mathcal{H}^{2}(\mathfrak{bms}_{4};\mathfrak{T})$. Nonetheless, as we will see, direct analysis of $\mathcal{H}^{2}(\mathfrak{bms}_{4};(\mathfrak{witt}\oplus\mathfrak{witt}))$ and $ \mathcal{H}^{2}(\mathfrak{bms}_{4};\mathfrak{T})$  {may carry information about the} structure of $\mathcal{H}^{2}(\mathfrak{bms}_{4};\mathfrak{bms}_{4})$.    

To this end, we follow the Hochschild-Serre spectral sequence method (\emph{cf}. subsection \ref{sec:HS-seq}) and consider the following short exact sequence of $\mathfrak{bms}_{4}$ algebra 
\begin{equation}
    0\longrightarrow \mathfrak{T}_{ab}\longrightarrow \mathfrak{bms}_4 \longrightarrow \mathfrak{bms}_4/\mathfrak{T}_{ab}\cong (\mathfrak{witt}\oplus\mathfrak{witt})\longrightarrow 0,\label{short-exact-bms}
\end{equation}
where $\mathfrak{T}_{ab}$ is the abelian ideal of the $\mathfrak{bms}_4$ algebra which is spanned by $T_{m,n}$ generators.
As in the case of $\mathfrak{bms}_{3}$ in chapter \ref{ch4}, we compute $\mathcal{H}^{2}(\mathfrak{bms}_{4};\mathfrak{witt}\oplus\mathfrak{witt})$ and $\mathcal{H}^{2}(\mathfrak{bms}_{4};\mathfrak{T}_{ab})$ separately. 

\paragraph{Computation of  $\mathcal{H}^{2}(\mathfrak{bms}_{4};\mathfrak{T})$.}
As is reviewed in subsection \ref{sec:HS-seq} and from \eqref{E2-dec} and \eqref{dec2} one can compute the $\mathcal{H}^{2}(\mathfrak{bms}_{4};\mathfrak{T})$ as 
\begin{equation}
\begin{split}
    \mathcal{H}^{2}(\mathfrak{bms}_{4};\mathfrak{T})&=\oplus_{p+q=2}E_{2;\mathfrak{T}}^{p,q}=E_{2;\mathfrak{T}}^{2,0}\oplus E_{2;\mathfrak{T}}^{1,1}\oplus E_{2;\mathfrak{T}}^{0,2}\\
    &=\mathcal{H}^{2}((\mathfrak{witt}\oplus\mathfrak{witt});\mathcal{H}^{0}(\mathfrak{T};\mathfrak{T}))\oplus \mathcal{H}^{1}((\mathfrak{witt}\oplus\mathfrak{witt});\mathcal{H}^{1}(\mathfrak{T};\mathfrak{T}))\cr &\oplus \mathcal{H}^{0}((\mathfrak{witt}\oplus\mathfrak{witt});\mathcal{H}^{2}(\mathfrak{T};\mathfrak{T})),\label{bmsT}
\end{split}
\end{equation}
where the subscript $\mathfrak{T}$ in $E_{2;\mathfrak{T}}^{p,q}$ means we are computing $\mathcal{H}^{2}(\mathfrak{bms}_{4};\mathfrak{T})$.
We compute the  three terms  above separately. 
$\mathcal{H}^{2}((\mathfrak{witt}\oplus\mathfrak{witt});\mathcal{H}^{0}(\mathfrak{T};\mathfrak{T}))$ contains $\mathcal{H}^{0}(\mathfrak{T};\mathfrak{T})$ which by its definition \eqref{H01} and the fact that the action of $\mathfrak{T}$ on $\mathfrak{T}$ is trivial, one concludes that $\mathcal{H}^{0}(\mathfrak{T};\mathfrak{T})=\mathfrak{T}$ then $\mathcal{H}^{2}((\mathfrak{witt}\oplus\mathfrak{witt});\mathcal{H}^{0}(\mathfrak{T};\mathfrak{T}))=\mathcal{H}^{2}((\mathfrak{witt}\oplus\mathfrak{witt});\mathfrak{T})$. On the other hand, by the direct computations in subsection \ref{Witt-deformations-sec} we have shown that two Witt subalgebras in \bmsf\ do not admit any nontrivial deformation with terms by coefficients in $T$. So one concludes that $\mathcal{H}^{2}((\mathfrak{witt}\oplus\mathfrak{witt});\mathfrak{T})=0$. 
Therefore the two first commutators in \eqref{bms4} remains intact by deformation procedure, in accord with results of subsection \ref{Witt-deformations-sec}.

Next we  analyze $\mathcal{H}^{1}((\mathfrak{witt}\oplus\mathfrak{witt});\mathcal{H}^{1}(\mathfrak{T};\mathfrak{T}))$. It  is constructed by 1-cocycle $\varphi_{1}$ which is $1-$cocycle defined as a function $ \varphi_{1}: (\mathfrak{witt}\oplus\mathfrak{witt})\longrightarrow \mathcal{H}^{1}(\mathfrak{T};\mathfrak{T})$.
The expression of $\varphi_{1}(\mathcal{L}_{m},\bar{\mathcal{L}}_{m})(T_{p,q})$ can hence be expanded in terms of $T$'s as $\varphi_{1}(\mathcal{L}_{m},\bar{\mathcal{L}}_{m})(T_{p,q})=\tilde{K}(m,p)T_{m+p,q}$\\$+ \tilde{\bar{K}}(m,q)T_{p,m+q},\label{Cjpp1}$
where $\tilde{K}(m,p)$ and $\tilde{\bar{K}}(m,q)$ are arbitrary functions. 
The deformation of $[\mathcal{L},T]$ part corresponding to $\varphi_{1}$ is $[\mathcal{L}_m,T_{p,q}]=(\frac{m+1}{2}-p)T_{m+p,q}+\tilde{K}(m,p){T}_{m+p,q}$.
The Jacobi identity for the above bracket imposes restraints on $\tilde{K}(m,n)$ exactly like the ones on $K(m,n)$ in \eqref{eq-K}, so one finds the same result as $\tilde{K}(m,n)=\alpha +\beta m$ and the same result is obtained for $\tilde{\bar{K}}(m,n)$. 

\vspace{0.1cm}
The last term we study is  $\mathcal{H}^{0}((\mathfrak{witt}\oplus\mathfrak{witt});\mathcal{H}^{2}(\mathfrak{T};\mathfrak{T}))$. We use the definition of $\mathcal{H}^{0}$ as
\begin{multline}
      \mathcal{H}^{0}((\mathfrak{witt}\oplus\mathfrak{witt});\mathcal{H}^{2}(\mathfrak{T};\mathfrak{T}))=\\
      \{ \psi  \in \mathcal{H}^{2}(\mathfrak{T};\mathfrak{T})| (\mathcal{L}\oplus\bar{\mathcal{L}})\circ \psi=0 ,\,\,\, \forall\,\, \mathcal{L}\,\, ,\,\, \bar{\mathcal{L}}\in (\mathfrak{witt}\oplus\mathfrak{witt}) \},\label{H0T}
\end{multline}
 where $\psi$ is a $T$-valued  2-cocycle. 
 The action ``$\circ$'' of $\mathcal{L}$ on a 2-cocycle $\psi$ is defined as \cite{MR0054581}
\begin{equation}
  (\mathcal{L}_{r}\circ \psi)(T_{m,n},T_{p,q})=[\mathcal{L}_{r},\psi(T_{m,n},T_{p,q})]-\psi([\mathcal{L}_{r},T_{m,n}],T_{p,q})-\psi(T_{m,n},[\mathcal{L}_{r},T_{p,q}]),\label{ideal-jacobi}
\end{equation}
Expanding $\psi$ in terms of $T$s as $\psi(T_{m,n},T_{p,q})=G(m,n;p,q)T_{m+p,n+q}$, we get the same relation as \eqref{TTT} which has the solution $G(m,n;p,q)=0$. The same relation can be obtained by $\bar{\mathcal{L}}$. 
The above discussion leads to 
\begin{equation}
    \mathcal{H}^{2}(\mathfrak{bms}_{4};\mathfrak{T})=\mathcal{H}^{1}((\mathfrak{witt}\oplus\mathfrak{witt});\mathcal{H}^{1}(\mathfrak{T};\mathfrak{T})),\label{bmsT2}
\end{equation}
which means that turning on deformations with coefficients in $\mathfrak{T}$, we can only  deform the $[\mathcal{L},T]$ part by $\tilde{K}(m,n)$. This is exactly in agreement of our results of direct and explicit calculations in section \ref{sec:1.5}.

 \paragraph{Computation of  $\mathcal{H}^{2}(\mathfrak{bms}_{4};(\mathfrak{witt}\oplus\mathfrak{witt}))$.}
One can expand the latter as
\begin{equation}
\begin{split}
     \mathcal{H}^{2}(\mathfrak{bms}_{4};(\mathfrak{witt}\oplus\mathfrak{witt}))&=\oplus_{p+q=2}E_{2;(\mathfrak{w}\oplus\mathfrak{w})}^{p,q}=E_{2;(\mathfrak{w}\oplus\mathfrak{w})}^{2,0}\oplus E_{2;(\mathfrak{w}\oplus\mathfrak{w})}^{1,1}\oplus E_{2;(\mathfrak{w}\oplus\mathfrak{w})}^{0,2}\\
    &=\mathcal{H}^{2}((\mathfrak{witt}\oplus\mathfrak{witt});\mathcal{H}^{0}(\mathfrak{T};(\mathfrak{witt}\oplus\mathfrak{witt})))\cr& \oplus \mathcal{H}^{1}((\mathfrak{witt}\oplus\mathfrak{witt});\mathcal{H}^{1}(\mathfrak{T};(\mathfrak{witt}\oplus\mathfrak{witt})))\cr &\oplus \mathcal{H}^{0}((\mathfrak{witt}\oplus\mathfrak{witt});\mathcal{H}^{2}(\mathfrak{T};(\mathfrak{witt}\oplus\mathfrak{witt}))),\label{bms-LbarL}
\end{split}
\end{equation}
where the subscript $(\mathfrak{w}\oplus\mathfrak{w})$ denotes we are computing $  \mathcal{H}^{2}(\mathfrak{bms}_{4};(\mathfrak{witt}\oplus\mathfrak{witt}))$.

To compute  $\mathcal{H}^{2}((\mathfrak{witt}\oplus\mathfrak{witt});\mathcal{H}^{0}(\mathfrak{T};(\mathfrak{witt}\oplus\mathfrak{witt})))$, we recall that the action  of $\mathfrak{T}$ on $(\mathfrak{witt}\oplus\mathfrak{witt})$ (which is induced via the short exact sequence \eqref{short-exact-bms},  is trivial and  hence $\mathcal{H}^{0}(\mathfrak{T};(\mathfrak{witt}\oplus\mathfrak{witt}))\cong (\mathfrak{witt}\oplus\mathfrak{witt})$. We then conclude 
\be\label{H2-witt-witt}
\mathcal{H}^{2}((\mathfrak{witt}\oplus\mathfrak{witt});\mathcal{H}^{0}(\mathfrak{T};(\mathfrak{witt}\oplus\mathfrak{witt})))\cong  \mathcal{H}^{2}((\mathfrak{witt}\oplus\mathfrak{witt});(\mathfrak{witt}\oplus\mathfrak{witt}))\cong 0,
\ee 
where in the last step we used the fact that $(\mathfrak{witt}\oplus\mathfrak{witt})$ algebra is rigid as we showed in previous chapter.
For more details see appendix \ref{witt+witt-rigid}.

Next, we consider the second term in \eqref{bms-LbarL}, $\mathcal{H}^{1}((\mathfrak{witt}\oplus\mathfrak{witt});\mathcal{H}^{1}(\mathfrak{T};(\mathfrak{witt}\oplus\mathfrak{witt})))$ which is constructed by 1-cocycle $\varphi_{2}$ as $\varphi_{2}: (\mathfrak{witt}\oplus\mathfrak{witt})\longrightarrow \mathcal{H}^{1}(\mathfrak{T};(\mathfrak{witt}\oplus\mathfrak{witt}))$.
A similar analysis as the previous case tells us that $\varphi_2$ deforms the commutator $[\mathcal{L},T]$ part as $[\mathcal{L}_{m},T_{p,q}]=(\frac{m+1}{2}-p)T_{m+p,q}+f(m,p)\mathcal{L}_{p+m-1}\delta_{q,0}+g(m)\bar{\mathcal{L}}_{q-1}\delta_{m+p,0}$ (and a similar relation can be obtained for $[\bar{\mathcal{L}},T]$). Recalling the arguments of previous section, 
one concludes 
\be
\mathcal{H}^{1}((\mathfrak{witt}\oplus\mathfrak{witt});\mathcal{H}^{1}(\mathfrak{T};(\mathfrak{witt}\oplus\mathfrak{witt})))=0.
\ee
This means that the $[\mathcal{L},T]$ and $[\bar{\mathcal{L}},T]$ commutators cannot be deformed by  terms with coefficients {in} $\mathcal{L}$ and $\bar{\mathcal{L}}$.

We finally compute the last term in \eqref{bms-LbarL}, $\mathcal{H}^{0}((\mathfrak{witt}\oplus\mathfrak{witt});\mathcal{H}^{2}(\mathfrak{T};(\mathfrak{witt}\oplus\mathfrak{witt})))$.  One can repeat the procedure exactly the same as the previous case to get 
\begin{equation}
  (\mathcal{L}_{m}\circ \psi)(T_{p,q},T_{r,s})=[\mathcal{L}_{m},\psi(T_{p,q},T_{r,s})]-\psi([\mathcal{L}_{m},T_{p,q}],T_{r,s})-\psi(T_{r,s},[\mathcal{L}_{m},T_{p,q}]),\label{hjacobi-2}
\end{equation}
and
\begin{equation}
  (\bar{\mathcal{L}}_{m}\circ \psi)(T_{p,q},T_{m,n})=[\bar{\mathcal{L}}_{m},\psi(T_{p,q},T_{m,n})]-\psi([\bar{\mathcal{L}}_{m},T_{p,q}],T_{m,n})-\psi(T_{m,n},[\bar{\mathcal{L}}_{m},T_{p,q}]),\label{hjacobi-3}
\end{equation}
with $\psi(T_{p,q},T_{m,n})=A(m,n;p,q)\mathcal{L}_{m+p-1}+B(m,n;p,q)\bar{\mathcal{L}}_{n+q-1}$ where $A(m,n;p,q)$ and \\$B(m,n;p,q)$ are arbitrary antisymmetric functions. Inserting the latter into \eqref{hjacobi-2} and \eqref{hjacobi-3}  we get the same relation as \eqref{B-eq0} and its analogue for $A$, leading to $A(m,n;p,q)=B(m,n;p,q)=0$. Therefore,  $\mathcal{H}^{0}((\mathfrak{witt}\oplus\mathfrak{witt});\mathcal{H}^{2}(\mathfrak{T};(\mathfrak{witt}\oplus\mathfrak{witt})))=0$ for  \bmsf\ case. This is in contrast to the \bms\ case \ref{sec:cohomology-bms}.

As summary of the above discussions one concludes that 
\be\mathcal{H}^{2}(\mathfrak{bms}_{4},\mathfrak{bms}_{4})=\mathcal{H}^{1}((\mathfrak{witt}\oplus\mathfrak{witt});\mathcal{H}^{1}(\mathfrak{T};\mathfrak{T})).\label{bms4-witt}
\end{equation}
That is,  deformations of $\mathfrak{bms}_{4}$ are just those that deform the $[\mathcal{L},T]$ and $[\bar{\mathcal{L}},T]$ commutators by terms with coefficients in $T$. In other words, \bmsf\ algebra can only be deformed into ${W(a,b;\bar{a},\bar{b})}$, in accord with our direct Jacobi identity analysis of previous section.

\subsection*{Cohomological consideration for \texorpdfstring{$\widehat{\mathfrak{bms}}_{4}$}{BMS4}  algebra}

As discussed in subsection \ref{sec:3.3} and \cite{Oblak:2016eij} for a given algebra $\mathfrak{g}$, (1) central extensions are classified by Gel'fand-Fucks 2-cocycles, or by $\mathcal{H}^{2}(\mathfrak{g}; \mathbb{R})$ and, (2)
to deal with the central extensions in the cohomological analysis we need to extend the algebra by {the addition of the generators of abelian $u(1)$ Lie algebras, one $u(1)$ algebra for each possible central term.} Explicitly, we need to consider $\hat{\mathfrak{g}}=\mathfrak{g}\oplus \mathfrak{u}(1)\oplus\cdots \oplus \mathfrak{u}(1)$, where the number of $\mathfrak{u}(1)$ factors is equal to the number of independent commutators in the algebra, i.e. for \bms\ it is three and for \bmsf\ it is six. Of course closure condition may not allow to turn on all these central terms.\footnote{{For instance, the Jacobi identity allows \bms\ to have two central terms in the $[\mathcal{J},\mathcal{J}]$ and $[\mathcal{J},\mathcal{P}]$ commutators, for the $\mathfrak{KM}_{\mathfrak{u}(1)}$ we can also have central term in $[\mathcal{P},\mathcal{P}]$ commutators, altogether three central extensions \ref{sec:4.2} \cite{gao2011low}. The \bmsf, however, just admits two central terms in its two Witt subalgebras.}} In our analysis, we adopt the viewpoint that each central extension is a deformation in the $\mathfrak{u}(1)$ extended algebra and that turning on a central charge is like a deformation in $\hat{\mathfrak{g}}$. Therefore, we need not first study $\mathcal{H}^{2}(\mathfrak{g}; \mathbb{R})$ and then analyze the deformation, we may directly focus on $\mathcal{H}^{2}(\hat{\mathfrak{g}}; \hat{\mathfrak{g}})$. 

{As in the $\mathfrak{bms}_{4}$ case, $(\mathfrak{vir}\oplus\mathfrak{vir})$ is not a $\widehat{\mathfrak{bms}}_{4}$ module by the adjoint action and $\mathcal{H}^{2}(\widehat{\mathfrak{bms}}_{4};(\mathfrak{vir}$\\$\oplus\mathfrak{vir}))$ is defined by the action induced from the short exact sequence \eqref{short-exact-vir-vir}, as was discussed, we should not expect
$\mathcal{H}^{2}(\widehat{\mathfrak{bms}}_{4};\widehat{\mathfrak{bms}}_{4})$ to be equal to $\mathcal{H}^{2}(\widehat{\mathfrak{bms}}_{4};(\mathfrak{vir}\oplus\mathfrak{vir}))\oplus \mathcal{H}^{2}(\widehat{\mathfrak{bms}}_{4};\mathfrak{T}_{ab})$.} 
As in the \bmsf\ case, computation of each of these two factors would be helpful in computing the former.
To this end we employ the Hochschild-Serre spectral sequence procedure. The $\widehat{\mathfrak{bms}}_{4}$ has semi-direct sum structure  $\widehat{\mathfrak{bms}}_{4}\cong(\mathfrak{vir}\oplus\mathfrak{vir})\inplus\mathfrak{T}_{ab}$ where $(\mathfrak{vir}\oplus\mathfrak{vir})$ is spanned by  $\mathcal{L}$ and $\bar{\mathcal{L}}$ generators plus two {unit elements as} central generators and $\mathfrak{T}_{ab}$ is ideal part which is spanned by $T$.
The short exact sequence for the above is  
\begin{equation}
    0\longrightarrow \mathfrak{T}\longrightarrow \widehat{\mathfrak{bms}}_{4} \longrightarrow \widehat{\mathfrak{bms}}_{4}/\mathfrak{T}\cong (\mathfrak{vir}\oplus\mathfrak{vir})\longrightarrow 0.\label{short-exact-vir-vir}
\end{equation}

\paragraph{Computation of $\mathcal{H}^{2}(\widehat{\mathfrak{bms}}_{4};(\mathfrak{vir}\oplus\mathfrak{vir}))$.}
Using the Theorem 1.2 in \cite{degrijse2009cohomology} and Hochschild-Serre spectral sequence method we get
\begin{equation}
\begin{split}
     \hspace*{-5mm}\mathcal{H}^{2}(\widehat{\mathfrak{bms}}_{4};(\mathfrak{v}\oplus\mathfrak{v}))&=\oplus_{p+q=2}E_{2;(\mathfrak{v}\oplus\mathfrak{v})}^{p,q}=E_{2;(\mathfrak{v}\oplus\mathfrak{v})}^{2,0}\oplus E_{2;(\mathfrak{v}\oplus\mathfrak{v})}^{1,1}\oplus E_{2;(\mathfrak{v}\oplus\mathfrak{v})}^{0,2}\\
     &= \mathcal{H}^{2}((\mathfrak{vir}\oplus\mathfrak{vir});\mathcal{H}^{0}(\mathfrak{T},(\mathfrak{vir}\oplus\mathfrak{vir})))\cr &\oplus \mathcal{H}^{1}((\mathfrak{vir}\oplus\mathfrak{vir});\mathcal{H}^{1}(\mathfrak{T},(\mathfrak{vir}\oplus\mathfrak{vir})))\cr &\oplus\mathcal{H}^{0}((\mathfrak{vir}\oplus\mathfrak{vir});\mathcal{H}^{2}(\mathfrak{T},(\mathfrak{vir}\oplus\mathfrak{vir}))),
\end{split}
\end{equation}
where $E_{2;(\mathfrak{v}\oplus\mathfrak{v})}^{p,q}\equiv\mathcal{H}^{p}((\mathfrak{vir}\oplus\mathfrak{vir});\mathcal{H}^{q}(\mathfrak{T},(\mathfrak{vir}\oplus\mathfrak{vir})))$.

The first term we have to consider is $E_{2;(\mathfrak{v}\oplus\mathfrak{v})}^{2,0}=\mathcal{H}^{2}((\mathfrak{vir}\oplus\mathfrak{vir});\mathcal{H}^{0}(\mathfrak{T},(\mathfrak{vir}\oplus\mathfrak{vir})))$. From the definition of $\mathcal{H}^{0}$ one gets $\mathcal{H}^{0}(\mathfrak{T},(\mathfrak{vir}\oplus\mathfrak{vir}))=(\mathfrak{vir}\oplus\mathfrak{vir})$ because the action of $\mathfrak{T}$ as an ideal part of the algebra, on $(\mathfrak{vir}\oplus\mathfrak{vir})$, {induced by the short exact sequence \eqref{short-exact-vir-vir},} is trivial. Then, recalling the fact that $(\mathfrak{vir}\oplus\mathfrak{vir})$ algebra is rigid, see appendix \ref{rigid-vir+vir}, one concludes that $E_{2;(\mathfrak{v}\oplus\mathfrak{v})}^{2,0}=\mathcal{H}^{2}((\mathfrak{vir}\oplus\mathfrak{vir});(\mathfrak{vir}\oplus\mathfrak{vir}))=0$.

We should next consider $E_{2;(\mathfrak{v}\oplus\mathfrak{v})}^{1,1}=\mathcal{H}^{1}((\mathfrak{vir}\oplus\mathfrak{vir});\mathcal{H}^{1}(\mathfrak{T},(\mathfrak{vir}\oplus\mathfrak{vir})))$. One may generalize discussions of the case without the central elements to conclude
$\mathcal{H}^{1}((\mathfrak{vir}\oplus\mathfrak{vir});\mathcal{H}^{1}(\mathfrak{T},(\mathfrak{vir}$\\$\oplus\mathfrak{vir})))=0$. Therefore,  the $[\mathcal{L},T]$ and $[\bar{\mathcal{L}},T]$ commutators cannot be deformed by the terms with coefficient of $\hat{\mathcal{L}}$ and $\hat{\bar{\mathcal{L}}}$ where the hatted objects, such as $\hat{\mathcal{L}}_{m}$, denote generators of the Virasoro algebra, i.e. Witt algebra plus central element hence $E_{2;(\mathfrak{v}\oplus\mathfrak{v})}^{1,1}=0$.

The last term we compute is $E_{2;(\mathfrak{v}\oplus\mathfrak{v})}^{0,2}=\mathcal{H}^{0}((\mathfrak{vir}\oplus\mathfrak{vir});\mathcal{H}^{2}(\mathfrak{T},(\mathfrak{vir}\oplus\mathfrak{vir})))$. Using  definition of $\mathcal{H}^{0}$, one observes that its elements are solutions of
\begin{equation}
  (\hat{\mathcal{L}}_{m}\circ \hat{\psi})(T_{p,q},T_{r,s})=[\hat{\mathcal{L}}_{m},\hat{\psi}(T_{p,q},T_{r,s})]-\hat{\psi}([\hat{\mathcal{L}}_{m},T_{p,q}],T_{r,s})-\hat{\psi}(T_{p,q},[\hat{\mathcal{L}}_{m},T_{r,s}]),\label{psi-hat-2}
\end{equation}         
and
\begin{equation}
  (\bar{\hat{\mathcal{L}}}_{m}\circ \hat{\psi})(T_{p,q},T_{r,s})=[\bar{\hat{\mathcal{L}}}_{m},\hat{\psi}(T_{p,q},T_{r,s})]-\hat{\psi}([\bar{\hat{\mathcal{L}}}_{m},T_{p,q}],T_{r,s})-\hat{\psi}(T_{p,q},[\bar{\hat{\mathcal{L}}}_{m},T_{r,s}]),\label{psi-hat1}
\end{equation}    
where $\hat{\psi}(T_{p,q},T_{r,s})$ is a $2-$cocycle. 
The linear expansion of $\hat{\psi}$ in terms of generators is \[\hat{\psi}(T_{m,n},T_{p,q})=\tilde{A}(m,n;p,q)\mathcal{L}_{m+p-1}+\tilde{B}(m,n;p,q)\bar{\mathcal{L}}_{n+q-1}+\tilde{Z}(m,n;p,q).\] 
Inserting the expansion of $\hat{\psi}$ into \eqref{psi-hat-2} and \eqref{psi-hat1} we reach to some relations which force all of the above functions to be zero so $E_{2;(\mathfrak{v}\oplus\mathfrak{v})}^{0,2}=\mathcal{H}^{0}((\mathfrak{vir}\oplus\mathfrak{vir});\mathcal{H}^{2}(\mathfrak{T},(\mathfrak{vir}\oplus\mathfrak{vir})))=0$.

To conclude this part, we have shown that $\mathcal{H}^{2}(\widehat{\mathfrak{bms}}_{4};(\mathfrak{vir}\oplus\mathfrak{vir}))=0$. 

\paragraph{Computation of $\mathcal{H}^{2}(\widehat{\mathfrak{bms}}_{4};\mathfrak{T})$.}

Using Theorem 1.2 in \cite{degrijse2009cohomology} and Hochschild-Serre spectral sequence method we get
\begin{multline}
         \mathcal{H}^{2}(\widehat{\mathfrak{bms}}_{4};\mathfrak{T})=\oplus_{p+q=2}E_{2;\mathfrak{T}}^{p,q}=E_{2;\mathfrak{T}}^{2,0}\oplus E_{2;\mathfrak{T}}^{1,1}\oplus E_{2;\mathfrak{T}}^{0,2}\cr
     = \mathcal{H}^{2}((\mathfrak{vir}\oplus\mathfrak{vir});\mathcal{H}^{0}(\mathfrak{T},\mathfrak{T}))\oplus \mathcal{H}^{1}((\mathfrak{vir}\oplus\mathfrak{vir});\mathcal{H}^{1}(\mathfrak{T},\mathfrak{T}))\oplus\mathcal{H}^{0}((\mathfrak{vir}\oplus\mathfrak{vir});\mathcal{H}^{2}(\mathfrak{T},\mathfrak{T})),\nonumber
\end{multline}
where we defined $E_{2;\mathfrak{T}}^{p,q}=\mathcal{H}^{p}((\mathfrak{vir}\oplus\mathfrak{vir});\mathcal{H}^{q}(\mathfrak{T},\mathfrak{T}))$. Unlike the $\widehat{\mathfrak{bms}}_{3}$ case, the ideal part of the $\widehat{\mathfrak{bms}}_{4}$ does not admit any central generator in its ideal part, so the results of the case without central extension can be generalized to this case too. 

In summary, we conclude that 
\be\mathcal{H}^{2}(\widehat{\mathfrak{bms}}_{4},\widehat{\mathfrak{bms}}_{4})=\mathcal{H}^{1}((\mathfrak{vir}\oplus\mathfrak{vir});\mathcal{H}^{1}(\mathfrak{T};\mathfrak{T})).\label{bms-vir}
\end{equation}
That is, in accord with our closure condition analysis, $\widehat{\mathfrak{bms}}_4$ algebra can only be deformed to $\widehat{W}(a,b;\bar{a},\bar{b})$.

\subsection*{Cohomological consideration of \texorpdfstring{$W(a,b;\bar{a},\bar{b})$}{W}  algebra} 


$W(a,b;\bar{a},\bar{b})$ which is introduces by \eqref{W4-algebra} can be considered as semi-direct sum of $(\mathfrak{witt}\oplus\mathfrak{witt})\inplus \mathfrak{T}$ which it has the following short exact sequence 
\begin{equation}
     0\longrightarrow \mathfrak{T}\longrightarrow W(a,b;\bar{a},\bar{b})\longrightarrow W(a,b;\bar{a},\bar{b})/\mathfrak{T}\cong (\mathfrak{witt}\oplus\mathfrak{witt})\longrightarrow 0,
\end{equation}   
where $\mathfrak{T}$  and $(\mathfrak{witt}\oplus\mathfrak{witt})$, respectively, denote the ideal part and subalgebra of $W(a,b;\bar{a},\bar{b})$. {Note that since $(\mathfrak{witt}\oplus\mathfrak{witt})$ is not a $W(a,b;\bar{a},\bar{b})$ module by the adjoint action, $\mathcal{H}^2(W(a,b;\bar{a},\bar{b});$\\$(\mathfrak{witt}\oplus\mathfrak{witt}))$ is defined by the action induced from the above short exact sequence. The relevant second adjoint cohomology of  $\mathcal{H}^2(W(a,b;\bar{a},\bar{b});W(a,b;\bar{a},\bar{b}))$ may hence be computed much the same as the \bmsf\ case discussed earlier, with the same result, namely,

$\mathcal{H}^2(W(a,b;\bar{a},\bar{b}); W(a,b;\bar{a},\bar{b}))=0$ for the \emph{family of ${W(a,b;\bar{a},\bar{b})}$ algebras}; i.e. ${W(a,b;\bar{a},\bar{b})}$ family  for generic value of parameters is infinitesimally and formally rigid.

\paragraph{Cohomological consideration of \texorpdfstring{$\widehat{W}(a,b;\bar{a},\bar{b})$}{W}  algebra.}

The case $\widehat{W}(a,b;\bar{a},\bar{b})$ is exactly the same as \hbmsf\ from cohomological point of view: It just admits two nontrivial central terms in its two Witt subalgebras and that $\mathcal{H}^2(\widehat{W}(a,b;\bar{a},\bar{b}); \widehat{W}(a,b;\bar{a},\bar{b}))=0$ for the family of $\widehat{W}$ algebras.









\section{Physical realization of obtained algebras through deformation}

The first aim of this chapter was possibly obtaining infinite dimensional Lie algebra, including $\mathfrak{so}(3,2)$ as its global part, through deformation of \bmsf\ algebra. Our computations showed that the answer of this question is negative. So the asymptotic symmetry algebra of $AdS_{4}$ will remain as an open problem which needs more considerations. Deposit this result, we concluded that \bmsf\ algebra is not a rigid algebra and, the same as \bms case, it can be deformed into a four parameters family algebra \wf. Now we are going to explore physical realisation of deformation of \bmsf to \wf. 

\begin{itemize}
    \item  $W(0,0;0,0)$ algebra, at first, was obtained as near horizon symmetry algebra of nonextremal black holes by Donnay and her collaborators in \cite{Donnay:2015abr}. In this way the deformation of \bmsf, which is equivalent to $W(-\frac{1}{2},-\frac{1}{2};-\frac{1}{2},-\frac{1}{2})$, may be viewed as procedure which connects the asymptotic and near horizon regions to each other through their symmetries algebras. For possibly dual field theories with mentioned symmetries algebra, deformation can be viewed as RG flow which relates the IR and UV regions.
    
    \item In recent work by Grumiller and his collaborators \cite{Grumiller:2019fmp} has been shown that the near horizon symmetry algebra for $4d$ nonextremal black holes is $W(a,a;a,a)$ family algebra which \bmsf\  is an element of this family with specific values of parameters as $W(-\frac{1}{2},-\frac{1}{2};-\frac{1}{2},-\frac{1}{2})$. Beside the interpretation we mentioned in prior case, deformation of \bmsf\ can be considered as way which changes \bmsf\ to \wf\ family algebra with arbitrary values for its parameters. 
    
    \item The specific kind of GCA which is known as semi-Galilean algebra introduced in \cite{Alishahiha:2009np}, is exactly the same as $\mathfrak{bms}_{4}$ algebra where $d=1$ and is also discussed in \cite{Bagchi:2010eg}. In this context, one may look for the physical interpretation for $W(a,b;,\bar{a},\bar{b})$ family algebra, as the same as $W(0,b)$ case, to possibly find some thing like $\ell$-conformal Galilei algebra such as $\ell-{\bar{\ell}}$-conformal semi Galilei symmetry algebra. \footnote{Since the parameter $b$ has the same role as $\ell$ in CGA, we suggested to find a $\ell-{\bar{\ell}}$-conformal semi Galilei symmetry where the parameters $\ell$ and ${\bar{\ell}}$ are associated to $b$ and $\bar{b}$.}
    
\section{Summary of the chapter}
In this chapter we considered formal deformations of \bmsf\ and its central extensions \hbmsf. We explored that \bmsf\ can not be deformed into an infinite dimensional Lie algebra which contains $\mathfrak{so}(3,2)$ as its global part. This shows that the asymptotic symmetry of flat spacetime \bmsf\ and its possibly corresponding in AdS$_{4}$ are not related to each other through standard deformation theory of Lie algebras. This is unlike the 3d case where \hbms\ and $\mathfrak{vir}\oplus\mathfrak{vir}$ are related through deformation/contraction, which prompts  a possible formulation of field theory holographically dual to gravity on 3d flat space \cite{Barnich:2010eb, Bagchi:2010eg, Hartong:2015usd, Bagchi:2016bcd}.

Despite the latter, we have proved that \bmsf\ algebra and its central extension are not rigid and can be deformed into a four parameters family algebra \wf\ and its central extension \wfh. We discussed the various physical interpretations of \wf\ family algebra. We also considered deformation of \wf\ algebra and its central extensions and showed that for generic values of their parameters they are rigid, as the same as \w\ algebra in previous chapter, in the sense that they just can be deformed into another $W(\tilde{a},\tilde{b};\tilde{\bar{a}},\tilde{\bar{b}})$ algebra with shifted parameters. 
However, we showed that for specific points such as $W(0,0;0,0)$ and $W(0,-1;0,0)$, there are other possibilities for deformation which takes us out of the $W$ family algebra. For instance $W(0,-1;0,0)$, which has $\mathfrak{bms}_{3}\oplus\mathfrak{witt}$ subalgebra, can be deformed into a new algebra in its ideal part which includes three copies of \wit\ as subalgebra.

\end{itemize}

\chapter{Deformation of Maxwell-BMS algebra and its physical realization} \label{ch6}
This chapter consists of two parts which are subjectwise related but are essentially independent. 
In the first part, the first and second sections, we study formal deformations of Maxwell-BMS algebra \Max, which is infinite dimensional enhancement of $2+1$ Maxwell algebra. We find that \Max\ algebra can be deformed into $\mathfrak{bms}_{3}\oplus\mathfrak{witt}$ and three copies of \wit\ algebra in its ideal part. We also show that there are some formal deformations in other commutators leading to new families of algebras. The latter provide other examples of deviations from the Hochschild-Serre factorization theorem. We then explore possibly central extensions of obtained algebras through deformations.  

In the second part, the  third and fourth sections, we try to give a physical interpretation for $\mathfrak{bms}_{3}\oplus\mathfrak{witt}$ algebra obtained in the first part. We construct a Chern-Simons gravity theory invariant under $\mathfrak{iso}(2,1)\oplus\mathfrak{so}(2,1)$. By choosing an appropriate boundary fall-off condition we explore surface charges associated to asymptotic Killing vector fields which keep the fall-off condition intact. We show that the surface charge algebra is central extension of $\mathfrak{bms}_{3}\oplus\mathfrak{witt}$ with three independent central charges. 

\paragraph{Remark.} Since \Max\ algebra has different commutators, consideration of separate commutators deformation lead to long computations. So to avoid tedious calculations, in this chapter we just mention the most general deformation of \Max\ algebra and their non trivial cases.

\section{The most general deformation of \texorpdfstring{$\mathfrak{Max}_{3}$}{Max3} algebra}\label{sec:1.6}

Here we explore all possible deformations of the commutators of the \Max\ algebra which may be parametrized as
\begin{equation} 
\begin{split}\label{most-general-deform}
&i[\mathcal{J}_{m},\mathcal{J}_{n}]=(m-n)\mathcal{J}_{m+n}+(m-n)F(m,n)\mathcal{P}_{m+n}+(m-n)G(m,n)\mathcal{M}_{m+n},\\
&i[\mathcal{J}_{m},\mathcal{P}_{n}]=(m-n)\mathcal{P}_{m+n}+K(m,n)\mathcal{P}_{m+n}+I(m,n)\mathcal{M}_{m+n}+O(m,n)\mathcal{J}_{m+n},\\
&i[\mathcal{J}_{m},\mathcal{M}_{n}]=(m-n)\mathcal{M}_{m+n}+\tilde{K}(m,n)\mathcal{M}_{m+n}+\tilde{I}(m,n)\mathcal{P}_{m+n}+\tilde{O}(m,n)\mathcal{J}_{m+n},\\
 &i[\mathcal{P}_m,\mathcal{P}_n]=(m-n)\mathcal{M}_{m+n}+(m-n)f_{1}(m,n)\mathcal{P}_{m+n}+(m-n)h_{1}(m,n)\mathcal{J}_{m+n}, \\
 &i[\mathcal{P}_m,\mathcal{M}_n]=f_{2}(m,n)\mathcal{P}_{m+n}+g_{2}(m,n)\mathcal{M}_{m+n}+h_{2}(m,n)\mathcal{J}_{m+n},\\
 &i[\mathcal{M}_m,\mathcal{M}_n]=(m-n)f_{3}(m,n)\mathcal{P}_{m+n}+(m-n)g_{3}(m,n)\mathcal{M}_{m+n}+(m-n)h_{3}(m,n)\mathcal{J}_{m+n},
\end{split}
\end{equation}
where the arbitrary functions can be fixed from the Jacobi identities leading to diverse deformations. It is important to emphasize that throughout this chapter the indices of the generators $\mathcal{J}$, $\mathcal{P}$ and $\mathcal{M}$ which appear in the right-hand-side are fixed to be $m+n$. On the other hand, the functions have a polynomial expansion in terms of their arguments. Furthermore, we shall not write the deformation term as $(m-n)g_{1}(m,n)\mathcal{M}_{m+n}$ which just rescales the term $(m-n)\mathcal{M}_{m+n}$ by a constant parameter as $\alpha(m-n)\mathcal{M}_{m+n}$. Of course this can be absorbed into a redefinition of generators. In what follows we study each Jacobi identity and its respective implications.

\subsection*{Infinitesimal deformations}
In this part we study infinitesimal deformation in which we consider the constraints coming from the Jacobi identities in first order of the functions. 
Let us consider first the Jacobi identity $[\mathcal{J},[\mathcal{J},\mathcal{J}]]+\text{cyclic permutations}=0$ which implies,
\begin{multline} \label{F,G-eq}
(n-l)(m-n-l)[G(m,l+n)+G(n,l)]+ (l-m)(n-l-m)[G(n,l+m)+G(l,m)]+\\
 (m-n)(l-m-n)[G(l,m+n)+G(m,n)]=0.
\end{multline}
Analogously, the same relation will be obtained for $F(m,n)$. There is no other constraint for $G$ and 
\begin{equation}\label{JJ-Z}
        G(m,n)=Z(m)+Z(n)-Z(m+n),
\end{equation}
for any arbitrary function $Z$, provides the most general solution to \eqref{F,G-eq}. Nevertheless, it is possible to show that the deformations of the form \eqref{JJ-Z} are trivial deformations since they can be reabsorbed by redefining the generators\footnote{One can show that this result is true when different deformations (functions) are turned on simultaneously.} as: 
\begin{equation} 
\begin{split}\label{ZY-redefinition}
  \mathcal{J}_{m}:=\tilde{\mathcal{J}}_{m}+Z(m)\tilde{\mathcal{M}}_{m}, \qquad \mathcal{P}_{m}:=\tilde{\mathcal{P}}_{m}, \qquad \mathcal{M}_{m}:=\tilde{\mathcal{M}}_{m},
\end{split}
\end{equation}
where $\tilde{\mathcal{J}}_{m}$, $\tilde{\mathcal{P}}_{m}$ and $\tilde{\mathcal{M}}_{m}$ satisfy the commutation relations of the \Max\ algebra \eqref{maxwell}.

On the other hand, one finds from the Jacobi identity $[\mathcal{J},[\mathcal{J},\mathcal{P}]]+\text{cyclic permutations}=0$ the following relation at the first order for $K$:
  \begin{multline}\label{firsteq-K}
(n-l) K(m,l+n)+(m-n-l)K(n,l)+(l-m) K(n,l+m) +\\
+(l+m-n)K(m,l)+(n-m)K(m+n,l)=0,
\end{multline}
which can be solved, as was discussed in section \ref{sec:4.1}, by
\begin{equation} \label{KIO-solution}
 K(m,n)=\alpha+\beta m+\gamma m(m-n)+\cdots.
\end{equation}
The same relation and solution is found for $O$. One can see that the Jacobi identity $[\mathcal{J},[\mathcal{J},\mathcal{M}]]+\text{cyclic permutations}=0$ also leads to the same relation and solution as \eqref{firsteq-K} and \eqref{KIO-solution} for $\tilde{K},\tilde{I}$ and $\tilde{O}$.

The Jacobi identity $[\mathcal{J},[\mathcal{J},\mathcal{P}]]+\text{cyclic permutations}=0$ also leads to a relation for functions $F$ and $I$ as
 \begin{multline}\label{I-F}
   (n-l) I(m,l+n)+(m-n-l)I(n,l)+(l-m)I(n,l+m) +\\
+(l+m-n)I(m,l)+(n-m)I(m+n,l)+(m-n)(l-m-n)F(m,n)=0.
\end{multline}
which is solved by $I(m,n)=\bar{\alpha}+\bar{\beta}m-\bar{\nu} n+(\bar{\gamma} mn^{2}+\frac{1}{2}(\bar{\lambda}-\bar{\gamma})nm^{2}+\frac{1}{2}(-\bar{\lambda}-\bar{\gamma})m^{3})+\cdots$ and $F(m,n)=\bar{\nu}+\bar{\lambda} mn+\cdots$.

One can see that three independent relations appear by considering the Jacobi identity $[\mathcal{J},[\mathcal{P},\mathcal{M}]]$  $+\text{cyclic permutations}=0$ in the first order in functions.  In particular, we have the following relation for $h_{2}$:
\begin{equation}\label{f,g,h-2(1)}
(n-l)h_{2}(m,l+n)+(m-l)h_{2}(m+l,n)+(l-m-n)h_{2}(m,n)=0.
\end{equation}
By setting $m=n=l$ we obtain $mh_{2}(m,m)=0$. Then we have that $h_{2}(m,m)=0$ for $m\neq 0$. This means that we can write $h_{2}(m,n)=(m-n)\bar{h}_{2}(m,n)$ where $\bar{h}_{2}(m,n)$ is a symmetric function. By inserting the latter into \eqref{f,g,h-2(1)} one gets
\begin{equation}\label{f,g,h-2(2)}
(n-l)(m-l-n)\bar{h}_{2}(m,l+n)+(l-m)(n-m-l)\bar{h}_{2}(m+l,n)+(l-m-n)(m-n)\bar{h}_{2}(m,n)=0,  
\end{equation}
which is solved for $h_{2}(m,n)=\alpha(m-n)$ where $\alpha$ is arbitrary constant.
For the functions $f_{2}$ and $\tilde{O}$ one obtains
\begin{equation}\label{f2-tildeO}
 (n-l)f_{2}(m,l+n)+(m-l)f_{2}(m+l,n)+(l-m-n)f_{2}(m,n)-(m-n-l)\tilde{O}(l,n)=0.
  \end{equation}
Then, by replacing $m=n+l$ one finds the same relation as \eqref{f,g,h-2(1)} leading to $f_{2}(m,n)=\beta(m-n)$ and $\tilde{O}(m,n)=0$.
Furthermore the Jacobi identity $[\mathcal{J},[\mathcal{P},\mathcal{M}]]+\text{cyclic permutations}=0$ gives rise to a relation for $g_{2}, O$ and $\tilde{I}$ as follows
\begin{multline}\label{g2-O-tildeI}
    (n-l)g_{2}(m,l+n)+(m-l)g_{2}(m+l,n)+(l-m-n)g_{2}(m,n)+\\ 
    +(n+l-m)\tilde{I}(l,n)+(n-l-m)O(l,m)=0.
\end{multline}

By studying the Jacobi identity $[\mathcal{J},[\mathcal{P},\mathcal{P}]]+\text{cyclic permutations}=0$ it is possible to see that such identity puts not only the following constraints on the functions
\begin{equation} 
\begin{split}\label{K-tildeK,O-tildeI}
     &(n+l-m)K(l,n)+(n-l-m)K(l,m)+(m-n)\tilde{K}(l,m+n)=0,\\
& (n+l-m)O(l,n)+(n-l-m)O(l,m)+(m-n)\tilde{I}(l,m+n)=0.
\end{split}
\end{equation}
but also leads to a new relation for $f_{1}, O$ and $\tilde{I}$ as
\begin{multline}\label{f1-O-tildeI}
    (n-l)(m-n-l)f_{1}(m,l+n)+ (l-m)(n-l-m)f_{1}(n,l+m)+(m-n)(l-m-n)f_{1}(m,n)+ \\ 
    +(n+l-m)O(l,n)+(n-l-m)O(l,m)+(m-n)\tilde{I}(l,m+n)=0.
\end{multline}

Nnote that the relation \eqref{f1-O-tildeI} is linear in $f_{1}, O$ and $\tilde{I}$ and the coefficients of the latter two are first order in $m,n,l$ while the coefficients of the $f_{1}$ terms are second order in $m,n,l$. We expect that these functions are polynomials of positive powers in their arguments, so one concludes that if $O$ and $\tilde{I}$ are monomials of degree $p$, $f_{1}$ should be a monomial of degree $p+1$. Since the solutions of $O$ and $\tilde{I}$ are similar to the ones of \eqref{KIO-solution}, we have that \eqref{f1-O-tildeI} is satisfied considering $f_{1}(m,n)=constant$, $O(m,n)=\alpha+\beta m+\gamma m(m-n)$ and $\tilde{I}=2\alpha+2\beta m +\tilde{\gamma}m(m-n)$. On the other hand, one finds that \eqref{g2-O-tildeI} is linear in all functions so they should appear as monomial with the same degree. Then one can insert the solutions $O(m,n)=\alpha+\beta m+\gamma m(m-n)$ and $\tilde{I}(m,n)=2\alpha+2\beta m +\tilde{\gamma}m(m-n)$ into \eqref{g2-O-tildeI} and finds that there is no solution for $g_{2}(m,n)$ for none of them. Thus we have to set $g_{2}(m,n)=0$, which implies that $O(m,n)=\tilde{I}(m,n)=0$.

In the case of $h_{1}$, one can find a relation for such function from the Jacobi identity $[\mathcal{P},[\mathcal{P},\mathcal{P}]]+\text{cyclic permutations}=0$ which implies at first order in function the following relation:
\begin{equation}\label{f1h1-2}
(n-l)(m-n-l)h_{1}(n,l)+ (l-m)(n-l-m)h_{1}(l,m)+(m-n)(l-m-n)h_{1}(m,n) =0,
\end{equation}
which is solved for $h_{1}(m,n)=constant$. 

Following the same procedure, it is possible to show from the Jacobi identity $[\mathcal{P},[\mathcal{P},\mathcal{M}]]+\text{cyclic permutations}=0$ that the functions $g_{3}$, $f_{3}$ and $h_{3}$ have to satisfy:
 \begin{multline}\label{f2-g3-h1}
(m-n-l)f_{2}(n,l)-(n-m-l)f_{2}(m,l)+(m-n)(l-m-n)g_{3}(l,m+n)+
\\ (m-n)(l-m-n)h_{1}(m,n)=0,
\end{multline}
 \begin{equation} 
\label{h2-f3}
 (m-n-l)h_{2}(n,l)-(n-m-l)h_{2}(m,l)+(m-n)(l-m-n)f_{3}(l,m+n)=0,\\
\end{equation} 
and 
\begin{equation}\label{h3}
    (m-n)(l-m-n)h_{3}(l,m+n)=0,
\end{equation}
which imply $h_{3}(m,n)=0,g_{3}(m,n)=\text{constant}$ and $f_{3}(m,n)=\text{constant}$.

Finally, one can show that the Jacobi identity $[\mathcal{J},[\mathcal{M},\mathcal{M}]]+ \text{cyclic permutations}=0$  leads to the same results as before for $f_{3}, g_{3}$ and $h_{3}$, while the Jacobi identity $[\mathcal{M},[\mathcal{M},\mathcal{M}]]+ \text{cyclic permutations}=0$ does not lead to any new constraint in first order of functions.



\subsection*{Formal deformations}

Until here we have obtained non trivial solutions for different functions which led to simultaneous infinitesimal deformations. 
In fact, we can turn on the functions $f_{1}$, $f_{2}$, $f_{3}$, $g_{2}$, $g_{3}$, $h_{1}$, $h_{2}$, $K$, $\tilde{K}$, $I$, $G$ and $F$ at the same time. However all of these infinitesimal deformations can not be extended to formal deformations. To obtain a formal deformation, the functions should satisfy the Jacobi identities for all orders in functions \ref{sec:3.3}. Here without entering the details, we discuss possible formal deformations.

As summary, one can see from the Jacobi identities that the non-trivial formal deformations of the \Max\ algebra can be classified in four different algebras. As we shall see, two of the deformed algebras can be written as the direct sum of known structures. The others deformed algebras are new family infinite dimensional algebras. In particular, a new family algebra reproduces, for particular values, interesting results already known in the literature. In what follows we discuss the diverse deformations obtained induced by one of several functions simultaneously. One can show that there is no additional formal deformations when we consider other possible infinitesimal deformations induced by the present functions. 

\section*{Deformations in ideal part} 
The formal deformations of \Max\ algebra in its ideal part lead to obtain two new non isomorphic algebras:

\subsection*{1) \texorpdfstring{$\mathfrak{bms}_{3}\oplus\mathfrak{witt}$}{bms3+witt} algebra} 

A new formal deformation appears by studying the deformation of commutator $[\mathcal{P}_{m},\mathcal{P}_{m}]$ without modifying the other commutation relations. Indeed, as we have previously discussed, the Jacobi identity $[\mathcal{P},[\mathcal{P},\mathcal{P}]]$ $+\text{cyclic permutations}=0$ leads to relations \eqref{f1-O-tildeI} which is linear in functions. A non linear relation also appears from such Jacobi identity as
 \begin{equation}
     \begin{split}
         &(n-l)(m-n-l)f_{1}(n,l)f_{1}(m,l+n) + (l-m)(n-l-m)f_{1}(l,m)f_{1}(n,l+m)+\\
         &(m-n)(l-m-n)f_{1}(m,n)f_{1}(l,m+n) = 0. 
     \end{split}
 \end{equation}
with the same solution as $f_1(m,n)=\text{constant}$. The complete analysis to solve the equation for $f_{1}$ in the relation
\begin{multline}\label{f1h1}
(n-l)(m-n-l)f_{1}(m,l+n)+ (l-m)(n-l-m)f_{1}(n,l+m)+\\+(m-n)(l-m-n)f_{1}(m,n)=0,
\end{multline}can be found in section \ref{sec:4.1}.
One can show that the new algebra 
\begin{equation} 
\begin{split}\label{f1-deform-maxwell}
 & i[\mathcal{J}_m,\mathcal{J}_n]=(m-n)\mathcal{J}_{m+n}, \\
 &i[\mathcal{J}_m,\mathcal{P}_n]=(m-n)\mathcal{P}_{m+n},\\
 &i[\mathcal{J}_m,\mathcal{M}_n]=(m-n)\mathcal{M}_{m+n},\\
 &i[\mathcal{P}_m,\mathcal{P}_n]=(m-n)\mathcal{M}_{m+n}+\varepsilon(m-n)\mathcal{P}_{m+n}.\\
\end{split}
\end{equation}
 obtained by $f_{1}(m,n)=\varepsilon$, with $\varepsilon$ being an arbitrary constant, is not isomorphic to the original algebra and hence the deformation is non trivial. By a redefinition of generators \footnote{The parameter $\varepsilon$ can be removed by an appropriate redefinition as $\mathcal{P}\equiv\varepsilon\mathcal{P}$ and $\mathcal{M}\equiv\varepsilon^{2}\mathcal{M}$. } as \begin{equation} 
\begin{split}\label{redefine-pp}
  \mathcal{J}_m\equiv L_{m}+S_{m}, \qquad &\mathcal{P}_m\equiv T_{m}+S_{m}, \qquad \mathcal{M}_m\equiv - T_{m},
\end{split}
\end{equation}
one reaches to the new algebra with non vanishing commutators
\begin{equation} 
\begin{split}\label{bms3+witt-2}
 & i[{L}_m,{L}_n]=(m-n){L}_{m+n}, \\
 &i[{L}_m,{T}_n]=(m-n){T}_{m+n},\\
 &i[{S}_m,{S}_n]=(m-n){S}_{m+n}.
\end{split}
\end{equation}
The new algebra \eqref{bms3+witt-2} has the direct sum structure as $\mathfrak{bms}_{3}\oplus\mathfrak{witt}$. 
The global part of the algebra \eqref{bms3+witt-2} corresponds to the $\mathfrak{iso}(2,1)\oplus\mathfrak{so}(2,1)$ algebra when we restrict ourselves to $m,n=\pm1,0$ which is the direct sum of the $3d$ Poincar\'{e}  and the $3d$ Lorentz algebras. A similar  structure has also been obtained as a deformation of the $d=2+1$ Maxwell algebra in \cite{Gomis:2009dm} but not in the same basis as \eqref{f1-deform-maxwell}. Note also that this algebra is a subalgebra of $W(0,-1;0,0)$, which is obtained as deformation of $\mathfrak{bms}_{4}$ algebra in section \ref{sec:5.3}.

Interestingly the same structure can be obtained by turning on $f_{1}$ and $g_{2}$ simultaneously. In fact we have from the Jacobi identity $[\mathcal{P},[\mathcal{P},\mathcal{M}]]+\text{cyclic permutations}=0$ the following relation
\begin{equation}\label{g2f1-formal}
g_{2}(n,l)g_{2}(m,n+l)-g_{2}(m,l)g_{2}(n,l+m)-(m-n)g_{2}(m+n,l)f_{1}(m,n)=0,
\end{equation}
which is solved for $g_{2}(m,n)=\varepsilon(m-n)$ and $f_{1}(m,n)=\varepsilon$. Let us note that $g_{2}(m,n)=\varepsilon(m-n)$ comes directly from a relation similar to \eqref{f,g,h-2(1)} as a consequence of the Jacobi identity $[\mathcal{P},[\mathcal{M},\mathcal{J}]]+\text{cyclic permutations}=0$. So the commutators of the new algebra obtained through this deformation are
\begin{equation} 
\begin{split}\label{f1g2-deform-maxwell}
 & i[\mathcal{J}_m,\mathcal{J}_n]=(m-n)\mathcal{J}_{m+n}, \\
 &i[\mathcal{J}_m,\mathcal{P}_n]=(m-n)\mathcal{P}_{m+n},\\
 &i[\mathcal{J}_m,\mathcal{M}_n]=(m-n)\mathcal{M}_{m+n},\\
 &i[\mathcal{P}_m,\mathcal{P}_n]=(m-n)\mathcal{M}_{m+n}+\varepsilon(m-n)\mathcal{P}_{m+n},\\
 &i[\mathcal{P}_m,\mathcal{M}_n]=\varepsilon(m-n)\mathcal{M}_{m+n},\\
 &i[\mathcal{M}_m,\mathcal{M}_n]=0. 
\end{split}
\end{equation}
One can show that the $\mathfrak{bms}_{3}\oplus\mathfrak{witt}$ algebra appears by considering an appropriate redefinition of the generators as\footnote{For convenience we drop the parameter $\varepsilon$ in our redefinition since it can be absorbed by an appropriate redefinition of the generators.}
\begin{equation} 
\begin{split}\label{redefine-f1g2}
  \mathcal{J}_m\equiv L_{m}+S_{m}, \qquad \mathcal{P}_m\equiv L_{m}+T_{m}, \qquad \mathcal{M}_m\equiv T_{m}.
\end{split}
\end{equation}

\subsection*{2) \texorpdfstring{$\mathfrak{witt}\oplus\mathfrak{witt}\oplus\mathfrak{witt}$}{witt+witt+witt} algebra}

Three copies of the Witt algebra can be obtained through deformations induced by two or more functions simultaneously and after an appropriate redefinition of the generators. Here, based on \eqref{f2-tildeO} for $f_2$ and using \eqref{f2-g3-h1} for $g_{3}$, we shall turn on two functions simultaneously. Indeed, the Jacobi identity $[\mathcal{M},[\mathcal{M},\mathcal{P}]]+\text{cyclic permutations}=0$ and $[\mathcal{M},[\mathcal{M},\mathcal{M}]]+\text{cyclic permutations}=0$ gives rise to non linear relations as
\begin{equation}\label{f2-g3-formal}
 f_{2}(l,n)f_{2}(l+n,m)- f_{2}(l,m)f_{2}(l+m,n)+(m-n)g_{3}(m,n)f_{2}(l,m+n)=0,
\end{equation}
and 
\begin{multline}\label{g3-formal}
 (n-l)(m-n-l)g_{3}(n,l)g_{3}(m,l+n)+ (l-m)(n-l-m)g_{3}(l,m)g_{3}(n,l+m)+\\ +
  (m-n)(l-m-n)g_{3}(m,n)g_{3}(l,m+n)=0,
\end{multline}
where is also satisfied with the solutions $f_2(m,n)=\lambda (m-n)$ and $g_3(m,n)=\lambda$. Then we find the following formal deformation of the \Max\ algebra
 \begin{equation} 
\begin{split}\label{PM-deform-maxwell4}
 & i[\mathcal{J}_m,\mathcal{J}_n]=(m-n)\mathcal{J}_{m+n}, \\
 &i[\mathcal{J}_m,\mathcal{P}_n]=(m-n)\mathcal{P}_{m+n},\\
 &i[\mathcal{J}_m,\mathcal{M}_n]=(m-n)\mathcal{M}_{m+n},\\
 &i[\mathcal{P}_m,\mathcal{P}_n]=(m-n)\mathcal{M}_{m+n},\\
 &i[\mathcal{P}_m,\mathcal{M}_n]=\lambda(m-n)\mathcal{P}_{m+n},\\
 &i[\mathcal{M}_m,\mathcal{M}_n]=\lambda(m-n)\mathcal{M}_{m+n}. 
\end{split}
\end{equation}
Upon the following redefinition of the generators,
\begin{equation} 
\begin{split}\label{redefine-f2g3}
  \mathcal{J}_m\equiv L_{m}+T_{m}+S_{m}, \qquad \mathcal{P}_m\equiv L_{m}-T_{m}, \qquad \mathcal{M}_m\equiv L_{m}+T_{m}.
\end{split}
\end{equation}
the above algebra reproduces three copies of the Witt algebra
\begin{equation} 
\begin{split}\label{3witt}
 & i[L_m,L_n]=(m-n)L_{m+n}, \\
 &i[T_m,T_n]=(m-n)T_{m+n},\\
 &i[S_m,S_n]=(m-n)S_{m+n}.
\end{split}
\end{equation}
This result is the infinite dimensional generalization of the one obtained in \cite{Gomis:2009dm} for the $2+1$ Maxwell algebra which was called $k-$deformation. In particular, they showed that the $k-$deformation leads to one of $\mathfrak{so}(2,2)\oplus\mathfrak{so}(2,1)$ or $\mathfrak{so}(3,1)\oplus\mathfrak{so}(2,1)$ algebras depending on the sign of the deformation parameter. On the other hand, the three copies of the Witt algebra have three $\mathfrak{sl}(2,\mathbb{R})$ algebras as their global part. In this specific basis both $\mathfrak{so}(2,2)$ and $\mathfrak{so}(3,1)$ are written as $\mathfrak{sl}(2,\mathbb{R})\oplus\mathfrak{sl}(2,\mathbb{R})$, while $\mathfrak{so}(2,1)$ is written as $\mathfrak{sl}(2,\mathbb{R})$. At the gravity level, the so-called AdS-Lorentz algebra, which can be written as three $\mathfrak{so}(2,1)$, allows to accommodate a cosmological constant to the three-dimensional Maxwell Chern-Simons gravity action \cite{Hoseinzadeh:2014bla, Diaz:2012zza, Concha:2018jjj}. 

It is interesting to note that three copies of the Witt algebra can be alternatively obtained by turning on other functions. Indeed one can easily verify that $f_1(m,n)=g_3(m,n)=\delta$ and $f_2=\delta(m-n)$ also reproduces such structure. The formal deformations induced by two functions simultaneously as $h_1$ and $g_3$ or $h_2$ and $f_3$ also reproduce the three copies of the Witt algebra after an appropriate redefinition of the generators. It is important to clarify that such deformations with coefficients being not in the ideal part can be obtained as a redefinition of a deformed \Max\ algebra with coefficients in the ideal part such that the conjecture presented in chapter \ref{ch4} about a possible extension of the Hochschild-Serre factorization theorem is still valid.

One could conjecture that, based on the analysis done for the direct sum of two Witt algebras, \emph{cf.} chapter \ref{ch4}, the direct sum of three Witt algebra is rigid. Furthermore, one could expect to recover the $\mathfrak{witt}\oplus\mathfrak{witt}\oplus\mathfrak{witt}$ algebra as a deformation of the $\mathfrak{bms}_3\oplus\mathfrak{witt}$ algebra since we know that the $\mathfrak{bms}_3$ algebra is not stable and can be deformed to two copies of the Witt algebra.

We can express our results for deformations of \Max\ ideal part into a theorem as 

\begin{tcolorbox}[colback=red!3!white]
\paragraph{Theorem 6.1} {\it The most general formal deformations of \Max\ algebra in its ideal part is either $\mathfrak{bms}_3\oplus\mathfrak{witt}$ or $\mathfrak{witt}\oplus\mathfrak{witt}\oplus\mathfrak{witt}$.}
\end{tcolorbox}

\section*{Deformations in non ideal part}
We considered formal deformations in non ideal commutators of \Max\ algebra and found that they lead to two separate family algebras as 

\subsection*{The \texorpdfstring{$\bar{M}(\bar{\alpha},\bar{\beta};\bar{\nu})$}{M(bar-alpha,bar-beta)} algebra}

One of the new formal deformations obtained is induced by the functions $F(m,n)=\bar{\nu}$ and $I(m,n)=\bar{\alpha}+\bar{\beta}m-\bar{\nu} n$ coming from \eqref{F-I-JJJ} such that the new algebra satisfies the following non vanishing commutation relations:
\begin{equation} 
\begin{split}\label{new-family-IF}
 & i[\mathcal{J}_m,\mathcal{J}_n]=(m-n)\mathcal{J}_{m+n}+\bar{\nu}(m-n)\mathcal{P}_{m+n}, \\
 &i[\mathcal{J}_m,\mathcal{P}_n]=(m-n)\mathcal{P}_{m+n}+(\bar{\alpha}+\bar{\beta}m-\bar{\nu} n)\mathcal{M}_{m+n},\\
 &i[\mathcal{J}_m,\mathcal{M}_n]=(m-n)\mathcal{M}_{m+n},\\
 &i[\mathcal{P}_m,\mathcal{P}_n]=(m-n)\mathcal{M}_{m+n}.
\end{split}
\end{equation}
We call this new family algebra as $\bar{M}(\bar{\alpha},\bar{\beta};\bar{\nu})$. 
 One can check that the functions $F(m,n)$ and $I(m,n)$ are fixed by the Jacobi identity $[\mathcal{J},[\mathcal{J},\mathcal{J}]]+\text{cyclic permutations}=0$ which implies the non linear relation in deformation parameter as
\begin{multline}\label{F-I-JJJ}
    (n-l)F(n,l)I(m,n+l)+(l-m)F(l,m)I(n,l+m)+(m-n)F(m,n)I(l,m+n)=0,
\end{multline}
whose solution is given by $F(m,n)=\bar{\nu}$ and $I(m,n)=\bar{\alpha}+\bar{\beta}m-\bar{\nu}n$.

To our knowledge, this is a novel structure whose global part has not been explored yet. It would be interesting to study the implication of such symmetry and analyze it for different  values of $\bar{\alpha}$, $\bar{\beta}$ and $\bar{\nu}$.

It is interesting to note that $\bar{\nu}=0$ reproduces a deformed algebra induced by $I=\bar{\alpha}-\bar{\beta}m$. The particular case $\bar{M}(\bar{\alpha},\bar{\beta};0)$ can be recovered by deforming only the commutator $[\mathcal{J}_m,\mathcal{P}_n]$ which implies $I=\bar{\alpha}+\bar{\beta}m+\bar{\gamma}m(m-n)+\cdots$ from the Jacobi identity $[\mathcal{J},[\mathcal{J},\mathcal{P}]]+\text{cyclic permutations}=0$ as we have previously discussed. A specific redefinition of the generators can be considered as
\begin{equation} 
\begin{split}\label{redefine-I}
  \mathcal{J}_m\equiv \tilde{\mathcal{J}}_m, \qquad \mathcal{P}_m\equiv \tilde{\mathcal{P}}_m+F(m)\tilde{\mathcal{M}}_m, \qquad \mathcal{M}_m\equiv \tilde{\mathcal{M}}_m.
\end{split}
\end{equation}
This redefinition does not change the ideal part and yields to the following relation:
\begin{equation}
    (m-n)(F(n)-F(m+n))\tilde{\mathcal{M}}_{m+n}=I(m,n)\tilde{\mathcal{M}}_{m+n}.
\end{equation}
One can then check that the solutions given by $I(m,n)=\bar{\gamma} m(m-n)+...$ can be absorbed by the above redefinition when $F(m)=a_{0} +a_{1} m+a_{2}m^{2}+\cdots$. In this way, the only non trivial formal deformation induced by $I(m,n)$ is 
\begin{equation}\label{barM-1}
     [\mathcal{J}_m,\mathcal{P}_n]=(m-n)\mathcal{P}_{m+n}+(\bar{\alpha}+\bar{\beta} m)\mathcal{M}_{m+n}.
\end{equation}
An interesting feature of the $\bar{M}(\bar{\alpha},\bar{\beta};\bar{\nu})$ algebra is that such symmetry is obtained by deforming the commutators $[\mathcal{J},\mathcal{J}]$ and $[\mathcal{J},\mathcal{P}]$ which are not the ideal part of the infinite dimensional algebra. 


\subsection*{The \texorpdfstring{$M(a,b;c,d)$}{M(a,b;c,d)} algebra}
Another formal deformation is obtained by turning on simultaneously the functions $K$ and $\tilde{K}$. One can show from the Jacobi identities $[\mathcal{J},[\mathcal{J},\mathcal{P}]]+\text{cyclic permutations}=0$ and $[\mathcal{J},[\mathcal{J},\mathcal{M}]]+\text{cyclic permutations}=0$ that (see \eqref{firsteq-K} and \eqref{K-tildeK,O-tildeI})
\begin{equation}
K(m,n)=\alpha +\beta m, \qquad \tilde{K}(m,n)=2\alpha +2\beta m, \label{K-generic}
\end{equation}
which is the only solution to  all orders in functions.

The new algebra, which we name it as $M(a,b;c,d)$ algebra, has the following non vanishing commutation relations
\begin{equation} 
\begin{split}\label{new-Walgebra}
 & i[\mathcal{J}_m,\mathcal{J}_n]=(m-n)\mathcal{J}_{m+n}, \\
 &i[\mathcal{J}_m,\mathcal{P}_n]=-(bm+n+a)\mathcal{P}_{m+n},\\
 &i[\mathcal{J}_m,\mathcal{M}_n]=-(dm+n+c)\mathcal{M}_{m+n},\\
 &i[\mathcal{P}_m,\mathcal{P}_n]=(m-n)\mathcal{M}_{m+n},
\end{split}
\end{equation}
 where $c=2a=-2\alpha$ and $d=b-\beta=-2\beta-1$. One can show that such formal deformation can alternatively be obtained by considering the functions $K$ $\tilde{K}$ and $I$ simultaneously. Indeed, from the Jacobi identity $[\mathcal{J},[\mathcal{J},\mathcal{P}]]+\text{cyclic permutations}=0$ we have $K$ and $\tilde{K}$ are given by \eqref{K-generic} and $I=\xi(\alpha+\beta m)$. 
 Although such functions seems to induce a new formal deformation, one can use the same redefinition as in \eqref{redefine-I} to obtain
 \begin{equation}
     [\tilde{\mathcal{J}}_m,\tilde{\mathcal{P}}_n+F(n)\tilde{\mathcal{M}}_n]=(m-n+\alpha+\beta m)\left(\tilde{\mathcal{P}}_{m+n}+F(m+n)\tilde{\mathcal{M}}_{m+n}\right)+\xi(\alpha+\beta m)\tilde{\mathcal{M}}_{m+n},
\end{equation}
which reproduces the same algebra as \eqref{new-Walgebra} when $F(m)=constant=\xi$.

As it is discussed in section \ref{sec:4.1}, in the context of $2d$ conformal field theory the parameters $b$ and $d$ are related to $h$ and $\tilde{h}$, which are the conformal weight of $\mathcal{P}$ and $\mathcal{M}$ respectively, through $b=1-h$ and $d=1-\tilde{h}$. On the other hand, the parameters $a$ and $c$ are related to the periodicity  of primary field $\mathcal{P}(\varphi)$ (or $\mathcal{M}(\varphi)$) through
\begin{equation*}
\mathcal{P}(\varphi+2\pi)=e^{2\pi i a}\mathcal{P}(\varphi),\qquad \mathcal{P}(\varphi)=\sum_n \mathcal{P}_n e^{i(n+a)\varphi}.
\end{equation*}
It is interesting to point out that various infinite dimensional structures appears for specific values of $a,b,c$ and $d$. In particular, let us suppose that $a=c=0$ in \eqref{new-Walgebra} and let us consider different values of $b,d$. First we set $b=0, d=1$ which leads to the algebra $M(0,0;0,1)$ with the following commutators
\begin{equation} 
\begin{split}\label{new-Walgebra-1}
 & i[\mathcal{J}_m,\mathcal{J}_n]=(m-n)\mathcal{J}_{m+n}, \\
 &i[\mathcal{J}_m,\mathcal{P}_n]=(-n)\mathcal{P}_{m+n},\\
 &i[\mathcal{J}_m,\mathcal{M}_n]=(-m-n)\mathcal{M}_{m+n},\\
 &i[\mathcal{P}_m,\mathcal{P}_n]=(m-n)\mathcal{M}_{m+n}.
\end{split}
\end{equation}
The generators $\mathcal{P}$ and $\mathcal{M}$ can be seen as a $U(1)$ current and a primary operator with conformal weight $h=0$, respectively. The infinite dimensional algebra \eqref{new-Walgebra-1} corresponds to a Maxwellian version of the so-called Virasoro-Kac-Moody algebra.
A different choice is $b=-\frac{1}{2},d=0$ which leads to a new algebra $M(0,-\frac{1}{2};0,0)$ whose non vanishing commutators are given by
\begin{equation} 
\begin{split}\label{twisted-virasoro-schrodinger}
 & i[\mathcal{J}_m,\mathcal{J}_n]=(m-n)\mathcal{J}_{m+n}, \\
 &i[\mathcal{J}_m,\mathcal{P}_n]=(\frac{m}{2}-n)\mathcal{P}_{m+n},\\
 &i[\mathcal{J}_m,\mathcal{M}_n]=(-n)\mathcal{M}_{m+n},\\
 &i[\mathcal{P}_m,\mathcal{P}_n]=(m-n)\mathcal{M}_{m+n},
\end{split}
\end{equation}
in which the generators $\mathcal{P}$ and $\mathcal{M}$ can be seen as a primary operator with conformal weight $h=\frac{3}{2}$ and a $U(1)$ current, respectively. This algebra is known as {\it twisted Schr\"{o}dinger-Virasoro algebra} \cite{Unterberger:2011yya}. In this reference the infinite enhancement of $2+1$ Maxwell algebra, which is called $\mathfrak{sv}_{1}(0)$, is obtained as a deformation of the twisted Schr\"{o}dinger-Virasoro algebra.

When the indices of the generator $\mathcal{P}$ are half integer valued the algebra corresponds to the so-called {\it Schr\"{o}dinger-Virasoro algebra} with spatial dimension $d=1$. The Schr\"{o}dinger-Virasoro algebra has a global part which is spanned by $6$ generators $\mathcal{J}_{0,\pm 1}$, $\mathcal{P}_{\pm\frac{1}{2}}$ and $\mathcal{M}_{0}$ where the latter appears as a central term. There are different work, for instance \cite{Alishahiha:2009nm, Compere:2009qm}, in which the authors have tried to find the Schr\"{o}dinger-Virasoro algebra as asymptotic symmetry of some spacetimes. 

An interesting feature of the $M(a,b;c,d)$ algebra is that, as the $\bar{M}(\bar{\alpha},\bar{\beta};\bar{\nu})$ algebra, it obeys our  conjecture made in chapter \ref{ch4}. 

Let us note that the family algebra $M(a,b;c,d)$, for some specific values of its parameters, can be deformed into new algebras out of this family. For example the \Max\ algebra given by $M(0,-1;0,-1)$  can be deformed in its ideal part into $\mathfrak{bms}_{3} \oplus \mathfrak{witt}$ as we shall see in the next section. Furthermore, the Schr\"{o}dinger-Virasoro algebra given by $M(0,\frac{1}{2};0,0)$ can be deformed in its $[\mathcal{J},\mathcal{J}]$ commutator. Despite this, it seems that the family algebra $M(a,b;c,d)$ is stable in the sense that for generic values of its parameters it can just be deformed into another family algebra $M(\bar{a},\bar{b};\bar{c},\bar{d})$ with shifted parameters. The latter should however be proved by direct computations.

As an ending remark, let us note that in
\cite{Compere:2009qm} they introduced the algebra with the same structure as $M(a,b;c,d)$. This algebra which is obtained with specific values of parameters as $M(\frac{z-2}{2z},\frac{-1}{z};\frac{z-2}{z},\frac{z-2}{z})$, is introduced as asymptotic symmetry algebra of Schr\"{o}dinger spacetimes.

The formal deformations of \Max\ algebra are classified into four mentioned algebras. This result can be stated as following theorem

\begin{tcolorbox}[colback=red!3!white]
\paragraph{Theorem 6.2} {\it The most general deformations of \Max\ algebra are  $\mathfrak{bms}_{3} \oplus \mathfrak{witt}$ and $\mathfrak{witt} \oplus \mathfrak{witt}\oplus\mathfrak{witt}$ in its ideal part and two family algebras $\bar{M}(\bar{\alpha},\bar{\beta};\bar{\nu})$ and $M(a,b;c,d)$.}
\end{tcolorbox}

\paragraph{Remark.} Although the deformations leading to family algebras $\bar{M}(\bar{\alpha},\bar{\beta};\bar{\nu})$ and $M(a,b;c,d)$ constitute obvious deviations from the Hochschild-Serre factorization theorem, they reinforce our conjecture presented in chapter \ref{ch4} which may be viewed as an extension of this theorem for infinite dimensional Lie algebras. Also, it is important to note that the deformation leading to three copies of \wit\ algebra with coefficients being not in the ideal part, can be obtained as a redefinition of a deformed \Max\ algebra with coefficients in the ideal part such that the conjecture is still valid.

\subsection*{Algebraic cohomology argument}\label{sec:cohomology}

Until now we have classified all possible nontrivial infinitesimal and formal deformations of the \Max\ algebra by studying the Jacobi identities. As discussed in sections \ref{sec:cohomology-bms} and \ref{sec:5.5}, one can approach and analyze such issue by cohomology consideration. Indeed one can classify all infinitesimal deformations of the \Max\ algebra by computing $\mathcal{H}^{2}(\mathfrak{Max}_{3};\mathfrak{Max}_{3})$. In our previous works, in which we tackled Lie algebras with abelian ideal, we used the theorem 2.1 of \cite{degrijse2009cohomology} which is  crucial for cohomological consideration. Nonetheless, we cannot use this theorem here since \Max\ does not have abelian ideal. We shall only state our result in cohomological language. As we can see from the our results in previous part, we have just four formal deformations for the \Max\ algebra. It is obvious that both $M(a,b;c,d)$ and $\bar{M}(\bar{\alpha},\bar{\beta};\bar{\nu})$ family algebras are deformed by the $K, \tilde{K}, I$ and $F$ terms, with coefficients from ideal part, $\mathcal{P}$ and $\mathcal{M}$. The same argument is true for the new algebra  $\mathfrak{bms}_{3} \oplus \mathfrak{witt}$ which is obtained through deformation induced by $f_{1}$ with coefficient in $\mathcal{P}$. The three copies of the Witt algebra can be obtained via deformation induced by $h_{1}, g_{3}$ or $h_{2}, f_{3}$  and also by $f_{2},g_{3}$, which means that the two first cases are just a redefinition of the latter. As summary, we have shown that, unlike the Hochschild-Serre factorization theorem of finite Lie algebras, other commutators of \Max\ algebra, except the ideal part, can also be deformed but only by terms with coefficients from the ideal part. As it has been discussed in chapters \ref{ch4} and \ref{ch5} this result can be viewed as an extension of the Hochschild-Serre factorization theorem for infinite dimensional algebras. \footnote{Here we are tackling infinite dimensional Lie algebras which are extensions of the Witt algebra.} 
 
 In the cohomological language our results for the \Max\ algebra can be written as
\begin{equation}
    \mathcal{H}^{2}(\mathfrak{Max}_{3};\mathfrak{Max}_{3})\cong \mathcal{H}^{2}(\mathfrak{Max}_{3};\mathfrak{h}).
\end{equation}
where $\mathfrak{h}$ denotes the ideal part of \Max\ algebra spanned by generators $\mathcal{P}$ and $\mathcal{M}$.



\section{Central extensions of the deformed \texorpdfstring{$\mathfrak{Max}_{3}$}{Max3} algebras}\label{sec:6.4}

In this section, we present explicit central extensions of the infinite-dimensional algebras obtained as a deformation of the $\mathfrak{Max}_3$ algebra introduced previously. In particular, one of the central extension reproduces a known asymptotic symmetry of a three-dimensional gravity theory.

\subsection*{Central extensions of the \texorpdfstring{$\mathfrak{bms}_{3} \oplus \mathfrak{witt}$}{bms3+witt} and the \texorpdfstring{$\mathfrak{witt}\oplus\mathfrak{witt}\oplus\mathfrak{witt}$}{witt+witt+witt} algebra}

We have shown that ones of the deformations of the \Max\ algebra are given by the $\mathfrak{bms}_{3} \oplus \mathfrak{witt}$ and three copies of the Witt algebra. In this section we briefly review the known central extensions of the $\mathfrak{bms}_3$ and the Witt algebra.

The most general central extension of the $\mathfrak{bms}_{3} \oplus \mathfrak{witt}$ is given by
\begin{equation} 
\begin{split}\label{bms3+witt_CE}
 & i[{L}_m,{L}_n]=(m-n){L}_{m+n}+\frac{c_{LL}}{12}m^{3}\delta_{m+n,0}, \\
 &i[{L}_m,{T}_n]=(m-n){T}_{m+n}+\frac{c_{LT}}{12}m^{3}\delta_{m+n,0},\\
 &i[{S}_m,{S}_n]=(m-n){S}_{m+n}+\frac{c_{SS}}{12}m^{3}\delta_{m+n,0},
\end{split}
\end{equation}
where the central charges $c_{LL}$, $c_{LT}$ and $c_{SS}$ can be related to three independent terms of the Chern-Simons $\mathfrak{iso}(2,1)\oplus \mathfrak{so}(2,1)$ gravity action \eqref{CS} as follows 

\begin{equation}
\begin{split}\label{CentralTerms4}
c_{LL}=12k(\alpha_{0}-\alpha_{1}-\alpha_{2}), \qquad c_{LT}=-12k\alpha_{2}, \qquad c_{SS}=12k(\alpha_{1}+\alpha_{2}),
\end{split}
\end{equation}
where we used the redefinition \eqref{cLL,cSS,cLT,ci} and relation \eqref{Ci-kalphai} when we set $\varepsilon=\ell=1$.
   In section \ref{sec:3.6} we explore the possibility that the central extensions of the infinite-dimensional algebra \eqref{f1-deform-maxwell} or \eqref{bms3+witt-2}, with the above central terms, appear as the asymptotic symmetries of three-dimensional gravity theory invariant under deformations of the Maxwell algebra.

On the other hand, centrally extended $\mathfrak{witt}\oplus\mathfrak{witt}\oplus\mathfrak{witt}$ algebra is  given by\begin{equation} 
\begin{split}\label{3Vir}
 & i[L_m,L_n]=(m-n)L_{m+n}+\frac{c_{LL}}{12}m^{3}\delta_{m+n,0}, \\
 &i[T_m,T_n]=(m-n)T_{m+n}+\frac{c_{TT}}{12}m^{3}\delta_{m+n,0},\\
 &i[S_m,S_n]=(m-n)S_{m+n}+\frac{c_{SS}}{12}m^{3}\delta_{m+n,0}.
\end{split}
\end{equation}
Interestingly, considering the following redefinition of the generators
\begin{equation} 
\begin{split}\label{redefine-AdSL}
L_m\equiv \frac{1}{2}(\mathcal{M}_{m}+\mathcal{P}_{m}),\quad T_m\equiv \frac{1}{2}(\mathcal{M}_{m}-\mathcal{P}_{m}), \quad S_m\equiv \mathcal{J}_{m}-\mathcal{M}_{m},
\end{split}
\end{equation}
 and the following redefinition of the central terms
\begin{equation} 
\begin{split}\label{redefine-centralterms}
 c_{LL}\equiv \frac{1}{2}(c_{JM}+c_{JP}), \quad c_{TT}\equiv \frac{1}{2}(c_{JM}-c_{JP}), \quad c_{SS}\equiv (c_{JJ}-c_{JM}).
\end{split}
\end{equation}
we recover the asymptotic symmetry of the Chern-Simons gravity theory invariant under the so-called AdS-Lorentz algebra \cite{Concha:2018jjj}. Such symmetry has been previously studied in \cite{Soroka:2006aj, Gomis:2009dm, Hoseinzadeh:2014bla, Diaz:2012zza} and extended to higher dimensions in \cite{Concha:2016kdz, Concha:2016tms, Concha:2017nca} in Lovelock theory.

\subsection*{Central extensions of \texorpdfstring{$M(a,b;c,d)$}{M(a,b;c,d)}}
Here we shall classify the central terms of the $M(a,b;c,d)$ algebra. One can easily find that the $M(a,b;c,d)$ algebra for generic values of parameter space $a,b,c$ and $d$ admits only one central term in its Witt subalgebra. However there are some specific points in which it is possible to have other non-trivial central terms. We follow the results of the work \cite{gao2011low} which classifies the central terms of $W(a,b)$ algebra.

\subsubsection*{Central terms for specific points in parameters space of \texorpdfstring{$M(a,b;c,d)$}{M(a,b;c,d)}}
\paragraph{ \texorpdfstring{$M(0,0;0,1)$}{M(d=1,a,b,c=0)} case.}
By setting the parameters as $a,b,c=0,d=1$ we obtain a new algebra with non vanishing commutators as in  \eqref{new-Walgebra-1}. One can readily check that there is a central term in the Witt subalgebra given by $c_{JJ}m^{3}$ 
so we shall take it in account in what follows.
Let us consider now the central term as $[\mathcal{J}_m,\mathcal{P}_n]=(-n)\mathcal{P}_{m+n}+S(m,n)$ where $S(m,n)$ is an arbitrary function. One can see that the Jacobi identity $[\mathcal{J},[\mathcal{J},\mathcal{P}]]+\text{cyclic permutations}=0$ implies the following constraint
\begin{equation}\label{eq-S}
   -lS(m,n+l)+lS(n,l+m)+(n-m)S(m+n,l)=0,
\end{equation}
If the function $S(m,n)$ is a symmetric function by setting $l=0$ one obtains $S(m+n,0)=0$. Then the only solution is $S(m,n)=c_{JP}m^{2}\delta_{m+n,0}$ in which $c_{JP}$ is an arbitrary constant as expected from central extension of the Virasoro-Kac-Moody algebra, \emph{cf.} section \ref{sec:4.2}. On the other hand, one can show that there is no solution for $S(m,n)$ being an anti symmetric function. The rest of the Jacobi identities do not put additional constraints on $S(m,n)$ reproducing a non trivial central extension. 
Another central term can appear as $[\mathcal{J}_m,\mathcal{M}_n]=(-m-n)\mathcal{M}_{m+n}+T(m,n)$ where $T(m,n)$ is an arbitrary function. The Jacobi identity $[\mathcal{J},[\mathcal{J},\mathcal{M}]]+\text{cyclic permutations}=0$ leads to
\begin{equation}\label{eq-T}
    (-n-l)T(m,n+l)+(m+l)T(n,l+m)+(n-m)T(m+n,l)=0.
\end{equation}
If the function $T(m,n)$ is a symmetric function one obtains $T(m,n)=T(m+n,0)=\bar{T}(m+n)$. Then we have $T(m,n)=(c_{JM1}m+c_{JM2})\delta_{m+n,0}$ where $c_{JM1,2}$ are arbitrary constants. On the other hand the Jacobi identity $[\mathcal{P},[\mathcal{P},\mathcal{J}]]+\text{cyclic permutations}=0$ implies $T(m,n)=0$. One can also see that there is no solution for $T(m,n)$ being an anti symmetric function. 
Let us consider now the presence of central terms in both $[\mathcal{J}_m,\mathcal{M}_n]=(-m-n)\mathcal{M}_{m+n}+T(m,n)\delta_{m+n+l,0}$ and $[\mathcal{P}_m,\mathcal{P}_n]=(m-n)\mathcal{M}_{m+n}+U(m,n)\delta_{m+n,0}$ simultaneously. The Jacobi identity  $[\mathcal{P},[\mathcal{P},\mathcal{J}]]+\text{cyclic permutations}=0$ leads to
\begin{equation}\label{eq-T-U}
    \left((n)U(m,n+l)-(m)U(n,l+m)+(m-n)T(l,m+n)\right)\delta_{m+n+l,0}=0,
\end{equation}
which does not have a non zero solution for $U(m,n)$ when $T(m,n)=c_{JM1}m$. However when we consider $T(m,n)=c_{JM2}$, one finds $U(m,n)=c_{JM2}$ which represents another non trivial central extension. An additional central term can appear in $[\mathcal{P}_m,\mathcal{P}_n]=(m-n)\mathcal{M}_{m+n}+U(m,n)\delta_{m+n,0}$ when other central terms are turned off. The Jacobi identities $[\mathcal{P},[\mathcal{P},\mathcal{P}]]+\text{cyclic permutations}=0$ and $[\mathcal{P},[\mathcal{P},\mathcal{M}]]+\text{cyclic permutations}=0$ do not constrain $U(m,n)$. The only remaining Jacobi identity is $[\mathcal{P},[\mathcal{P},\mathcal{J}]]+\text{cyclic permutations}=0$ which implies
\begin{equation}\label{eq-U}
    \left((n)U(m,n+l)-(m)U(n,l+m)\right)\delta_{m+n+l,0}=0,
\end{equation}
with $U(m,n)=c_{PP}m$. However, it is possible to see that the following redefinition 
\begin{equation}
    \mathcal{M}_{m}\equiv \tilde{\mathcal{M}}_{m}+c\delta_{m,0},
\end{equation}
do not reproduce a non trivial central extension for $c=-\frac{c_{PP}}{2}$ since the central term $c_{PP}$ can be absorbed.

To summarize, the most general central extension of $M(0,0;0,1)$ is
\begin{equation} 
\begin{split}\label{beta=-1}
 & i[\mathcal{J}_m,\mathcal{J}_n]=(m-n)\mathcal{J}_{m+n}+\frac{c_{JJ}}{12}m^{3}\delta_{m+n,0}, \\
 &i[\mathcal{J}_m,\mathcal{P}_n]=(-n)\mathcal{P}_{m+n}+c_{JP}m^{2}\delta_{m+n,0},\\
 &i[\mathcal{J}_m,\mathcal{M}_n]=(-m-n)\mathcal{M}_{m+n}+c_{JM}\delta_{m+n,0},\\
 &i[\mathcal{P}_m,\mathcal{P}_n]=(m-n)\mathcal{M}_{m+n}+c_{JM}\delta_{m+n,0}.
\end{split}
\end{equation}

\paragraph{\texorpdfstring{$M(0,-2;0,-3)$}{a=c=0,b=-2,d=-3} case.}
The next values of the parameters which we will consider is $a=c=0,b=-2,d=-3$ for which we obtain a new algebra with non vanishing commutators as
\begin{equation} 
\begin{split}\label{beta=1,1}
 & i[\mathcal{J}_m,\mathcal{J}_n]=(m-n)\mathcal{J}_{m+n}, \\
 &i[\mathcal{J}_m,\mathcal{P}_n]=(-m-n)\mathcal{P}_{m+n},\\
 &i[\mathcal{J}_m,\mathcal{M}_n]=(-3m-n)\mathcal{M}_{m+n},\\
 &i[\mathcal{P}_m,\mathcal{P}_n]=(m-n)\mathcal{M}_{m+n}.
\end{split}
\end{equation}
Let us consider first the central term in $[\mathcal{J}_m,\mathcal{P}_n]=(-m-n)\mathcal{P}_{m+n}+S(m,n)$. The Jacobi identity $[\mathcal{J},[\mathcal{J},\mathcal{P}]]+\text{cyclic permutations}=0$ reproduces the same constraint as \eqref{eq-T} on $S(m,n)$. So we obtain $S(m,n)=(c_{JP1}m+c_{JP2})\delta_{m+n,0}$. One can turn on a central term as $[\mathcal{J}_m,\mathcal{M}_n]=(-3m-n)\mathcal{M}_{m+n}+T(m,n)$. The Jacobi identity $[\mathcal{M},[\mathcal{M},\mathcal{J}]]+\text{cyclic permutations}=0$ implies
\begin{equation}
   -(3n+l)S(m,l+n)+(3m+l)S(n,l+m)+(n-m)S(m+n,l)=0,
\end{equation}
which has no non trivial solution leading to $T(m,n)=0$. On the other hand one may consider the central term as $[\mathcal{P}_m,\mathcal{P}_n]=(m-n)\mathcal{M}_{m+n}+U(m,n)\delta_{m+n,0}$ however this does not lead to a non trivial central term.  
Therefore, there is no further central extensions for $M(a=c=0,b=-2,d=-3)$ and the most general central extension of this 
algebra is given by
\begin{equation} 
\begin{split}\label{beta=1}
 & i[\mathcal{J}_m,\mathcal{J}_n]=(m-n)\mathcal{J}_{m+n}+\frac{c_{JJ}}{12}m^{3}\delta_{m+n,0}, \\
 &i[\mathcal{J}_m,\mathcal{P}_n]=(-m-n)\mathcal{P}_{m+n}+(c_{JP1}m+c_{JP2})\delta_{m+n,0},\\
 &i[\mathcal{J}_m,\mathcal{M}_n]=(-3m-n)\mathcal{M}_{m+n},\\
 &i[\mathcal{P}_m,\mathcal{P}_n]=(m-n)\mathcal{M}_{m+n}.
\end{split}
\end{equation}
As we can see this is in contradiction with the result of theorem 5.7. of \cite{Roger:2006rz} in which they did not mention the term $c_{JP1}\delta_{m+n,0}$ in \eqref{beta=1}.

\paragraph{\texorpdfstring{$M(0,-\frac{1}{2};0,0)$}{beta=-1,alpha=0} case.}
Another value of the parameters that one could explore is $a=c=0,b=-\frac{1}{2},d=0$ which leads to the new algebra \eqref{twisted-virasoro-schrodinger}. As  mentioned before this algebra is known as the twisted Schr\"{o}dinger-Virasoro algebra. According to the theorem 2.2 in \cite{gao2011low} we know that there is no central term in the $[\mathcal{J}_m,\mathcal{P}_n]$ commutator.\footnote{This can be easily checked by adding a central term like $S(m,n)$ to this commutator and considering the Jacobi identity $[\mathcal{J},[\mathcal{J},\mathcal{P}]]+\text{cyclic permutations}=0$.} One can indeed show from the Jacobi identity that the only central extension for the twisted Schrödinger-Virasoro algebra appears in its Witt subalgebra:
\begin{equation} 
\begin{split}\label{beta=-1/2}
 & i[\mathcal{J}_m,\mathcal{J}_n]=(m-n)\mathcal{J}_{m+n}+\frac{c_{JJ}}{12}m^{3}\delta_{m+n,0}, \\
 &i[\mathcal{J}_m,\mathcal{P}_n]=(\frac{m}{2}-n)\mathcal{P}_{m+n},\\
 &i[\mathcal{J}_m,\mathcal{M}_n]=(-n)\mathcal{M}_{m+n},\\
 &i[\mathcal{P}_m,\mathcal{P}_n]=(m-n)\mathcal{M}_{m+n}.
\end{split}
\end{equation} 



\subsection*{Central extensions of \texorpdfstring{$\bar{M}(\bar{\alpha},\bar{\beta};\bar{\nu})$}{M(bar-alpha,bar-beta)}}

As we have mentioned the functions $I(m,n)$ and $F(m,n)$ are just constrained by the Jacobi identities $[\mathcal{J},[\mathcal{J},\mathcal{J}]]+\text{cyclic permutations}=0$ and $[\mathcal{J},[\mathcal{J},\mathcal{P}]]+\text{cyclic permutations}=0$. Let us then  consider the central terms constrained by these Jacobi identities. In particular, let us first consider the central term as 
$[\mathcal{J}_m,\mathcal{J}_n]=(m-n)\mathcal{J}_{m+n}+\bar{\nu}(m-n)\mathcal{P}_{m+n}+R(m,n)\delta_{m+n,0}$. From the Jacobi identity $[\mathcal{J},[\mathcal{J},\mathcal{J}]]+\text{cyclic permutations}=0$ we find the solution $R(m,n)=c_{JJ}m^{3}$. Let $S(m,n)$ be an arbitrary functions which appears in $[\mathcal{J}_m,\mathcal{P}_n]=(m-n)\mathcal{P}_{m+n}+(\bar{\alpha}+\bar{\beta}m)\mathcal{M}_{m+n}+S(m,n)$ and satisfy the following constraint
\begin{equation}
   (n-l)S(m,l+n)+(l-m)S(n,l+m)+(n-m)S(m+n,l)=0.
\end{equation}
The Jacobi identities $[\mathcal{J},[\mathcal{J},\mathcal{J}]]+\text{cyclic permutations}=0$ and $[\mathcal{J},[\mathcal{J},\mathcal{P}]]+\text{cyclic permutations}=0$, as expected, indicate the existence of a central term $S(m,n)=c_{JP}m^{3}\delta_{m+n,0}$. One can see that a central term can also appear in the commutator $[\mathcal{J}_m,\mathcal{M}_n]=(m-n)\mathcal{M}_{m+n}+T(m,n)$ where $T(m,n)$ is an arbitrary function. From the Jacobi identity $[\mathcal{J},[\mathcal{J},\mathcal{M}]]+\text{cyclic permutations}=0$ we find that the function is fixed as $T(m,n)=c_{JM}m^{3}\delta_{m+n,0}$ if we also turn on the same central term in $[\mathcal{P}_m,\mathcal{P}_n]=(m-n)\mathcal{M}_{m+n}+U(m,n)$ with $U(m,n)=c_{JM}\delta_{m+n,0}$. However one should also consider the Jacobi identity $[\mathcal{J},[\mathcal{J},\mathcal{P}]]+\text{cyclic permutations}=0$ which leads to
\begin{equation}
  c_{JM} \left((\bar{\alpha}+\bar{\beta}n-\bar{\nu}l)m^{3}-(\bar{\alpha}+\bar{\beta}m+\bar{\nu}l)n^{3}+\bar{\nu}(m-n)l^{3}\right)\delta_{m+n+l,0}=0.
\end{equation}
Let us note that since the three parameters $\bar{\alpha}, \bar{\beta}$ and $\bar{\nu}$ are independent, there is no solution for the above expression for $\bar{\alpha},\bar{\beta},\bar{\nu}\neq 0$. Nevertheless for $\bar{\alpha}=\bar{\nu}=0$, we have the non trivial central extension $T(m,n)=U(m,n)=c_{JM}m^{3}\delta_{m+n,0}$. Thus, we conclude that the most general central extension for the $\bar{M}(0,\bar{\beta};0)$ algebra is given by
\begin{equation} 
\begin{split}
 & i[\mathcal{J}_m,\mathcal{J}_n]=(m-n)\mathcal{J}_{m+n}+\frac{c_{JJ}}{12}m^{3}\delta_{m+n,0}, \\
 &i[\mathcal{J}_m,\mathcal{P}_n]=(m-n)\mathcal{P}_{m+n}+\bar{\beta}m\mathcal{M}_{m+n}+\frac{c_{JP}}{12}m^{3}\delta_{m+n,0},\\
 &i[\mathcal{J}_m,\mathcal{M}_n]=(-n)\mathcal{M}_{m+n}+\frac{c_{JM}}{12}m^{3}\delta_{m+n,0},\\
 &i[\mathcal{P}_m,\mathcal{P}_n]=(m-n)\mathcal{M}_{m+n}+\frac{c_{JM}}{12}m^{3}\delta_{m+n,0}.
\end{split}
\end{equation}

\section{Maxwell Chern-Simons gravity theory with torsion}\label{sec:3-6}

Using the CS formalism, we present the three-dimensional gravity theory based on a particular deformation of the Maxwell algebra. Unlike the Maxwell case, such deformation leads to a non-vanishing torsion as equation of motion. The deformed Maxwell algebra is spanned by the generators $\{J_a,P_a,M_a\}$, which satisfy the following non-vanishing commutation relations:
\begin{equation}\label{algebra01} 
\begin{split}
 & [J_a,J_b]=\epsilon_{ab}^{\hspace{3 mm}c} J_{c} \,, \\
 &[J_a,P_b]=\epsilon^{\hspace{3 mm}c}_{ab}P_{c} \,,\\
 &[J_a,M_b]=\epsilon^{\hspace{3 mm}c}_{ab} M_{c} \,,\\
 &[P_a,P_b]=\epsilon^{\hspace{3 mm}c}_{ab} (M_{c}+\frac{1}{\ell}P_{c}) \,,
\end{split}
\end{equation}
where $\epsilon_{abc}$ is the three-dimensional Levi-Civita tensor and $a,b=0,1,2$ are the Lorentz indices which are lowered and raised with the Minkowski metric $\eta_{ab}$.  
Let us note that the Hietarinta-Maxwell algebra \cite{Hietarinta:1975fu,Bansal:2018qyz,Chernyavsky:2020fqs} is recovered in the limit $\ell\rightarrow\infty$ when the role of the $P_{a}$ and $M_{a}$ generators is interchanged. One can see that $J_a$ and $P_a$ are not the generators of a Poincaré subalgebra. However, as it is pointed out
in section \ref{sec:1.6}, \eqref{algebra01} can be rewritten as the $\mathfrak{iso}(2,1)\oplus\mathfrak{so}(2,1)$ algebra. This can be seen by a redefinition of the generators,

\begin{equation} 
\begin{split}\label{redefine-pp1}
 & L_{a}\equiv J_{a} - \ell P_{a}- \ell^2 M_{a} \,,\\
 & S_a\equiv \ell P_{a}+ \ell^2 M_{a} \,,\\
 & T_a\equiv - \ell \,M_{a} \,,
\end{split}
\end{equation}
where $L_{a}$ and $T_{a}$ are the respective generators of the $\mathfrak{iso}(2,1)$ algebra, while $S_{a}$ is a $\mathfrak{so}(2,1)$ generator. Then, the Lie algebra \eqref{algebra01} can be rewritten as
\begin{equation} 
\begin{split}
 & [L_a,L_b]=\epsilon_{ab}^{\hspace{3 mm}c} L_{c} \,, \\
 & [L_a,T_b]=\epsilon_{ab}^{\hspace{3 mm}c} T_{c} \,, \\
 &[S_a,S_b]=\epsilon^{\hspace{3 mm}c}_{ab} S_{c} \,.
\end{split}
\end{equation}

It is important to point out that \eqref{algebra01} is not the unique way of deforming the Maxwell algebra. 
It is shown in \cite{Gomis:2009dm} that
the Maxwell algebra can be deformed into two different algebras: $\mathfrak{so}(2,2)\oplus\mathfrak{so}(2,1)$ and $\mathfrak{iso}(2,1)\oplus\mathfrak{so}(2,1)$. The former has been largely studied in \cite{Diaz:2012zza,Hoseinzadeh:2014bla,Concha:2018jjj,Concha:2019lhn} whose asymptotic symmetry is described by three copies of the Virasoro algebra \cite{Caroca:2018obf,Concha:2018jjj}, while the latter has only been approached through a deformation process \cite{Gomis:2009dm,Concha:2019eip}. In the present work, using the basis $\left\{ J_a,P_a,M_a\right\}$, we find asymptotic symmetry of the CS gravity theory based on the $\mathfrak{iso}(2,1)\oplus\mathfrak{so}(2,1)$ algebra.
The motivation to use such basis is twofold. First, it allows us to recover the Maxwell CS gravity theory in a particular limit. Second, as we shall see, it reproduces the Maxwell field equations with a non-vanishing torsion.

A three-dimensional gravity can be formulated as a CS theory described by the action
\begin{equation}\label{CS-Action}
    S_{\text{CS}}[A]= \frac{k}{4 \pi} \int_{\mathcal{M}} \langle A \di A+ \frac{2}{3} A^{3}  \rangle \,,
\end{equation}
with a given Lie algebra on a manifold $\mathcal{M}$, where $A$ is the gauge connection, $\langle \, , \, \rangle$ denotes the invariant trace and $k=1/(4G)$ is the CS level. For the sake of simplicity we have omitted writing the wedge product. 
The gauge connection one-form $A$ for the deformed Maxwell algebra reads
\begin{equation}\label{one-form}
    A= e^{a} P_{a}+\omega^{a}J_{a}+f^{a} M_{a} \,,
\end{equation}
where $e^{a}$, $\omega^{a}$ and $f^{a}$ are the dreibein, the (dualized) spin connection and an auxiliary one-form field, respectively. 
The associated field strength $F=\di A+\frac{1}{2}[A,A]$ can be written as
\begin{equation}\label{two-form}
    F= K^{a} P_{a}+R^{a}J_{a}+W^{a} M_{a},
\end{equation}
where
\begin{subequations}
\begin{align}\label{curvatures}
   & R^{a}=\di \omega^{a}+\frac{1}{2}\epsilon^a_{ \, \, \,  bc}\omega^{b}\omega^{c} \,,\\
   & K^{a}= T^{a}+\frac{1}{2\ell}\epsilon^a_{ \, \, \,  bc} e^{b}e^{c} \,, \\ 
   & W^{a}=D(\omega) f^{a} +\frac{1}{2}\epsilon^a_{ \, \, \,  bc} e^{b}e^{c} \,.
    \end{align}
\end{subequations}
Here, $T^ {a}=D(\omega) e^{a}$ is torsion two-form, $R^{a}$ is curvature two-form, and $D(\omega)\Phi^{a}=\di \Phi^{a}+\epsilon^a_{ \, \, \,  bc}\omega^{b}\Phi^{c}$ is the exterior covariant derivative. Naturally, the 
limit $\ell\rightarrow\infty$ reproduces the Maxwell field strength \cite{Concha:2018zeb}. On the other hand, the non-degenerate bilinear form of the algebra \eqref{algebra01} reads
\begin{equation}
\begin{split}\label{invtensor}
     &\langle J_{a} J_{b} \rangle = \alpha_0 \eta_{a b}, \hspace{0.7 cm} \langle P_{a} P_{b} \rangle = (\frac{\alpha_1}{\ell} + \alpha_2) \eta_{a b},  \\
     & \langle J_{a} P_{b} \rangle = \alpha_1 \eta_{a b}, \hspace{0.7 cm} \langle P_{a} M_{b} \rangle = 0,\\
     &\langle J_{a} M_{b} \rangle = \alpha_2 \eta_{a b}, \hspace{0.7 cm} \langle M_{a} M_{b} \rangle = 0,
 \end{split}  
\end{equation}
where $\alpha_0, \alpha_1$ and $\alpha_2$ are arbitrary constants satisfying $\alpha_2\neq0$ and $\alpha_1\neq-\ell\alpha_2$. Both conditions are required to ensure the non-degeneracy of the invariant tensor \eqref{invtensor}. One can see that the 
limit $\ell\rightarrow\infty$ leads to the non-vanishing components of the invariant tensor for the Maxwell algebra \cite{Concha:2018zeb}.

Considering the one-form gauge potential \eqref{one-form} and the non-vanishing components of the invariant tensor \eqref{invtensor}, one can rewrite the CS action \eqref{CS-Action} as
\begin{equation}
\begin{split}\label{CS}
    S_{CS} = \frac{k}{4 \pi} \int_{\mathcal{M}} \bigg\{ &\alpha_0 \left( \omega^a \di \omega_a + \frac{1}{3}\epsilon^{abc}\omega_a \omega_b \omega_c
    \right) + \alpha_1 \left( 2 R_a e^a + \frac{1}{3\ell^{2}} \epsilon^{abc} e_a e_b e_c + \frac{1}{\ell}T^a e_a \right) \\
    &+ \alpha_2 \left( T^a e_a +2R^a f_a + \frac{1}{3\ell} \epsilon^{abc} e_a e_b e_c \right) \bigg\}
\end{split}
\end{equation}
up to a surface term. One can see that the CS action is proportional to three independent sectors each one with its respective coupling constant $\alpha_i$. In particular, the first term is the so-called exotic Lagrangian \cite{Witten:1988hc}. The second term contains the usual Einstein Lagrangian with cosmological constant term plus a torsional term related to the so-called Nieh-Yan invariant density.

Interestingly, the CS action \eqref{CS} can be seen as a Maxwell version of a particular case of the Mielke-Baekler (MB) gravity theory \cite{Mielke:1991nn} which describes a three-dimensional gravity model in presence of non-vanishing torsion. The MB action is given by
\begin{equation}\label{MB}
    I_{MB}=aI_{1}+\Lambda I_{2}+\beta_{3} I_{3}+\beta_{4} I_{4}
\end{equation}
where $a,\Lambda,\beta_{3}$ and $\beta_{4}$ are constants and
\begin{eqnarray}
I_{1} & = & 2\int e_{a}R^{a}\,,\nonumber\\
I_{2} & = & -\frac{1}{3}\int \epsilon_{abc}e^{a}e^{b}e^{c}\,,\nonumber\\
I_{3} & = & \int \omega^a \di \omega_a + \frac{1}{3}\epsilon^{abc}\omega_a \omega_b \omega_c\,,\label{MBterms}\\
I_{4} & = &\int e_{a}T^{a}\,.\nonumber
\end{eqnarray}
Particularly, in the absence of the auxiliary field $f^a$ in \eqref{CS}, the constants appearing in the MB gravity can be identified with those of the   the deformed Maxwell algebra CS theory as
\begin{equation}\label{MBcts}
    a=\frac{k}{4\pi}\alpha_{1}\,, \qquad \Lambda=-\frac{k}{4\pi \ell}\left( \frac{\alpha_{1}}{\ell}+\alpha_2\right)\,, \qquad \beta_{3}=\frac{k}{4\pi}\alpha_{0}\, \qquad \beta_{4}=\frac{k}{4\pi }\left( \frac{\alpha_{1}}{\ell}+\alpha_2\right)\,.
\end{equation}
Thus, the CS gravity action \eqref{CS} can be interpreted as the Maxwellian version of a particular case of the MB gravity action when the MB's constants satisfy \eqref{MBcts}. Further studies of the MB gravity have been subsequently developed in \cite{Baekler:1992ab,Blagojevic:2003wn,Giacomini:2006dr,Santamaria:2011cz,Cvetkovic:2018ati,Peleteiro:2020ubv}.

It is important to point out that one can accommodate a generalized cosmological constant in the Maxwell gravity theory using another deformation of the Maxwell algebra, known as AdS-Lorentz algebra \cite{Soroka:2004fj}. However, in the AdS-Lorentz case, besides the Einstein-Hilbert term there is an additional gauge field $f_a$, while there is no  torsion term $\alpha_{1}$ \cite{Diaz:2012zza,Hoseinzadeh:2014bla,Concha:2018jjj,Concha:2019lhn}. In the present case, the deformed Maxwell symmetry allows us to introduce both a cosmological constant and  a torsion term. In the flat limit $\ell\rightarrow\infty$ the CS action reproduces the Maxwell CS gravity action which contains pure GR as sub-case.  Dynamics of the $f_a$ gauge field is completely determined by the last term with coupling constant $\alpha_2$. In particular, the equations of motion appear by considering the variation of the action \eqref{CS} under the respective gauge fields:
\begin{eqnarray}
\delta e^{a} & : & \qquad0=\alpha_{1}\left(R^{a}+\frac{1}{\ell}K^{a}\right)+\alpha_{2}K^{a},\nonumber \\
\delta\omega^{a} & : & \qquad0=\alpha_{0}R^{a}+\alpha_{1}K^{a}+\alpha_{2}W^{a},\label{eom}\\
\delta f^{a} & : & \qquad0=\alpha_{2}R^{a},\nonumber
\end{eqnarray}
Then, when $\alpha_2\neq 0$ we find the curvature two-forms \eqref{two-form} should vanish,
\begin{equation}\label{EOM}
    R^{a}= 0\,, \qquad 
    K^{a}=0\,, \qquad
    W^{a}=0 \, .
\end{equation}
Indeed, from last equation in \eqref{eom}, we find $R^{a}=0$. Nevertheless, it is important to emphasize that $\alpha_1=-\ell\alpha_2$, which solves the first equation of \eqref{eom} and would imply a relation between $K^{a}$ and $W^{a}$, cannot be considered as a solution of the theory. As was previously mentioned, the non-degeneracy of the invariant tensor implies $\alpha_2\neq0$ and $\alpha_1\neq -\ell\alpha_2$. In particular, in the three-dimensional CS formalism, the non-degeneracy of the bilinear form ensures that the CS action involves a kinematical term for each gauge field and the equations of motion imply that all curvature two-forms vanish as in \eqref{EOM}. Note that the CS gravity theory \eqref{CS} describes the Maxwell CS gravity theory in presence of a non-vanishing torsion $T^{a}\neq 0$. In particular, the first two equations $R^{a}=0$ and $T^{a}=-\frac{1}{2 \ell}\epsilon^{a}_{\,\,\, bc}e^{b}e^{c}$ correspond to the three-dimensional teleparallel theory in which the cosmological constant can be seen as a source for the torsion.  On the other hand, the vanishing of $W^a$ implies
that the exterior covariant derivative of the auxiliary field $f_a$
is constant. In particular, in the flat limit $\ell\rightarrow\infty$ the field equation for $f_{a}$ remains untouched and is analogue to the constancy of the electromagnetic field in flat spacetime.

One can see that each term of the action \eqref{CS} is invariant under the gauge transformation laws of the algebra \eqref{algebra01}. Indeed, considering
\begin{equation}
    \Lambda=\varepsilon^{a}P_{a}+\rho^{a}J_{a}+\chi^{a}M_{a},
\end{equation}
we have that the gauge transformations $\delta A =\di \Lambda +[A,\Lambda]$ of the theory are given by
\begin{equation}
   \begin{split}
        &\delta_{\Lambda} e^{a}=D(\omega)\varepsilon^{a}-\epsilon^{abc}\rho_{b}e_{c}+\frac{1}{\ell}\epsilon^{abc}e_{b}\varepsilon_{c}, \\
        &\delta_{\Lambda}\omega^{a}=D(\omega)\rho^{a}, \\
        &\delta_{\Lambda}f^{a}=D(\omega)\chi^{a}+\epsilon^{abc}e_{b}\varepsilon_{c}-\epsilon^{abc}\rho_{b}f_{c}.
    \end{split}
\end{equation}

In this work, we  analyze the consequences of this particular deformation of the Maxwell symmetry at the level of the 
asymptotic structure. In the Maxwell case, as was shown in \cite{Concha:2018zeb}, the presence of the additional gauge field $f^{a}$ leads to new effects compared to GR and the asymptotic symmetries is found to be a new algebra,  denoted as $\mathfrak{Max}_{3}$ in section \ref{sec:1.6}. Here, we explore the implications of deforming the Maxwell algebra as in \eqref{algebra01}.


Let us recall that given an action there are two ways to render it having a well-posed variation principle. One of them is to add boundary terms to the action and the other is imposing suitable boundary conditions on fields. Let us consider the variation of the action \eqref{CS-Action},
\begin{equation}
    \delta S_{\text{CS}}[A]= \frac{k}{2 \pi} \int_{\mathcal{M}} \langle \delta A F  \rangle  + \frac{k}{4 \pi} \int_{\partial\mathcal{M}} \langle \delta A\, A \rangle \,,
\end{equation}
where $\partial\mathcal{M}$ is the boundary of $\mathcal{M}$. The field equations require vanishing of the field strength and the on-shell boundary contribution to the action is the surface term
\begin{equation}\label{bdy}
    \delta S_{\text{CS}}[A]\big|_{\text{bdy}} = - \frac{k}{4 \pi} \int_{\partial\mathcal{M}} \langle  A  \delta A \rangle\, .
\end{equation} 
In our case this term reads as
\begin{equation}
    \delta S_{\text{CS}}\big|_{\text{bdy}} = \frac{k}{4 \pi} \int_{\partial\mathcal{M}}  \left[ \delta\omega^{a}\left(\alpha_{0}\omega_{a}+\alpha_{1} e_{a}+\alpha_{2}f_{a}\right)+\delta e^{a}\left(\alpha_{1}\omega_{a}+\left(\frac{\alpha_1}{l}+\alpha_{2}\right)e_{a}\right)+\alpha_{2}\delta f^{a}\omega_{a}\right] \, .
\end{equation}
We will see later that for spacetimes with null boundary, where the boundary is located at $r=const\rightarrow\infty$, the action principle is satisfied without addition of boundary terms.

\subsection*{BMS-like solution}
In this section we analyze the field equations \eqref{EOM}. We consider spacetimes with null boundary, which can be described in the three-dimensional BMS gauge. We parametrize spacetime by the local coordinates $x^{\mu}=\left(u,r,\phi\right)$, where $-\infty < u <\infty$ is the retarded time coordinate,  $\phi \sim \phi + 2 \pi$ is the angular coordinate and the boundary is located at $r=\it{const}$. Then, the metric can be written as \cite{Barnich:2014cwa}
\begin{equation}
    \di s^{2}=\mathcal{M}\di u^{2}-2\di u \di r +\mathcal{N}\di\phi \di u+r^{2}\di \phi^{2}\,.
\end{equation}
where $\mathcal{M}$ and $\mathcal{N}$ are two arbitrary functions of $u,\phi$. As was previously discussed, the deformed Maxwell symmetry \eqref{algebra01} can be written in a certain basis as the direct sum $\mathfrak{iso}(2,1)\oplus\mathfrak{so}(2,1)$. Following the same trick considered in \cite{Concha:2018jjj}, we can find  solutions of the present theory by first working in the direct sum basis and then, go to the basis we are interested in. In the direct sum basis, the fields $(\Tilde{\omega}^a,\Tilde{e}^a)$ associated to the $\mathfrak{iso}(2,1)$ generators obey the very well-known GR boundary conditions, and the field $\Tilde{f}^a$ associated to the $\mathfrak{so}(2,1)$ generator can be set as a flat connection.

Furthermore, in the aforementioned direct sum basis, the functions $\mathcal{M}$ and $\mathcal{N}$ are given for the known results in asymptotically flat gravity in three dimensions
\begin{equation}\label{M,N}
    \mathcal{M}=\mathcal{M}(\phi)\,,\qquad \mathcal{N}=\mathcal{J}(\phi)+u \mathcal{M}^{\prime}(\phi)\,.
\end{equation}
 The spacetime line element can be written in terms of the dreibein as $\di s^2 = \eta_{ab} \Tilde{e}^{a} \Tilde{e}^{b}$, where
\begin{equation}
    \eta_{ab} = \begin{pmatrix}
0 & 1 & 0\\
1 & 0 & 0\\
0 & 0 & 1
\end{pmatrix}
\end{equation}
is the Minkowski metric in null coordinate system. Then, the dreibein and the torsionless spin connection one-forms are written as
\begin{align}
     \Tilde{e}^{0}&=-\di r+\frac{1}{2}\mathcal{M}\di u+\frac{1}{2}\mathcal{N}\di \phi\,, & \Tilde{e}^{1}&= \di u\,, & \Tilde{e}^{2}&= r\di\phi \,, \nonumber \\ 
     \Tilde{\omega}^{0}&=\frac{1}{2}\mathcal{M}\di \phi\,, & \Tilde{\omega}^{1}&= \di \phi\,, & \Tilde{\omega}^{2}&=0\,.
\end{align}
The field $\Tilde{f}^a$ can then be chosen as a Lorentz flat connection, 
\begin{equation}
    \Tilde{f}^{0}=\frac{1}{2}\mathcal{L} \di \phi , \qquad\Tilde{f}^{1}= \di \phi, \qquad\Tilde{f}^{2}= 0\,.
\end{equation}
where $\mathcal{L}=\mathcal{L}(\phi)$. In this way, we have found the solutions in the BMS gauge for the fields $(\Tilde{e}^a, \Tilde{\omega}^a, \Tilde{f}^a)$. 

As mentioned before, we are interested in the basis where the Poincaré-Lorentz symmetry appears as a deformation of the Maxwell algebra. From \eqref{redefine-pp1}, it is possible to show that the fields $({e}^a, {\omega}^a, {f}^a)$ are related to those in the direct sum basis $(\Tilde{e}^a, \Tilde{\omega}^a, \Tilde{f}^a)$ as follows:  \begin{equation}
    e^a=\ell\left(\Tilde{f}^a-\Tilde{\omega}^a\right)\,,\qquad \omega^a=\Tilde{\omega}^a\,,\qquad f^a=\ell^{2}\left(\Tilde{f}^a-\Tilde{\omega}^a\right)-\ell\Tilde{e}^a\,.
    \end{equation}
Consequently, the field equations \eqref{EOM} are solved by the following components of the gauge fields
\begin{align}
    e ^{0} & =\frac{1}{2}\mathcal{P} \di \phi\,,  &
\omega ^{0} & =\frac{1}{2}\mathcal{M}\di \phi\,, & f ^{0} & =\ell \di r+\frac{1}{2}\mathcal{F}\di \phi-\frac{\ell}{2}\mathcal{M}\di u\,,\nonumber \\ \label{ewf}
e ^{1} & =0\,, & \omega ^{1} & =\di \phi\,, & f ^{1} & =-\ell \di u\,, \\ \nonumber
e ^{2} & =0\,, & \omega ^{2} & =0\,, & f^{2} &
=-\ell r \di \phi\,,
\end{align}
where, for later convenience, we have defined the functions $\mathcal{P}=\ell (\mathcal{L}-\mathcal{M})$ and $\mathcal{F}=\ell(\mathcal{P}-\mathcal{N})$.

As discussed we need to ensure  vanishing of the boundary term in the variation of the action when suitable boundary conditions on the fields are imposed. 
The radial dependence of the connection $A$ can be gauged away by the gauge transformation $A=h^{-1}\di h+h^{-1}ah$, where the asymptotic field $a=a_{u}(u,\phi)\di u+a_{\phi}(u,\phi)\di \phi$ does not depend on $r$ and $h=e^{\ell r M_{0}}$. Then, at the boundary $r=const.\rightarrow \infty$, the on-shell action \eqref{bdy} takes the form
\begin{align}
   \delta S_{\text{CS}}\big|_{\text{bdy}}&=-\frac{k}{4\pi}\int_{\partial\mathcal{M}} \langle a\delta a\rangle \nonumber\\
&=\frac{k}{4\pi}\int_{\partial\mathcal{M}}\di u \di \phi\,\left[\delta e^{a}_{u}\left(\alpha_{1}\omega_{a\phi}+\left(\frac{\alpha_{1}}{\ell}+\alpha_{2}\right)e_{a\phi}\right)-\delta e^{a}_{\phi}\left(\alpha_{1}\omega_{a u}+\left(\frac{\alpha_{1}}{\ell}+\alpha_{2}\right)e_{a u}\right)\right.  \nonumber \\
     & +\left.\delta \omega^{a}_{u}\left(\alpha_{0}\omega_{a\phi}+\alpha_{1}e_{a\phi}+\alpha_{2}f_{a\phi}\right)-\delta \omega^{a}_{\phi}\left(\alpha_{0}\omega_{a u}+\alpha_{1}e_{a u}+\alpha_{2}f_{a u}\right) \right.\\
     &\left. +  \alpha_2\delta f^{a}_{u}\omega_{a\phi}-\alpha_2\delta f^{a}_{\phi}\omega_{a u} \right] \,.\nonumber
\end{align}
 Furthermore, from \eqref{ewf} we find the following boundary conditions for the gauge fields
\begin{equation}\label{bdy-condition}
    e^{a}_{u}=0\,, \qquad \omega^{a}_{u}=0\,, \qquad \omega^{a}_{\phi}=-\ell f^{a}_u\,,
\end{equation}
upon the first two,  the variation of the action reduces to
\begin{equation}
    \delta S_{\text{CS}}\big|_{\text{bdy}}=\frac{k \alpha_2}{4\pi}\int_{\partial\mathcal{M}}\di u \di \phi\,\left(\delta \omega^{a}_{\phi}f_{au}-\delta f^{a}_{u} \omega_{a\phi} \right)\,.
\end{equation}
Finally,  applying the last condition $\omega^{a}_{\phi}=-\ell f^{a}_u$, we arrive at
\begin{equation}
    \delta S_{\text{CS}}\big|_{\text{bdy}}=0\,,
\end{equation}
for any value of $\alpha_2$. Thus, in space-time with
 boundary conditions \eqref{bdy-condition}, the action principle is well-posed.


\section{Asymptotic symmetries and surface charges algebras}\label{sec:3.6}
The aim of this section is to find the asymptotic symmetry algebra for the Maxwell CS gravity with torsion which was  previously constructed. To start with, we provide the suitable fall-off conditions for the gauge fields at infinity and the gauge transformations which preserve our boundary conditions. Then, the charge algebra is found using the Regge-Teitelboim method \cite{REGGE1974286}.
\subsection{Boundary conditions}
 Inspired by the results obtained in the previous section, we consider the gauge connection evaluated in the BMS gauge as follows
\begin{equation}
\begin{split}\label{Connection}
       A & = \frac{1}{2}\mathcal{M}\left(u,\phi\right) \di \phi J_0 + \di \phi J_1 + \frac{1}{2}\mathcal{P}\left(u,\phi\right) \di \phi P_0 \\ & + \left[\ell \di r+\frac{1}{2}\mathcal{F}\left(u,\phi\right)\di\phi-\frac{\ell}{2}\mathcal{M}\left(u,\phi\right)\di u\right]M_0
      -\ell \di u M_1-r\ell \di\phi M_2\,.
\end{split}
\end{equation}
The radial dependence can be gauged away by an appropriate gauge transformation on the connection
\begin{equation}
    A = h^{-1} \di h +h^{-1} a h\,,
\end{equation}
where the group element is given by $h=e^{\ell r M_0}$. Then, if we use the identity $h^{-1}\di h=\ell \di r M_0$, and the Baker-Campbell-Hausdorff formula, we find
\begin{equation}
   h^{-1} a h = a-\ell r \di\phi M_2\,. 
\end{equation}
Therefore, once we have dropped out the radial dependence from the gauge field $A$, we are left with the asymptotic field $a=a_{u}\di u+a_{\phi}\di \phi$, whose components are given by
\begin{equation}
    \begin{split}
        &a_{u}=-\frac{\ell}{2}\mathcal{M}M_0-\ell M_1\,,\\
        &a_{\phi}=\frac{1}{2}\mathcal{M} J_0+J_1+\frac{1}{2} \mathcal{P} P_0+\frac{1}{2}\mathcal{F}M_0\,,
    \end{split}
\end{equation}
which depend only on time and the angular coordinate. The equations of motion, which are required to hold in the asymptotic region, 
imply that
\begin{equation}\label{MLN}
   \mathcal{M}=\mathcal{M}\left(\phi\right)\,, \qquad \mathcal{P}=\mathcal{P}\left(\phi\right)\,, \qquad \mathcal{F}=\mathcal{Z}\left(\phi\right)-u \ell \mathcal{M^{\prime}}\left(\phi\right)\,.
\end{equation}


\subsection*{Residual gauge transformations}
Asymptotic symmetries correspond to residual gauge transformations $\delta_{\Lambda}A = \di{\Lambda}+[A,\Lambda]$ which preserve boundary conditions \eqref{Connection}. We consider the following gauge parameters
\begin{equation}
    \Lambda=h^{-1}\lambda h\,, \qquad \lambda=\lambda^{a}_{(J)}\left(u,\phi\right) J_a
+\lambda^{a}_{(P)}\left(u,\phi\right) P_a+\lambda^{a}_{(M)}\left(u,\phi\right) M_a\,.
\end{equation}
Then, gauge transformations of the connection $A$ with gauge parameter $\Lambda$, lead to $r$-independent gauge transformations of the connection $a$ with gauge parameter $\lambda$, i.e.
\begin{equation}
    \delta_{\Lambda}a \equiv \delta_{\lambda}a=\di\lambda+[a,\lambda] \, .
\end{equation}
The gauge transformations that preserve the boundary conditions \eqref{Connection} with \eqref{MLN} for $\lambda_{(J)}^{a}$ and $\lambda_{(P)}^{a}$ are given by
\begin{align}\label{deltaa-1}
    \lambda_{(J)}^{0} &=\frac{\mathcal{M}}{2}\,\varepsilon-\varepsilon^{\prime \prime }\,,    &  \lambda_{(P)}^{0} & =\frac{1}{2}\left[\mathcal{P}\left(\frac{\chi}{\ell}+ \varepsilon\right)+ \mathcal{M} \chi\right] -\chi^{\prime \prime }\,, \nonumber \\ 
    \lambda_{(J)}^{1} & =\varepsilon\,,   & \lambda_{(P)} ^{1} & =\chi \,, \\
    \lambda_{(J)}^{2} & =-\varepsilon^{\prime }\,, & \lambda_{(P)}^{2} & =-\chi^{\prime} \nonumber\,,
\end{align}
while for $\lambda_{(M)}^{a}$ we get
\begin{align}\label{deltaa-2}
    \lambda_{(M)}^{0}& =\frac{\mathcal{M}}{2}\gamma+\frac{\mathcal{P}}{2} \chi+\frac{ \mathcal{F}}{2}\varepsilon-\frac{\ell \mathcal{M}}{2}u \varepsilon^{\prime }+u\ell  \varepsilon^{\prime\prime\prime }-\gamma^{\prime\prime}\,, \nonumber\\
    \lambda_{(M)}^{1} & =\gamma-u \ell \varepsilon^{\prime }\,, \\
    \lambda_{(M)}^{2} & =-\gamma^{\prime }+u \ell \varepsilon^{\prime \prime } \nonumber\,.
    \end{align}
where $\varepsilon$, $\chi$ and $\gamma$ are three arbitrary, periodic functions of the angular coordinate $\phi$. Under the given gauge transformation, the dynamical fields transform as
\begin{eqnarray}
\delta_\lambda \mathcal{M} &=&\mathcal{M}^{\prime }\varepsilon+2\mathcal{M}\varepsilon^{\prime
}-2\varepsilon{}^{\prime \prime \prime }\,,  \notag \\
\delta_\lambda \mathcal{P} &=&\mathcal{P}^{\prime }\left(\frac{\chi}{\ell}+\varepsilon\right)+2\mathcal{P}\left(\varepsilon^{\prime}+\frac{\chi^{\prime
}}{\ell}\right)+\mathcal{M}^{\prime }\chi+2\mathcal{M}\chi^{\prime}-2\chi^{\prime \prime \prime }\,,  \label{translaw} \\
\delta_\lambda \mathcal{Z} &=&\mathcal{Z}^{\prime}\varepsilon+2\mathcal{Z}\varepsilon^{\prime}+\mathcal{M}^{\prime}\gamma+2\mathcal{M}\gamma^{\prime}+\mathcal{P}^{\prime}\chi+2\mathcal{P}\chi^{\prime}-2\gamma^{\prime \prime \prime}\,.  \notag
\end{eqnarray}


\subsection*{Canonical surface charges and asymptotic symmetry algebra}
 Asymptotic symmetries of the Maxwell gravity theory with non-vanishing torsion can be found in the canonical approach \cite{REGGE1974286}. In particular, in the case of a three-dimensional Chern-Simons theory, the variation of the canonical generators is given by \cite{Banados:1998gg,Banados:1994tn}
\begin{equation}
    \delta \mathcal{Q}[\lambda]=\int d\phi \langle \lambda \delta a_\phi\rangle\,.
\end{equation}
Therefore one can show that surface charge variation associated with \eqref{deltaa-1} and \eqref{deltaa-2} is

\begin{equation}\label{charge-variation-01}
    \delta \mathcal{Q} (\varepsilon,\chi, \gamma )= \int_0^{2\pi} \di \phi \left( \varepsilon \delta \mathbf{J}+ \chi \delta \mathbf{P} + \gamma \delta \mathbf{M} \right)
\end{equation}
with
\begin{subequations}
\begin{align}
   \mathbf{J} =& \frac{k}{4\pi}\left(\alpha_2\mathcal{Z}+\alpha_0\mathcal{M}+\alpha_1\mathcal{P}\right)\, , \\
   \mathbf{P}= & \frac{k}{4\pi}\left[\left(\frac{\alpha_{1}}{\ell}+ \alpha_{2}\right)\mathcal{P}+ \alpha_{1}\mathcal{M}\right]\, , \\
   \mathbf{M}= & \frac{k}{4\pi} \alpha_{2} \mathcal{M} \, .
\end{align}
\end{subequations}
One can take $\varepsilon$, $\chi$ and $\gamma$ to be state-independent and then the charge variation \eqref{charge-variation-01} is integrable on the phase space.

There are three independent surface charges,
\begin{equation}
    J(\varepsilon) = \mathcal{Q}(\varepsilon,0,0) \, , \qquad P(\chi) = \mathcal{Q}(0,\chi,0) \, , \qquad M(\gamma) = \mathcal{Q}(0,0,\gamma) \, , \qquad 
\end{equation}
associated with three independent symmetry generators $\varepsilon$, $\chi$ and $\gamma$.
It is shown that the algebra among surface charges is given by \cite{REGGE1974286,brown1986poisson,compere2019advanced}
\begin{equation}
    \{ \mathcal{Q}(\Lambda_{1}), \mathcal{Q}(\Lambda_{2}) \} = \mathcal{Q}([\Lambda_{1},\Lambda_{2}])+ \mathcal{C}(\Lambda_{1},\Lambda_{2})
\end{equation}
where Dirac bracket is defined as $\{ \mathcal{Q}(\Lambda_{1}), \mathcal{Q}(\Lambda_{2}) \}:= \delta_{\Lambda_{2}}\mathcal{Q}(\Lambda_1)$ and $\mathcal{C}(\Lambda_{1},\Lambda_{2})$ is central extension term. Therefore, using the transformation laws \eqref{translaw}, one can show that the algebra of charges \eqref{charge-variation-01} is
\begin{eqnarray}
\left\lbrace J(\varepsilon_1),J(\varepsilon_2)\right\rbrace&=&J([\varepsilon_1,\varepsilon_2])-\frac{k\alpha_0}{2\pi}\int \di \phi \, \varepsilon_1 \varepsilon^{\prime\prime\prime}_2\,, \nonumber\\
\left\lbrace J(\varepsilon),P(\chi)\right\rbrace&=&P([\varepsilon,\chi])-\frac{k\alpha_1}{2\pi} \int \di \phi
\, \varepsilon \chi^{\prime\prime\prime}\,, \nonumber\\
\left\lbrace J(\varepsilon),M(\gamma)\right\rbrace&=&M([\varepsilon,\gamma])-\frac{k\alpha_2}{2\pi} \int \di \phi
\, \varepsilon \gamma^{\prime\prime\prime}\,,\\
\left\lbrace P(\chi_1),P(\chi_2)\right\rbrace&=&M([\chi_1,\chi_2])+\frac{1}{\ell} P([\chi_1,\chi_2])-\frac{k}{2\pi}\left(\frac{\alpha_1}{\ell}+\alpha_2\right)\int \di \phi \chi_1 \chi^{\prime\prime\prime}_2\,, \nonumber\\
\left\lbrace P(\chi),M(\gamma)\right\rbrace&=&0\,, \nonumber\\
\left\lbrace M(\gamma_1),M(\gamma_2)\right\rbrace&=&0\,,  \nonumber
\end{eqnarray}

where $[x,y]=x y^{\prime}-y x^{\prime}$.
The resulting algebra corresponds to an infinite-dimensional lift of the deformed Maxwell algebra.  Now we can express the above algebra in Fourier modes,
\begin{equation}
    J_n := J(e^{in\phi})\,, \qquad P_n := P(e^{in\phi})\,, \qquad M_n := M(e^{in\phi})\,,
\end{equation}
which give rise to the following centrally extended algebra
\begin{eqnarray}
i\left\lbrace J_m,J_n\right\rbrace &=& \left(m-n\right)J_{m+n}+\frac{c_1}{12}m^3\delta_{m+n,0}\,,\nonumber\\
i\left\lbrace J_m,P_n\right\rbrace &=& \left(m-n\right)P_{m+n}+\frac{c_2}{12}m^3\delta_{m+n,0}\,,\nonumber\\
i\left\lbrace J_m,M_n\right\rbrace &=& \left(m-n\right)M_{m+n}+\frac{c_3}{12}m^3\delta_{m+n,0}\,,\nonumber\\
i\left\lbrace P_m,P_n\right\rbrace &=&
\left(m-n\right)M_{m+n} +\frac{1}{\ell}\left(m-n\right)P_{m+n}+\frac{1}{12}\left(\frac{c_2}{\ell} +c_3\right)m^3\delta_{m+n,0}\,,\label{inf-algebra}\\
i\left\lbrace P_m,M_n\right\rbrace &=&0\,,\nonumber\\
i\left\lbrace M_m,M_n\right\rbrace &=&0\,.\nonumber
\end{eqnarray}
The central charges $c_{1}$, $c_{2}$ and $c_{3}$ are related to the CS level $k$ and the arbitrary constant appearing in the invariant tensor \eqref{invtensor} as
\begin{equation}\label{Ci-kalphai}
    c_{i}=12 k\alpha_{i-1}\,.
\end{equation}
It should be noticed that \eqref{inf-algebra} is central extension of \eqref{f1-deform-maxwell} where $\varepsilon$ is replaced by $\frac{1}{\ell}$. 
\subsection*{Change of basis}

The infinite-dimensional algebra \eqref{inf-algebra} can be seen as the infinite-dimensional enhancement of the deformed Maxwell algebra \eqref{algebra01}. In particular, in the flat limit $\ell\rightarrow\infty$ we recover the asymptotic symmetry of the three-dimensional Maxwell CS gravity theory introduced in \cite{Concha:2018zeb}. Interestingly, as was shown in section \ref{sec:6.4}, the algebra \eqref{inf-algebra} is isomorphic to the $\widehat{\mathfrak{bms}}_3\oplus \mathfrak{vir}$ algebra.
To observe this at the level of charge algebra, we use the change of basis proposed in \cite{Grumiller:2019fmp}.

Suppose $\varepsilon$, $\chi$ and $\gamma$ are now state-dependent, i.e. they are functions of dynamical fields. We require that charge variation be integrable which leads to
\begin{equation}
    \varepsilon = \frac{\delta \mathcal{G}}{\delta \mathbf{J}}\, , \qquad \chi = \frac{\delta \mathcal{G}}{\delta \mathbf{P}}\, , \qquad \gamma = \frac{\delta \mathcal{G}}{\delta \mathbf{M}}\, ,
\end{equation}
for some functional
\begin{equation}
    \mathcal{G}[\mathbf{J},\mathbf{P},\mathbf{M}]= \int_0^{2\pi} \di \phi \, \mathbf{G}(\mathbf{J},\mathbf{P},\mathbf{M})\, .
\end{equation}
By choosing
\begin{equation}
    \mathbf{G}= \tilde{\varepsilon} \left( \mathbf{J}- \ell \mathbf{P}- \ell^2 \mathbf{M} \right) +\tilde{\chi} \left( \ell \mathbf{P}+ \ell^2 \mathbf{M}\right)- \ell \tilde{\gamma} \,\mathbf{M}
\end{equation}
where $\tilde{\varepsilon},\tilde{\chi}, \tilde{\gamma}$ are state-independent functions, one can show that the charge variation \eqref{charge-variation-01} can be written as
\begin{equation}\label{charge-variation-02}
    \delta \mathcal{Q} (\tilde{\varepsilon},\tilde{\chi}, \tilde{\gamma} )= \int_0^{2\pi} \di \phi \left( \tilde{\varepsilon} \delta \mathbf{L}+ \tilde{\chi} \delta \mathbf{S} + \tilde{\gamma} \delta \mathbf{T} \right)
\end{equation}
with
\begin{subequations}
\begin{align}
   \mathbf{L} =& \mathbf{J}- \ell \mathbf{P}- \ell^2 \mathbf{M}\, , \\
   \mathbf{S}= & \ell \mathbf{P}+ \ell^2 \mathbf{M}\, , \\
   \mathbf{T}= &  - \ell \,\mathbf{M}\, .
\end{align}
\end{subequations}
Now, by introducing Fourier modes
\begin{equation}
   L_m := \mathcal{Q} (\tilde{\varepsilon}=e^{in\phi},0,0 ),\qquad S_m:= \mathcal{Q} (0,\tilde{\chi}=e^{in\phi},0 ) ,\qquad T_{m}:=\mathcal{Q} (0,0, \tilde{\gamma}=e^{in\phi} ),
\end{equation}
one finds the direct sum of the $\widehat{\mathfrak{bms}}_3$ and Virasoro algebra:
\begin{eqnarray}\label{bms3+vir}
i\left\lbrace L_m,L_n\right\rbrace &=& \left(m-n\right)L_{m+n}+\frac{c_{LL}}{12}m^3\delta_{m+n,0}\,,\nonumber\\
i\left\lbrace L_m,T_n\right\rbrace &=& \left(m-n\right)T_{m+n}+\frac{c_{LT}}{12}m^3\delta_{m+n,0}\,,\label{bmsVir}\\
i\left\lbrace S_m,S_n\right\rbrace &=& \left(m-n\right)S_{m+n}+\frac{c_{SS}}{12}m^3\delta_{m+n,0}\,,\nonumber
\end{eqnarray} where the central charges are related to those appearing in \eqref{inf-algebra} as
\begin{equation}\label{cLL,cSS,cLT,ci}
    c_{LL}\equiv c_{1}-\ell c_{2}-\ell^2 c_{3}\,, \qquad c_{SS}\equiv \ell c_{2}+\ell^2 c_{3}\,, \qquad c_{LT}\equiv -\ell c_{3}\,.
\end{equation} 
The central charges $\left(c_{LL},c_{LT},c_{SS}\right)$ of the $\widehat{\mathfrak{bms}}_3\oplus\mathfrak{vir}$ algebra are related to the three independent terms of the CS gravity action for the $\mathfrak{iso}(2,1)\oplus\mathfrak{so}(2,1)$ algebra.

Let us note that the algebra \eqref{inf-algebra} can be obtained alternatively as a central extension of the deformed \Max\ \ref{sec:6.4} and deformed $\mathfrak{bms}_{4}$ \ref{sec:5.3}. It is also worth  pointing out that by ignoring the generators $T_{m}$ and central charge $c_{LT}$ in \eqref{bms3+vir}, which is equivalent to ignoring the Maxwell generators $M_{m}$ and central charge $c_{3}$ in \eqref{inf-algebra}, one obtains two copies of Virasoro algebra, which is the asymptotic symmetry of the AdS$_{3}$ CS gravity (or Teleparallel theory \cite{Blagojevic:2003uc}) with two central charges as 
\begin{equation}
    c_{LL}\equiv c_{1}-\ell c_{2}\,, \qquad c_{SS}\equiv \ell c_{2}\,.
\end{equation}

\section{Summary of the chapter}
In the first part of this chapter we have considered the deformation and stability of \Max\ algebra which is the infinite enhancement of the $2+1$ dimensional Maxwell algebra and describes the asymptotic symmetry of the Chern-Simons gravity theory invariant under the Maxwell algebra \cite{Concha:2018zeb}. We have shown that the \Max\ algebra is not stable and can be deformed to four possible formal deformations. The \Max\ algebra can be formally deformed into $\mathfrak{bms}_{3}\oplus\mathfrak{witt}$ or three copies of the Witt algebra in its ideal part. Furthermore, the \Max\ algebra can be formally deformed into two new families of algebras when we consider deformations of other commutators. The new infinite dimensional algebras obtained have been denoted as $M(a,b;c,d)$ and $\bar{M}(\bar{\alpha},\bar{\beta};\bar{\nu})$.
In particular, the \Max\ algebra can be formally deformed to the (twisted) Schrödinger-Virasoro algebra for the specific values of parameters $a=c=d=0$ and $b=-\frac{1}{2}$, which can be seen as the asymptotic symmetry algebra of the spacetimes invariant under Schrödinger symmetry \cite{Alishahiha:2009nm, Compere:2009qm}. 

We have then explored possible central terms for the obtained algebras through deformation procedure. We have first briefly review the well-known central extensions of the  $\mathfrak{bms}_{3}$ and the $\mathfrak{witt}$ algebra. We also explored the central extensions of $M(a,b;c,d)$ and $\bar{M}(\bar{\alpha},\bar{\beta};\bar{\nu})$ in some specific points of their parameters space. For a generic point in the parameter space $M(a,b;c,d)$ algebra  admits only one central term in its Witt subalgebra. For specific values of parameters it can admit more central terms which means that the deformation procedure can change the number of possible non trivial central terms. On the other hand the algebra $\bar{M}(\bar{\alpha},\bar{\beta};\bar{\nu})$ in general admits two non trivial central terms and a third central terms can appear for $\bar{\alpha}=\bar{\nu}=0$ in $\bar{M}(\bar{\alpha},\bar{\beta};\bar{\nu})$ as in the \Max\ algebra.

In the rest of the chapter, we have studied a Chern-Simon gravity theory with a deformed Maxwell algebra $\mathfrak{iso}(2,1)\oplus\mathfrak{so}(2,1)$ as the gauge group, which allows us to introduce a non-vanishing torsion to the Maxwell CS gravity. In particular, the CS action can be seen as a Maxwell generalization of a particular case of the MB gravity \cite{Mielke:1991nn} whose field equations correspond to those of the Maxwell gravity theory but in presence of a non-vanishing torsion. Motivated by the fact that the deformed Maxwell is isomorphic to the $\mathfrak{iso}(2,1)\oplus\mathfrak{so}(2,1)$ algebra, we considered asymptotically flat geometries and discuss the BMS-like solution. We then explored the implications of the deformed Maxwell algebra \eqref{algebra01} at the level of the asymptotic symmetry. In particular, we have shown that the asymptotic symmetry for the Maxwell algebra with torsion is described by an infinite-enhancement of the deformed Maxwell algebra which can be written as the $\widehat{\mathfrak{bms}}_3\oplus\mathfrak{vir}$ algebra with three independent central charges.

\newpage
\thispagestyle{empty}
\mbox{}

\chapter{Discussion and outlook} \label{ch7}
 In this thesis we worked out deformations and stabilizations of infinite dimensional Lie algebras which are obtained as asymptotic symmetries algebras and not subject to Hochschild-Serre factorization theorem. In particular, we showed that \bms, \bmsf\ and \Max\ algebras are not rigid and can be deformed into some new non isomorphic algebras. Some are deformations can be viewed as deviation of 
Hochschild-Serre factorization theorem. We found that the deformation of mentioned algebras may have various physical interpretations which can be viewed as specific properties of these family algebras. We also worked out deformations of obtained algebra and introduced a new notion of rigidity for infinite dimensional family algebras. In mathematical setting, we presented a conjecture as a new extended version of Hochschild-Serre factorization theorem for infinite dimensional lie algebras and also introduced a new notion of rigidity for infinite dimensional family algebras. 

In $3d$ we showed that \bms\ can be deformed into either two copies of \wit\ algebra or \w\ family algebra. The centrally extended \hbms\ also can be deformed into either two copies of \vir\ algebra or \wh\ when $c_{JP}=0$. We also found that \kac\ algebra can be deformed into either \w\ family  or $\mathfrak{KM}(a,\nu)$ family algebras and that its central extension \kach\ with three independent central charges is rigid. Among different algebras obtained through deformation in $3d$, \w\ algebra is more interesting. First of all, two specific elements of this family $W(0,0)$ and $W(0,-1)$ are matched on \kac\ and \bms\ respectively. So one concludes that these two can be deformed to each other through \w\ algebra. If \bms\ is considered as asymptotic and \kac\ as near horizon symmetry algebras, the deformation can be viewed as 
the procedure which connects these two regions. If one considers a dual field theory description in $3d$ flat space time, this connection may be viewed as RG flow of IR and UV region of corresponding field theory. On the other hand, it has been recently shown that $W(0,b)$ family algebra can be obtained as $3d$ near non-extremal horizons. In this picture, \bms\ may be viewed as one of element of $W(0,b)$ family and its deformation is considered as a procedure which modifies $-1 \rightarrow b$. We also discussed how deformation of \hbms\ can be considered as quantum corrections which leads to turning on the central charge $c_{JJ}$. In mathematical point of view, we presented a new notion of rigidity for \w\ family algebra which \w\ algebra with generic value of its parameters just can be deformed to another $W(\bar{a},\bar{b})$ algebra with shifted parameters.

In $4d$ case we explored deformations of \bmsf\ algebra and its central extension \hbmsf. We showed that \bmsf\ can not be deformed into an infinite dimensional Lie algebra including $\mathfrak{so}(3,2)$ as global part. So finding asymptotic symmetry algebra for AdS$_{4}$ will remain an open question. Although we showed that the ideal part of \bmsf\ is rigid, it can be deformed in its other commutators. Similar to \bms\ case, \bmsf\ can be deformed into \wf\ family algebra. Recently $W(a,a;a,a)$ family algebra is obtained as near horizon symmetry algebra of $4d$ black holes. So as the same as $3d$ case, there are two different interpretations for deformation: Deformation as procedure which connects the asymptotic and near horizon regions OR as procedure which takes $W(-\frac{1}{2},-\frac{1}{2};-\frac{1}{2},-\frac{1}{2})$ as one elements of near horizon algebras to $W(a,a;a,a)$. The same notion of rigidity, as \w\ algebra in $3d$ case, for \wf\ is considered and sowed that in this sense it is rigid. We also found deformations in some specific points as $W(0,0;0,0)$ and $W(0,-1;0,0)$ which takes us out of \wf\ family algebra. In particular, $W(0,-1;0,0)$ which has $\mathfrak{bms}_{3}\oplus\mathfrak{wit}$ subalgebra can be deformed into a new non isomorphic algebra with there copies of \wit\ algebra as subalgebra. 

As the last problem we carried out deformation and stabilization of \Max\ algebra which is infinite enhancement of $2+1$ Maxwell algebra. We showed that \Max\ algebra can be deformed in its ideal part to $ \mathfrak{bms}_{3}\oplus\mathfrak{wit}$ and three copies of \wit\ algebra. \Max\ algebra and its deformation, three copies of \wit\ algebra, have been obtained as asymptotic symmetry of two Chern-Simons gravity theories. We also found deformations in other commutators of \Max\ algebra which are $M(a,b;,c,d)$ and $\bar{M}$ family algebras. For specific value of the parameters as $(0,-\frac{1}{2};0,0)$, $M$ family algebra is the same as twisted Schr\"{o}dinger-Virasoro algebra. It is also noteworthy to mention that the $M(\frac{z-2}{2z},\frac{-1}{z};\frac{z-2}{z},\frac{z-2}{z})$ family algebra, which is introduced as asymptotic symmetry algebra of Schr\"{o}dinger spacetimes \cite{Compere:2009qm}, has the same structure as $M(a,b;c,d)$ family algebra. We also constructed a Chern-Simons gravity theory with torsion invariant under $\mathfrak{iso}(2,1)\oplus\mathfrak{so}(2,1)$ and by choosing an appropriate boundary conditions, we obtained $\mathfrak{bms}_{3}\oplus\mathfrak{wit}$ algebra as its asymptotic symmetry algebra.  

In continuation of the problems which we addressed here, one can search for other problems/questions may be enriched our knowledge about asymptotic symmetry groups, their corresponding algebras and also infinite dimensional algebras in high energy physics. The main questions are as follows:
 
   \paragraph{Deformation of Heisenberg-like algebras.}
The asymptotic symmetry algebra analysis depends very much on the choice of boundary fall-off behavior on metric fluctuations.
As mentioned in recent and notable paper \cite{Grumiller:2019fmp}, it has been studied near non-extremal horizons specifically for $3d$ and $4d$ and should that different boundary conditions are specified by a functional of the dynamical variables, describing inequivalent interactions at the horizon with a thermal bath. As another result, they showed that the near horizon algebra of a set of boundary conditions, is given by the semi-direct sum of diffeomorphisms at the horizon with “spins supertranslations” which is exactly the family algebra $W(0,b)$. They also explored another choice of boundary conditions and found a non-linear extension of the Heisenberg algebra, generalizing recent results in $3d$. In another recent paper\cite{Adami:2020uwd}, it has been considered that, the soft part of the symmetry algebra is independent of the radius and in the near horizon is just the Heisenberg-like algebra. 

As the combination of these two, one may propose that the fundamental symmetry algebra in the near horizon of Black holes is just Heisenberg algebra and other algebra, for instance $W(0,b)$ in $3d$, might be obtained through its deformation. To this end, we are going to study deformation of Heisenberg algebra in three and four dimension. If this conjecture is true, with combination of these recent works, it shows that  1- There is no relation between the near horizon symmetry algebra and symmetry algebras in other radii. 2- All obtained algebras as near horizon algebras with different choices of boundary condition can be obtained through deformation of Heisenberg algebra. For instance in $3d$, one can start with Heisenberg-like algebra and by deformation obtain \wh\ family algebra.

 \paragraph{Deformation of ``original'' \bmsf\ algebra.}
 As we showed that here, deformation of \bmsf\ can not be led to obtain an infinite dimensional Lie algebra containing $\mathfrak{so}(3,2)$ as its global part. It means that asymptotic symmetry algebra of AdS$_{4}$ can not be an infinite dimensional Lie algebra which is contracted to \bmsf. But as it is known, there three various version algebras which are obtained as asymptotic symmetry algebra of $4d$ flat space time. The natural continuation of our consideration is to explore possible deformation of original \bmsf\ introduced by \cite{Bondi:1962px, Sachs:1962zza} which is a semi-direct sum of Lorentz algebra and Abelian ideal spanned by supertranslations. If one finds the ideal part of the original \bmsf\ is rigid, the conclusion is that the infinite dimensional asymptotic symmetry algebra of AdS$_{4}$ is not a deformation of this algebra.  The one may need to look for new mathematical frameworks other than Lie algebra such as Lie algebroid as it is proposed in \cite{Compere:2019bua,Compere:2020lrt}. Then the interesting question is how one can extend the concepts of deformation/contraction of Lie algebras to Lie algebroids.

 \paragraph{Black hole solutions' thermodynamics of Chern-Simons gravity theories invariant under $2+1$ Maxwell algebra and its deformations.}

It was pointed that there are three Chern-Simons gravity theories which are invariant under $2+1$ Maxwell algebra and its deformations which are $\mathfrak{so(2,2)\oplus\mathfrak{so(2,1)}}$ and $\mathfrak{iso(2,1)\oplus\mathfrak{so(2,1)}}$. It has been shown that asymptotic symmetries algebras of first and second theories are infinite dimensional enhancement of $2+1$ Maxwell and three copies of Virasoro algebra respectively \cite{Concha:2018zeb,Concha:2018jjj}. In a recent work we have constructed a Chern-Simons gravity theory with torsion which is invariant under $\mathfrak{iso(2,1)\oplus\mathfrak{so(2,1)}}$ and showed that the associated asymptotic symmetries algebra is $\mathfrak{bms}_{3}\oplus \mathfrak{witt}$ with independent central charges \cite{Adami:2020xkm}. As the first problem we should search for solutions of the last theory. Then one may ask what is relation of physical quantities of these theories which are connected to each other through deformation of algebras. In particular, one may consider thermodynamic of black hole solutions, specifically entropy, in these three theories to find meaning of deformation in this level. The similar notion of deformation may be existed when one explores Cardy like formula for thermodynamics of mentioned solutions. 

 \paragraph{Deformation at the level of group and coadjoint orbits.}
 Here we focused on the algebras and their deformations. We know that there are groups associated with \bms\ and \bmsf\ and their central extensions \hbms\ and \hbmsf\ algebras \cite{Oblak:2016eij,Barnich:2016lyg}. One may ask how the deformation of algebras appear in the associated groups, e.g. whether there are groups associated with \w, $W(a,b;\bar{a},\bar{b})$ and also $M(a,b;c,d)$ family algebras and their central extensions. Another related question is analyzing the (co)adjoint orbits of these groups and algebras and how the deformations affect the unitary coadjoint orbits which is crucial for building Hilbert space of physical theories invariant under the deformed symmetry algebras.

 \paragraph{Hochschild-Serre factorization (HSF) theorem for infinite dimensional algebras.} We discussed that the HSF theorem states that a non-rigid \emph{finite dimensional} algebra can only be deformed in its ideal part and other parts of the algebra cannot be deformed. It is, however, known that this theorem does not apply to infinite dimensional algebras. Our analysis in the case for the \bms,  $\mathfrak{KM}_{u(1)}$ and $W(a,b)$ in $3d$, \bmsf\ and \wf\ in $4d$ and also \Max\ algebra led us to a proposal for a version of 
HSF which works for the infinite dimensional algebras: 

\emph{For infinite dimensional algebras with countable basis the deformations may appear in ideal and non-ideal parts, however, the deformations are always by coefficients of term in the ideal part.} 

For instance the $W(a,b)$ class with generators ${\cal J}_n, {\cal P}_n$ the deformations may appear in $[{\cal J}, {\cal J}],\ [{\cal J}, {\cal P}],\ [{\cal P}, {\cal P}],$ nonetheless it is proportional to ${\cal P}$ generators. It would be desirable to attempt proving this proposal, maybe using cohomological arguments.


\appendix 

\chapter {Noether theorems, gauge symmetries and surface charges algebra}\label{appendix-A}
  There are long standing debates around finding the meaningful conserved charges for gauge symmetries such as energy and momentum in gravity and electrical charge in Maxwell theory \cite{Barnich:2001jy, Compere:2007az, cogliati2012noether, Avery:2015rga, Banados:2016zim, Seraj:2016jxi}. The problem arises when one computes a Noether current associated to a gauge symmetry and finds that it vanishes on-shell up to divergence of an antisymmetric rank-2 tensor known as superpotential. This is a direct consequence of Noether's second theorem which is mentioned by her in \cite{Noether:1918zz}. The attempts to solve the conserved charge puzzle in gauge theories give rise to study asymptotic symmetries and define surface charges. Here we are going to briefly reviewe the main problem to associate a conserved charge to gauge symmetries and then explain the solution for it. Also we describe the notions of asymptotic symmetries and surface charges for gauge theories and then consider how the infinite dimensional Lie algebras obtained through this construction.  
  It should be mentioned that although there are different approaches and descriptions about Noether's second theorem, surface charges, asymptotic symmetries and algebras, in this review we follow mainly the similar description as \cite{Barnich:2001jy, Ruzziconi:2019pzd}.

\subsection*{Noether's theorems}

 The description of a field theory on the $n$ dimensional manifold $(\mathcal{M},g)$ is given by the action 
\begin{equation}
    S[\phi(x)]=\int_{\mathcal{M}}  \mathbf{L}[\phi(x), \partial_{\mu} \phi (x),...],
\end{equation}
 where $\phi$ denotes arbitrary field and $\mathbf{L}=Ld^{n}x$ is the Lagrangian. 
 
 \subsubsection*{Global and gauge symmetries}
 Consider a Lagrangian theory with Lagrange density $\mathbf{L}[\phi,\partial_{\mu}\phi,...]$ and a transformation of fields as $\delta_{\chi}\phi=\chi$ where $\chi$ is a local function. Then the symmetry of theory is defined as
 \begin{equation}\label{symmetry-definition}
     \delta_{\chi}\mathbf{L}=d\mathbf{B}_{\chi}
 \end{equation}
 where the boundary term $\mathbf{B}$ is a $n-1$-form as $\mathbf{B}_{\chi}=B^{\mu}_{\chi}(d^{n-1}x_{\mu})$.  
  
  An infinitesimal gauge transformation on fields which are depend on parameters $\Xi=(\xi^{\alpha})$ is defined as 
  \begin{equation}\label{infinit-gauge transfom}
      \delta_{\Xi}\phi=G[\Xi]= G_{\alpha}\xi^{\alpha}+G^{\mu}_{\alpha}\partial_{\mu}\xi^{\alpha}+G^{\mu\nu}_{\alpha}\partial_{\mu}\partial_{\nu}\xi^{\alpha}+...=\sum_{k \geq0}G^{\mu_{1},...,\mu_{k}}_{\alpha}\partial_{\mu_{1}}...\partial_{\mu_{k}}\xi^{\alpha},
  \end{equation}
 where $\xi^{\alpha}$ are arbitrary functions of space-time coordinates and $G^{\mu_{1}...\mu_{k}}$ are local functions. 
 A gauge symmetry similar to definition \eqref{symmetry-definition}, which is for arbitrary symmetry, is defined as
 \begin{equation}\label{gaugesymmetry-definition}
     \delta_{\Xi}\mathbf{L}=d\mathbf{B}_{\Xi}.
 \end{equation}
 So one concludes that a gauge symmetry is just a symmetry when its parameters are arbitrary functions of space-time coordinates $\chi=G[\Xi]$.  
 
 \subsubsection*{Examples}
 For the first example we consider the free Maxwell theory with the action
 \begin{equation}
     S[A]=\int_{\mathcal{M}} \mathbf{F}\wedge \star \mathbf{F},
 \end{equation}
 where the $2$-form $\mathbf{F}=d\mathbf{A}$ is field strength and $\mathbf{A}$ is the $1$-form potential. One checks that the Maxwell Lagrangian is invariant under the gauge transformation $\delta_{\Lambda}\mathbf{A}=d\lambda(x)$ where $\lambda(x)$ is an arbitrary function of coordinates. In comparison with \eqref{infinit-gauge transfom}, it is obvious that the form of gauge transformation in Maxwell theory is like $ \delta_{\Xi}\phi= G^{\mu}_{\alpha}\partial_{\mu}\xi^{\alpha}$ which is involved just with first derivative of the parameter. 
 
 As the second example we study the action of Einstein-Hilbert in General Relativity which is 
 \begin{equation}
     S[A]=\int_{\mathcal{M}}d^{n}x \,\sqrt{-g}(R-2\Lambda),
 \end{equation}
 where $\Lambda$ and $R$ are cosmological constant and Ricci scalar respectively.
 It is straightforward to show that the gauge transformation induced by diffeomorphism $\xi$ as 
 \begin{equation}
     \delta_{\xi}g_{\mu\nu}=\mathcal{L}_{\xi}g_{\mu\nu}=\xi^{\alpha}\partial_{\alpha}g_{\mu\nu}+g_{\mu\alpha}\partial_{\nu}\xi^{\alpha}+g_{\nu\alpha}\partial_{\mu}\xi^{\alpha},
 \end{equation}
 is a symmetry of theory. One finds that this gauge transformation is in the form of $\delta_{\Xi}\phi= G_{\alpha}\xi^{\alpha}+G^{\mu}_{\alpha}\partial_{\mu}\xi^{\alpha}$ which is only deal with first order derivatives of the parameters.

 Now we define an equivalence relation $\sim$ between symmetry transformations: Two transformations are belong to the same equivalence class if they differ, on shell, by a gauge transformation as
 \begin{equation}
     \chi\sim\chi+G[\Xi].
 \end{equation}
 The equivalence class of symmetries $[\chi]$ is called a global symmetry. 
 We define a conserved current as a $n-1$ form $J$ which is closed on shell as $dJ\approx 0$. The on shell equivalence relation between the currents is defined as 
 \begin{equation}
     J\sim J+dK,
 \end{equation}
 where $K$ is a $n-2$ form called superpotential. The equivalence class $[J]$ is called Noether current.

 \subsubsection*{Noether's first theorem} 
 There is one to one relation between global symmetries $[\chi]$ and Noether currents $[J]$ as
 \begin{equation}
     [\chi]\leftrightarrow[J].
 \end{equation}
 This implies that the Noether currents associated to gauge symmetries are trivial. In the other words, as mentioned, the Noether current vanishes on shell up to divergence of a $n-2$ form called superpotential.  
 To see the relation between global symmetry and Noether current one considers the explicit variation of Lagrangian as 
 \begin{equation}
 \begin{split}
          \delta_{\chi}L&=\delta_{\chi}\phi\frac{\partial L}{\partial \phi}+\delta_{\chi}\partial_{\mu}\phi\frac{\partial L}{\partial(\partial_{\mu}\phi)}+...\\
     &= \chi\frac{\partial L}{\partial \phi}+\partial_{\mu}\chi\frac{\partial L}{\partial(\partial_{\mu}\phi)}+...\\
     &= \chi\big(\frac{\partial L}{\partial \phi}+\partial_{\mu}\frac{\partial L}{\partial(\partial_{\mu}\phi)}+...\big)+ \partial_{\mu}(\chi\frac{\partial L}{\partial(\partial_{\mu}\phi)}+...)\\
     &= \chi \frac{\delta L}{\delta \phi}+\partial_{\mu}(\chi\frac{\partial L}{\partial(\partial_{\mu}\phi)}+...)
      \end{split}
 \end{equation}
where we used $[\delta,\partial]=0$ and $\chi=\delta_{\chi}\phi$. With comparison with relation \eqref{symmetry-definition} one concludes that 
\begin{equation}
    \chi \frac{\delta L}{\delta \phi}=\partial_{\mu}\big(B-\chi\frac{\partial L}{\partial(\partial_{\mu}\phi)}+...\big)\equiv\partial_{\mu}J_{\chi}^{\mu}.
\end{equation}
 or equivalently
 \begin{equation}\label{current-EOM}
      \chi \frac{\delta \mathbf{L}}{\delta \phi}=d\mathbf{J}_{\chi},
 \end{equation}
 where $\mathbf{J}_{\chi}=J_{\chi}^{\mu}(d^{n-1}x)_{\mu}$. By applying the equation of motion $\frac{\delta \mathbf{L}}{\delta \phi}$ one obtains
 \begin{equation}
     d\mathbf{J}_{\chi}\approx 0,
 \end{equation}
 which $\approx$ shows this relation holds on-shell. 
 
 \subsubsection*{Noether charge} The Noether charge is defined as integrating the Noether current $[J]$ on a $n-1$ dimensional space like surface $\Sigma$ as
 \begin{equation}\label{Q-intJ}
      \mathcal{Q}_{\chi}=\int_{\Sigma} J.
 \end{equation}
 As we defined the Noether current as equivalence class $J\sim J+dK$ one obtains
 \begin{equation}
     \hat{\mathcal{Q}}_{\chi}=\mathcal{Q}_{\chi}+\int_{\partial\Sigma}K,
 \end{equation}
 in which the Stokes' theorem is applied. To avoid inconsistency, it is supposed that the Noether current and the superpotential $K$ vanishes at infinity so $\int_{\partial\Sigma}K=0$ then one obtains
 \begin{equation}
     \hat{\mathcal{Q}}_{\chi}=\mathcal{Q}_{\chi}.
 \end{equation}
 As we will see this assumption that the superpotential $K$ vanishes at infinity is not true and the surface charges associated to gauge symmetries are computed by using $K$. 
 
 \subsubsection*{Noether charge conservation}
 The Noether charge is conserved in time in the sense that its time integration is zero on shell as
 \begin{equation}
     \frac{d}{dt}\mathcal{Q}_{\chi}\approx 0.
 \end{equation}
 
  \subsubsection*{Noether current, charge and representation theory}
  The Noether currents form a representation of symmetries as 
  \begin{equation}
      \{J_{\chi_{1}},J_{\chi_{2}}\}\approx J_{[\chi_{1},\chi_{2}]},
  \end{equation}
where we the bracket of Noether currents and symmetries are defined as  $\{J_{\chi_{1}},J_{\chi_{2}}\}=\delta_{\chi_{1}}J_{\chi_{2}}$ and $[\chi_{1},\chi_{2}]=\delta_{\chi_{1}}\chi_{2}-\delta_{\chi_{2}}\chi_{1}$ respectively. 
It can be also shown that the Noether charges form a representation of the global symmetry algebra as
\begin{equation}
          \{\mathcal{Q}_{\chi_{1}},\mathcal{Q}_{\chi_{2}}\}\approx \mathcal{Q}_{[\chi_{1},\chi_{2}]},
\end{equation}
 where the bracket of charges is defined as $ \{\mathcal{Q}_{\chi_{1}},\mathcal{Q}_{\chi_{2}}\}=\delta_{\chi_{1}}\mathcal{Q}_{\chi_{2}}=\int \delta_{\chi_{1}}J_{\chi_{2}}$.

\subsubsection*{Gauge symmetries and lower degree conservation law}
One may consider the same relation as \eqref{current-EOM} for gauge symmetries as
\begin{equation}\label{gaugecurrent-EOM}
    G[\Xi]\frac{\delta L}{\delta \phi}=\partial_{\mu}J_{\Xi}^{\mu}.
\end{equation}
By using \eqref{infinit-gauge transfom} one works out the LHS as
\begin{equation}
    \begin{split}
        G[\Xi]\frac{\delta L}{\delta \phi}&=\big(G_{\alpha}\xi^{\alpha}+G^{\mu}_{\alpha}\partial_{\mu}\xi^{\alpha}+G^{\mu\nu}_{\alpha}\partial_{\mu}\partial_{\nu}\xi^{\alpha}+...\big)\frac{\delta L}{\delta \phi}\\
        &= \xi^{\alpha}\big[G_{\alpha}\frac{\delta L}{\delta \phi}-\partial_{\mu}\big(G^{\mu}_{\alpha}\frac{\delta L}{\delta \phi}\big)+...\big]\\
        &+\partial_{\mu}\big[G^{\mu}_{\alpha}\xi^{\alpha}\frac{\delta L}{\delta \phi}-\xi^{\alpha}\partial_{\nu}\big(G^{\mu\nu}_{\alpha}\frac{\delta L}{\delta \phi}\big)+...\big].
        \end{split}
\end{equation}
 The last term in RHS is called $S_{\mu}^{G}=G^{\mu}_{\alpha}\xi^{\alpha}\frac{\delta L}{\delta \phi}-\xi^{\alpha}\partial_{\nu}\big(G^{\mu\nu}_{\alpha}\frac{\delta L}{\delta \phi}\big)+...$. By using the latter, relation \eqref{gaugecurrent-EOM} is written as
 \begin{equation}\label{NoethercurrentJS}
     \xi^{\alpha}\Tilde{G}_{\alpha}\frac{\delta L}{\delta \phi}=\partial_{\mu}(J_{G}^{\mu}-S_{G}^{\mu}),
 \end{equation}
 where $\Tilde{G}_{\alpha}\frac{\delta L}{\delta \phi}=G_{\alpha}\frac{\delta L}{\delta \phi}-\partial_{\mu}\big(G^{\mu}_{\alpha}\frac{\delta L}{\delta \phi}\big)+...$. Since the RHS of \eqref{NoethercurrentJS} is a total derivative, it has been proved that 
 \begin{equation}\label{Noether-identity}
      \xi^{\alpha}\Tilde{G}_{\alpha}\frac{\delta L}{\delta \phi}=0,
 \end{equation}
 which is called the Noether identity. For more details about the proof we refer the reader to \cite{Ruzziconi:2019pzd, Barnich:2018gdh}. 
 
 \subsubsection*{Example}
 It is straight forward to show that the Noether identity in the case of Maxwell theory and gravity is just the Bianchi identity as 
 \begin{equation}
     \nabla_{\mu}\nabla_{\nu}F^{\mu\nu}=0,\,\,\, \nabla_{\mu}G^{\mu\nu}=0,
 \end{equation}
 respectively. 
 
 \subsubsection*{Noether's second theorem}
 For a gauge symmetry $G[\Xi]$ we have 
 \begin{equation}\label{Noether-second}
     G[\Xi]\frac{\delta \mathbf{L}}{\delta \phi}=d\mathbf{S}_{G}\big[\frac{\delta \mathbf{L}}{\delta \phi}\big],
 \end{equation}
 where $\mathbf{S}=S_{G}^{\mu}(dx^(n-1))_{\mu}$ is defined as the above and called the weakly vanishing Noether current which means that it vanishes on-shell $S_{G}^{\mu}\approx0$. In the other word, Noether's second theorem states that for a gauge symmetry the multiplication of Euler-Lagrange equation with gauge transformation is equal to a total derivative. 
 
 One can also consider relations \eqref{NoethercurrentJS} and \eqref{Noether-identity} to conclude that 
 \begin{equation}
     d(\mathbf{J}_{G}-\mathbf{S}_{G})=0.
 \end{equation}
 By using the Poincaré lemma one obtains
 \begin{equation}
     \mathbf{J}_{G}=\mathbf{S}_{G}+d\mathbf{K}_{G},
 \end{equation}
 where $\mathbf{K}_{G}$ is a $n-2$ form as mentioned is called superpotential. The Euler-Lagrange equations leads to 
 \begin{equation}\label{J-dK}
     \mathbf{J}_{G}\approx d\mathbf{K}_{G}.
 \end{equation}
 As we mentioned, \eqref{J-dK} states that the Noether current associated to a gauge symmetry is trivial on-shell up to an exact $n-1$ form $d\mathbf{K}_{G}$. One may follow the previous procedure which led to obtain an expression for Noether charge from Noether current \eqref{Q-intJ} and define a charge associated to a Noether current of a gauge symmetry as 
 \begin{equation}\label{Q-Jgauge-K}
     \mathcal{Q}_{G}=\int_{\Sigma}\mathbf{J}_{G}\approx \int_{\partial\Sigma}\mathbf{K},
 \end{equation}
 where we used \eqref{J-dK} and Stokes' theorem.  Since the charge for gauge theories receives only contribution from surface region, we call it surface charge. One can show that since $d\mathbf{J}_{G}\approx0$ the charge associated to this current is conserved on-shell $\partial_{\mu}K^{\mu\nu}_{G}\approx 0$. In fact, relation \eqref{Q-Jgauge-K} suggests that instead of conserved current one can associate a superpotential to a gauge symmetry.   
 The main problem to compute the conserved charge \eqref{Q-Jgauge-K} is that there is no constraint on superpotential $\mathbf{K}_{G}$ so the integral \eqref{Q-Jgauge-K} may take any value. To avoid this problem, one should look for a procedure to consider only a certain $\mathbf{K}_{G}$.  
 
 There are different methods to compute surface charges associated to gauge transformations. Two of main methods are known as the covariant phase space methid which is introduced and extended by Wald, Zoupas and Iyer in \cite{Wald:1993nt,Iyer:1994ys,Wald:1999wa} and the Barnich-Brandt prescription worked out by Barnich and Brandt \cite{Barnich:2001jy, Barnich:2003xg, Barnich:2007bf}. For more details about these two techniques and their advantages-disadvantages we refer the reader to \cite{Ruzziconi:2019pzd}.

 \subsubsection*{Asymptotic symmetry and surface charge algebras}
 As mentioned, finding a conserved charge corresponding to a gauge symmetry is more involved than global symmetries. Here without details, we are going to review the main ideas which lead to obtain asymptotic symmetry group and its corresponding algebra. We will mainly follow \cite{Oblak:2016eij}. 
 
 To find asymptotic symmetry group and surface charge algebra one should work out some main separate steps as follows
 \begin{itemize}
     \item First of all, one should specify the theory by choosing the bulk action, the field content of the theory with appropriate fall-off conditions and if needed to add a boundary term to the action such that it remains differentiable. In some cases as general relativity the fall-off boundary conditions imposing on field content of the theory is specified after fixing an appropriate gauge.

     \item One then looks for the gauge transformations which keep the fall-off conditions intact. The gauge transformations which preserves fall-off conditions are called allowed gauge transformations. In contrast there are gauge transformations who do not respect fall-off condition so called forbidden gauge transformations. Allowed gauge transformations are thought the transformations which lead to symmetries (global or gauge) of the theory.

     \item One finds a conserved superpotential $\mathbf{K}_{G}$ corresponding to each allowed gauge transformation. It can be checked that the superpotential is a linear function of gauge parameter but an arbitrary function of field content depending on details of the theory under consideration. For instance, the superpotential in Maxwell theory is given by 
     \begin{equation}
         K_{\mu\nu}= F_{\mu\nu}\lambda(x),
     \end{equation}
     where $F_{\mu\nu}$ is field strength and $\lambda(x)$ is an arbitrary function of spacetime which induces the gauge transformation as $A_{\mu}\rightarrow A_{\mu}+\partial_{\mu}\lambda(x)$.  
     
     \item Finally, one computes the surface charges associated to allowed gauge transformations and identifies two separate classifications: 1- If the value of obtained surface charge is zero, the corresponding allowed gauge transformation is called trivial gauge transformation. 2- If the value of obtained surface charge is finite and non zero, the corresponding allowed gauge transformation is called nontrivial gauge transformation. If the value of all surface charges are infinite it shows that the boundary conditions are not consistence. For instance, the surface charges associated to allowed gauge transformation of Maxwell theory is given by 
     \begin{equation}
         \mathcal{Q}_{\lambda}=\oint_{\partial\Sigma} d\Sigma_{\mu\nu}F_{\mu\nu}\lambda(x).
     \end{equation}
     For particular value $\lambda=1$ the surface charge will coincide with electric charge. For more details about computation of surface charge for Maxwell theory we refer the reader to \cite{Seraj:2016jxi}.

 \end{itemize}
 
  \paragraph{Remark} It would be worth noting that there are mainly three approaches to define asymptotic of the space in the case of gauge theories, specifically general relativity, which can be classified as 1- Geometric approach \cite{Penrose:1962ij, Ashtekar:1981bq,Ashtekar:1991vb} 2- Gauge fixing approach \cite{Barnich:2017ubf, Barnich:2009se} 3- Hamiltonain approach \cite{regge:1974zd, henneaux:2018cst}. For more details about these approaches and their difference we refer to \cite{Ruzziconi:2019pzd}.

 As result one can separate three different categories of gauge transformations: 1- Trivial gauge transformations which the associated surface charges are zero. This family of gauge transformations are truly redundancies of the theory and do not contain physical information. 2- Non trivial gauge transformation with associated surface charges are finite and non zero. The electric charge in Maxwell theory is one example which shows that non trivial gauge transformation may include important information of theory. 3- Forbidden gauge transformation which are neither trivial, nor non trivial gauge transformation so should be excluded from the theory since they do not leave the phase space invariant.   
     
     We mentioned before that the infinitesimal global transformations are endowed with Lie bracket so they satisfies a Lie algebra structure. One can show that the non trivial gauge transformation and their associated surface charges satisfy the same structure so it leads us to this terminology:
     \begin{tcolorbox}
         \paragraph{Definition.} {\it Asymptotic symmetries algebra (group) of a theory is defined as quotient of the algebra (group) of all allowed gauge transformations to its ideal part including trivial gauge transformations. }
          \end{tcolorbox}

     \paragraph{Remark} In gravitational physics, the gauge transformations are diffeomorphisms $x_{\mu}\rightarrow x_{\mu}+\xi_{\mu}$ on spacetime manifold $\Sigma$ which are generated by vector fields $\xi_{\mu}(x)$. Allowed gauge transformations are called asymptotic Killing vector fields which are endowed with usual Lie bracket of vector fields.  
     
     \subsubsection*{Surface charge algebra and its central extension}
     The surface charge may be seen as generator of allowed non trivial gauge transformation $\chi$ which acts on field content of the theory through Poison bracket as
     \begin{equation}
         \{\mathcal{Q}_{\chi},\phi\}=\delta_{\xi}\phi,
     \end{equation}
     where $\mathcal{Q}_{\chi}$ is the surface charge associated to gauge transformation $\chi$. In comparison with representation theorem for Noether current and charges, one shows that surface charges satisfy the poison bracket as
     \begin{equation}\label{surfacalgebra-central}
         \{\mathcal{Q}_{\xi_{1}},\mathcal{Q}_{\xi_{2}}\}=\mathcal{Q}_{[\xi_{1},\xi_{2}]}+c(\xi_{1},\xi_{2}),
     \end{equation}
     where $[\chi_{1},\chi_{2}]=\delta_{\chi_{1}}\xi_{2}-\delta_{\chi_{2}}\chi_{1}$ and $c(\chi_{1},\chi_{2})$ is real-valued two co-cycle. 
     The last term in RHS in \eqref{surfacalgebra-central} is responsible for central extensions of surface charge algebra. It should be noticed that in gravity the bracket $[\chi_{1},\chi_{2}]$ is the usual bracket between vector fields. The surface charge associated to asymptotic Killing vector fields satisfy the same algebra with an extra central term. For more details see e.g. \cite{Barnich:2001jy,Ruzziconi:2019pzd}.

\chapter{On rigidity of  \texorpdfstring{$\mathfrak{witt}\oplus \mathfrak{witt}$}{wittpluswitt}, \texorpdfstring{$\mathfrak{vir}\oplus \mathfrak{vir}$}{virplusvir}, \texorpdfstring{$W(a,b)$}{Wab} and \texorpdfstring{$\widehat{W(a,b)}$}{hWab} algebras}\label{appendix-B}

As discussed and summarized in table \hyperlink{table1}{4.1} formal deformations of the \bms, \hbms, ${\mathfrak{KM}_{\mathfrak{u}(1)}}$ and $\widehat{\mathfrak{KM}_{\mathfrak{u}(1)}}$ algebras yields $\mathfrak{witt}\oplus \mathfrak{witt}$, $W(a,b)$, $\mathfrak{vir}\oplus \mathfrak{vir}$ and $\widehat{W(a,b)}$ algebras repectively. As such one expects these latter algebras to be rigid. We explore this explicitly in this appendix.

\section{\texorpdfstring{$\mathfrak{witt}_{L}\oplus \mathfrak{witt}_{R}$}{virplusvir} is rigid.}\label{witt+witt-rigid}
Consider the most general infinitesimal deformation of $\mathfrak{witt}_{L}\oplus \mathfrak{witt}_{R}$ algebra:
\begin{equation} 
\begin{split}
 & i[L_{m},L_{n}]=(m-n)L_{m+n}+\eta (m-n)h(m,n)\bar{L}_{m+n}, \\
 &i[L_{m},\bar{L}_{n}]=\varepsilon_{1}K(m,n)L_{m+n}+\varepsilon_{2} I(m,n)\bar{L}_{m+n},\\
 &i[\bar{L}_{m},\bar{L}_{n}]=(m-n)\bar{L}_{m+n}+\zeta(m-n)f(m,n)L_{m+n},
\end{split}
\end{equation}
where $f(m,n),h(m,n)$ and $K(m,n), I(m,n)$ are symmetric and arbitrary functions. Also ${\varepsilon}_{i}, \eta$ and $\zeta$ are deformation parameters. The Jacobi identities lead to some independent relations for each of the above functions as well as two equations relating $I(m,n)$ and $h(m,n)$, and $K(m,n)$ and $f(m,n)$ to each other. If we turn on each one of the deformations individually, one can see that there are only trivial solutions.

There are solutions involving simultaneous deformation by two parameters. If we turn on $I,h$ together we find solutions  $I(m,n)=\alpha(n-m)m^{r}$ and $h(m,n)=\alpha(m+n)^{r}$ and if we turn on  $K,f$ together one finds $K(m,n)=\beta(n-m)n^{s}$ and $f(m,n)=-\beta(m+n)^{s}$,  where $r,s$ are arbitrary integers.  Let us consider deformations by $I(m,n)$ and $h(m,n)$. The Jacobi identity leads to  $\tilde{I}(m+n,l)=-h(m,n)$, where  $I(m,n):=(m-n)\tilde{I}(m,n)$. Therefore, $\tilde{I}(m+n,l)=-h(m,n)=-R(m+n)$. Redefining generators as
\begin{equation} 
\begin{split}
   L_{m}:=\tilde{L}_{m}+R(m)\tilde{\bar{L}}_{m}, \qquad 
 \bar{L}_{m}:=\tilde{\bar{L}}_{m},
\end{split}\label{A-redefinition}
\end{equation}
one can show that $\tilde{L}_{m}$ and $\tilde{\bar{L}}_{m}$ satisfy the commutation relations of  $\mathfrak{witt}_{L}\oplus \mathfrak{witt}_{R}$ algebra. This shows that all of the solutions we have derived for $I(m,n)$ and $h(m,n)$ are just trivial deformations so $\mathcal{H}^2(\mathfrak{witt}_{L}\oplus \mathfrak{witt}_{R}; \mathfrak{witt}_{L})=0$. The same result can be obtained for $K(m,n)$ and $f(m,n)$). So we conclude that $\mathcal{H}^2(\mathfrak{witt}_{L}\oplus \mathfrak{witt}_{R}; \mathfrak{witt}_{L}\oplus \mathfrak{witt}_{R})=0$ and therefore $\mathfrak{witt}_{L}\oplus \mathfrak{witt}_{R}$ is infinitesimally and formaly rigid.

\paragraph{Cohomological considerations.} Theorem \ref{2.1} states that an infinite dimensional Lie algebra $\mathfrak{g}$ is called formally and infinitesimally rigid when we have $\mathcal{H}^2(\mathfrak{g};\mathfrak{g})=0$. So we consider the second cohomology of $\mathfrak{witt}_{L}\oplus \mathfrak{witt}_{R}$  algebra with values in adjoint module, namely, 
\begin{equation}
    \mathcal{H}^2(\mathfrak{witt}_{L}\oplus \mathfrak{witt}_{R};\mathfrak{witt}_{L}\oplus \mathfrak{witt}_{R}).\label{2-ads3-cohomo}
\end{equation}
Then one can decompose the relation \eqref{2-ads3-cohomo} as \cite{ChevalleyEilenberg}, 
\begin{multline}\label{2-ads3-cohomo-decom}
     \mathcal{H}^2(\mathfrak{witt}_{L}\oplus \mathfrak{witt}_{R};\mathfrak{witt}_{L}\oplus \mathfrak{witt}_{R})=
    \mathcal{H}^2(\mathfrak{witt}_{L}\oplus \mathfrak{witt}_{R}; \mathfrak{witt}_{L})\oplus\mathcal{H}^2(\mathfrak{witt}_{L}\oplus \mathfrak{witt}_{R}; \mathfrak{witt}_{R}).
\end{multline}
{We should note that unlike $\mathfrak{bms}_{3}$ and $\mathfrak{KM}_{\mathfrak{u}(1)}$ and their central extensions in this case both of $\mathfrak{witt}_{R,L}$ are $\mathfrak{witt}_{L} \oplus \mathfrak{witt}_{R}$ module by the adjoint action.} Since the first and second terms in the above are essentially the same, it suffices just to show that
\begin{equation}
     \mathcal{H}^2(\mathfrak{witt}_{L}\oplus \mathfrak{witt}_{R}; \mathfrak{witt}_{L})=0.\label{coho-leftwitt}
\end{equation}
We consider the following exact sequence of Lie algebras:
\begin{equation}
     0\longrightarrow\mathfrak{witt}_{L}\longrightarrow\mathfrak{witt}_{L}\oplus \mathfrak{witt}_{R}\longrightarrow \mathfrak{witt}_{R}\longrightarrow 0.
\end{equation}
The Hochschild-Serre spectral sequence associated to the above exact sequence with coefficients in $\mathfrak{witt}_{L}$ is convergent with the following property \cite{degrijse2009cohomology}:
\begin{equation}
     E_{\infty}^{p,q}\Longrightarrow\mathcal{H}^{p+q}(\mathfrak{witt}_{L}\oplus \mathfrak{witt}_{R}; \mathfrak{witt}_{L}).
\end{equation}
On the other hand, we know (see \cite{Ecker:2018iqn} for more detail)
\begin{equation}
\begin{split}
   & \mathcal{H}^{0}(\mathfrak{witt}_{L};\mathfrak{witt}_{L})=\text{Cent}(\mathfrak{witt}_{L})=0,\\
   &\mathcal{H}^{1,2,3}(\mathfrak{witt}_{L};\mathfrak{witt}_{L})=0.
\end{split}
\end{equation}
In view of the Hochschild-Serre theorem \cite{MR0054581}, this implies that 
\begin{equation}
    E_{2}^{p,q}=\mathcal{H}^{p}(\mathfrak{witt}_{R};\mathcal{H}^{q}(\mathfrak{witt}_{L};\mathfrak{witt}_{L}))=0\,\,\,\text{for} \,\,q=0,1,2,3    \Longrightarrow \ \ E_{\infty}^{p,q}=0\,\,\text{for} \,\,q=0,1,2,3.
\end{equation}
Recalling that
\begin{equation}
     \mathcal{H}^2(\mathfrak{witt}_{L}\oplus \mathfrak{witt}_{R}; \mathfrak{witt}_{L})=\oplus_{p+q=2} E_{\infty}^{p,q}=E_{\infty}^{0,2}\oplus E_{\infty}^{1,1}\oplus E_{\infty}^{2,0}=0.\label{coho-decompos}
\end{equation}

The same result can be obtained for $ \mathcal{H}^2(\mathfrak{witt}_{L}\oplus \mathfrak{witt}_{R}; \mathfrak{witt}_{R})$. So we conclude that
the centerless asymptotic symmetries algebra of (A)dS$_3$ spacetime in $3d$, the direct sum of two Witt algebras $\mathfrak{witt}_{L}\oplus\mathfrak{witt}_{R}$, is formaly and infinitesimally {\it rigid}.

The same result can also be reached through the long exact sequence \eqref{long-exact} and \eqref{coho-leftwitt} without using \eqref{2-ads3-cohomo-decom}. From  \eqref{coho-leftwitt} and the same result for the second term in \eqref{2-ads3-cohomo-decom} which are  the second and fourth terms in \eqref{long-exact}, one concludes $ \mathcal{H}^2(\mathfrak{witt}_{L}\oplus \mathfrak{witt}_{R};\mathfrak{witt}_{L}\oplus \mathfrak{witt}_{R})=0$ .  

\section{On rigidity of $\mathfrak{vir}_{L}\oplus \mathfrak{vir}_{R}$}\label{rigid-vir+vir}
To establish rigidity of $\mathfrak{vir}_{L}\oplus \mathfrak{vir}_{R}$, we show that $\mathcal{H}^2(\mathfrak{vir}_{L}\oplus \mathfrak{vir}_{R};\mathfrak{vir}_{L}\oplus \mathfrak{vir}_{R})=0$. To this end, { as in the previous case}, we use decomposition, 
\begin{equation}
\begin{split}
    \mathcal{H}^2(\mathfrak{vir}_{L}\oplus \mathfrak{vir}_{R};\mathfrak{vir}_{L}\oplus \mathfrak{vir}_{R})=
    \mathcal{H}^2(\mathfrak{vir}_{L}\oplus \mathfrak{vir}_{R}; \mathfrak{vir}_{L})\oplus\mathcal{H}^2(\mathfrak{vir}_{L}\oplus \mathfrak{vir}_{R}; \mathfrak{vir}_{R}).
\end{split}
\end{equation}
     It just suffices to show that
\begin{equation}
     \mathcal{H}^2(\mathfrak{vir}_{L}\oplus \mathfrak{vir}_{R}; \mathfrak{vir}_{L})=0,
\end{equation}
for which we have the short exact sequence
\begin{equation}
     0\longrightarrow\mathfrak{vir}_{L}\longrightarrow\mathfrak{vir}_{L}\oplus \mathfrak{vir}_{R}\longrightarrow \mathfrak{vir}_{R}\longrightarrow 0.
\end{equation}

The Hochschild-Serre spectral sequence associated to the above exact sequence with coefficients in $\mathfrak{vir}_{L}$ is convergent with the following property \cite{degrijse2009cohomology}:
\begin{equation}
     E_{\infty}^{p,q}\Longrightarrow\mathcal{H}^{p+q}(\mathfrak{vir}_{L}\oplus \mathfrak{vir}_{R}; \mathfrak{vir}_{L}).
\end{equation}
On the other hand, we know (see \cite{Ecker:2017sen, Ecker:2018iqn} for more detail)
\begin{equation}
\begin{split}
   & \mathcal{H}^{0}(\mathfrak{vir}_{L};\mathfrak{vir}_{L})=\mathbb{R}c,\\
   &\mathcal{H}^{1,2}(\mathfrak{vir}_{L};\mathfrak{vir}_{L})=0,
\end{split}
\end{equation}
where $c$ is a central element.
In view of the Hochschild-Serre theorem \cite{MR0054581}, this implies that 
\begin{equation}
    E_{2}^{p,q}=\mathcal{H}^{p}(\mathfrak{vir}_{R};\mathcal{H}^{q}(\mathfrak{vir}_{L};\mathfrak{vir}_{L}))=0\,\,\,\text{for} \,\,q=0,1,2    \Longrightarrow \ \ E_{\infty}^{p,q}=0\,\,\text{for} \,\,q=0,1,2.
\end{equation}
We should just explain one point. In the case $E_{2}^{2,0}=\mathcal{H}^{2}(\mathfrak{vir}_{R};\mathcal{H}^{0}(\mathfrak{vir}_{L}, \mathfrak{vir}_{L}))$ the above leads to  $E_{2}^{2,0}=\mathcal{H}^{2}(\mathfrak{vir}_{R};\mathbb{R})$ but we know that $\mathcal{H}^{2}(\mathfrak{vir}_{R};\mathbb{R})=0$ then we conclude that $E_{2}^{2,0}=0$. Recalling that
\begin{equation}
     \mathcal{H}^2(\mathfrak{vir}_{L}\oplus \mathfrak{vir}_{R}; \mathfrak{vir}_{L})=\oplus_{p+q=2} E_{\infty}^{p,q}=E_{\infty}^{0,2}\oplus E_{\infty}^{1,1}\oplus E_{\infty}^{2,0},\label{coho-decompos-vir}
\end{equation}
we obtain
\begin{equation}
     \mathcal{H}^2(\mathfrak{vir}_{L}\oplus \mathfrak{vir}_{R}; \mathfrak{vir}_{L})=0,
\end{equation}
and consequently 
\begin{equation}
    \mathcal{H}^2(\mathfrak{vir}_{L}\oplus \mathfrak{vir}_{R}; \mathfrak{vir}_{L})\oplus\mathcal{H}^2(\mathfrak{vir}_{L}\oplus \mathfrak{vir}_{R}; \mathfrak{vir}_{R})=0.
\end{equation}
That is, the asymptotic symmetry algebra of (A)dS$_3$, the direct sum of two Virasoro algebras $\mathfrak{vir}_{L}\oplus\mathfrak{vir}_{R}$, is formaly and infinitesimally {\it rigid}.


\section{On rigidity of  \texorpdfstring{$W(a,b)$}{Wab}} 
Here we discuss that $W(a,b)$ algebra for generic $a,b$ is expected to be  rigid in the sense that it does not admit a nontrivial deformation. {As we mentioned in footnote \ref{footnote3}  and also in caption of table \hyperlink{table1}{4.2}, here we are considering the $W(a,b)$ family as the two-parameter family of the algebras. However, the $W(a,b)$ algebra as single algebra with two distinct parameters $a$ and $b$ is not rigid and can be deformed through $(a,b)$ space.} In what follows we show that all members of the $W(a,b)$ family for generic values of $a,b$ parameters are cohomologous to each other. To this end and as before, we deform each commutator separately. 
   \begin{equation} 
\begin{split}
 & i[\mathcal{J}_{m},\mathcal{J}_{n}]=(m-n)\mathcal{J}_{m+n}+\eta (m-n)h(m,n)\mathcal{P}_{m+n}, \\
 &i[\mathcal{J}_{m},\mathcal{P}_{n}]=-(n+bm+a)\mathcal{P}_{m+n}+\zeta K(m,n)\mathcal{P}_{m+n}+\kappa I(m,n) {\cal J}_{m+n},\\
 &i[\mathcal{P}_{m},\mathcal{P}_{n}]=\varepsilon_{1}(m-n)f(m,n)\mathcal{J}_{m+n}+\varepsilon_{2}(m-n)g(m,n)\mathcal{P}_{m+n},\label{W-general-deformation}
\end{split}
\end{equation}
where $h(m,n),f(m,n)$ and $g(m,n)$ are symmetric and $K(m,n),I(m,n)$ are arbitrary functions and $\eta,\zeta,\kappa $and $\varepsilon_{i}$ are arbitrary deformation parameters. We then consider Jacobi identities for generic $a,b$. 

The Jacobi $[\mathcal{P}_{m},[\mathcal{P}_{n},\mathcal{J}_{l}]]$ leads to two different relations one of them for just $f(m,n)$ and the second one which relates $g(m,n)$ and $I(m,n)$ which are respectively
\begin{equation}
\begin{split}
    &(l+m-n)(m+bl+a)f(n,m+l)+(m-n-l)(n+bl+a)f(m,n+l)+\\
    &(m-n)(l-m-n)f(m,n)=0,
\end{split}\label{ff-appendix}
\end{equation}
and 
\begin{equation}
\begin{split}
    \hspace*{-16mm}&(m+b(n+l)+a)I(l,n)-(n+b(l+m)+a)I(l,m)+(n-m)(m+n+bl+a)g(m,n)\\ &+(l+m-n)(m+bl+a)g(n,m+l)+(m-n-l)(n+bl+a)g(m,n+l)=0,\label{W-I,g}
\end{split}
\end{equation}
Consider \eqref{ff-appendix}. Since the parameter $a$ and $b$ are arbitrary and independent, the coefficients of $a$ and $b$ should be equal to zero separately. The relation for $b$ coefficient is 
\begin{equation}
    (l+m-n)f(n,m+l)+(m-n-l)f(m,n+l)=0.
\end{equation}
Choosing $m=n+l$ we get $f(n,2l+n)=0$. Since $l$ is an arbitrary integer ($l\neq 0$), we conclude $f(m,n)=0$. For the second relation we need to have more information about $I(m,n)$, for which we consider the Jacobi identity $[\mathcal{J}_{m},[\mathcal{J}_{n},\mathcal{P}_{l}]]$. This leads to two independent relations for $K(m,n)$ and $I(m,n)$ as
\begin{equation}
\begin{split}
    &-(l+bn+a)K(m,n+l)-(n+l+bm+a)K(n,l)+\\
    &(l+bm+a)K(n,m+l)+(m+l+bn+a)K(m,l)+(n-m)K(m+n,l)=0,
\end{split}\label{K-Jacobi-W}
\end{equation}
and 
\begin{equation}
\begin{split}
   &-(l+bn+a)I(m,n+l)+(m-n-l)I(n,l)+\\
    &(l+bm+a)I(n,m+l)+(m+l-n)I(m,l)+
    (n-m)I(m+n,l)=0,
\end{split}\label{I-Jacobi-W}
\end{equation}
\eqref{K-Jacobi-W} states that $K(m,n)=\alpha +\beta m$ and \eqref{I-Jacobi-W} leads to $I(m,n)=0$, and hence  \eqref{W-I,g} yields $g(m,n)=0$. 

Finally one can check the Jacobi  $[\mathcal{J}_{m},[\mathcal{J}_{n},\mathcal{J}_{l}]]$ which leads to the equation for $h(m,n)$,
\begin{multline}
    (n-l)(m-n-l)h(m,n+l)+(l-n)(n+l+bm+a)h(n,l)\\
    +(l-m)(n-l-m)h(n,l+m)+(m-l)(l+m+bn+a)h(l,m)\\
    +(m-n)(l-m-n)h(l,m+n)+(n-m)(m+n+bl+a)h(m,n)=0.\label{h-Jacobi-W}
\end{multline}
If $h(m,n)$ is expanded as power series as \eqref{power series}, the only solution for general $a,b$ is $h(m,n)=constant=h$. This deformation is, however, a trivial one, as it can be absorbed into redefinition of generators as \eqref{Z-redefinition} which leads to the relation $Z(m)(m+bn+a)-Z(n)(n+bm+a)+(n-m)Z(m+n)=\nu(m-n)$. One can then check that by choosing $Z(m)=-\nu/b$ the above relation is satisfied. So we conclude that for generic $a,b$ we cannot deform the $W(a,b)$ algebra; i.e. the family of $W(a,b)$ algebras is rigid. 

\paragraph{Cohomological considerations.} Unlike the case $\mathfrak{witt}\oplus \mathfrak{witt}$ algebra, $W(a,b)$ is not direct sum of two algebras and 
the Witt part is not an ideal in $W(a,b)$. However, like the \bms\ or $\mathfrak{KM}_{\mathfrak{u}(1)}$ case we can still use the Hochschild-Serre spectral sequence for the $W(a,b)$ algebra. We have the following short exact sequence 
\begin{equation}
     0\longrightarrow \mathcal{P}\longrightarrow W(a,b)\longrightarrow W(a,b)/\mathcal{P}\cong \mathfrak{witt}\longrightarrow 0,
\end{equation}   
where $\mathcal{P}$  and $\mathfrak{witt}$, respectively, denote the ideal part and subalgebra of $W(a,b)$. {Note that since $\mathfrak{witt}$ is not a $W(a,b)$ module by the adjoint action, $\mathcal{H}^2(W(a,b); \mathfrak{witt})$ is defined by the action induced from the above short exact sequence. So, the second adjoint cohomology of $W(a,b)$, $\mathcal{H}^2(W(a,b);W(a,b))$, similarly to the $\mathfrak{bms}_{3}$ and $\mathfrak{KM}_{u(1)}$ cases,  may not be decomposed as the sum of two cohomologies, i.e. $\mathcal{H}^2(W(a,b); \mathcal{P})\oplus\mathcal{H}^2(W(a,b); \mathfrak{witt})$.} 
We will argue below the first factor is zero and hence the computations reduce to finding the second factor, which may again be argued to be zero.

We first consider $\mathcal{H}^2(W(a,b); \mathcal{P})$. One can decompose the latter as
\begin{equation}
\begin{split}
    \mathcal{H}^2(W(a,b); \mathcal{P})&=\oplus_{p+q=2} E_{2;\mathcal{P}}^{p,q}=E_{2;\mathcal{P}}^{0,2}\oplus E_{2;\mathcal{P}}^{1,1}\oplus E_{2;\mathcal{P}}^{2,0}\\
    &=\mathcal{H}^{2}(\mathfrak{witt};\mathcal{H}^{0}(\mathcal{P};\mathcal{P}))\oplus \mathcal{H}^{1}(\mathfrak{witt};\mathcal{H}^{1}(\mathcal{P};\mathcal{P}))\oplus \mathcal{H}^{0}(\mathfrak{witt};\mathcal{H}^{2}(\mathcal{P};\mathcal{P})) ,\label{coho-decompos-W}
\end{split}
\end{equation}
where the subscript $\mathcal{P}$ denotes we are considering $\mathcal{H}^2(W(a,b); \mathcal{P})$. Because of trivial action of $\mathcal{P}$ on $\mathcal{P}$, the first term of the above is $\mathcal{H}^{2}(\mathfrak{witt};\mathcal{P})$ which is exactly determined by the nontrivial solutions of \eqref{h-Jacobi-W}. The latter has just one solution as $h(m,n)=constant$ which can be absorbed by a proper redefinition as we mentioned in previous part. Therefore, $\mathcal{H}^{2}(\mathfrak{witt};\mathcal{P})=0$. Next we  consider $\mathcal{H}^{1}(\mathfrak{witt};\mathcal{H}^{1}(\mathcal{P};\mathcal{P}))$. Its elements are  solutions of  \eqref{K-Jacobi-W} which just leads to a shift in $a$ and $b$ and hence these solutions are trivial deformations in the family of $W(a,b)$ algebras, so $\mathcal{H}^{1}(\mathfrak{witt};\mathcal{H}^{1}(\mathcal{P};\mathcal{P}))=0$. The last term $\mathcal{H}^{0}(\mathfrak{witt};\mathcal{H}^{2}(\mathcal{P};\mathcal{P}))$ just contains solutions of \eqref{W-I,g}, when $I=0$. It does not have any nontrivial solution, so $\mathcal{H}^{0}(\mathfrak{witt};\mathcal{H}^{2}(\mathcal{P};\mathcal{P}))=0$. From the above discussions one gets $\mathcal{H}^2(W(a,b); \mathcal{P})=0$. One can readily repeat the above procedure and using the relations of previous part to find that all terms in $\mathcal{H}^2(W(a,b); \mathfrak{witt})$ are also equal to zero. To arrive the same result one can use the long exact sequence \eqref{long-exact} and the above results; the latter are the second and the fourth term in \eqref{long-exact}. 
As summary we conclude that $\mathcal{H}^2(W(a,b); W(a,b))=0$, i.e. $W(a,b)$ for generic $a,b$ is infinitesimally and formaly rigid. 

\paragraph{On rigidity of $\widehat{W(a,b)}$. }
As it is discussed in \cite{gao2011low} $W(a,b)$ algebra, for generic $a,b$, just admits one central extension whose commutation relations are 
\begin{equation} 
\begin{split}
 & i[\mathcal{J}_{m},\mathcal{J}_{n}]=(m-n)\mathcal{J}_{m+n}+{\frac{c_{JJ}}{12}m^{3}\delta_{m+n,0}\mathcal{J}_{n}}, \\
 &i[\mathcal{J}_{m},\mathcal{P}_{n}]=-(n+bm+a)\mathcal{P}_{m+n},\\
 &i[\mathcal{P}_{m},\mathcal{P}_{n}]=0.\label{W-central}
\end{split}
\end{equation}
In the other words, the space of $\mathcal{H}^2(W(a,b); \mathbb{R})$ for generic $a,b$ is one dimensional. $\widehat{W(a,b)}$ has 
the short exact sequence,
\begin{equation}
     0\longrightarrow\mathcal{P}\longrightarrow \widehat{W(a,b)}\longrightarrow \widehat{W(a,b)}/\mathcal{P}\cong \mathfrak{vir}\longrightarrow 0,
\end{equation}  
where $\mathcal{P}$ and $\mathfrak{vir}$ are respectively ideal part and sub algebra of $\widehat{W(a,b)}$. As the previous case, $W(a,b)$, to compute
$\mathcal{H}^2(\widehat{W(a,b)};\widehat{W(a,b)})$ we focus on $\mathcal{H}^2(\widehat{W(a,b)}; \mathcal{P})$ and $\mathcal{H}^2(\widehat{W(a,b)}; \mathfrak{vir})$. 

One can readily observe that   $\mathcal{H}^2(\widehat{W(a,b)}; \mathcal{P})$  is exactly the same as $\mathcal{H}^2(W(a,b); \mathcal{P})$, which is equal to zero. For $\mathcal{H}^2(\widehat{W(a,b)}; \mathfrak{vir})$, we have found $f(m,n)=I(m,n)=0$. The combination of the latter and using the fact that $\mathfrak{vir}$ algebra is rigid yields $\mathcal{H}^2(W(a,b); \mathfrak{vir})=0$. 
{On the other hand one can show, by direct calculations, that the simultaneous  infinitesimal deformations of \eqref{W-central} do not lead to any new nontrivial infinitesimal deformation.
Therefore, $\mathcal{H}^2(\widehat{W(a,b)};\widehat{W(a,b)})=0$ which means that  $\widehat{W(a,b)}$ algebra is infinitesimally and formaly rigid. 

\chapter{Algebra generators as functions on celestial two sphere}\label{appendix-C}

The \bms\ and \bmsf\ algebras may be obtained as asymptotic symmetry algebra 3d and 4d flat spacetimes, respectively. As such the generators of these algebras are given through co-dimension two surface integrals which is an integral over a circle for \bms\ case and over a two-sphere for the \bmsf\ case.
In other words, the generators of \bms\ and \bmsf\ algebras respectively may be viewed as functions on an $S^1$ or $S^2$. In fact, it is known that $\mathfrak{witt}$ (or Virasoro algebra) is nothing but algebra of diffeomorphisms on an $S^1$. In this appendix we explore this viewpoint and its implications.

\paragraph{$\mathfrak{witt}$ algebra case, a warm up example.} Let us first analyze the case of $\mathfrak{witt}$. The case of \bms\ was analyzed in chapter \ref{ch4}. Generators of the $\mathfrak{witt}$ are Fourier modes of a (periodic) function on $S^1$:
\begin{equation}
    \mathcal{L}_{n}=\frac{1}{2\pi}\int^{2\pi}_{0} d\varphi \mathcal{L}(\varphi)\exp{(in\varphi)}.
\end{equation}
The index $n$ on the generators, hence, depends on the Fourier basis used. Explicitly, one may use 
\begin{equation}\begin{split}
    \tilde{\mathcal{L}}_{n}&=\frac{1}{2\pi}\int^{2\pi}_{0}  d\varphi \mathcal{L}(\varphi)\exp{(in\varphi)}{\Phi'(\varphi)}\cr
    &= \frac{1}{2\pi}\int^{2\pi}_{0}  d\Phi   \tilde{\mathcal{L}}(\Phi)\exp{(in\varphi(\Phi))},
    \end{split}
\end{equation}
where  $\Phi(\varphi)$ is some periodic function, $\varphi(\Phi)$ is its inverse and $\tilde{\mathcal{L}}(\Phi)=\mathcal{L}(\varphi)$. For example, if we choose 
$\Phi=\frac{K}{d_{0}} e^{id_{0}\varphi}$, then  $\tilde{\mathcal{L}}_{n}=\sum_{d}C_{d}\mathcal{L}_{n+d}$ with $C_{d}=(\frac{K}{d_{0}})^{l}\frac{1}{l!}$ in which $K$ is a constant number. 

The above simple analysis shows that one have the freedom to shift the index $n$ on ${\cal L}_n$ generators and go to another equivalent basis; this change of basis geometrically corresponds to  a diffeomorphism on circle.

\paragraph{$\mathfrak{bms}_{4}$ algebra case.} We are now ready to make a similar analysis for \bmsf\ case where the generators are function on an $S^2$ \cite{Barnich:2009se, Barnich:2011ct}. To this end, let adopt Poincar\'e coordinates for the $S^2$,
$ds^{2}=\frac{1}{(1+\zeta\bar{\zeta})^{2}}d\zeta d\bar{\zeta}$. In this coordinates the standard basis for expansion of generators are
\begin{equation}
\begin{split}
{\cal L}(\zeta)=\sum {\cal L}_n \zeta^{n+1}&,\qquad \bar{\mathcal{L}}(\zeta)=\sum \bar{\mathcal{L}}_n \bar{\zeta}^{n+1},\cr
    T(\zeta,\bar{\zeta})&=\sum T_{m,n}\zeta^{n}\bar{\zeta}^{m}
\end{split}
\end{equation}
where $\zeta,\bar{\zeta}$ are coordinates on the sphere. A change of coordinates on the $S^2$, for example like the one discuss above for the case of a circle, yields a change in the indices on ${\cal L}_n, \bar{\mathcal{L}}_n, T_{m,n}$. Fixing this coordinates, fixes the conventions on the indices.\footnote{Of course, recalling that the global part of the supertranslations $T_{00},T_{01},T_{10}, T_{11}$ are in the $(2,2)$ representation of the Lorentz group $\mathfrak{su}(2)_L\times \mathfrak{su}(2)_R$, it is also natural to choose the indices to be half-integer valued, as suggested in \cite{Barnich:2017ubf}. } 
Analyzing the commutators and hence in what can appear on the Right-Hand-Side of the deformed commutators,  we are dealing with product of these functions. Let us e.g. consider $T^{(1)}(\zeta,\bar{\zeta})T^{(2)}(\zeta,\bar{\zeta})$,
\begin{equation}
    T^{(1)}(\zeta,\bar{\zeta})T^{(2)}(\zeta,\bar{\zeta})=\sum T^{(1)}_{m,n}T^{(2)}_{p,q}\zeta^{m+p}\bar{\zeta}^{n+q}\equiv \sum T_{k,l}\zeta^{k}\bar{\zeta}^{l}\ \Longrightarrow\ 
    T_{k,l}=\sum T^{(1)}_{m,n}T^{(2)}_{k-m,l-n},
\end{equation}
so one finds that the index $k,l$ are fixed to be sum indices of $T^{(1)}_{m,n}$ and $T^{(2)}_{k-m,l-n}$. In a similar way, the indices in deformations of $[{\cal L}_n,{\cal L}_m], [{\cal L}_n, \bar{\mathcal{L}}_m], [{\cal L}_n, T_{p,q}], [{\cal L}_n, \bar{\mathcal{L}}_m], [\bar{\mathcal{L}}_m,\bar{\mathcal{L}}_n]$ and $[\bar{\mathcal{L}}_m, T_{p,q}]$ are fixed.


\bibliographystyle{utphys}
\bibliography{references}


\end{document}